# Issues in Gravitational Clustering and Cosmology

A Thesis submitted to the
UNIVERSITY OF PUNE
for the degree of
**Doctor of Philosophy**
in Physics

by

## JATUSH V. SHETH

INTER-UNIVERSITY CENTRE FOR
ASTRONOMY AND ASTROPHYSICS
PUNE

July 2004

# Declaration

CERTIFIED that the work incorporated in the thesis **Issues in Gravitational Clustering and Cosmology** submitted by **Jatush V. Sheth** was carried out by the cadidate under my supervision. Such material as has been obtained from other sources has been duly acknowledged in the thesis.

Place: **IUCAA**, Pune                    **Professor Varun Sahni**

Date : July, 2004                    Thesis Supervisor

To my parents.....

# Acknowledgments

Carrying out a thesis-work is such an elaborate process ! Over years of hard work, one evolves both as a researcher as well as a person. Many a times, it is a journey in solitude; a journey into uncharted territories. It is a funful adventure to seek goals amidst uncertainties.

To me, the experience of carrying out this thesis-work has been this, and much more. It is a pleasure to have reached this destination successfully. There are numerous people who have directly or indirectly helped me find my way, have nurtured my personal and/or academic growth. You all have been my fellow-travellers, friends and at times, my guiding light! Today I want to thank you all for being with me in this journey.

It was during my course work that I developed a sense of fascination for large-scale structure. Through SURFGEN, I could give an objective meaning to this nameless, somewhat poetic sense of fascination. It is gladdening that this pursuit has served as a link to my further scientific explorations. Thanks Varun, for offering me to work on this challenging problem! You encouraged me to develop myself as an independent researcher; a quality which I am sure, I will cherish forever. You truly *supervised* my progress, corrected me when I went wrong and were always available in tough times, giving me a friendly shoulder when needed. Thanks for everything!

It has been a rewarding experience to collaborate with Sergei Shandarin. I have learnt much from him. His unique insight into large-scale structure studies have often helped me give an objective meaning to my sense of wonder for cosmic structure.

Through stories which one hears during graduate school, I had developed a sense of regard towards Sathya (B.S.Sathyaprakash) for his exceptional programming skills. My efforts to venture into C programming language and later to utilise it in writing SURFGEN, I think, were guided by Sathya's inspiring example of similar sort during his postdoctoral tenure here at IUCAA. It has been a learning experience to work with Sathya, Somnath Bharadwaj and Suketu Bhavsar. Working with Tolia (Anatoly Klypin) on a recent collaboration was similarly quite rewarding. He has been an elderly friend, from whom I have learnt to be careful in work, enthusiastic and hard-working. I would forever be grateful both to Sathya and Tolia for supporting my case while I was applying for jobs. Further, I want to thank my instructors, teachers and colleagues at IUCAA, whose lectures and seminars have been a source of much of my understanding of cosmology and astrophysics.

Numerous friends have made my stay at IUCAA cherishable and those lonely months of struggle, bearable! Dinners and treks with Ranjeev-Rama will always remain fresh in my memory. It always paid to discuss problems with Ranjeev. Jasjeet and Harvinder have been so caring! I have received much warmth and guidance from them. Thanks to both of them for being such wonderful friends! I warmly remember another wonderful couple, Archana and Shanki, who lent me a friendly ear, stood by me in tough times, and received me with understanding. I heartily recall great times I have had with Tarun Saini, Arun Thampan, Sunu, Srini, Niranjan, Suman Datta, Tirthankar, Anand Sankar, Ujjaini, Atul, Amrit, Arman, Sanjit, Amir, Humchand, Abhishek, Saumyadeep, Tapan, Sushmita and Gaurang. It is also a pleasure to recall many discussions with dear friends C.D.Ravikumar and Sudhanshu Barway. Similarly, I recall my friends *from across the road* − Dharam, Poonam, Ayesha and Neeraj whose friendship I have cherished! Special thanks to Dharam for providing me with thesis style files, and helping me out with many related problems.

Thanks to the administrative staff at IUCAA for their support and cooperation in solving bureaucratic glitches as and when they arose. I would specially like to acknowledge Chella, Rajesh Pardeshi, Lata Shankar, Ratna Rao, Neelima, Deepika, S.L.Gaekwad, Mr.Modak and Mr.Sahasrabuddhe. Neelima and her team (Hemant, Sandeep, Vijay and Shashikant) at the library were always very helpful. The computational work in this thesis was carried out with the help of excellent computational facility provided by the computer centre. Sarah, Anjali, Manisha and Neelam have always helped me meet my computing needs. Thanks to you all!

In tough times, it is difficult to maintain sanity. I am blessed with some of the most loving friends from whom I have received pure, selfless love and understanding. But for them, it would have been difficult to keep calm and remain balanced in toughest hours of my stay at IUCAA. Some of those whom I want to fondly recall at this happy hour are Jignesh and Umeshbhai; Naushad, Maitreya and Vikram. Thanks to you all for being with me. I am sure, we shall continue treading the same path, and shall always have each others' company to rejoice in!

Mummy-Papa, I dedicate this thesis to you! You both are my ever-lasting source of inspiration and energy, my guiding light! But for your love and the faith you put in me, it would have been impossible for me to survive these long years of academic pursuit! It is difficult to say it in person, but let me steal this opportunity to say that *this* is the tribute I found most fitting for all your love, support and understanding! I wish and pray, I can offer you much more in days to come. My dear brother Chirag, my Bhabhi and my cute, little nephew Uddhava are too near to be thanked! They are but a part of my life. Today I am warmly thinking of them!

IUCAA, Pune                                                    **Jatush V.Sheth**

July, 2004

# Abstract

The present era marks a remarkable milestone in the history of cosmology. While values of the cosmological parameters have been obtained to a great precision through measurement of microwave background anisotropies (e.g., COBE, BOOMERANG, WMAP) and type Ia supernovae, dedicated telescopes have been yielding information about the distribution of galaxies going back in time to when the Universe was only a fraction of its current age. These observations allow us to understand the history of the Universe. We have recently witnessed the culmination of one of the two most ambitious surveys − the 2 degree Field Galaxy Redshift Survey (2dFGRS). It maps out the 3−dimensional distribution of galaxies over 1500 steradians of the sky going as deep as 600 Mpc[*]. As a result we know the 3−dimensional locations of a quarter of a million galaxies. Another even more ambitious survey, the Sloan Digital Sky Survey (SDSS) is underway, and when completed in a few more years, it would have covered a quarter of the sky and would give us positions of a million galaxies.

Thus, the time is ripe enough to confront predictions for large-scale structure (hereafter, LSS) arising from theoretical models, with observations. This thesis reports the development of a unique methodology which will help the cosmology community in comparing theoretical predictions of LSS with observations in a detailed and reliable manner.

The LSS data from either numerical experiments or observations consist of a large number of *galaxies* distributed over a cosmic volume, with mean inter-galactic separation of $\simeq 5$Mpc. Before comparing theory with observations, we first need to quantify the basic facts concerning LSS. The traditional way of doing this has been to utilise the two-point correlation function and its Fourier-space counterpart, the power spectrum. It turns out however that quantifying LSS using the $2-$point correlation function alone can result in a considerable information loss. For example, the correlation length-scale for galaxies is estimated to be $\simeq 5$Mpc, signifying that the galaxy-distribution is highly clustered compared to one that is random. But this information is of little help in understanding the existence of large-scale superclusters $50-100$ Mpc in length as well as large voids which exist in the universe which reflect the fact that the underlying density field is quite non-Gaussian. It is now widely believed that the non-Gaussian features in the LSS formed as a result of the gravitational clustering of a density field which was initially distributed in the manner of a Gaussian random field (hereafter, GRF). Evidence of Gaussian initial conditions come primarily from observations of cosmic microwave background (CMB). Gravitational clustering is characterised by the transfer of power from larger to smaller scales, which leads to phase$-$coupling between different Fourier wavemodes. Although the two$-$point correlation function is sufficient to describe the properties of a GRF, the density field which we see in the Universe today is significantly non-Gaussian. To fully specify its statistical properties, one in principle requires to know the entire hierarchy of correlation functions. However, the task of estimating $n-$point correlation functions is highly cumbersome and offers limited physical insight into the gravitational clustering process. Hence, the need for alternative statistics, which would be more effective in quantifying LSS, is highly essential.

---

[*]1Mpc $\simeq 3 \times 10^6$ light-years

Minkowski Functionals (hereafter, MFs) provide us with the needed alternative to traditional approach of utilising correlation functions. This is chiefly due to two reasons:

- MFs have been shown to depend on the entire hierarchy of correlation functions of a system of N points.

- MFs give us physically relevant interpretation involving geometry and topology of the LSS. Thus, for example in 3−dimensions, there are 4 MFs, namely (1) Volume $V$, (2) surface area $S$, (3) integrated extrinsic curvature $C$ and (4) genus $G$.

The first three MFs tell us about the geometry of the surface (such as an isodensity contour), and the fourth MF conveys information about the connectivity (topology) of the density field.

An added advantage of utilising MFs is in their ability to quantify the morphology of LSS. As first evidenced by the famous CfA-slice and later confirmed by slices of LSS from deeper redshift surveys (2dFGRS and SDSS) and N−body simulations, the LSS is rich in texture. Its visual impression is that of filaments and sheet-like structures running through space, separated by large, empty and quasi-spherical regions − voids. The large-scale structure often reminds us of the web of cosmic spider, a reason why it is dubbed as the *Cosmic Web* or *Cosmic Froth*. The problem of quantifying our visual impression of LSS has been pursued for a long time, and pattern-specific statistics have been found to be preferable over the traditional $n$−point correlation functions. In this regard Shapefinders, a set of morphological statistics derived from MFs, provide us with an excellent probe of the shapes and sizes of coherent large-scale superclusters and voids. Thus by using MFs in conjunction with Shapefinders, we can systematically compare cosmic web due to two or more theoretical models and can point out the relevant differences between models. Thus, the MF-based approach of quantifying LSS provides us with an efficient way to compare the theoretical models with the observations.

This thesis is chiefly devoted to the accurate estimation of MFs for cosmic density fields. The power and promise of this method is illustrated by applying it to a class of N−body simulations and to mock galaxy catalogues.

The last part of the thesis concerns the analysis of the distribution of galaxies in the Las Campanas Redshift Survey (LCRS). The survey consists of 6 quasi 3−dimensional slices containing about 25,000 galaxies with a depth similar to that of 2dFGRS. The issue of the scale of homogeneity of the universe has been addressed using data from these slices. In addition, the size of the longest coherent superclusters has also been derived. These results will be reported in this thesis.

The overall structure of the thesis is as follows:

**Chapter 1** will briefly summarise the observational evidence for LSS and the properties of the supercluster-void network in the universe. We will discuss the importance of various statistics including the $n$-point correlation functions in quantifying the LSS and the role these play in interpretation of cosmic datasets. This will be motivated by summarising the current status of our knowledge about bias between baryonic and dark matter. Minkowski Functionals shall be introduced and their expected utility in quantifying the LSS shall be stressed. MFs can also be used to independently study the morphology of the large-scale superclusters and voids. The related morphological measures, the Shapefinders, shall be introduced. Finally, an overall layout of how the MFs can be used to confront various models with each other shall be presented.

**Chapter 2** is devoted to the techniques developed in this thesis for determining the MFs. Minkowski Functionals are geometric and topological measures, and for a scalar field (for example, the cosmic density field) in 3-dimensional space, the four MFs are defined on a 2−dimensional surface which could represent, say an isodensity contour. Thus, in order to estimate MFs one first needs to have a mathematically robust prescription with which to define an isodensity contour for a chosen threshold of density. Further, this prescription is to be complemented by an algorithm which enables us to compute MFs for the resulting surface defined at this threshold.

This thesis reports a unique method which unites both the above aspects into a software called SURFGEN (short for "surface generator"). SURFGEN accepts at its input a density field defined on a cubic grid, and builds a surface for an overdense (underdense) region with respect to a user-specified threshold of density. SURFGEN simultaneously computes MFs for that surface. Such a surface could refer to overdense regions including clusters and superclusters of galaxies and to underdense regions (voids). Thus SURFGEN enables us to study superclusters and voids individually as well as the entire supercluster-void network.

This chapter describes in detail the methodology and algorithms which form the foundation of SURFGEN. This is followed by tests of SURFGEN which serve to illustrate its accuracy.

**Chapter 3** will be devoted to a comprehensive morphological study of LSS resulting from three rival cosmological models. An important test for statistical diagnostics of LSS is that they be capable of discriminating between rival theoretical models of structure formation. This chapter is devoted to MF-based comparison of three rival cosmological models, SCDM, $\tau$CDM and $\Lambda$CDM. It will be shown that MFs for the three models are considerably different and hence their MFs can distinguish between rival structure-formation models. Further the MFs are used to glean information about relative shapes and sizes of superclusters occurring in the density fields corresponding to the three cosmological models. Thus, this chapter will provide valuable insight into the morphology of LSS. Earlier in this chapter we will gain deeper insights into the morphology of the cosmic web resulting from the $\Lambda$CDM model. Properties of both superclusters and voids will be quantified using MFs. This analysis will provide precious information, the first of its kind, about the shapes and sizes of voids which may conceivably refer to the real universe.

There is ample observational evidence to suggest that the dark matter is almost 10 times as abundant as the baryonic or visible component of the universe. Furthermore, dark matter is believed to be collisionless and to interact only through gravity. For this reason, the clustering of dark matter is very well understood theoretically, and gravitational clustering in dark matter has been simulated to great precision using cosmological N−body simulations. However, galaxies are composed of baryons and it is still a subject of some debate as to how galaxies form and evolve together with the evolving dark matter. Most theoretical studies assume that baryons follow dark matter on large scales, i.e., on the scales of superclusters. However, on smaller scales, it is not clear how the galaxies may be distributed relative to the dominant dark matter component. In the absence of a complete understanding of baryonic physics, one can try and derive the *galaxy distribution* from the (simulated) dark matter distribution using a well-motivated ansatz. Such an ansatz is expected to provide the observed bias between the dark matter and galaxies and to satisfy observational constraints on the galaxy-distribution such as the two-point correlation function and the power spectrum.

**Chapter 4** will report the analysis of mock SDSS *galaxy catalogues* constructed from two rival cosmological models, $\tau$CDM and $\Lambda$CDM. For both models, *galaxies* are selected as biased tracers of the underlying dark matter mass in such a way that the resulting *galaxy distributions* reproduce the observed two−point correlation function and the power spectrum of galaxies derived from the APM galaxy-catalogue.

Because mock catalogues of both $\tau$CDM and $\Lambda$CDM have identical 2−point correlation functions and power spectra, discriminating between these models is an excellent challenge for new statistical diagnostics like MFs.

In this chapter, we compare 10 realizations of $\tau$CDM mock SDSS catalogues with a given realization of its $\Lambda$CDM counterpart. We show that MFs are very well constrained statistics and are remarkably different for the two models. We further note an interesting effect of bias. A scale-dependent bias is found to lead to different clustering properties for the galaxies compared to the underlying dark matter-distribution. The effect is such that the dark matter distribution due to $\tau$CDM is less clustered compared to its $\Lambda$CDM counterpart, whereas the galaxy-distributions show a completely reverse trend, i.e., $\tau$CDM galaxy distribution is found to be *more* clustered compared to the $\Lambda$CDM galaxy distribution. Thus MFs help us gain deeper insight into clustering properties of the galaxies, an effect which would have been otherwise missed because the galaxy-catalogues share the same $2-$point correlation functions. We do however stress that a relatively simple treatment of having a scale-independent uniform bias should be complemented by more thorough treatment which incorporates complex physics of galaxy formation in preparing galaxy catalogues before confronting our theoretical models with observations.

A general impression of the large-scale structure is that of a percolating web of galaxies. Are the superclusters of galaxies *actually* so long as to percolate the entire available volume of the universe? Or is it that the cosmic web arises because of a chance alignment of superclusters of much smaller size?

The two$-$point correlation function of galaxies fluctuates around zero beyond about 20 Mpc. But we do see structures spanning across much longer length scales. Clearly, detection of the longer structures can be attributed to nonzero higher order correlation functions and correlations among phases. Much like in the case of $2-$point correlation function, we can assign an effective length to all the correlation functions, beyond which their contribution is negligible. A situation may be envisaged wherein there is a maximum length-scale in the cosmic density field beyond which the phase correlations are negligibly small. Naturally this length-scale refers to the largest coherent structures in the universe. Further, the universe can be interpreted as composed of representative volume elements of this size. Thus, this scale is also a scale of homogeneity.

In **Chapter 5**, we report measurement of the scale of homogeneity, and of the size of the longest superclusters using 6 slices of the Las Campanas Redshift Survey. Here we have adopted an MF-based morphological shapefinder, *Filamentarity* to measure the shapes and sizes of structures. Further we employed a statistical method of *Cos-* *mic Shuffle* to establish the scale of homogeneity. This chapter discusses the method employed in our analysis and presents the main results of this exercise.

Finally, **Chapter 6** presents the conclusions of the thesis and offers an integrated discussion based on the previous chapters.

# List of Publications

This thesis is based on the following publications.

- *Measuring Geometry and Topology of Large Scale Structure using SURFGEN: Methodology and Prelimenary Results*

  **Jatush V. Sheth**, Varun Sahni, Sergei F. Shandarin and B.S.Sathyaprakash, 2003, MNRAS, **343**, 22

- *Morphology of the Supercluster-Void Network in $\Lambda CDM$ Cosmology*

  Sergei F. Shandarin, **Jatush V. Sheth** and Varun Sahni, 2004, Accepted for publication in MNRAS

- *Morphology of mock SDSS Catalogues*

  **Jatush V. Sheth**, 2004, Accepted for publication in MNRAS

- *Size of the Longest Filaments in the Universe*

  Somnath Bharadwaj, Suketu P. Bhavsar and **Jatush V. Sheth**, 2004, ApJ, **606**, 25

# Contents

# List of Tables

# List of Figures

# List of Symbols

$\alpha$ : spectral index defines as $S_\nu \propto \nu^{-\alpha}$, p. 55

$\alpha_{ff}$ : significance of Spearman's rank correlation,

for 'flux density−flux density' scatter diagram, p. 124

$\alpha_{ll}$ : significance of Spearman's rank correlation

for 'luminosity−luminosity' scatter diagram, p. 124

$\beta$ : $v/c$, $v$ is the velocity, p. 105

$\Gamma$ : X-ray power law photon index, p. 133

$\gamma$ : Lorentz factor, $\frac{1}{\sqrt{1-\beta^2}}$, p. 105

$\gamma_{ff}$ : Spearman's rank correlation coefficient

for 'flux density−flux density' scatter diagram, p. 124

$\gamma_{ll}$ : Spearman's rank correlation coefficient

for 'luminosity−luminosity' scatter diagram, p. 124

$\eta_s$ : VLBI system efficiency, p. 36

$\theta$ : angle between point source and line joining two antennas, p. 22

$\lambda$ : emission wavelength, p. 22

$\nu$ : emission frequency, p. 105

$\nu_{SN}$ : supernova rate, p. 92

$\tau$ : geometric delay, p. 22

$\phi_{ij}$ : phase angle on the baseline formed by stations i and j, p. 30

$\phi_{123}$ : closure phases, p. 30

$\Omega$ : solid angle, p. 31

$\omega$ : angular frequency of radiation, p. 22

$A_{1234}$ : closure amplitudes, p. 31

$B(x, y)$ : synthesised beam, p. 25

$B_T^c$ : total observed optical luminosity in the $B$ band (corrected), p. 39

$D$ : Doppler factor, p. 105

$E$ : electric field, p. 22

$F_{5007}$ : [O III] $\lambda 5007$ emission line flux, p. 41

$F_{H\beta}$ : $H\beta$ emission line flux, p. 41

$g_i$ : complex antenna gain, p. 31

$H_0$ : Hubble constant, p. 15

$M_B^{total}$ : absolute magnitude of the host galaxy corrected for nuclear

non-stellar emission, p. 40

$(M_B)_{bulge}$ : absolute bulge magnitude, p. 42

$M_\odot$ yr$^{-1}$ : star-formation rate in solar mass year$^{-1}$, p. 92

$P$ : probability, p. 132

$q_0$ : the cosmological deceleration parameter;

assumed to be 0 unless stated otherwise, p. 15

$\Re$ : real part of a complex quantity, p. 24

$R$ : ratio of observed radio emission from the compact feature

to that from the outer components, p. 106

$S(u, v)$ : sampling function, p. 25

$S_\nu$ : flux density, p. 55

$T$ : Hubble type of the galaxy, p. 13

$T_A$ : antenna temperature, p. 31

$T_s$ : system temperature, p. 31

$t_{int}$ : integration time, p. 36

$V$ : voltage, p. 22

$V(b)$ : complex visibility of the source, p. 24

$z$ : redshift, p. 4

# List of Constants

| | | |
|---|---|---|
| c | Speed of light in vacuum | $2.99792458 \times 10^{10}$ cm sec$^{-1}$ |
| au | Astronomical unit | $1.49597892(1) \times 10^{13}$ cm |
| pc | Parsec | $3.0856(1) \times 10^{18}$ cm |
| ly | Light year | $9.4605 \times 10^{17}$ cm $= 6.324 \times 10^{4}$ au |
| $M_{\odot}$ | Solar mass | $1.989(2) \times 10^{33}$ g |
| $R_{\odot}$ | Solar radius | $6.9598(7) \times 10^{10}$ cm |
| $L_{\odot}$ | Solar luminosity | $3.826(8) \times 10^{33}$ erg sec$^{-1}$ |

# Chapter 1

# Introduction

This thesis is devoted to a study of the large scale structure (hereafter, LSS) of the Universe. By *large*, we mean length-scales which light covers only over a time-scale comparable to the age of the Universe. The basic constituent of LSS is a *galaxy*. In the present study of LSS, we are mainly concerned with the largest coherent structures; the superclusters of galaxies, and their complements voids, which are practically devoid of galaxies. Superclusters and voids are dynamically evolving systems *even today*, and there is a lot to learn about the basic parameters governing the evolution of our Universe and about the complex interplay between dark matter and baryons by studying properties of these systems. A study such as this, is also motivated by remarkable, relatively recent improvements in the extragalactic database. Over last less than a decade, the data have grown to be impressively richer; giving us access to a near-complete sample of 3-dimensional distribution of $\simeq 10^{5-6}$ galaxies over a cosmological volume as large as $0.1-1$ [Gpc]$^3$. The distribution of these galaxies is far from being uniform. In fact, the galaxies appear to show strong clustering in an anisotropic manner; giving a phenomenally rich, textural appearance to LSS.

As in other branches of science, the study of LSS can be perceived in terms of three basic directions:

1. Observations of LSS

2. Effective description of the observed LSS

3. Explanation of observations (of LSS) in terms of theoretical models

Of these, (2) quantifies LSS through a set of statistics and discriminating measures which describe properties of LSS. These measures are also applied to theoretical models (of LSS) to compare theory with observations.

Our currently favoured models of large-scale structure formation are based on the principle of gravitational instability. According to this principle, tiny fluctuations in the matter distribution present in the early universe are amplified by gravity, resulting in the formation of significantly overdense structures (clusters and superclusters of galaxies) which we see today. Such models have been worked out in great detail, and they lend us valuable insights into the process of gravitational clustering. On the observational front, we have recently witnessed culmination of one of the most ambitious redshift surveys, the 2 degree field redshift survey of galaxies (2dFGRS). In addition, several impressive redshift surveys (e.g., SDSS, 2MASS etc.) are in progress. In trying to make a connection between the theoretical predictions and such large body of observational data on LSS, one needs to develop unbiased and well-behaved statistics with which to quantify the large-scale distribution of galaxies. The main thrust of this thesis is in *this* direction. We report in this thesis, statistical and diagnostic measures which characterise the complex patterns made by superclusters and voids belonging to LSS. We then go on to present algorithms and software developed for efficient numerical determination of these diagnostics. Such statistics/diagnostics could eventually be utilised to quantify LSS and to compare theoretical predictions for LSS with the observational data.

To understand the importance of such a study, it will be beneficial to briefly review the three items listed above concerning theoretical and observational study of LSS. The work reported subsequently in this thesis will thereby be placed in its proper global context.

## 1.1    Theoretical Preliminaries

The birth of modern cosmology was marked by the classic discovery of the expansion of the Universe (Hubble, 1929). Hubble established the linear relation between the velocity of recession of a given galaxy and its distance from our galaxy. Robertson (1928) showed that Einstein's general theory of relativity admitted solutions for such *expanding world* models. Hubble's law was soon interpreted by Friedman, and independently by Robertson and Walker and by Lemaitre in terms of FRWL (Friedman, Robertson, Walker, Lemaitre) cosmological solutions of the Einstein Field Equations for gravitation. These solutions described a homogeneous and isotropic Universe emerging from a singular state of infinite density. The hypotheses of homogeneity and isotropy of the universe were later supported by the near-uniformity on large scales, of the galaxy counts in the sky recorded by the Lick Survey (Shane & Wirtanen 1967; Seldner et al., 1977). Groth & Peebles (1977) used this survey and demonstrated the power-law form of the two-point galaxy-galaxy correlation function, which pointed out that galaxies are strongly clustered on small scales, and are *not* randomly distributed. Explaining such clustering of galaxies in accordance with the large-scale homogeneity and isotropy in an expanding Universe thus became the fundamental question to be answered by the theories of structure formation.

The discovery of Cosmic Microwave Background Radiation (CMBR) by Penzias and Wilson (1965) provided firm support to the Big Bang paradigm for early history and evolution of the Universe, and established the cosmological framework in which structure formation could be studied. Two and a half decades later, CMBR-observations by COBE satellite provided precious information about the initial conditions for structure formation. During the past decade, there has been a tremendous refinement in the CMBR-data. Surveys like DASI, MAXIMA-1, BOOMERANG, WMAP, CBI, ACBAR etc. have together mapped out a large range of angular scales ($7' \leq \theta \leq 10^o$), from which the matter-fluctuations at the epoch of matter-radiation decoupling can be deduced (Leitch et al. 2002; Pryke et al. 2002; Balbi et al. 2000; de Barnardis et al. 2000; Bennett et al. 2003; Mason et al. 2003; Kuo et al. 2004). These fluctuations reflect the structure in the gravitational potential that eventually leads to the formation of galaxies, their clustering into groups and clusters and ultimately to the formation of LSS in the Universe.

We are thus faced with a classical initial value problem: having set up the initial density field in accordance with the CMBR-maps, the task is to derive the clustering properties of galaxies as observed in redshift surveys (about which we shall say more in the next section). To this end, several analytical and numerical approaches have been developed. We shall summarise them in this section. Detailed treatment can be found in several excellent reviews (for example, Jones (1992); Sahni & Coles (1995); Bertschinger (1998); Peacock (2003); Jones et al.(2004)) and in books such as Peebles(1980), Padmanabhan(1993) and Peacock(1998).

## 1.1.1   Linear Theory of Structure Formation

As mentioned earlier, *gravitational instability* is the central physical mechanism believed to be responsible for structure formation in the Universe. The domain of linear theory is during the earliest phase of the evolution of the Universe when the density fluctuations are believed to have been very small (or on the very large, $\geq 100h^{-1}$Mpc

scales)[*]. Zel'dovich (and independently Harrison and Peebles) invoked in 70s, the argument of self-similar evolution of the Universe to predict that the power spectrum of the primordial density fluctuations should be a power-law ($P(k) \sim k^n$), with the spectral index $n = 1$. The exact nature of density fluctuations remained a mystery until *Inflation* was introduced as a phenomenological model for the early Universe (Guth 1981). Inflation, in addition to solving problems like *horizon problem* and *flatness problem*, naturally predicts a form for density fluctuations generated quantum mechanically during Inflation. These fluctuations have a Gaussian, random phase behaviour. The statistical properties of a Gaussian, random phase density field (hereafter, GRF) are completely determined in terms of its power spectrum. For such a GRF, the spectral index of the power spectrum, predicted by Inflation is close to being scale-invariant ($P(k) = A \times k^n$; $n \approx 1$) (Starobinsky 1982; Hawking 1982; Guth & Pi 1982). The amplitude $A$ of the power spectrum however, remained unfixed until COBE measured the quadrupole anisotropy in the CMBR in 1992. The fluctuations in CMBR (and hence in matter-density) as measured by COBE, were found to be indeed quite small, with deviations of 1 part in $10^5$ about the mean density, i.e., $\delta \simeq 10^{-5}$.

The above developments gave us a picture of the matter-density field at the epoch of matter-radiation decoupling ($z \approx 1100$). To summarise, this density field is believed to be a Gaussian, random phase field, which is nearly homogeneous with departures from the mean of $\simeq \pm 10^{-5}$ and is described by a power law power spectrum, $P(k) \sim k$. GRF is featureless, with complete symmetry between overdense and underdense regions. It is a puzzle hence, to understand as to how the featureless GRF evolves into a complex, nonlinear network of galaxies with rich textural properties[†]. The theoretical pursuit hence, is to understand the growth of density fluctuations from these initial conditions under the influence of gravitational instability in an expanding Universe.

It turns out that so long as $\delta$, the density contrast remains small ($\delta \ll 1$), the equations of motion can be solved perturbatively. A natural consequence of this fact is that density perturbations at all scales evolve independently (Lifshitz 1946; Sahni & Coles, 1995). As a result, the random-phase nature of the initial density field is preserved and the density field remains featureless. The amplitude of density fluctuations grow as the Universe expands (e.g., $\delta_m \sim t^{2/3}$ in the matter-dominated, flat Universe), and Linear Theory breaks down when $\delta \sim 1$. After this, approximation schemes such as Zel'dovich

---

[*]The density fluctuations are described in terms of the dimensionless scalar field, the density contrast $\delta(\mathbf{x}, t) = \rho(\mathbf{x}, t)/\bar{\rho}(t) - 1$, where $\bar{\rho}(t)$ is the mean background density.

[†]The morphology of LSS is indeed tantalizing, and gives us the appearance of network of sheets and filaments of galaxies. We shall have more to say about this in the next section.

Approximation and Adhesion Model or N−body simulations have to be employed in order to further probe the growth of structures.

## 1.1.2 Zel'dovich Approximation and the Adhesion Model

In an effort to understand the nature of the *first collapsed* objects in a gravitationally unstable, generic density field, Zel'dovich (1970) proposed an approximation (the so called, Zel'dovich Approximation − ZA). This model is based on the result from Linear Theory, that the gravitational potential is a much smoother function in space (compared to the density contrast $\delta(\mathbf{x}; t)$), and evolves very slowly with time. Further, in an expanding Universe, it is possible to express the velocity field as a gradient of a scalar *velocity-potential* $\phi_v(\mathbf{q})$. The potential $\phi_v(\mathbf{q})$ can be shown to be proportional to the gravitational potential $\phi_g(\mathbf{q})$. Due to this, $\phi_v(\mathbf{q})$ evolves slowly with time. Utilising such weak temporal dependence of the *velocity-potential*, Zel'dovich proposed that the motion of particles may be completely prescribed in terms of their *initial* velocities as follows the gravitational potential $\phi_g(\mathbf{q})$, and that this may well approximate the evolution of an N−body system in its early phase up to a time when the density field grows to become mildly nonlinear ($\delta \rightarrow 1$). Thus in this approximation, the initial configuration of particles are allowed to move according to their initial velocities. The final (Eulerian) distribution of particles $\mathbf{r}$ can be obtained from the initial (Lagrangian) distribution of coordinates $\mathbf{q}$ by using the following mapping relation:

$$\mathbf{r} = a(t)[\mathbf{q} + D(t) \times \mathbf{V}(\mathbf{q})] = a(t)[\mathbf{q} + D(t) \times \nabla\phi_g(\mathbf{q})], \qquad (1.1)$$

where a(t) is the scale factor and D(t) is the growth rate of density fluctuations prescribed by Linear Theory. Using this approximation, Zel'dovich predicted that in case of a generic initial density field within an expanding Universe, the first non-linear structures to form would be due to a unidirectional collapse, and would exhibit predominantly pancake-like morphology (although higher order singularities in the density field would also be present; see Arnold, Shandarin & Zel'dovich 1982). Indeed when Geller & Huchra (1989) reported the existence of a *Great Wall* in the CfA-II redshift survey (de Lapparent et al., 1986), a generic prediction of the ZA seemed to be confirmed.

Zel'dovich approximation was the first to give some glimpse into how the large-scale structure may look like: according to this approximation, the large scale objects should be sheet-like, and should intersect into denser filaments; compartmentalising the empty regions; the voids. Today when we have access to galaxies which are four orders of magnitude larger in number, the LSS appears to exhibit a network of sheets and filaments of galaxies interspersed with large voids.  In their visual appearance, filaments stand out remarkably well in our impression of LSS. However, the occasional occurrence of *vast*, sheet-like superclusters such as the *SDSS Great Wall* (which is almost twice as large as the CfA Great Wall; Gott et al., 2003; Tegmark et al., 2004) raises questions as to how significant are sheets with respect to filaments on large scales. In fact, the morphology of superclusters (their sheet-like or filamentary character as well as their sizes) needs to be objectively defined and quantified, so that more refined theoretical predictions about LSS (based on N−body simulations) can be tested against the redshift surveys.  This is a theme which is *central* to this thesis.  In future, if the sheet-like superclusters in redshift surveys are confirmed and their counterpart structures *not* found in N−body simulations, it may present a clear dichotomy between theory and observations.  In *this* thesis, we report methods which we shall eventually use to make such a comparison between theory and observations possible by investigating the morphology of LSS [‡]. We shall introduce our methods and the statistics used for our purpose a little later (see Section 1.4).

The Zel'dovich approximation is not applicable beyond the regime of first shell-crossing (Shandarin & Zel'dovich 1989).  Nevertheless, it correctly predicts the formation of filaments at the intersection of sheets; a feature which has inspired many (e.g., Klypin and Shandarin, 1993) to look for Zel'dovich pancakes at a density level lower than that for filaments. Such filaments can be made to survive, by attributing an artificial "stickiness" to the particles. Gurbatov, Saichev & Shandarin (1989) introduced such an approach by proposing the "Adhesion Model (AM)". AM is shown to reproduce the overall patterns in LSS (on scales $\geq 2h^{-1}$Mpc) remarkably well, in agreement with simulations (Weinberg & Gunn, 1990; Kofman et al., 1990; Kofman et al., 1992). Further, AM can be used to study the evolution of voids (Sahni et al., 1994) and to probe the evolution of lower-order correlation functions into the weakly nonlinear clustering regime.  Thus, AM is a useful analytic approximation to study dynamics of LSS in the weakly nonlinear regime of gravitational clustering.

### 1.1.3   N−Body Simulations

The best way to understand the evolution of a gravitating system consisting of N particles would be to *actually* solve for the force acting on a given particle due to the re-maining particles. In a rather different form, by solving for the Poisson's equation, this is precisely the goal which modern versions of N−body simulation codes achieve. (For an excellent introduction to N−body simulations, see Bertschinger (1998) and Klypin (2000a, 2000b, 2000c). As an illustration of a typical N−body code, see Springel, Yoshida and White (2001), who present GADGET, a code which has become a standard for many further add-on developments.)

---

[‡]Morphology is a combined study of sizes and shapes of structures.  Connectivity, geometry and the overall shape of the structures form the subject matter of such a study.

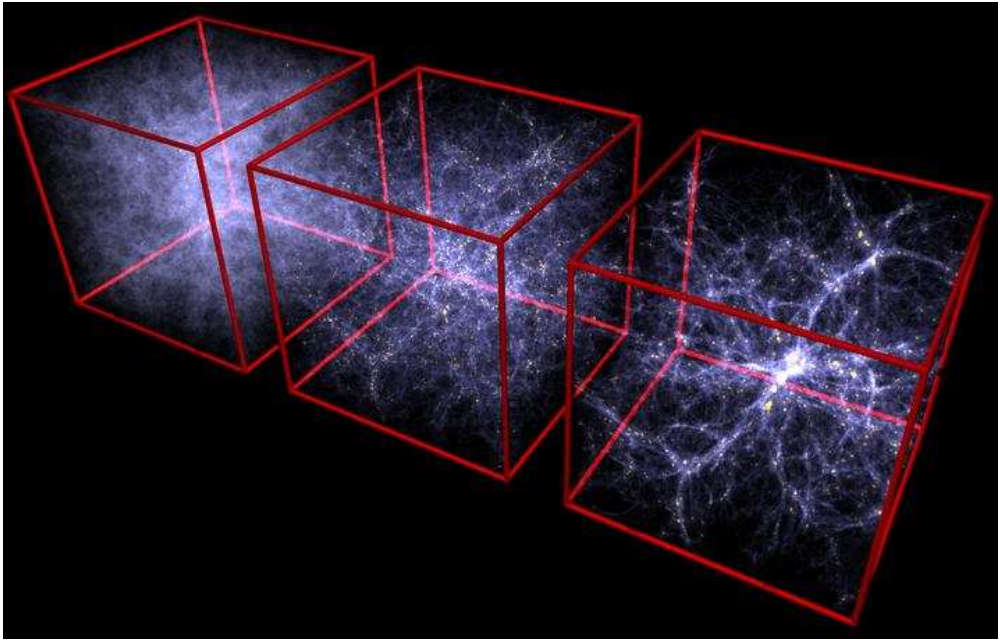

Figure 1.1: The figure illustrates the development of nonlinear structures in an N−body simulation from $z = 6$ (leftmost cube) to $z = 0$ (rightmost cube) via $z = 2$ (middle cube). A near featureless density field evolves to produce filamentary and sheet-like superclusters which percolate through the box-volume. These are separated by large voids. Due to its web-like appearance, LSS is dubbed as the *Cosmic Web*. [Figure courtesy: Volker Springel; see *http://www.mpa-garching.mpg.de/galform/data_vis/index.shtml*]

Simulations allow us to probe the nonlinear regime of the dark matter density field, and can provide us with detailed predictions of the *dark side* of the Universe at today's epoch. One can further study the development and growth of structures in a cosmic density field using these simulations. A fully nonlinear LSS in dark matter may look like that shown in Figure 1.1. We note from this figure, how the matter condenses along filamentary/sheet-like ridges between clusters and how voids get progressively more evacuated with the passage of time. The appearance of LSS reminds us of the web of a spider; a reason why LSS has often been dubbed as "Cosmic Web" (Bond et al., 1996). Due to its frothy appearance, LSS is also termed as "Cosmic Froth" (van de Weygaert 2002).

Studying the growth of rich texture of the Cosmic Web (made evident by N−body simulations) is an important aim in Cosmology. This can be achieved with the help of efficient quantifying statistics. Peebles(1980) discusses application of the traditional quantifiers of LSS such as correlation functions. These measures give us a direct handle on the clustering properties of the matter distribution, but at a substantial cost of computing time. Further, it is difficult to use these diagnostics to have a physically appealing interpretation of the data. It is desirable that we try to isolate the very features in LSS − the clusters and superclusters of galaxies and voids − which stand out in the visual appearance of the Cosmic Web, and study them individually. This is the approach which we shall take in this thesis. The methods reported here may thus provide a correct mathematical description of the complex morpholgy of LSS, and would pave a way for confronting simulations with data.

During the past 2 decades, advent of parallel supercomputers have ensured a phenomenal improvement in the computing power available to tackle complex astrophysical processes. Hence, it has become possible to include complex gas-dynamical processes into simulations. Various recipes for cooling, star formation and feedback due to supernovae and galactic winds have been incorporated to mimic the formation and evolution of galaxies in a cosmological setup with simulated volumes ranging from 50 to 100 $h^{-1}$Mpc (e.g., Weinberg et al., 2004; Weinberg, Hernquist and Katz, 2002). This has made it possible to develop *mock* catalogues of galaxies inspired by a given cosmological model. Relatively less time-consuming semianalytic methods have also been proposed which hope to achieve similar goals while bypassing the complex baryonic physics (Benson et al., 2002; Cole et al., 2000; Kauffmann et al., 1997). Such mock catalogues can eventually be compared with clustering properties of galaxies derived from redshift surveys with the help of methods developed here.

Before we begin to tackle the important question of how best to compare two distributions of galaxies, it is necessary to develop insight into what we *already* know about LSS in light of the recent redshift surveys. This is the subject of the next section.

## 1.2   Observations of LSS: Galaxy Redshift Surveys

Ever since sky maps of galaxy counts, e.g., Lick catalogue (Shane and Wirtanen, 1967) revealed a rich pattern of LSS of galaxies in projection, it was realised that comple-menting this information with the redshifts of individual galaxies was essential in order to get a fully 3-dimensional view of the Universe. In this exercise, one turns redshift of the galaxy into its distance from us with the help of Hubble's law, and is thus able to study *redshift space* distribution of galaxies.

Various small surveys were carried out in the late 80s and mid 90s in a controlled fashion; their positive scientific outcome providing impetus for the initiation of bigger ones. One of the most noticeable among these is the Center for Astrophysics Survey (de Lapparent et al.1986). This survey had an angular coverage of $6^o \times 120^o$ and depth of $\simeq 150h^{-1}$Mpc. The survey measured redshifts of about 2400 galaxies. The famous *Great Wall* centered around the Coma cluster was the most striking feature of this survey. The slice showed very clearly the "bubbly" nature of LSS with voids having sharp boundaries; the largest void having a diameter of $\simeq 50h^{-1}$Mpc. Subsequent surveys, the following CfA slices and the ESO Southern Sky Survey (da Costa et al., 1991) amply confirmed the impression given by the CfA slice. The Southern Sky Redshift Survey (SSRS) discovered a second Great Wall in the southern sky. This was the first serius challenge to cosmologists to seek an explanation for the observed LSS. The Optical Redshift Survey (ORS) (Santiago et al., 1995) had a depth of $80h^{-1}$Mpc, but attempted a complete sky coverage of the sky (except for the *zone of avoidance*). This survey measured 8500 redshifts in total, and was heavily used to estimate the luminosity functions, galaxy correlations, velocity dispersions etc. In mid 90s, the Stromlo-APM redshift survey (Loveday et al., 1995) and the Durham/UKST redshift survey (Ratcliffe et al., 1998) were focussed towards the southern sky and led to fruitful results on correlation functions in real and redshift space, power spectra, redshift distortions, cosmological parameters, bias etc. However, the galaxy redshifts in all these surveys never exceeded the count of 10000.

The Las Campanas Redshift Survey (Shectman et al., 1996) of six $1.5^o \times 80^o$ slices (3 in the northern galactic hemisphere and 3 in the southern hemisphere) went up to the depth of $\simeq 750 \ h^{-1}$Mpc ($z \sim 0.25$) and recorded redshifts of about 25000 galaxies. This was an order-of-magnitude improvement over earlier surveys. This was the first deep survey of sufficient volume so as to test whether our knowledge of the local Universe can be sufficient to describe the more distant regions. The usual tests included the luminosity function, second and third-order correlation functions, power spectra etc. We shall see below the important conclusions which can be derived about the cosmological model(s) of our Universe by making use of some of these measurements.

### 1.2.1   Two Degree Field Galaxy Redshift Survey (2dFGRS)

The 2dF multi-fiber spectrograph on the 3.9m Anglo-Australian Telescope can measure up to 400 objects simultaneously over a field of view with $2^o$ diameter. This facility was put to use to sample two contiguous constant-declination strips of the sky, one in the northern galactic hemisphere (with angular coverage of $10^o \times 90^o$ in $(\delta, \alpha)$) and the other in the southern galactic hemisphere (with angular coverage $\sim 15^o \times 90^o$ in $(\delta, \alpha)$). Working in the $B_J$ band, the survey reached a depth of $z = 0.3$ with median redshift of 0.11, and has compiled redshifts of about $2.5 \times 10^5$ galaxies which are brighter than an extinction corrected magnitude of $b_J = 19.45$ (Colless et al., 2003). The survey is already complete and the resulting correlation functions, redshift distortions and pairwise velocity dispersions (Hawkins et al., 2003) demonstrate the superb quality of the data set. The science output of this survey and its overall cosmological relevance have been beautifully reviewed by Peacock (2003). We shall review here some of the results which are most relevant to our present discussion.

One of the main science goals of 2dFGRS was to get improved estimates of the 3-dimensional power spectrum and two-point galaxy-galaxy correlation function[§]. The first data release of 2dFGRS (Colless et al., 2001) was large enough to measure 3-dimensional power spectrum over the length scales $0.01 \leq k \leq 0.4h$ Mpc$^{-1}$ (Percival et al., 2001; Tegmark et al., 2002). Percival et al.(2001) assumed a linear, constant bias between the dark matter and baryons based on the justification provided by N$-$body simulations (Cole et al., 1998). Based on this assumption, one would expect the nonlinear, redshift space power spectrum to have the *same* shape as the linear, real space power spectrum on length scales corresponding to $k \leq 0.15h$ Mpc$^{-1}$ (on scales smaller than this, the nonlinear redshift distortions may distort the power spectrum in a nontrivial way). It would hence, be fair to compare the measured power spectrum with that predicted by linear theory. Since the latter explicitly depends on the cosmological parameters such as (1) the power spectrum index $n$, (2) the total matter density $\Omega_m$, (3) the baryon-to-matter ratio, or the baryon fraction $\Omega_B/\Omega_m$ and (4) the Hubble constant $h$, a likelihood analysis was carried out to constrain the values of $\Omega_m$ and $\Omega_B/\Omega_m$. The best fitting values are $\Omega_B/\Omega_m = 0.17 \pm 0.06$ and $\Omega_m.h = 0.18 \pm 0.02$. With a prior of $h = 0.7 \pm 10\%$, this implies $\Omega_m = 0.26 \pm 0.03$. One of the remarkable feats of this analysis is that it constrains the density of dark matter. The lack of large amplitude fluctuations in the power spectrum further implies that the dark matter should be nonbaryonic and collisionless.

---

[§]Earlier these were already estimated by Baugh and Efstathiou (1993) and Baugh (1996), respectively by working with the 2-dimensional galaxy-counts recorded in the APM catalogue (also see, Maddox, Efstathiou and Sutherland, 1996).

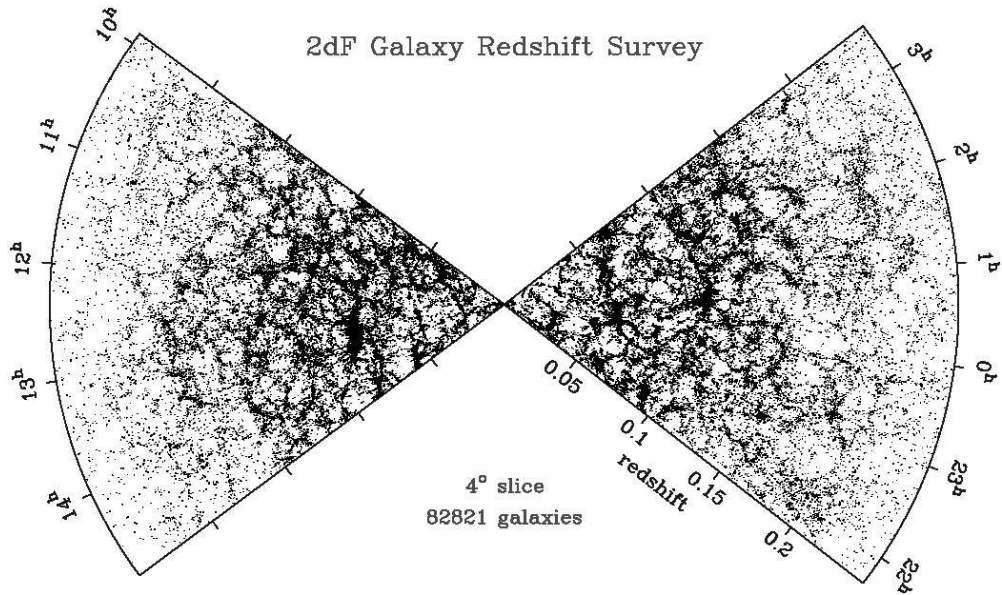

Figure 1.2: Flux-limited 2dFGRS slices in northern and southern galactic hemisphere are shown. The supercluster-void network visually stands out, and can be properly understood by computing either a hierarchy of correlation functions or by alternate, geometric and topological quantifiers. In this thesis, we develop methods to delineate individual superclusters and voids and describe their properties by using topological and geometrical diagnostics such as Minkowski Functionals and Shapefinders.

Cosmological parameter estimation is clearly one of the key results gleaned from large redshift surveys such as 2dFGRS. However, in our view several assumptions made in the process remain to be adequately justified, for instance, the assumption of linear, constant bias used in the above exercise is clearly inadequate. In fact, the study of biasing of galaxies with respect to the underlying mass distribution has its own history, and the issue has not yet settled completely (see Section 1.3). Figure 1.2 shows the distribution of galaxies in the flux-limited 2dFGRS slices. The 3−dimensional power-spectrum mentioned above, in a more appropriate form $\Delta^2(k) = dP(k)/dln(k)$, measures the amount of power per unit interval in the logrithmic bin around wavenumber $k$. In simple terms, this is a statistical measure of the density fluctuations within spheres of radius $R \sim 1/k$. In a measure such as this, clearly, the information at length-scales $r < R$ is averaged out. Hence, one learns relatively little about the nature of the *actual* distribution of galaxies. To the eyes, galaxies are distributed anisotropically, with sharp qualitative similarities with the distribution of dark matter in N−body simulations (compare the galaxy distribution in Figure 1.2 with that of dark matter in Figure 1.1). Describing (and explaining) the visually rich filamentary pattern of superclusters of galaxies and voids is another, *equally* important application where redshift surveys can be used. This problem has not been adequately addressed so far despite recent considerable progress (Hoyle and Vogeley 2004; Hoyle et al., 2002a, 2002b; Hikage et al., 2003; Basilakos 2003; Doroshkevich et al., 2003; Hikage et al., 2002) ¶. Clearly, the contribution of higher order correlation functions is an important ingredient in such a program. This will be very usefully complemented if we can quantify the shapes and sizes of individual superclusters and voids as well as assess the extent of non-Gaussianity in LSS as a whole. Our method, proposed and developed in the present thesis, is intended to complement more traditional, correlation function based approaches in developing a quantifiable picture of the cosmic web and its properties. The geometrical and topological methods which we employ will enable us to test theoretical predictions against observational results in the same integrated manner.

## 1.2.2   Sloan Digital Sky Survey

The Sloan Digital Sky Survey (SDSS) is derived from a dedicated 2.5m telescope. The initial photometric program has been to measure the positions and brightness of about $10^8$ objects in $\pi$ steradian (almost $1/4^{\text{th}}$) of the sky. The survey is centered on the northern galactic pole and has an elliptical angular coverage of $130^o \times 110^o$, where the semi-major axis runs along a line of constant right-ascension. Figure 1.3 shows the survey geometry of SDSS (large region bounded by solid line) and 2dFGRS (two shaded regions). Follow-up spectroscopy is planned to give redshifts of about $10^6$ galaxies and $10^5$ quasars. At the time of writing, the SDSS team has made its $2^{\text{nd}}$ data release, available for free download (Abazajian et al., 2004). It contains redshifts of $\simeq 3 \times 10^5$ galaxies, with the farthest galaxy at a redshift of ∼0.3. (See `http://cas.sdss.org/astro/en/tools/search/SQS.asp`.)

The limiting magnitude of the SDSS survey is in the red band ($r_{lim} = 17.77$), and its depth is similar to 2dFGRS. The survey will cover a cosmological volume as large as $1[h^{-1}\text{ Gpc}]^3$. Taking the length scale of 100 $h^{-1}$Mpc as a tentative scale of homogeneity, this is approximately thousand times a representative volume of the Universe. Hence when fully complete, SDSS can be fruitfully used to evaluate cosmic variance on many statistics associated with LSS. In Chapter 4 of this thesis, we study LSS in *mock* catalogues of SDSS where we discuss some of the applications of our method.

---

¶Recently, an attempt has been made to evaluate average $n$−point cummulants *up to* $6^{\text{th}}$ order using the method of *Counts in Cells* (Baugh et al., 2004; Croton et al., 2004). These authors test the hierarchical nature of gravitational clustering.

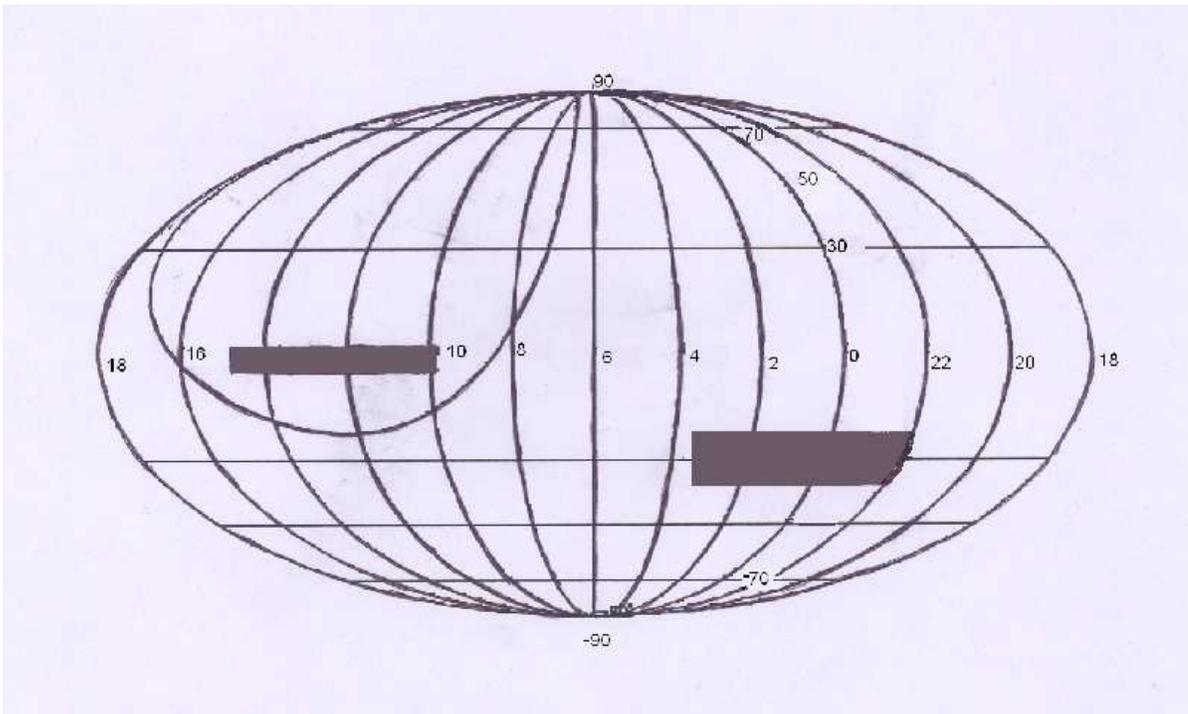

Figure 1.3: Survey geometry for SDSS (large region marked with solid line, to the left) and 2dFGRS (shaded regions) is shown on an Aitoff projection of the celestial sphere. The SDSS operates over an elliptical region which covers about $\pi$ steradian of the sky. By reaching a depth of $\simeq 10^3 h^{-1}$Mpc, this survey will record redshifts of $10^6$ galaxies, which is an improvement by a factor of 4, over the 2dFGRS.

In order to make full use of the information encoded in the observed LSS, it is essential to understand the relationship between the number density of galaxies and the mass density field. Since the 1980s, it has been realized that the *visible* density field due to galaxies may be *biased* relative to the underlying mass-density. Below we briefly review the state of our knowledge in this regard in order to appreciate the implications that bias may have in the program of confronting theory with observations.

## 1.3   Nature of Bias

There is compelling evidence that baryons (visible matter) constitute only 4% of the closure density of the Universe. Most remaining matter is likely to be nonbaryonic (and *dark*), interacting only gravitationally. This form of matter is ∼10 times more abundant than baryonic matter, and hence more important from a dynamical point of view. In Section 1.1, we summarised the theoretical efforts which are made to understand the clustering of dark matter.

From the predictive point of view, given the fully evolved nonlinear mass-density field, we would like to know the exact locations where galaxies are formed. An important question related to this is whether the clustering properties of the resulting galaxy distribution (quantified in terms of $n-$point correlation functions or some other alternate statistics) is *the same* as the underlying matter-field, or whether the two are different. The answer to this, clearly depends upon the recipe which we utilise in grafting galaxies on the density field. Our theoretical understanding in this regard is, time and again, supplemented by a wealth of observations of clustering at various scales: on the scale of Abel clusters of various richness classes (in early 80s), for radio galaxies, IRAS selected and optical galaxies (Peacock & Dodds, 1994) and lately for normal galaxies ($L = L_*$) and galaxies fainter than $L_*$ (Norberg et al., 2001, 2002). It is found that all these classes of objects show different clustering strengths; the more massive (and more luminous) objects showing more clustering. It is clearly important to understand the relative degree of clustering of these systems. Since these various visible tracers of LSS appear to cluster differently, their distributions are clearly biased relative to each other. There is hence, a reason to believe that the visible matter as a whole would be biased relative to the underlying dark matter. This phenomenon is termed as *bias*. Knowing the nature of bias would enable us to connect the two distributions − visible and dark.

If the galaxies *strictly* trace the mass, i.e., if the number density of galaxies is proportional to the mass contained in a given volume, we would expect the clustering properties of the two distributions to be *similar*. In this case, for example, the two-point correlation functions of the two components would differ from each other only by a constant factor. Groth & Peebles (1977) (using Lick Catalogue) and later Baugh (1996) (using the APM catalogue), established the perfect power-law form of the two-point galaxy-galaxy correlation function, $\xi_{gg}(r)$.

$$\xi_{gg}(r) = \left(\frac{r}{r_0}\right)^{\gamma}; \quad r_0 \simeq 5 \ h^{-1} \ Mpc; \quad \gamma \simeq -1.8. \tag{1.2}$$

However, the correlation function of *cold* dark matter $\xi_m(r)$ was shown to differ from $\xi_{gg}(r)$ in a complicated, scale-dependent fashion in case of various CDM cosmogonies (Klypin, Primack & Holtzman 1996; Peacock 1997; Jenkins et al., 1998). This at once ruled out the possibility of bias being constant and scale-independent.

During the late 1970s and early 1980s, clusters of galaxies were shown to cluster far more than galaxies. In an effort to understand this phenomenon, Kaiser (1984) proposed the *high-peak* model and pointed out that the strong clustering can be traced back to the structure of the primordial density field. The chief element of the high-peak model is to decompose the density field into a complete set of density-fluctuations of varying length. See Figure 1.4 where the dashed line represents a long wavelength mode; superimposed on this are density-fluctuations with smaller length-scale, which may collapse to form the objects of our interest (these may be galaxies or clusters of galaxies). At any given epoch, the dark matter halos with average density contrast $\delta$ exceeding the critical density $\delta_c$ =1.686 are prone to collapse. The high-peak model states that a rare high density fluctuation, corresponding to a massive object (solid line in Figure 1.4), collapses sooner if it lies in a region of large-scale overdensity (dashed line). This is because it is *aided* by the surrounding large-scale overdensity so that, $\delta > \delta_c$ criterion is fulfilled relatively easily for such a halo. This "helping hand" from the long-wavelength modes means that overdense regions contain an enhanced abundance of massive objects with respect to mean. Such objects form earlier than similar objects in the underdense regions, and they also participate in the subsequent gravitational dynamics. This model successfully explained the observation that Abell clusters are much more strongly clustered than galaxies in general: regions of large-scale overdensity systematically contain more high-mass halos than expected if the halos traced mass. The notion of what we call *large-scale overdensity* changes with time as gradually larger scales become nonlinear. It is possible to work out a relation between the mass of a halo and its relative degree of bias. Working in the Press-Schechter formalism (Press & Schechter 1974), one gets

$$b(\nu) = 1 + \frac{\nu^2 - 1}{\delta_c}, \tag{1.3}$$

where $\nu = \delta_c/\sigma(M)$ characterises the halos of mass M. Thus at every epoch, halos of mass $M_*$ ($\nu = 1$) are unbiased, those with $\nu > 1$, $M > M_*$ are positively biased and those with $\nu < 1$, $M < M_*$ are antibiased (see for example, Peacock 2003).

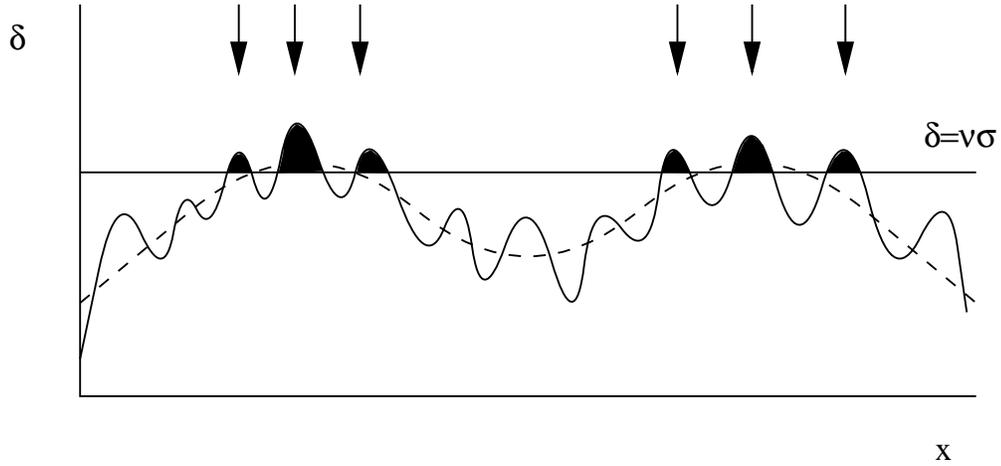

Figure 1.4: The essential features of *high-peak* model are illustrated. A density fluctuation *riding over* a large-scale overdensity of a longer wave mode is aided so that it can collapse earlier than similar objects in the underdense regions. This can explain the strong clustering of elliptical galaxies as well as that of Abell clusters. For galaxy-like objects, this model predicts high large-scale bias at high redshifts. The figure has been taken from Peacock(2003).

The above model, first worked out for clusters by Kaiser (1984), was later extended for objects of any mass by Cole & Kaiser (1989) (also see, Mo & White 1996 and Sheth et al., 2001). However, there is an intrinsic problem in relating predictions of this model for galaxies with the data from redshift surveys. The model heavily relies on the kinematic premises, and is applicable at early epochs when $\sigma(M)$ corresponding to the mass of normal galaxies was $\sim \delta_c$. In other words, the formalism may correctly predict the bias in galaxies at early epochs, when $M_*$, the typical mass of the halos most likely to collapse, is similar to the mass of a normal galaxy. Steidel (1997) employed the Lyman limit technique to select galaxies around redshifts $2.5 \leq z \leq 3.5$ and found their distribution to be highly inhomogeneous. The apparent value of $\sigma_8$ for these objects is of order unity (Adelberger et al., 1998), whereas the present value of $\sigma_8$ should have evolved to $\simeq 0.25$ in linear theory prediction of $\Lambda$CDM cosmology. This suggests a bias parameter $b \simeq 4$, i.e., $\nu(M) \simeq 2.5$ (Eq.1.3). The mass of each halo can be worked out to be $\simeq 10^{12}h^{-1}\mathrm{M_\odot}$. The baryon mass fraction in these objects was however estimated to be $\simeq 10^{10}h^{-1}\mathrm{M_\odot}$ which is $\sim$1% of the total mass. Since this is unusually small a number for baryon fraction, it was argued that each $10^{12}h^{-1}\mathrm{M_\odot}$ halo contains a number of Lyman-break galaxies. Thus the $\mathrm{M_*}$ galaxies (the Lyman break galaxies) may be unbiased ($\nu = 1$), but halos with $M = 10^{12}h^{-1}\mathrm{M_\odot} > \mathrm{M_*}$ may be positively biased (we noted above that their distribution is highly inhomogeneous).

We saw above that the kinematic bias formalism may correctly predict the relative population of galaxies at high redshifts. Starting from z$\simeq$3, up to the present, galaxies have possibly collapsed in all environments. The underlying mass-fluctuations have also grown. Thus, one predicts that the large bias which existed at high redshift should asymptotically approach unity on large scales (bias is defined as the ratio of r.m.s. fluctuations in the visible and dark matter). Lahav et al.(2002) have indeed shown that the 2dFGRS data are consistent with a scale-independent large scale bias which is of the order unity. Thus, broadly we seem to have effectively understood bias. However, we should see whether our original goal of *populating* the mass density field with galaxies can be achieved with the help of this knowledge. A recent approach which bypasses the complex route of following simulations or semianalytic calculations, and yet explains the shape and amplitude of the observed power spectrum of galaxies is to resort to *Halo Model*. The key feature of this model is to encode all the complications of galaxy formation via the halo occupation number: the number of galaxies found above some luminosity threshold in a virialized halo of a given mass. In a nut-shell, the halo model describes nonlinear structures as virialized halos of different mass, placing them in space according to the linear large-scale density field. For more details of this model see Cooray and Sheth (2002). Figure 1.5 compares the exact nonlinear matter distribution with the halo model representation. The halo model in the right panel is surprisingly fruitful, allowing us to reproduce correlation function, power spectrum etc. However, it is manifestly different from the fully evolved density field. This brings out an important aspect relevant to further study of LSS: the low-order correlations cannot be relied upon to provide us with a complete picture of clustering. In addition to the large halos of clusters and groups, today the rich filamentary/sheet like patterns in matter density field are also populated by galaxies $^\parallel$, and one must theoretically understand the emergence of halos along filaments/sheets in order to complete the picture. In fact, the following experiment can be carried out: the galaxies can be *selected* in $z = 0$ mass-density fields from two rival cosmogonies, so that the observed galaxy-galaxy correlation function $\xi_{gg}(r)$ and the power spectrum $P_k$ are reproduced. In comparing our predictions with the LSS-data, if we confine ourselves merely to two-point correlation function, such an algorithm would validate *both* models to be good representatives of our Universe. Evidently this is not true. Clustering properties of galaxies in the two models would of course be different from each other (and perhaps even from LSS in the Universe). This serves to establish that the large-scale distribution of galaxies *cannot* solely be described by the two-point correlation function alone. The resulting galaxy distributions can be shown to differ from each other by employing statistics which are sensitive to higher order correlation functions. Following this, we arrive at an important conclusion: the two distributions of galaxies *must be* compared with the help of higher order correlation functions or statistics derived therefrom. We shall explicitly address this problem in Chapter 4. Some nontrivial effects of simple scale-dependent bias shall be pointed out in this chapter.

---

$^\parallel$Indeed, about 70% of galaxies appear to be in the *field*, i.e., along filaments and sheets connecting the clusters of galaxies.

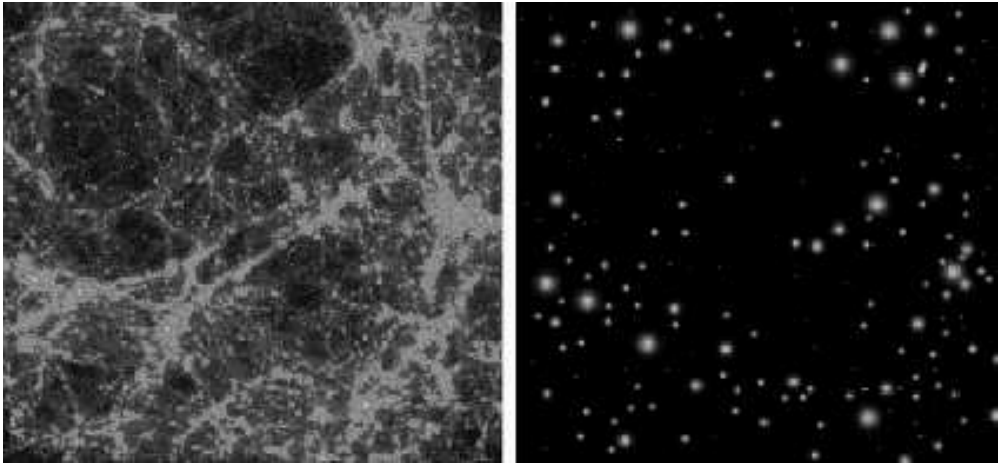

Figure 1.5: *Left panel*: Simulated dark matter distribution which shows complex morphology. *Right panel*: Halo Model corresponding to density field in the left panel, which successfully reproduces galaxy-galaxy correlation function and the power spectrum. The two are manifestly different, which points out the role which higher order correlation functions play in defining the cosmic structure. The figure is taken for illustrative purposes from Cooray and Sheth (2002).

To summarise, in this section we reviewed the state of our knowledge about the nature of bias. Although the *high peak model* (Kaiser 1984) gives us promising results for galaxies assembling at high redshift and for Abell clusters at present, the predictive power of *halo model* in populating mass density field with galaxies is clearly inadequate (in fact, due to a variety of reasons, the bias could also be stochastic; see Dekel & Lahav 1999). In particular, for a combination of a cosmological model and a biasing scheme to be proven correct, it is *not sufficient* that the two explain the two-point correlation function for galaxies *alone*. In order to effectively compare the theoretical predictions for galaxy-distribution with the data from redshift surveys, one should resort to higher order correlation functions, or alternate statistics which can effectively capture the anisotropies in the galaxy distribution. We introduce one such class of statistics in the next section, and explain the utility of these statistics in effectively quantifying the complex morphology of LSS.

## 1.4    Confronting Theories with Observations

It is well known that, unlike the primordial Gaussian random field, a fully evolved nonlinear density field cannot be fully quantified in terms of its two-point correlation function. In fact, it is the lowest and first of an infinite hierarchy of correlation functions describing the galaxy distribution. Furthermore, the bias in galaxies viz a viz dark matter can be nontrivial in nature, and can give rise to different clustering properties of galaxies and the underlying mass density field. A physical model, a random process or a combination of the two, generating correct correlation function $\xi_{gg}(r)$ do not on their own validate a given cosmological model unless other statistical measures of clustering provide independent support. It is hence, a challenge to compare the theoretical predictions for LSS with real data.

If we knew all n−point correlation functions, we would have a complete description of the galaxy clustering process. However, estimating $\xi_{gg}(r)$ for a sample of $N$ galaxies requires all pairs of galaxies in this sample, while calculating 3-point function requires taking all triples from $N$. The amount of computation escalates rapidly with $n$. Besides, it is difficult to extract intuitively useful information from these statistics. In this thesis, we propose to work with another class of statistics which complement the correlation function approach and which has the advantage of providing a physically appealing interpretation for the evolving density field. These are the *Minkowski Functionals*.

### 1.4.1    Minkowski Functionals

For an excursion set involving $N$ particles embedded in $n$−dimensions, the Minkowski Functionals (hereafter, MFs) are defined on an $(n-1)$-dimensional hypersurface. MFs number $(n + 1)$. In this thesis, we shall be concerned with 3-dimensional distribution of dark matter and/or galaxies. Hence, MFs will be defined on a 2-dimensional surface and they will reflect the physical properties of this surface (in the given instance, an isodensity contour). The four MFs (Blaschke 1936; Minkowski 1903; Mecke, Buchert and Wagner 1994) are, respectively

1. Volume V,

2. Surface area S,

3. Integrated Mean Curvature C,

$$C = \frac{1}{2} \oint \left( \frac{1}{R_1} + \frac{1}{R_2} \right) dS. \qquad (1.4)$$

   $R_1$ and $R_2$ are the two principal radii of curvature of the surface in a given local neighbourhood and the integral is taken over the entire closed surface.

4. Integrated Gaussian Curvature (or Euler characteristic) $\chi$,

$$\chi = \frac{1}{2\pi} \oint \left( \frac{1}{R_1 R_2} \right) dS. \qquad (1.5)$$

   A related quantity which is more popular in Cosmology is genus. The genus is related to $\chi$ by

$$G = 1 - \frac{\chi}{2}. \qquad (1.6)$$

   The 3-dimensional genus of an object is a topological invariant. It can be interpreted in terms of the connectivity of the surface. In simple terms it can be viewed as the number of independent cuts which one can make to the surface without breaking it into two pieces. A torus can be cut *once* and yet can be kept in one piece. Hence, its genus is 1. However, a sphere would be broken into two pieces if a cut is made. Hence, its genus is 0. Two objects with the same value of genus are topologically similar: one can be obtained by continuously deforming the other object. Thus, a sphere, a cube and a torus with one hole in it are topologically equivalent. More concretely, genus of an object is the number of handles that the object has, in excess of the number of holes which it encloses (e.g., see Matsubara 2003). Thus,

   G = [# of handles to the surface] − [# of holes enclosed by the surface].

According to this definition, a sphere has no handle, a torus has a single handle (equivalent to a sphere with one handle), and a pretzel has two handles. Introducing a hole or a bubble *inside* the surface *reduces* its genus by 1, whereas adding a handle to the surface *increases* its genus by 1.

An attractive feature of MFs is that they can be shown to depend on the entire hierarchy of correlation functions (Schmalzing 1999). Thus, MFs can indirectly help us capture the effect of $n-$point functions. Higher order $n-$point functions gradually become important as a primordial, featureless GRF evolves to develop nonlinear structures. This may very well reflect in the behaviour of MFs for the corresponding density field. MFs are additive in nature, i.e., they can be studied for individual objects (say, clusters or superclusters defined using some prescription) as well as for the entire ensemble of such objects.

Minkowski Functionals were discovered by Minkowski (1903) and later they were presented in their modern form by Blaschke (1936) (Minkowski's original article does *not* introduce the Minkowski Functionals, but rather related quantities known as mixed volumes; see Schmalzing 1999). and were introduced in Cosmology by Mecke, Buchert and Wagner (1994). Of the 4 MFs listed above, the genus G was already known to the cosmology community (Doroshkevich 1970, Gott, Melott and Dickinson 1986). Tools like percolation analysis which can also be related to MFs, had earlier been introduced in cosmology by Zel'dovich, Einasto and Shandarin (1982). There were strong reasons why this was the case.

The emerging picture of LSS showed us that galaxies are distributed along filaments and/or sheets which encircle vast voids. The distribution of galaxies appeared to occupy very little volume, and yet it would percolate through the entire volume. A percolating LSS would be such that one can visit *all* corners of the cosmic volume by following the filamentary/sheet-like ridges. This property reveals the connectedness of the LSS. Connectivity is a mathematical notion, and it is possible to precisely quantify connectivity of an object or a system by evaluating its percolation properties and its genus.

A Gaussian random field (GRF) is featureless and shows complete symmetry between overdense and underdense regions. Hence, at mean density level we expect that the overdense and underdense regions shall be interwoven and would lend as a visual impression of a sponge. When viewed at median density threshold, LSS gives us an impression of a sponge; a reason why genus was thought to be a promising quantifier of this visual appearance of LSS. Doroshkevich (1970) gave an important result for genus G of a GRF.

$$G(\nu) = A(1 - \nu^2) \exp\left(-\frac{\nu^2}{2}\right),                    (1.7)$$

where $\nu$ is the density contrast measured in units of the variance $\sigma$ of the density field. Later, grid-based algorithms were worked out in order to compute genus of LSS (Weinberg 1988; Gott, Melott & Dickinson 1986).  Because of the known analytical expression for the genus of a GRF, it can be employed to probe the Gaussianity of a density field or its departure from it.  In addition, the amplitude $A$ in the above equation explicitly involves the power spectrum of GRF. By measuring genus $G(\nu)$ for a galaxy-distribution smoothed at a variety of length-scales, one can have an independent estimate of the effective power spectrum index $n$ at a variety of length-scales of smoothing.  (For an excellent review, see Melott (1990).)

Although a study of the genus gives us a useful handle on the connectivity of a density field, we still lack information about the morphology of LSS. For example, if the domain of information is solely restricted to knowing genus, a filament with one handle would be considered to be identical to a sphere with one handle or a pancake with one handle.  Therefore, to gain objective insights into the nature of the supercluster-void network, we must complement genus of an object with quantities which in some way, characterise geometry of that object.  This is precisely the role played by the first three MFs, Volume V, Surface Area S and Integrated Mean Curvature C. These MFs change when local deformations are applied to the surface, and hence, these may be useful to glean information about the typical size of an object.  The final goal here would be to use this geometric information and give an objective meaning to generic sizes and shapes of superclusters and voids belonging to LSS. In this thesis we will utilise MFs to characterise the geometry and topology of the cosmic density field, and eventually use this information to probe the morphology of the supercluster-void network in LSS. Because MFs depend on the full hierarchy of correlation functions, we can use them to compare and distinguish two rival cosmological models.  Recently, Matsubara (2003) derived semianalytical expressions of MFs for weakly nonlinear cosmic density field by using a second-order perturbative formalism.  The methods developed by Matsubara and reported in this thesis can be used to obtain insight into the dynamics of the clustering process of a density field smoothed on sufficiently large scales.  The role of bias on large scales can also be directly probed using such a method.

There have been several attempts of evaluating MFs for a given point process. For example, the Boolean grain model (Mecke, Buchert and Wagner 1994; Schmalzing 1999) would decorate input set of points with spheres of varying radii, and would study the morphology of the structures which contain overlapping spheres.  Another approach was propounded by Schmalzing & Buchert (1997).  In this approach one smooths the distribution of points using a suitable window function, and defines the density field on a grid.  MFs can be evaluated for connected structures defined on a grid by using

Crofton's formulae or Koenderink Invariants (Schmalzing & Buchert 1997).

In this thesis, we have developed a *new* approach for determining MFs. Rather than working with a grid-approximation for connected structures, we go a step forward, and model surfaces for these objects. Minkowski Functionals are evaluated for the isodensity surfaces *online*, i.e., while these are being modelled using a surface triangulation technique. Chapter 2 describes the algorithms for surface construction and the evaluation of MFs on these surfaces. We also discuss here, our method of implementation of these algorithms which has culminated into a robust and accurate software SURFGEN. The development of SURFGEN forms a major component of this thesis.

Finally we briefly discuss how shapes and sizes of superclusters and voids can be determined using the Shapefinder diagnostics (Sahni, Sathyaprakash & Shandarin 1998).

## 1.4.2 Morphology of LSS with Shapefinders

Sahni, Sathyaprakash & Shandarin (1998) proposed to use differential geometric measures such as MFs to evaluate size and shape of a typical three-dimensional object. The size is given in terms of three Shapefinders which have dimensions of length. These are defined as ratios of MFs, and are conviniently termed as Length ($\mathcal{L}$), Breadth ($\mathcal{B}$) and Thickness ($\mathcal{T}$).

$$\mathcal{T} = \frac{3 \times V}{S} \tag{1.8}$$

$$\mathcal{B} = \frac{S}{C} \tag{1.9}$$

$$\mathcal{L} = \frac{C}{4\pi G}. \tag{1.10}$$

The three Shapefinders are defined using spherical normalization, so that for a sphere of radius $R$, $\mathcal{T} = \mathcal{B} = \mathcal{L} = R$. Later in Chapter 3, we shall see that for structures defined as connected overdense regions at or above the percolation threshold, the three Shapefinders convey the morphological information about superclusters in a generic cosmic density field remarkably well.

The above measures quantify the *size* of the object in question. To further quantify the *shape* of these objects, two dimensionless Shapefinders have been defined as follows. These are Planarity ($\mathcal{P}$) and Filamentarity ($\mathcal{F}$).

$$\mathcal{P} = \frac{\mathcal{B} - \mathcal{T}}{\mathcal{B} + \mathcal{T}} \tag{1.11}$$

$$\mathcal{F} = \frac{\mathcal{L} - \mathcal{B}}{\mathcal{L} + \mathcal{B}}. \tag{1.12}$$

Sahni et al.(1998) showed that $(\mathcal{P}, \mathcal{F})$ quantify the shape of fairly simple objects quite well. Thus, for an oblate ellipsoid its planarity is relatively large and its filamentarity is small. For a prolate spheroid, the reverse is true. The statistics respond monotonically to deformations of these surfaces. Further, for objects with toroidal topology, $(\mathcal{P}, \mathcal{F})$ statistics were again shown to give correct morphological prescription, so that a compact disc is found to have a sheet-like morphology, and a circular torus with negligible crosssection is detected as a filament.

Together with $(\mathcal{P}, \mathcal{F})$, genus G can define a three-dimensional *Shape-Space* which may be used to represent the distribution of shapes of superclusters and their topologies in a given cosmological density field. In Chapters 3 and 4, we shall demonstrate that the Shapefinders do indeed conform to our visual impression of LSS when evaluated for objects defined at or above the percolation threshold. Thus, Shapefinders will be proven to be unbiased and robust statistics with which to quantify the morphology of the Cosmic Web.

In summary, the major thrust of this thesis is to develop tools to determine accurately the MFs and Shapefinders. These tools have been developed and tested against Gaussian Random Fields (Chapter 2), N−body simulations of rival cosmogonies (Chapter 3) and mock SDSS catalogues of galaxies (Chapter 4). We report several successes of our method and conclude in Chapter 6. Chapter 5 is devoted to a morphological study of the 2−dimensional Las Campanas Redshift Survey slices.

# Chapter 2

# Estimating Minkowski Functionals on a two−dimensional surface

The last chapter was devoted to a historic account of the study of large scale structure and the current status of observations aimed at mapping the 3−dimensional distribution of matter in the universe. We motivated the need for quantification of large scale structure, and summarised various statistics which are used to meet this goal. In particular, we highlighted the importance of geometric and topological descriptors of large scale structure, and in this context we introduced the Minkowski Functionals (hereafter, MFs).

Since MFs are defined on a surface, it is important to develop a mathematically robust prescription to construct a surface *. Furthermore, it would be beneficial if we could evaluate MFs *online*, while constructing a surface out of input data. Both these demands are nicely met by SURFGEN − short for SURFace GENerator− a code which was developed as a part of this thesis. In the present chapter, we discuss the foundations on which SURFGEN is developed, and provide a detailed description of the method of implementation of SURFGEN. The last part of this chapter is devoted to various tests of the accuracy of SURFGEN.

## 2.1   Surface Generating Algorithm (SURFGEN)

SURFGEN is a versatile and powerful prescription which permits one to generate and study surfaces whose physical origin can be quite varied and different. Isophotes, isotherms and isodensity surfaces constructed from 3D data present us with examples of two dimensional surfaces which can be generated and studied using SURFGEN and the Minkowski functionals introduced in the previous chapter [†].  In this thesis, SURFGEN will mainly be applied to a density field $\rho(\mathbf{r})$ defined on a rectangular grid ($\rho(\mathbf{r})$ can itself be generated from particle positions using a Cloud-in-Cell method; see Hockney and Eastwood 1988).  The density-grid can be visualized as consisting of a large number of closely packed grid-cubes.  Any given grid-cube is characterised by the value of the density at its eight vertices.  SURFGEN determines a closed polyhedral surface at a preset value of the density threshold $\rho_{TH}$ by triangulating cubes while moving across the grid.  The process of triangulation is based on the following observation.

---

*In the context of cosmology, the fundamental physical entity is a cosmic density field and in this case, a surface may refer to an *isodensity contour*.

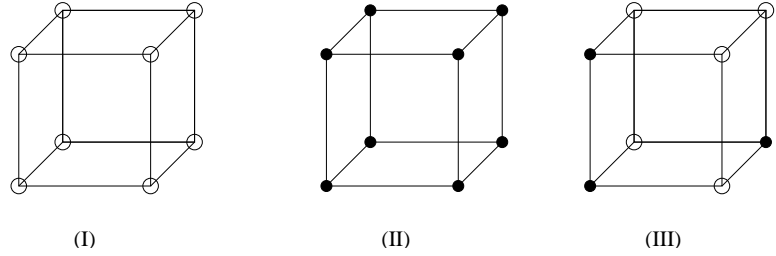

Figure 2.1: Given a density field defined on a grid, the grid-cubes fall into one of the above three classes:  (1) cubes which are completely underdense, (2) cubes which are completely overdense and finally (3) cubes which have both overdense and underdense vertices.  Cubes belonging to the 3$^{\mathrm{rd}}$ category are triangulated using SURFGEN to generate an isodensity surface.

For any arbitrary threshold of density $\rho_{TH}$, all grid cubes of the density field fall into one of the following three classes:

(i) those for which $\rho < \rho_{TH}$ at all eight vertices (*underdense cubes* – Fig. 2.1–I),

(ii) those for which $\rho > \rho_{TH}$ at all eight vertices (*overdense cubes* – Fig. 2.1–II),

(iii) cubes having both overdense and underdense vertices (*surface cubes* – Fig. 2.1–III).

When modelling clusters and superclusters overdense cubes will be *enclosed* by our surface while underdense cubes will be excluded by it.  (For voids the situation will be reversed.)  Clearly an isodensity surface will pass entirely *through the surface cubes*.  Therefore, the properties of *surface cubes* are vitally important for this surface-construction exercise.

We now describe how SURFGEN constructs a surface at the desired threshold $\rho_{TH}$ by triangulating *surface cubes*.  We work under the assumption that the underlying density field $\rho(\mathbf{r})$ is continuous, so that in moving from an overdense site to an underdense site (along a cube edge) we invariably encounter a point at which $\rho = \rho_{TH}$. SURFGEN classifies a given grid-cube in terms of the number of points (on the edges) where $\rho = \rho_{TH}$. The precise location of these points (found by interpolation) tells us exactly where our isodensity surface will intersect the cube. Since there are eight vertices and each vertex can either be overdense (1) or underdense (0), the number of possible configurations of the cube are $2^8 = 256$. Clearly the triangulation must be invariant upon interchanging 1's with 0's, and this reduces the number of independent configurations to 128. Upon invoking rotational symmetry this number further reduces to only 14. (For instance, although there are eight cubes which have a single overdense vertex, any two members of this family are related via the three dimensional rotation group. Therefore a single scheme of triangulation suffices to describe all members of this group.) These fourteen configurations of triangulated cubes are shown in Fig. 2.2 (also see Lorenson & Cline 1987).

---

[†]SURFGEN is a modified and extended version of the Marching Cubes Algorithm (MCA) which is used in the field of medical imaging to render high resolution images of internal organs by processing the output generated by X-ray tomography (see Lorenson & Cline 1987).

The surface intersecting a given *surface cube* clearly forms only a portion of the full isodensity surface which we are interested in constructing. By explicitly demanding that the density field be *continuous* across the isodensity surface one joins triangles across the faces of neighboring cubes, thus constructing the full continuous polyhedral surface. This then is the basic prescription which may be used to construct a closed triangulated surface at any desired value of the density threshold. However, a few comments about triangulation are in order: (i) the triangulation of the cubes having four contrasting vertices is *not unique*. To illustrate this let us consider Fig. 2.3 where cube # 10 of Fig. 2.2 is reproduced in the left panel. We note that this cube is symmetric with respect to an interchange of underdense sites with overdense sites. As a result, we could just as well triangulate it according to the prescription shown in the right panel of Fig. 2.3. This degeneracy in the way a given cube can be triangulated is inherent also in the 14[th] and 15[th] cube of Fig. 2.2. Noticeably, at all such places where we encounter one of these cubes, one scheme of triangulation has to be preferred over the other, otherwise the surface could become discontinuous. Thus to create a unique triangulated surface one must triangulate such cubes *in tandem* with their neighbors, ensuring that the continuity of the surface is maintained. (ii) There are instances when two neighboring cubes have a density configuration which does not lead to a complete closure of the surface in question. The resulting surface has a hole (Fig. 2.4a) which must be filled by constructing two additional triangles which close the surface (Fig. 2.4b). Hole formation can occur every time the common face shared by two cubes

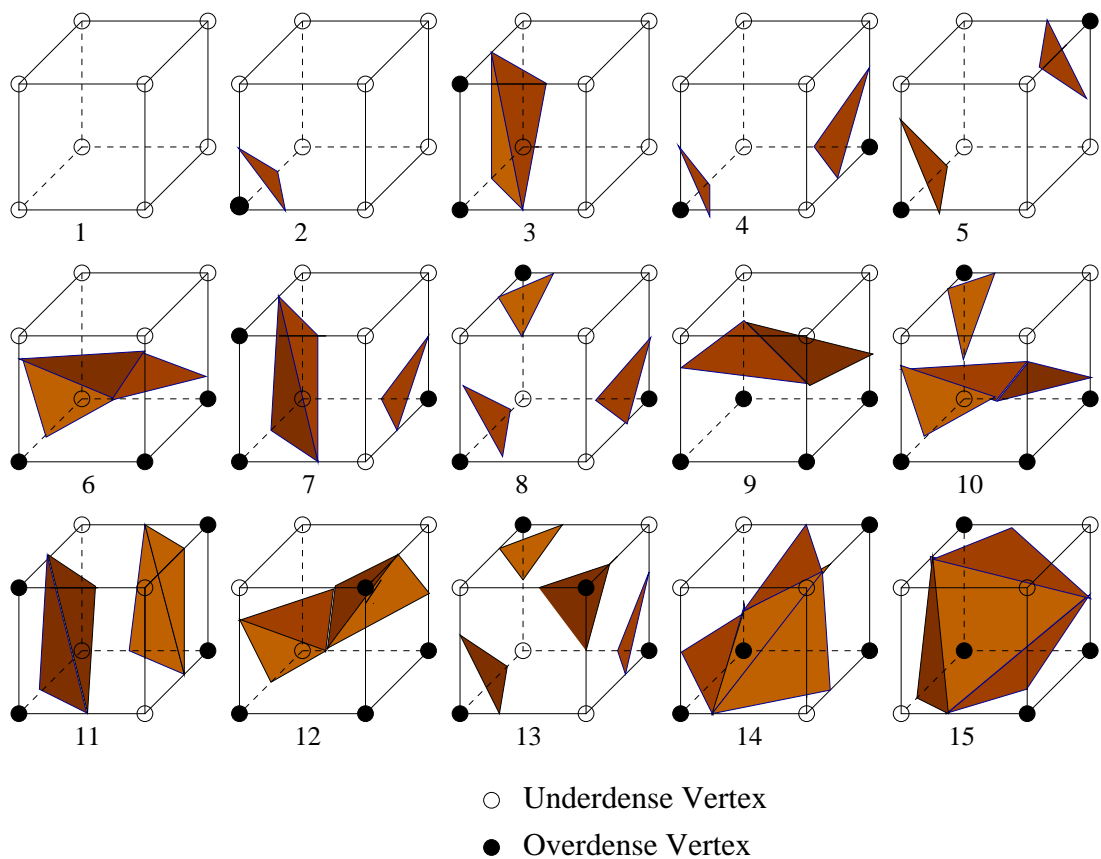

Figure 2.2: Depending upon the value of density at its vertices, a surface cube can be triangulated according to one of the fourteen independent configurations shown above. The black dots represent overdense vertices whereas underdense vertices are represented by open circles.

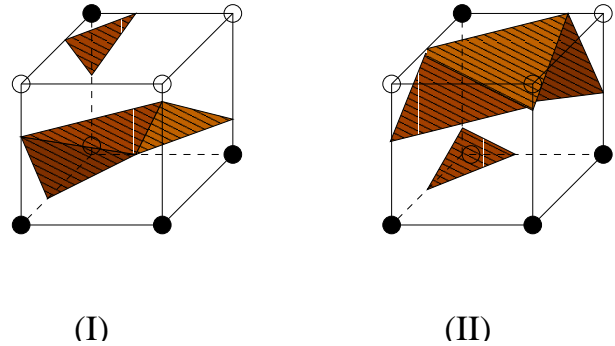

(I)                             (II)

Figure 2.3: The tenth cube of Figure 2.2 is shown in the left panel of this figure. It has four overdense and underdense vertices and, as shown above, it can be triangulated in two distinct ways. Two other cubes − numbered 14 and 15 in Figure 2.2 − also allow for this possibility (of being triangulated in two distinct ways) §. However, it can be shown that this degeneracy can be broken if we include information about neighboring cubes. Requiring that the density field be continuous across the neighboring cubes breaks an occasional degeneracy which may arise, and allows one to triangulate surfaces unambiguously.

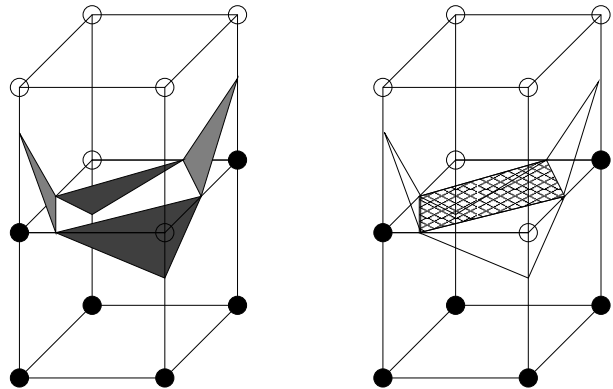

Figure 2.4: Formation of a 'hole' is illustrated. We 'fill' this hole by a pair of triangles as shown in the right panel of the figure. Since the original MCA algorithm was meant strictly for visualization purposes, this problem appears to have been overlooked by Lorenson & Cline (1987). We correct for this in our triangulation scheme by filling holes whenever they occur since its of prime importance to work with closed polyhedral surfaces when constructing MFs.

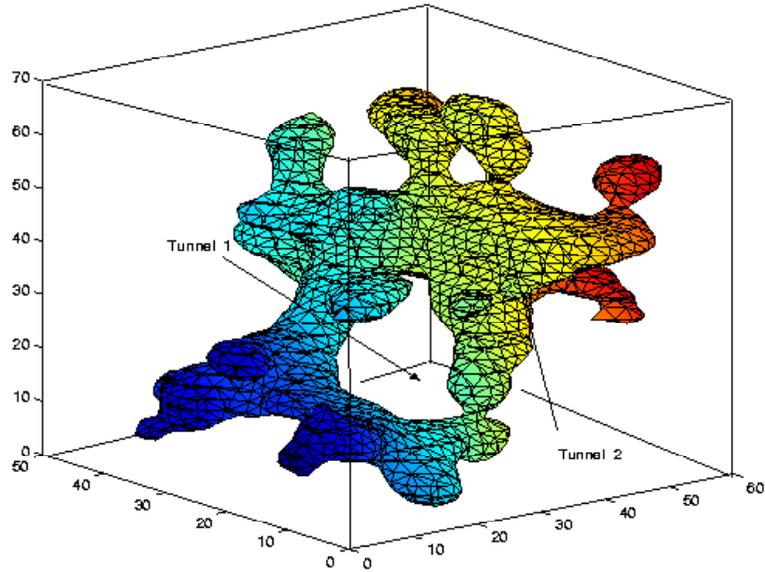

Figure 2.5: A cluster appearing in the CDM simulation is triangulated using SURF-GEN. The cluster is multiply connected with two tunnels and hence is described by a genus value of two.

shows two vertices of type 1 along one diagonal and of type 0 along the other diagonal.

Thus, in order to generate a closed polyhedral surface at a given value of the density threshold $\rho = \rho_{TH}$, SURFGEN uses the 14 independent triangulations of a grid cube in conjunction with an algorithm which ensures that the surface is continuous. We should emphasize that surfaces generated in this manner *need not be simply connected* (see Fig. 2.5). Indeed isodensity surfaces constructed at moderate density thresholds in N-body simulations as well as 3-dimensional galaxy surveys tend to show a large positive value for the genus. This issue will be discussed in detail in forthcoming chapters of this thesis. In the next subsection we shall discuss the implementation of the above algorithm which simultaneously allows us to generate isodensity contours and also calculates the four MFs.

## 2.2   Implementation of SURFGEN

In this section, we shall discuss in detail, how the algorithm for surface construction presented above, was implemented in the form of SURFGEN. It suffices here to clarify one major distinction between algorithm(s) and what we term as a *formalism*: algorithm is an abstract set of rules which guide us systematically from the input data to a result at the output. A formalism, which is usually specific to the programming language of implementation, is an efficient scheme to implement such an algorithm. Thus, it is the formalism underlying SURFGEN which will be discussed here.

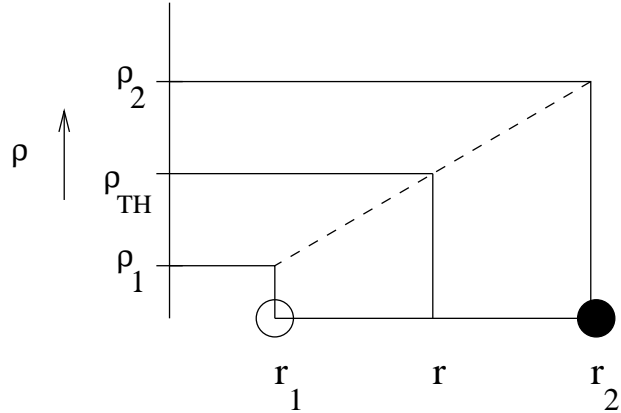

Figure 2.6: The process of finding point of intersection between a grid-edge and the surface of interest is illustrated.

Surface reconstruction on a cubic grid requires efficient handling of various geometric entities, like grid-vertices, grid-edges, faces of the cubic cells of the grid etc. Enabling the computers to carry out calculations utilising such geometric constructs, is an exercise which falls under the widely studied discipline of Computational Geometry. As will become clearer below, Computational Geometry is best practised in C; a reason why C was chosen for developing SURFGEN.

The principle underlying our implementation of SURFGEN is to start with a density grid and a threshold of density at the input and next to determine *all* points of intersection between the edges of the grid and the isodensity surface to be modelled. By definition, a grid-edge which is intersected by the isodensity surface *must have* one of its end-points overdense, and the other underdense. The point of intersection between a grid-edge and the isodensity contour is found by linear interpolation.

$$\mathbf{r} = \mathbf{r_2} - \frac{\rho(\mathbf{r_2}) - \rho_{\text{TH}}}{\rho(\mathbf{r_2}) - \rho(\mathbf{r_1})} \times (\mathbf{r_2} - \mathbf{r_1}), \tag{2.1}$$

where, $\mathbf{r_1}$ and $\mathbf{r_2}$ refer to the two neighbouring vertices of the grid and $\mathbf{r}$ refers to the point of intersection on the edge connecting these two vertices, at which $\rho = \rho_{\text{th}}$. In Figure 2.6 $\mathbf{r_1}$ refers to an underdense vertex, $\mathbf{r_2}$ refers to an overdense vertex and $\mathbf{r}$ refers to the point of intersection between the grid-edge and the surface.

Before proceeding further, let us take a brief discourse as to why is it so important to lay the foundations of SURFGEN on *this* principle. Notice that every grid-edge is shared by 4 cubic cells of the grid. Furthermore, a *surface cube* involves *at least* 3 intersected grid-edges and *at best* 12, i.e., all of them. As a rule, if a grid-edge is intersected, the four cubes sharing *that* edge, *will* exhibit 2 or more number of other intersected grid-edges, i.e., these cubes *will* be *surface cubes* and they will be eventually triangulated by our algorithm. We have mentioned it earlier that in order to build up the whole surface, we march across the entire grid, cell-by-cell, and triangulate the relevant cubes on the way. As can be appreciated, the triangulation of neighboring *surface cubes* belonging to, say, two neighboring $z = constant$ planes may take place separated by a considerable duration of time. See Figure 2.7 for an illustration. Here two neighboring cubic cells are shown to be sharing a common face □ABCD. The grid-edges $\overline{AB}$ and $\overline{BC}$ are respectively intersected by P1 and P2. Notice that the line-segment $\overline{P1P2}$ acts as a triangle-edge and participates in construction of triangles in both the cubes. In fact, $\overline{P1P2}$ connects the surface sections contributed by the two cubic cells. Recalling that SURFGEN will generate these two surface sections at different instants of time, evidently the information related to P1 and P2 is too precious to be evaluated repeatedly, i.e., while triangulating individual cells which happen to share the face □ABCD. The trick is *to create* this information *globally*. A globally created information will be accessible *throughout* the process of triangulation. To achieve this, we have to visualize the grid as consisting of properly aligned grid-edges. The *global* information is created by evaluating *all* points of intersection between the grid-edges and the surface to be modelled. After all the grid-edges being intersected have been identified, at a time 12 of them composing a cubic cell are considered. If a subset of the edges belonging to a cubic cell are intersected, the cube is classified as a *surface cube* and is considered further for triangulation according to the algorithm discussed above. While triangulating a given cube, the 12 *grid-edges* composing that cube are locally copied as *face_edges*. It is these *face_edges* which are manipulated for the purpose of triangulation of that cube (see below).

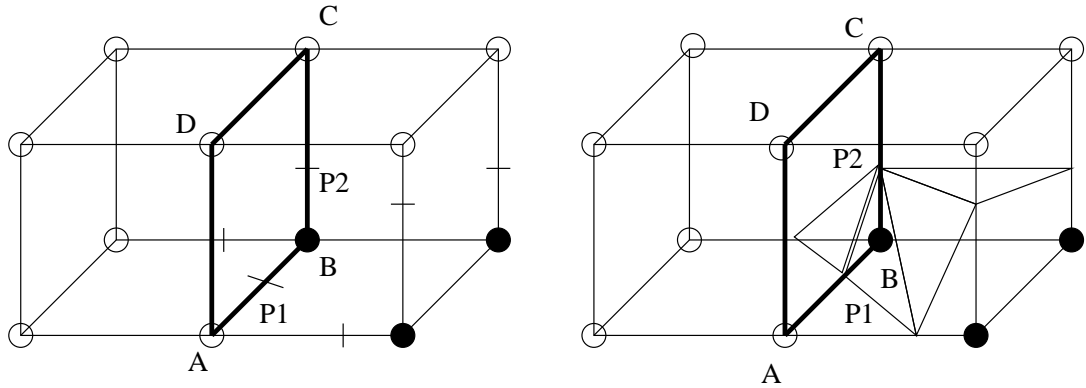

Figure 2.7: Two neighboring cubic cells are shown which share a face □ABCD com-
posed of 4 grid-edges. The two cubic cells *belong to* two neighboring $z = constant$
planes, and hence will be considered for triangulation at different times. Notice that
the grid-edges $\overline{AB}$ and $\overline{BC}$ are respectively intersected by points P1 and P2. In the
right panel of the figure, the two cubes are shown triangulated. Notice that the triangle-
edge $\overline{P1P2}$ connects the surface sections contributed by the two neighboring cubic cells.
Evidently, the information about P1 and P2 is too precious to be evaluated repeatedly,
i.e., while triangulating individual cubic cells which share □ABCD. As described in
the text, our entire formalism heavily relies on the fact that *this* information is stored
*globally*, so that one can retrieve it at any time during the course of triangulation.
While triangulating a given cube, the 12 *grid-edges* composing that cube are locally
copied as *face_edges*. These *face_edges* are used for the purpose of triangulation of that
cube. Please refer to the text for further details.

As it can now be appreciated, the step of finding *all* the sites of intersection between the grid and the surface turns out to be extremely helpful in the course of online triangulation. This is because a *globally* created information is used, as per need, in doing the surface modelling and computations of MFs *locally*, i.e., on a cell-by-cell basis.

In what follows, we shall systematically run through the formalism which implements the operations related to initialisation and triangulation of the grid. To do so, we have divided the process of surface-construction into three logical units.

- Initialisation of the grid

- Triangulation of cubes

- Calculation of Minkowski Functionals

## 2.2.1   Initialisation of the grid

In this subsection we shall discuss implementation of two concepts, which are intricately interwoven.

1. Finding the points of intersection between the grid-edges and the surface.

2. Incorporating a scheme to march across the grid.

In the course of this discussion, we shall introduce the necessary tools of programming and also the associated terminology. The tools will be used to do the next set of calculations and the related terminology will provide a language for describing the rest of the steps of implementation.

To begin with, let us motivate the relevance of implementing the above two tasks. Of these, the first has already been explained in terms of the principle underlying the formalism of SURFGEN. To understand the importance of developing the latter facility, let's imagine carrying out a simple-minded calculation involving a scalar field defined on a grid. In most of the numerical work required in solving physical problems, one is chiefly concerned with solving differential equations. A specific example which represents the general class of such routine scientific computations is to evaluate the gravitational acceleration field $\mathbf{g}(\mathbf{r})$ starting from a gravitational potential $\phi(\mathbf{r})$. Let's take $\phi(\mathbf{r})$ to be a discretised field defined on a grid. To obtain $\mathbf{g}(\mathbf{r})$ at an arbitrary location $(i, j, k)$, we would be evaluating the differences $\left[\frac{\phi(i+1,j,k)-\phi(i-1,j,k)}{2\Delta\ell_g}\right]$ etc., where $\Delta\ell_g$ is the resolution of the grid. Notice here that in all applications of this type, one only needs to invoke a single number from a given location $(i, j, k)$. Hence, it is quite straightforward to code such calculations. In the example considered here, we can trivially obtain $\mathbf{g}(\mathbf{r})$ by running through the $\phi-$grid using a simple triplet of `do-loops`.

It is easier to understand our computational needs in light of the above example. Let's note that our chief objective is to scan the grid and triangulate a cube at the upper-right corner of a given vertex, wherever appropriate. We shall need a mechanism which will give us complete control over the cube to be triangulated, in such a way that we can manipulate the geometric entities (grid-vertices, grid-edges, faces of the cube, triangles constructed within the cube, etc.) in a systematic manner, also ensuring proper book-keeping of relevant information. Scanning the grid, hence, acquires a different meaning in the present case. Contrary to the example considered above where we needed to retrieve a single value (of the scalar field) from a given location, here we need to retrieve the *entire* cube for the subsequent computations. In the earlier example, the only requirement before the beginning of calculations was to create a 3−dimensional array for the $\phi$−field, and later invoke it at various grid-locations. Similarly to this, we *too* will start with a density field $\rho(\mathbf{r})$ defined on a grid, but preparing a grid-defined density field *alone* will *not* sufficiently enable us to begin our calculations. In addition we shall need to *build* the entire information-network along the grid. It turns out that this can be wonderfully achieved with the help of `structures` for a variety of geometric quantities, such as a grid-vertex, a grid-edge, the faces of a cube etc. We shall shortly say more about the programming concept of a `structure`. Presently let us note that each of these structures will *physically* refer to a given grid-vertex, grid-edge or face of a cube. For example, at every vertex of the grid, we shall associate a structure, which will store more than one information attributes[¶]. All these structures not only need to be properly initialised and assigned to the grid-vertices, but they should be properly connected to one another so that our original goal of retrieving the entire cube starting from a given vertex, can be achieved. It turns out that this *same* mechanism also provides us a means to efficiently march across the grid. Thus (1) creating a facility to march across the grid, (2) initialising and connecting structures at the vertices of the grid, and (3) storing the points of intersection of the grid-edges with the surface to be modelled, are the three *basic* processes which are interwoven and which can be accomplished together. A successful implementation of these steps provides us a grid on which we can readily perform our calculations. In this section, we are concerned with the question as to how to implement these processes in an integrated manner.

Before proceeding further, we explain here the meaning of two terms which will frequently appear in our discussion : a `structure` and a `linked list`. In conventional terms, a `structure` is like a bucket containing several items (members). These members carry useful attributes of the structure in question, and store important information. A `structure` can be treated like any other variable of any allowed data-type (like a variable of an integer or a real data-type). Thus, one could have arrays of structures and linked lists of them, through which one can move from one member (say, a structure referring to a given grid-vertex), to the next (which will refer to the next ver-

---

[¶]Compare this with $\phi(i, j, k)$, a single value which we would store in solving the earlier example.

tex in the grid) with the help of pointers [∥],[**]. A `linked list` is a useful data-handling
facility. Linked lists come in two varieties: singly linked lists and doubly linked lists. In
general, a linked list refers to any of the allowed data-types, and consists of pointers to
the variables of that data-type (in our applications, these are pointers to structures).
Every member structure of `a singly linked list` has a pointer pointing to the next
member in the list. In case of `doubly linked lists`, every member structure has a
link both to the next and to the previous member in the list. Due to this property,
a doubly linked list is often called a *circular list*. A major difference between arrays
and linked lists must be evident from here: an array allows for a random access to
its members, whereas, any $n^{th}$ member of a linked list can be accessed only by step-
ping through the first $(n-1)$ members in the list. In general, geometric manipulations
involving a small number of members are best carried out using linked lists, chiefly
because (1) it is no longer needed to take care of the index of an array, and (2) a linked
list of say, six faces of a cube, couples well with an intuitive picture of the cube in a
geometrically more appealing manner whereas, handling an array of six faces of a cube
could be cumbersome. This fact will become clearer as we go along.

We now return to the topic of discussion, and show how the above facilities are
utilised in order to build up a grid. The chief aim is to store information so that it is
easily accessible during the course of surface-construction.

To illustrate the power and usage of the facility of structures, we show the structure
designed for the grid-vertex in the box below (see Box 2.1). This structure was specially
designed to develop a fully $3-$dimensional forward-moving linked list of grid-vertices.
The linked list itself was used as *a workhorse* in scanning the grid, from one vertex to
the next.

Notice in Box(2.1) that, in addition to the three coordinates $(i, j, k)$ of the vertex
and the value of the density field $\rho(i, j, k)$, the structure at $(i, j, k)$ also stores three

---

[∥]The facility of pointers is a unique feature of the C-language. Any variable in C can be pointed
to by a pointer of the same data-type. Thus, a pointer is just another variable situated in a specific
memory address. A special feature of using a pointer is that, the value of a variable can be accessed by
*dereferencing* the pointer pointing to that variable. The chief advantage offered by C is that this can
be done from any arbitrary function being called by the main program. Thus, the variable is indirectly
visible from everywhere; it can participate in calculations taking place in any arbitrary function, and
its value can be changed or updated from within any function by merely supplying the corresponding
pointer to that function. Due to this feature, C is termed as a *pass-by-reference* language. For a
detailed exposition of pointers see Kernighan & Ritchie (1988) and Balagurusamy (1992).

[**]Suppose, `V` is a structure of type `tsVertex` which has a real variable `x` as its member. In programs
using this structure, the value of `x` can be accessed in two ways: $x = $`V.x` and/or $x = $`vert->x`, where
`vert` is a pointer to `V`.

pointers of the same data-type as itself, i.e., `tsVertex`. These three pointers, respectively point to the structures at the 3 neighbouring vertices along the three principal axes, allowing us to retrieve information about the vertices along X, Y and Z-axis respectively from any given locality. For example, if the pointer `V` refers to a vertex $(i, j, k)$, then `V->xnext` is another pointer of type `tsVertex` which points to the next vertex along X-axis, i.e., $(i+1, j, k)$. The meaning of the 3−dimensional forward-moving linked list should now be clear : from every vertex, one can move a step forward along all the three axes, and also retrieve *all* the information pertaining to these vertices *at will*.

```
    /* ---------- Structure for the vertex of a grid ---------- */

typedef struct tVertexStructure tsVertex;   /* Type-declaration */

typedef tsVertex *tVertex;       /* tVertex is a pointer to the
                                    following structure */

struct tVertexStructure {
  float v[3];                    /* (i,j,k) coordinates of the
                                    vertex */

  double rho;                    /* Value of the density field at
                                    (i,j,k) */

  tVertex xnext, ynext, znext; /* Pointers to the three vertices,
                                    along X-axis, Y-axis and Z-axis */

  tEdge xedge,yedge,zedge;       /* Pointers to the three grid-edges
                                    which emanate from (i,j,k) along
                                    the three axes */
};
Memory allocation requirement = 72B
```

**Box 2.1: Structure for the vertex of a grid**

One of the special advantages of working with such a system of structures is that one can access the cube being triangulated *fully*. This is schematically depicted in

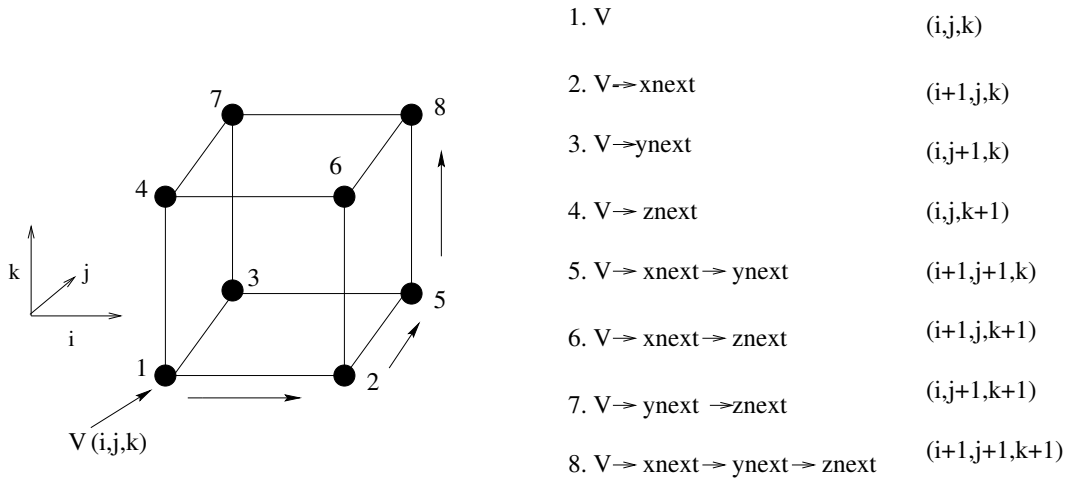

| | |
|---|---|
| 1. V | (i,j,k) |
| 2. V→xnext | (i+1,j,k) |
| 3. V→ynext | (i,j+1,k) |
| 4. V→znext | (i,j,k+1) |
| 5. V→xnext→ynext | (i+1,j+1,k) |
| 6. V→xnext→znext | (i+1,j,k+1) |
| 7. V→ynext→znext | (i,j+1,k+1) |
| 8. V→xnext→ynext→znext | (i+1,j+1,k+1) |

Figure 2.8: `V` is a pointer to a structure of data-type `tsVertex`. Here it is shown pointing to a grid-vertex with coordinates $(i, j, k)$. It suffices to add here that *all* the properties associated with vertex $(i, j, k)$ which are stored in the corresponding structure, can be accessed through pointer `V`. Remarkably, this also means that starting from the pointer `V`, we can completely recover the cube which is to be triangulated. This figure depicts the manner in which this can be achieved. Notice that *all* the vertices of the cube (actually, their corresponding structures) are pointed by pointers which are derivable from `V`. Please refer to the text for further clarifications.

Figure 2.8. Notice that every vertex of the cube can be reached by using a specific series of pointers emanating from `V`. For example, starting from vertex 1, if we want to arrive at vertex 8, we mentally carve out a path, $1 \rightarrow 2 \rightarrow 5 \rightarrow 8$ which is shown in the figure. In the code, this is equivalent to using the pointer `V→xnext→ynext→znext`. It is easy to convince one's self that this pointer indeed points to vertex 8. Thus, pointers allow us to successfully program the mental notion of hopping from one vertex of the cube to another.

The grid is *built up* by creating as many structures as there are vertices on the grid. Each structure is properly allocated the memory it requires (72B, see Box 2.1), and is assigned to a given vertex. Every such vertex can be accessed by a pointer to the corresponding structure. Further, from a structure referring to a given vertex, we pass pointers to relevant vertices and grid-edges as mentioned above. En route to building the grid, the grid-edges are properly initialised. The points of intersection of these grid-edges with the surface are obtained and are stored within their respective structures.

The task of building the grid is a major constituent of SURFGEN, and expands over 4 functions.

1. `Read_Vert_Init_Z.c` − A function which initialises all the vertices of the grid, and initialises all the grid-edges parallel to the Z−axis (alternatively the X or Y−axis).

   A successful run of this function ensures that all the vertices of the grid are initialised properly. From every vertex, a link is passed to each of the three neighbouring vertices along the three principal axes. Hence, the pointers to vertices can be used to march across the grid. The idea is to utilise this facility of scanning the grid, and initialise the rest of the grid-edges. This is achieved by calling the next three functions. The order of implementation of these functions is specific to our scheme, but grid-edges can also be initialised following a different route.

2. `Init_Y.c` − Initialises all the grid-edges parallel to Y−axis.

3. `Init_X_Y.c` − Initialises the grid-edges parallel to X-axis in the Z=0 plane.

4. `Init_X_Y_Z.c` − Initialises rest of the grid-edges parallel to X-axis.

We show below the structure of a grid-edge in Box 2.2. In the course of initialisation, the members of this structure are given relevant values.

Notice that the two end points of the grid-edge can be accessed through relevant pointers. In addition, all the grid-edges are put on a doubly linked circular list. The point of intersection of the grid-edge is stored in a structure of type `tsPtOfIntersect`[††].

In summary, the functions and concepts discussed in this section lead us to build the grid with all the necessary information stored. Further, our method of implementation utilises structures and linked lists which can help us use this information as desired. Presented next is the method used to triangulate an arbitrary cube. The power and beauty of the formalism discussed thus far, will be revealed in its full glory in the course of triangulating a cube.

## 2.2.2 Triangulation of cubes

Let us begin by summarising the three major tasks which we discussed in the last subsection.

---

[††]For illustrative purposes, consider `edge` to be a pointer to a structure `e` of type `tsEdge`. Suppose `e` is intersected by the surface and the point of intersection is $P(x, y, z)$. $P$ can be retrieved as follows:

$$x = \texttt{(edge -> pt).coord[0]};$$
$$y = \texttt{(edge -> pt).coord[1]};$$
$$z = \texttt{(edge -> pt).coord[2]};$$

```
  /* --------------- Structure for a grid-edge --------------- */

typedef struct tEdgeStructure tsEdge; /* Type-declaration */

typedef tsEdge *tEdge;  /* tEdge is a pointer to the following
                           structure */

struct tEdgeStructure {
  tVertex endpt[2];       /* Pointers to the two end-points */

  tsPtOfIntersect pt;    /* Member-structure containing point of
                            intersection of the grid-edge and the
                            surface */

  tEdge next, prev;      /* Pointer each, to the next and previous
                            grid-edge in a doubly lined circular list
                            of grid-edges */

  int status;            /* status = 1 iff the grid-edge is
                            intersected */
};
Memory allocation requirement = 64B

  /* --------- Structure for the point of intersection --------- */

struct tPtOfIntersectStructure {
  double coord[3];
};
Memory allocation requirement = 24B
```

Box 2.2: Structures for a grid-edge and its point of intersection with
the isodensity surface, respectively.

1. By introducing appropriate structures for the grid-vertices and grid-edges, we could build up the information network of a density grid in a fully recoverable fashion.

2. The grid can now be scanned in such a way that at every vertex of the grid, one could stop and if necessary, triangulate the corresponding cube.

3. The structures of the grid-vertex and grid-edge permit us to recover and handle the cube associated with a given vertex completely (see Figure 2.8).

In this subsection, we shall introduce the formalism which enables us to triangulate the cubes. It suffices to add here that this formalism allows us to triangulate *all* the 256 configurations of a *surface cube* which might occur in any random order in the course of scanning the grid [‡‡]. To this end, we shall make use of the already developed grid and its constituent structures and linked lists.

A cube to be triangulated can be viewed in many ways: it can be considered to be a collection of 8 vertices or of 12 grid-edges or as a cube of 6 faces. For reasons to be clarified below, in our formalism we prefer to visualise it in terms of its 6 faces. We take every face of the cube to contain *a local copy* of the four grid-edges which compose that face. We shall call them `face_edges`. The `face_edges` differ from the grid-edges in that, once an intersected `face_edge` has been fully utilised in triangulation, we choose to *delete it locally* and prevent it from further consideration while triangulation of the cube continues. However, as was stressed earlier, the corresponding grid-edge will be used in triangulating a neighbouring cube. Hence, we choose not to delete it, and all the information about it is preserved throughout the course of triangulation. Depending on the density field and the supplied values of $\rho_{\mathrm{TH}}$, 2 or 4 (or none) of the `face_edges` of a given face may be intersected by the isodensity surface which we aim to determine and construct. If the *strength* of a face is zero, it does not participate in triangulation. A tag acquires zero value in this case, and this means that we do not use this face any further. All this information pertaining to a face of the cube is stored in a structure, an example of which is shown below (Box 2.3). Notice that the structure has been designed so that a doubly linked, circular list of faces can be constructed. The facility of invoking the `next` and `prev` faces is introduced for this purpose. For convenience in the calculation of volume and integrated mean curvature, `face_edges` are arranged anticlockwise within the structure for a face.

---

[‡‡]Looking at the complexity and enormity of this job, our ability to achieve this goal is a reflection of the powerful programming techniques permitted by C in the domain of Computational Geometry.

We prefer to call this as the *strength of the face*. See the structure of the face.

For the sake of completeness, we also show the structure of a `face_edge` (Box 2.4). Notice that it contains a pointer to the grid-edge which it physically corresponds to. Moreover, the structure offers the option of *switching off* the `face_edge`, once it has been utilised completely. We shall elaborate on this concept when illustrating the process of triangulation.

```
 /* ------------- Structure for the face of a cube ------------- */

typedef struct tFaceStructure tsFace; /* Type-declaration */

typedef tsFace *tFace; /* tFace is a pointer to the following
                          structure */

struct tFaceStructure {
   tFaceEdge edge[4];    /* Pointers to the local copy of the four
                            edges */

   tFace next, prev;     /* Pointer to the next and previous
                            face in the list */

   int strength;         /* Counts the number of ON edges in a
                            face */

   int status;           /* ACTIVE iff strength > 0.  In that case,
                             strength can be 2 or 4 */
};
Memory allocation requirement = 56B
```

**Box 2.3: Structure for the face of a cube**

Having presented in detail, the design of structures of `faces` and `face_edges`, now we explain as to how do we *represent* a given cube in terms of its constituent faces. To illustrate the concept, we turn to Figure 2.9. The left part of this figure shows a pair of cubes; the first depicting our standard of labelling the vertices, and the second showing how we label various `face_edges`. For the sake of clarity, below the pair of cubes we explicitly specify as to which edges connect any two given vertices of the cube. To the right of the cubes, we show a table with six rows; each one referring to one face of the cube. Further, every row has 4 members, which correspond to the four `face_edges` bordering that face. In our scheme, we will complement the above table with a subtle aspect of the cubic symmetry: a cube perceived from any axis, or from any vertex looks the same ! In other words, no face is special; all the faces carry equal weightage. At this point, let us recall that *each* of the 14 independent density configurations of a *surface cube*, refers to a rotation group. For example, two cubes — each with only one overdense vertex, but situated differently on the cube — are members of a rotation group, since one can be obtained from the other cube through rotation. Therefore, both cubes can be triangulated following the same scheme.

```
        /* -------- Structure for the face_edge --------- */
        /* - It serves as a local copy of the grid-edge - */

typedef struct tFaceEdgeStructure tsFaceEdge;

tFaceEdge; /* tFaceEdge points to the following
                             structure */

struct tFaceEdgeStructure{
   tEdge e;                       /* Pointer to a grid-edge */

   int face_edge_status;          /* 1 if intersected and yet
                                     in use */
                                  /* Assigned 0 after being used
                                     fully */
};
Memory allocation requirement = 16B
```

**Box 2.4: Structure for a face_edge**

Figure 2.10 illustrates this concept. To achieve *full generality*, a scheme of implementing triangulation should utilise this rotational symmetry. In other words, one should be able to freely rotate a cube and construct triangle(s) enclosing the overdense vertices situated in any given order. Precisely such a mechanism is provided by a doubly linked *circular* list of faces. A doubly linked list allows us to rotate the cube and visit its different corners. This list (like all the lists) consists of a *head* of the list, pointed to by a specific pointer, and 5 other pointers referring to the remaining faces of the cube. All the search operations begin from the *head of the list*. The *head*, pointing at a time to a specific face of the cube, can be freely translated along the list, i.e., from one face of the cube to the other. Further, the structure of the `face` has been designed so that, from any given face one can move to the next and the previous face of the list. Thus, this procedure allows us to neatly translate the symmetry of a cube in 3-dimensions to a hexagonal symmetry on a 2-dimensional plane, with every side of the hexagon representing one of the faces of the cube. This aspect is depicted in the bottom-left part of Figure 2.9. Note that the circularity property allows us to visit the faces repeatedly, just as desired − an aspect which we shall be using tremendously to our benefit [†]. It is now possible to appreciate the elegance of our approach which consists in employing an hexagonal, *facial* representation of a cube: *all* the 256 configurations of a cube can be conveniently expressed in such a tabular form. Hence, the process of triangulation is founded on the efficient handling of the circular list of faces, which reflect the rotational symmetry between *surface cubes* corresponding to the same density configuration (see Figure 2.10).

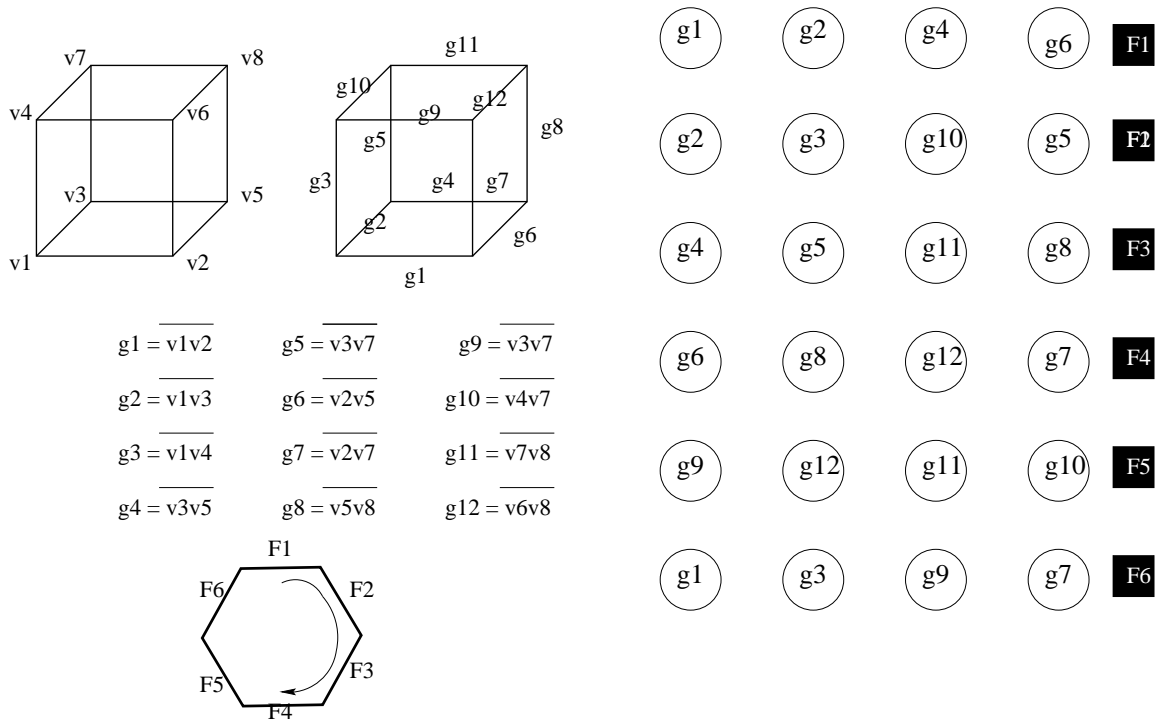

Figure 2.9: This figure contains several details. The left cube shows our convention of labelling the vertices of a cube. The right cube labels various grid-edges. To the right is a *tabular form* of the cube, where every row refers to one of the faces of the cube. The four elements of the row refer to the four constituent grid-edges composing a given face of the cube which arranged anticlockwise, when perceived from outside the cube. The faces are labelled in the same sequence from F1 to F6. On the bottom-left part of the figure, a hexagon is shown, which symbolically depicts the *doubly linked circular list* of the 6 faces. The circularity property of the linked list allows us to visit the six faces repeatedly, as per our need.

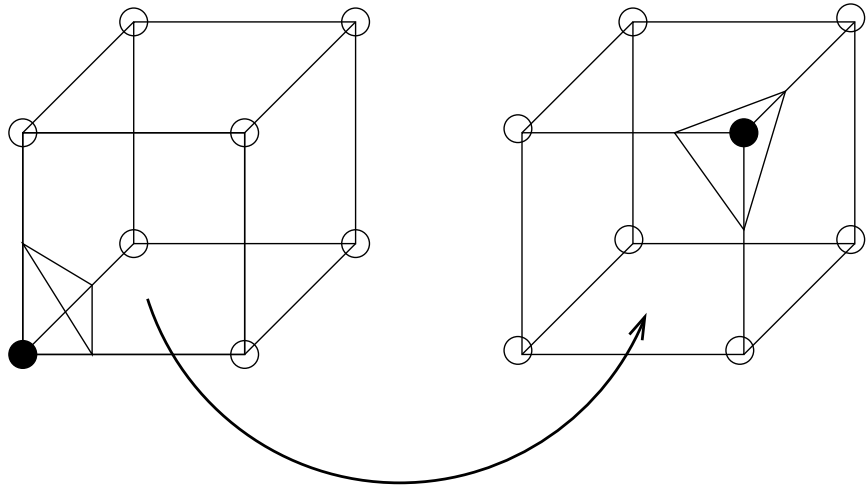

Figure 2.10: Two surface cubes, each with a single overdense vertex, are shown in the above figure. The cube to the right can be obtained from that to the left through rotation. Thus, these are members of a rotation group, and they can be triangulated using the *same* scheme. The triangulation is shown in the above figure. Our formalism achieves generality by utilising this inherent rotational symmetry, and eventually enables us to triangulate *all* the configurations of the cube.

Having established how we tackle a cube, let us now note some facts regard-ing a triangle. A triangle, which will be our primary unit in defining an isodensity surface, consists of three vertices and it has three edges at its boundary, which we call as `triangle-edges` or `t-edges`. Every `t-edge` connects a pair of points on two `face_edges`. Each of these points corresponds to the density threshold $\rho_{\mathrm{TH}}$ (see Figure 2.2). These two `face_edges` could either belong to a *single* face, or they may belong to two different faces. Thus, the process of constructing a triangle requires the setting up of a rule which helps us select three pairs of `face_edges`, each composing one `t-edge`. The rule should be general enough to handle all the configurations of a cube, and as we shall demonstrate below, the hexagonal linked list of faces helps us in achieving this. In what follows, we shall illustrate this rule by way of working out an example.

---

[†]Once again, this provides an instance of the power of using C in programming SURFGEN. The language provides such tools and machinery that, the programmer can put into practise all that (s)he visualises in one's imagination!

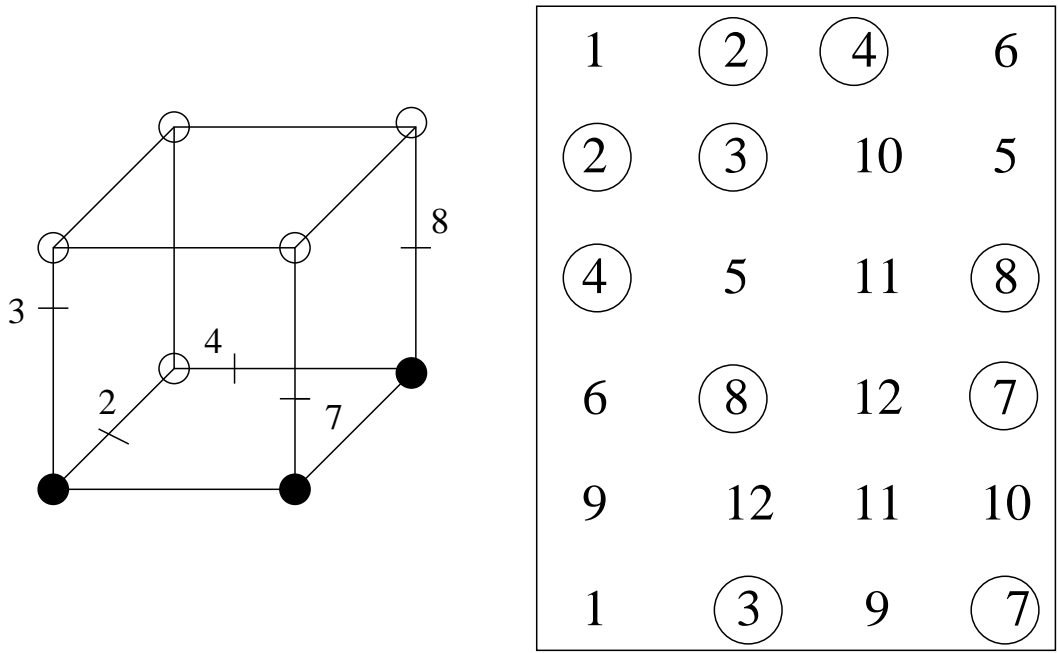

Figure 2.11: The density configuration of the sixth cube of Figure 2.2 has been reproduced here. Illustrated to the right is the tabular representation of this cube, where intersected edges have been marked by a circle. For convenience, the preceding $g$ has been omitted in representing the grid-edges and they are only marked by their respective numbers. Every *surface cube* can be represented in such a tabular form. As mentioned in the text, the tabular form is cast into a doubly linked, circular list of the 6 cube-faces, which can be conveniently handled in triangulating the cube.

Consider the sixth cube of Figure 2.2. The density configuration of this cube is reproduced in Figure 2.11. The grid-edges being intersected by the surface are marked accordingly. The right side of the figure shows a tabular representation of the linked list of faces, where the intersected `face_edges` have been marked with a circle [‡]. For convenience the preceding $g$ has been omitted and in stead, the grid-edges are now simply marked by their respective numbers.

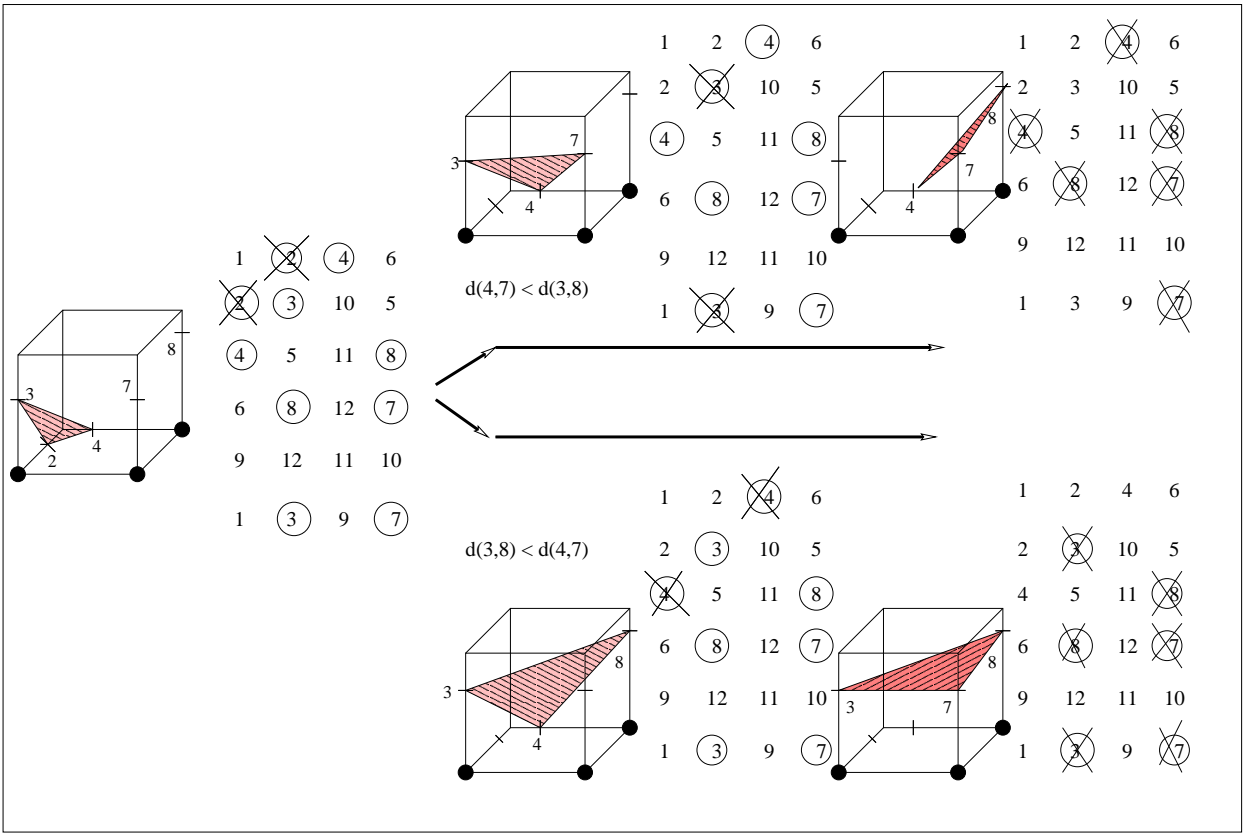

Figure 2.12: The process of triangulation is demonstrated for the cube shown in Figure 2.11. The scheme of implementation is discussed in the text. It suffices to add here that triangulation is done in a fully automated manner, and the algorithm SURFGEN proceeds to *chop off* the *surface cube* so that after completing the triangulation procedure, the cube resembles the sixth cube of Figure 2.2.

The triangulation of the cube in Figure 2.11 involves a series of search operations of linked list of faces. All search operations along the list are initiated from the *head* of the list. Hence, it is important to keep track of the *head*, and point it to the appropriate face of the cube. In Figure 2.11 the first face in the table is *active*, i.e., it has 2 intersected `face_edges` ($g_2$ and $g_4$). Hence, this face is a natural choice for being *head* of the list, and the search operations should commense from there. In other cases, *the*

---

[‡]Notice that each `face_edge` appearing twice in the table, is shared by two neighbouring faces. In future, we shall need to perform a search for a given `face_edge`. The *in-built* hexagonal symmetry in the system ensures that starting from a face containing a given `face_edge`, the same `face_edge` would be found by stepping two steps forward or backward along the linked list.

*first active face* in the table for a cube should be made the *head* of the list.

Coming back to Figure 2.11, notice that the strength of the head of the list is 2, i.e., two of the `face_edges` of F1 are active. These compose one `t-edge` of the triangle. To construct a triangle, we need another *active* `face_edge`. To this end, we now step forward along the linked list, and stop at the *first* face, which shares one of the 2 `face_edges` with the head of the list. Once obtained, we specialise to that face, and start to search this face for the first *active* `face_edge` other than the shared one. In our case, F2 shares the active `face_edge` labelled by 2 with F1. Hence, we perform a search within F2 for another active `face_edge` starting from `face_edge` 2. Here, the very next `face_edge`, labelled by 3, is active. Hence, we consider the triplet 2−4−3 for triangulation [§], which results in construction of the first triangle as shown in the left-most cube of Figure 2.12.

After constructing any given triangle, one *must delete* the *used up* `face_edge` from the list. In the present case, the `face_edge` 2 is *used up*, since it is shared by faces F1 and F2. Hence, it is deleted, i.e., its `face_edge_status` is set to 0, and it is no longer considered active while searching through the linked list. This is pictorially depicted in the table in the left of Figure 2.12.

Before proceeding further, it should be stated that a face with less than 2 *ON* `face_edges` is *switched off* and is not considered further in any search operations concerning triangulation. Thus, after having constructed △2-4-3 [¶], faces F1 and F2 go out of our search-list.

As we saw above, after every triangle is constructed, the circular list of faces undergoes a change; one or several faces are turned off, and they go out of the search-list. If there still exist *ON* faces, the triangulation of the cube proceeds further. Our very next step then is to appoint the *first* active face in the list to be the head of the list. In the present case, face F3 should be appointed the head of the list. The rest of the procedure follows the same route, with the pair of `face_edges` 2 and 3, replaced by `face_edges` 3 and 4. To construct a triangle, we need to select a third `face_edge`. Once again the search is confined to faces which contain `face_edges` 3 and 4 respectively. Unlike earlier, the triangle to be constructed this time is *inside* the cube. Hence, invariably we are left with a choice as to which `face_edge` should be selected, e.g., we can select one of the two `face_edges` − the one labelled 7 (sharing face F6 with face_edge labelled 3)

---

[§]The three vertices of the triangle are stored in the structure of a triangle (see below) in an anticlockwise order, so that the normal to the triangle always points outward from the surface.

[¶]Or △2-3-4, if that order of storing the vertices makes the normal to the triangle point outward from the surface

or 8 (sharing face F3 with face edge labelled 4) − to create a triangle. To break this degeneracy, we evaluate distances $d(3,8)$ and $d(4,7)$. We select edge 8 if $d(3,8) < d(4,7)$, and construct △3-8-4, and in the alternate case, we select edge 7 and construct △3-4-7. In the former case, edge 4 is deleted, whereas in the latter case, edge 3 is deleted. Appropriate face in the circular list is *turned off*, and the search operation continues because 2 faces in the cube are *still active*. In the present case, it is trivial to see that the third triangle could either be △3-8-7 (if edge 4 is deleted), or △4-8-7 (if edge 3 is deleted). When all the triangles have been created, *all* the faces of the cube are *switched off*, and the cube needs no triangulation. At this point, we move to the next cube.

The above scheme of triangulation can be *readily* applied to 6 out of 14 density configurations of the cube. These 6 configurations are characterised by 2 or more number of triangles which connect with one another inside the cube, e.g., the ninth and twelveth cubes in Figure 2.2. Seven out of the remaining density configurations of the cube are characterised by the presence of one or more number of *isolated* triangles‖. The above method, having no in-built prescription as to where to create an isolated triangle, cannot readily triangulate such cubes. In order to proceed with the triangulation of these cubes, we had to develop a clever piece of code. The main task of this code is to create an isolated triangle, wherever necessary. If the cube is still *active* after the creation of this triangle (e.g., the tenth cube of Figure 2.2), it is submitted to the above scheme of triangulation. Notice that in this particular case, after the construction of the isolated triangle, the cube resembles the first cube in the second row of Figure 2.2. Thus it is triangulated following exactly the same procedure discussed earlier. In passing, we should reiterate that degeneracy as offered by the cube in Figure 2.3 is also handled by the routine constructing the *first* isolated triangle.

One particular cubic configuration stands out, and cannot be triangulated by any of the schemes discussed above. This cube is shown at the beginning of the third row of Figure 2.2 (cube # 11). This configuration may arise when we have a narrow, sub-resolution tunnel passing through the surface. Notice that the cube can be triangulated in two ways: according to the scheme which is depicted in Figure 2.2, the triangles *enclose* the overdense vertices on either side of the cube and isolate the two pairs of overdense vertices from each other. This scheme leads to the formation of a tunnel.

---

‖Figure 2.2 shows 2 possible configurations of a triangle: (A) Cubes 2, 4, 5, 7, 8, 10 and 13 contain isolated triangles. These triangles have *all* their edges located on the faces of the cube. They do not share a common boundary with any other triangle belonging to this particular cube. (B) Cubes which contain joined triangles. The remaining cubes in Figure 2.2 contain 2 or more triangles which are joined together at a common edge which is *fully localised* within the cube.

However, the same cube can be triangulated so that the triangles enclose the 2 pairs of *underdense* vertices (not shown). In this case a tunnel does not result. Clearly, this cube offers an instance of degeneracy and has to be handled *in tandem* with the density distribution surrounding that cube. Fortunately, this degeneracy can be broken by constructing the first pair of triangles in such a way which maintains continuity of the surface. For this purpose, the code uses information from neighbouring cubes which have already been triangulated. For instance, if the cube corresponding to $P(i, j, k)$ is being triangulated, it shares its faces $x = i$, $y = j$ and $z = k$ with the neighbouring cubes which have been already triangulated. Notice that all these faces are *active*, and hence it is clear that these must have contributed 1 or 2 `t-edges` to the triangles constructed in the neighbouring cubes. In the next subsection we shall see that, for various computational benefits, a doubly linked circular list of `t-edges` is created and maintained. *All* the `t-edges` (before it is time when they are no longer needed and therefore deleted) are kept in this list. In the present context, this list plays a *crucial* role. Let us take an example. For our cube in Figure 2.2, the *strength* of face F1 is 4, and hence it must have contributed *two* `t-edges` to the cube corresponding to vertex $(i, j, k - 1)$. Out of the 2 choices, in which combination might these have occurred? Specifically, could the two `t-edges` be couplets $\overline{g_1 g_2}$ and $\overline{g_4 g_6}$, or could it be $\overline{g_1 g_4}$ and $\overline{g_2 g_6}$ ? The linked list of `t-edges` helps us answer this. We perform a search of this list, and figure out as to with which grid-edge has $g_1$ coupled itself. The corresponding `t-edge`, i.e., either $\overline{g_1 g_2}$ or $\overline{g_1 g_4}$, has already got one triangle from the cube below, and to ensure continuity, it must participate *again* in construction of another triangle within *this* cube. This then, is the method which helps us lift the degeneracy. Once the correct combination of `t-edges` has been identified, it is straightforward to construct the first pair of triangles. Later the cube is handled by the procedure already discussed above, which constructs the remaining two triangles.

Finally, Figure 2.4 draws our attention to a pathological case which appears to be quite common in surface-construction. When two neighbouring cubes share a face where two pairs of diagonally opposite vertices are 1s and 0s, formation of a hole is inevitable. We don't worry about these holes while scanning the grid in the normal course of triangulation. We *fill up* the holes *after* the scanning is complete.

Above we described the formalism which allows us to triangulate *all* 256 configurations of a cube while taking care of the pathological cases. We stress here that the entire formalism is based on the representation of a cube in terms of a doubly linked circular list of faces.

Surface-construction is *one* part of a larger exercise. An equally important task is to calculate Minkowski Functionals for the surface being modelled. As far as possible, this should be carried out *online*, i.e., alongside the process of surface construction, so as to minimise memory requirements. In the next subsection we focus on this important aspect, and show that with the help of efficient design for the structures of triangles and triangle-edges, we are able to meet this goal.

### 2.2.3  Calculation of Minkowski Functionals

The large scale structure of the universe can be studied on various scales by considering the geometry and topology of excursion sets of the density contrast $\delta(\mathbf{x})$ defined as $\mathcal{E}_{TH}^{+} = \{\mathbf{x}|\delta(\mathbf{x}) \geq \delta_{TH}\}$, for overdense regions (clusters, superclusters) and $\mathcal{E}_{TH}^{-} = \{\mathbf{x}|\delta(\mathbf{x}) \leq \delta_{TH}\}$, for underdense regions (voids). By specifying a given density threshold one effectively defines an isodensity surface for which the Minkowski functionals should be evaluated. Depending upon whether the surface encloses an overdense region of the space or an underdense region, the surface refers to a cluster/supercluster or a void. In the last two subsections we described how the process of triangulation has been implemented in SURFGEN in order to reliably construct an isodensity contour for a supercluster or a void. In the present subsection, we shall describe the calculation of Minkowski Functionals (hereafter, MFs) on these isodensity contours using SURFGEN. The MFs have already been introduced in Chapter 1. For completeness, let us list them out again. The four MFs describing the morphological properties of an isodensity surface in three dimensions are

- [1]*Area S* of the surface,

- [2]*Volume V* enclosed by the surface,

- [3]*Integrated mean curvature C* of the surface (or integrated extrinsic curvature),

$$C = \frac{1}{2} \oint \left( \frac{1}{R_1} + \frac{1}{R_2} \right) dS, \qquad (2.2)$$

  where $R_1$ and $R_2$ are the principal radii of curvature at a given point on the surface.

- [4] *Integrated intrinsic (or Gaussian) curvature $\chi$* of the surface – also called the *Euler characteristic*

$$\chi = \frac{1}{2\pi} \oint \left( \frac{1}{R_1 R_2} \right) dS. \qquad (2.3)$$

A related measure of topology is the genus defined as $G = 1 - \chi/2$. Multiply connected surfaces have $G > 0$ while $G = 0$ for a simply connected surface such as a sphere. (The topological properties of all orientable surfaces are equivalent to those of a sphere with one or more 'handles'. Thus a torus is homeomorphic to a sphere with one handle, while a pretzel is homeomorphic to a sphere with two handles etc.) While the genus provides information about the connectivity of a surface, the remaining three MFs are sensitive to local surface deformations and hence characterize the geometry and shape of large scale structure at varying thresholds of the density (Sahni, Sathyaprakash & Shandarin 1998).

In nature one seldom comes across surfaces that are perfectly smooth and differentiable. As a result expressions [1] - [4] which make perfect sense for manifolds $\mathcal{C}^n$, $n \geq 2$ are woefully inadequate when it comes to determining the Minkowski functionals for real data sets which are grainy and quite often sparse. Below we describe how one can determine the Minkowski functionals for isodensity, triangulated surfaces derived from real data.

Let us assume a polyhedral surface to be an assembly of triangles in which every triangle shares its edges with each of its three neighboring triangles. We have already seen how SURFGEN constructs such surfaces from a density field defined on a cubic grid.

- The area of such a triangulated surface is

$$S = \sum_{i=1}^{N_T} S_i, \tag{2.4}$$

  where $S_i$ is the area of the $i-$th triangle and $N_T$ is the total number of triangles which compose a given surface.

- The volume enclosed by this polyhedral surface is the summed contribution from $N_T$ tetrahedra

$$\begin{aligned} V &= \sum_{i=1}^{N_T} V_i, \\ V_i &= \frac{1}{3} S_i (n_j \bar{P}^j)_i. \end{aligned} \tag{2.5}$$

Here $V_i$ is the volume of an individual tetrahedron whose base is a triangle on the surface. $(n_j \bar{P}^j)$ is the scalar product between the outward pointing normal $\hat{n}$ to this triangle and the mean position vector of the three triangle vertices, for which the $j^{th}$ component is given by

$$\bar{\mathbf{P}}^{\mathbf{j}} = \frac{1}{3}(P_1^j + P_2^j + P_3^j).$$ (2.6)

The subscript $i$ in (4.3) refers to the $i-$th tetrahedron, while the vectors $P_1, P_2, P_3$ in (2.6) define the location of each of three triangle vertices defining the base of the tetrahedron relative to an (arbitrarily chosen) origin (Fig. 2.13). The vertices are ordered anticlockwise. This ensures that the contribution to the volume from tetrahedra whose base triangles lie on *opposite sides* of the origin add, while the volumes of tetrahedra with base triangles lying on the *same side* of the origin subtract out. Thus for both possibilities we get the correct value for the enclosed volume (Fig. 2.13).

Here we digress briefly and discuss the implementation of the above 2 algorithms. This once again takes us to the heart of the surface reconstruction exercise, which involves storing and manipulating the constituent triangles of the surface.

We show in Box 2.5, the structure which has been designed to store necessary information pertaining to a triangle. Among other quantities which we shall comment upon below, the structure of a triangle contains the three vertices of the triangle, pointers to three of its edges (the `t-edges`) and 3 components of the normal to the triangle which would point radially outward from the surface. Whenever a new triangle is to be constructed (according to the procedure laid out in the last subsection), first we allocate sufficient memory to store a structure for that triangle, and next initialise *all* the fields of that structure. The area and volume are *local* quantities, and are dependent solely on the triangle in question. These are hence trivially calculated once the right combination of `grid-edges e1, e2 and e3` have been supplied so that their points of intersection with the surface to be constructed $P_1(x_1, y_1, z_1), P_2(x_2, y_2, z_2)$ and $P_3(x_3, y_3, z_3)$ are arranged in an anticlockwise order. Area $A_i$ of $i-$th triangle is computed from

$$\mathbf{A_i} = A_i \times \hat{n}_i = \begin{vmatrix} \hat{i} & \hat{j} & \hat{k} \\ x_2 - x_1 & y_2 - y_1 & z_2 - z_1 \\ x_3 - x_1 & y_3 - y_1 & z_3 - z_1 \end{vmatrix}.$$ (2.7)

As can be seen, this formula also evaluates the normal $\hat{n}_i$ to the triangle, which is used to determine the total volume of the surface.

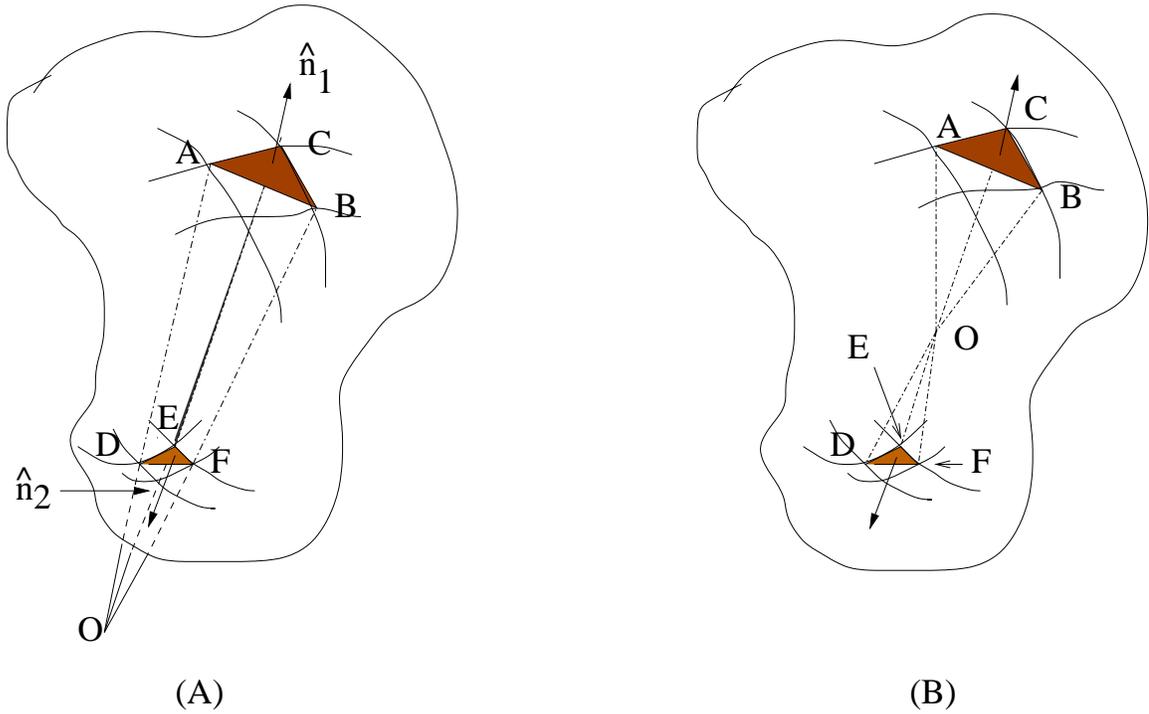

(A)                                                    (B)

Figure 2.13: Volume enclosed by the surface is estimated by *vectorially* summing over the volumes of individual tetrahedra; each having its base at one of the triangles composing the surface; and its apex at an arbitrarily chosen point which we call the origin. Note that the origin can lie both within as well as outside the surface. The left panel shows how SURFGEN estimates the volume when the origin lies *outside* the surface. In this case, the contribution (to the enclosed volume) from triangles falling in the same solid angle carries opposite sign for tetrahedra OABC and OFED. Clearly the volume of OFED must be subtracted from the volume OABC in order to give the true volume enclosed by the surface. Anticlockwise orientation of the vertices used in determining the normals of the base triangles helps us bring this about. The right panel shows the second possibility for which the two triangles ACB and DEF form tetrahedra whose fourth (common) vertex lies *within* the surface. In this case, the contributions from ODEF and OACB add to give the total volume.

```
    /* -------------- Structure for the triangle -------------- */

typedef struct tTriangStructure tsTriang; /* Type-declaration */

typedef tsTriang *tTriang; /* tTriang is a pointer to the following
                              structure */

struct tTriangStructure {
  double nx,ny,nz;           /* Components of the outward normal to the
                                triangle */

  float D1,D2,D3,D4;         /* Four determinants which define the
                                plane of the triangle */

  tsPtOfIntersect pt[3];     /* Three vertices of the triangle */

  tTriangEdge edge[3];       /* Pointers to the three triangle-edges */

  tTriang face[3];           /* Pointers to the three triangles */

  tTriang next,prev;         /* Pointers to in the linked list */
};
Memory allocation requirement = 176B
```

**Box 2.5: Structure for a triangle**

We have thus discussed the first three entries in the box for triangle-structure. The rest of the entries are useful in evaluating the contribution of the triangle to the total mean curvature, which will be discussed below.

- *The extrinsic curvature of a triangulated surface is localized in the triangle edges.* As a result the integrated mean curvature $C$ is determined by the formula

$$C = \frac{1}{2} \sum_{i,j} \ell_{ij}.\phi_{ij}.\epsilon, \tag{2.8}$$

where $\ell_{ij}$ is the edge common to adjacent triangles $i$ and $j$ and $\phi_{ij}$ is the angle between the normals to these triangles (see Figure 2.14).

$$\cos\phi_{ij} = \hat{n}_i.\hat{n}_j. \tag{2.9}$$

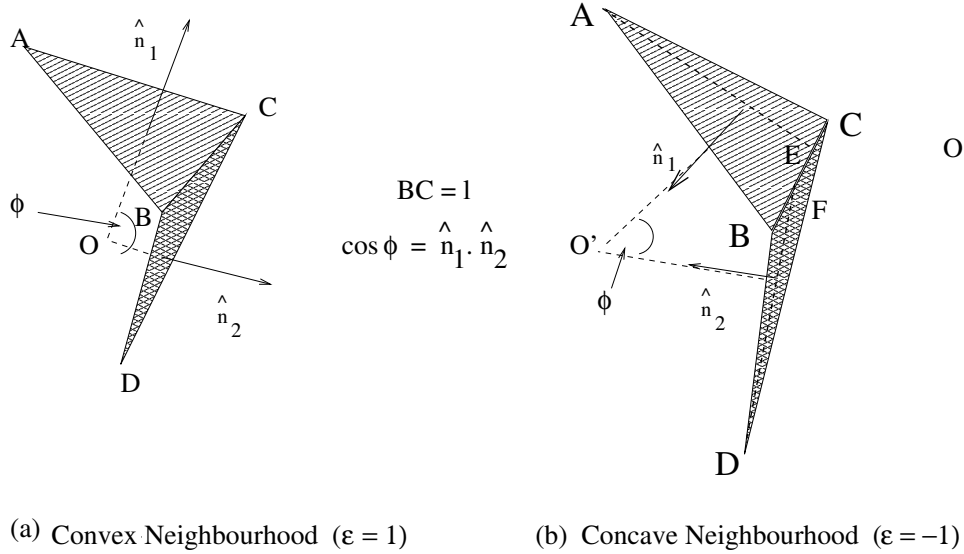

(a) Convex Neighbourhood  ($\epsilon = 1$)            (b) Concave Neighbourhood  ($\epsilon = -1$)

Figure 2.14: Illustrated here is the algorithm to compute the local contribution to the Integrated Mean Curvature: (a) $\triangle$ABC and $\triangle$CBD are two triangles sharing a side BC of length $\ell$. The normals $\hat{n}_1$ and $\hat{n}_2$ to the two triangles lie in the planes orthogonal to BC. Their projection in the plane of the paper diverges from a point O *inside* the surface. Thus the triangles correspond to a convex neighborhood and $\epsilon = 1$ in Eq.(2.8). The angle between the two triangles is denoted as $\phi$. (b) Shown here are a similar pair of triangles, but with inverted sense of the normals, so that the sense of the surface is opposite to that shown in (a). We note that in the present case, the two normals converge *outside* the surface, so that their projections in the plane of the paper would meet at O', which lies outside the surface. Thus this represents a concave neighborhood, for which $\epsilon = -1$ in Eq.(2.8). The angle $\phi$ represents the internal angle between $\hat{n}_1$ and $\hat{n}_2$.

The summation in (2.8) is carried out over all pairs of adjacent triangles. It should be noted that for a completely general surface, the extrinsic curvature can be positive at some (convex) points and negative at some other (concave) points on the surface. To accommodate this fact one associates a number $\epsilon = \pm 1$ with every triangle edge in (2.8). $\epsilon = 1$ if the normals on adjacent triangles diverge away from the surface, indicating a locally convex surface, while $\epsilon = -1$ if the normals converge towards each other outside the surface, which is indicative of a concave surface. In the former case the centre of curvature of the surface lies within the surface-body whereas in the latter case the centre of curvature lies outside of the surface-body (Fig. 2.14).

Unlike the previous two quantities − area and volume − where the object of interest was the triangle *itself*, the integrated mean curvature is to be evaluated visiting *every* triangle-edge. The curvature in the surface is mainly a cumulative effect of nonzero angle between the triangles sharing a given triangle-edge.

Notice that every triangle has three edges, and each one of them is shared by 2 triangles. Curvature-contribution from a given triangle-edge cannot be evaluated until *both* the triangles sharing that edge have been created. Furthermore, the information about the normal to the triangle is needed while computing the contribution to the curvature from *all* three edges of that triangle. However, storing every single triangle requires memory allocation = 176B (the memory-budget of each triangle-edge is 56B). Thus, a triangle requires the maximum space for storage. It is advisable, therefore, to free the memory allocated to a triangle once *all* its triangle-edges have been utilised for the calculation of mean curvature. In addition, every triangle-edge (regardless of whether the 2 triangles *bordering* it have been retained in the memory or deleted) can *always* be freed once it has been used for the calculation of curvature.

The structures for a triangle and a triangle-edge have been designed to take care of *all* the above requirements. While constructing any given triangle, a link is passed to all three of its edges. In turn, a link to this triangle is given from *each* of these edges, and the field `status`** in the structure for these triangle-edges is incremented by 1. For any triangle-edge, the `status` reaching the value of 2 signifies that the triangle *just* constructed is the second triangle which that triangle-edge required to calculate mean curvature using Eq.(2.8). While it is straightforward to compute $\ell_{ij}$ and $\phi_{ij}$, we need to be careful in choosing an algorithm to detect the concavity (convexity) of the local neighbourhood surrounding triangles $i$ and $j$. This in turn demands that we fix the value of $\epsilon$. We follow the following prescription for this.

Given a triangle with its vertices A$(x_1, y_1, z_1)$, B$(x_2, y_2, z_2)$ and C$(x_3, y_3, z_3)$ arranged in an anticlockwise order, Eq.(2.7) sets the direction of the normal to the triangle. Let us define four determinants as follows:

$$D_1 = \begin{vmatrix} 1 & y_1 & z_1 \\ 1 & y_2 & z_2 \\ 1 & y_3 & z_3 \end{vmatrix} ; D_2 = \begin{vmatrix} x_1 & 1 & z_1 \\ x_2 & 1 & z_2 \\ x_3 & 1 & z_3 \end{vmatrix} ; D_3 = \begin{vmatrix} x_1 & y_1 & 1 \\ x_2 & y_2 & 1 \\ x_3 & y_3 & 1 \end{vmatrix} ; D_4 = \begin{vmatrix} x_1 & y_1 & z_1 \\ x_2 & y_2 & z_2 \\ x_3 & y_3 & z_3 \end{vmatrix} .$$

$$(2.10)$$

---

**See the structure for the `triangle-edge`.

```
 /* -------------- Structure for the triangle-edge -------------- */

typedef struct tTriangEdgeStructure tsTriangEdge;

typedef tsTriangEdge *tTriangEdge;   /* tTriang is a pointer to the
                                        following structure */

struct tTriangEdgeStructure {
  tEdge edge[2];        /* Pointers to the pair of grid-edges which make
                           up this triangle-edge */

  int status;           /* How many triangles share this particular
                           triangle-edge?  During the running of the
                           code, this counter is either 1 or 2 */

  tTriang face[2];      /* Pointers to the two triangles which share
                           this triangle-edge */

  tTriangEdge next, prev;
};
Memory allocation requirement = 56B
```

**Box 2.6: Structure for a triangle-edge**

For $(x, y, z)$ to be an arbitrary point, let us further define an identity

$$I = D_1.x + D_2.y + D_3.z + D_4. \tag{2.11}$$

If $I < 0$, $(x, y, z)$ is towards the side of the normal (and hence, outside of the surface-body). On the other hand, if $I > 0$, the point is on the opposite side to the normal (and hence, inside the surface).

We use the above condition to find out whether a given locality is convex or concave. We consider the normals $\hat{n}_i$ and $\hat{n}_j$ with their respective base at $C_i$ and $C_j$, the centres of the two triangles $i$ and $j$. The points $P_i$ and $P_j$, which we define to be the heads of these two normals, are now subject to the above test. We check whether $P_i$ is outside the surface section defined by triangle $j$, and similarly for $P_j$ and triangle $i$. If $I_{i,P_j} < 0$ as well as $I_{j,P_i} < 0$, we conclude that the surface in the vicinity of triangles $i$ and $j$ is convex ($\epsilon$=1). If $I_{i,P_j} > 0$ and $I_{j,P_i} > 0$, we conclude that the surface in the vicinity of triangles $i$ and $j$ is concave ($\epsilon = -1$). The four determinants $\mathtt{D}_{1,2,3,4}$ are stored for *every* triangle at the time of creating it (see Box 2.5).

- The genus of a triangulated closed polyhedral surface is given by the convenient expression

$$\begin{aligned} G &= 1 - \frac{\chi}{2}, \\ \chi &= N_T - N_E + N_V, \end{aligned} \tag{2.12}$$

where $\chi$ is the Euler characteristic of the triangulated surface. $N_T, N_E, N_V$ are, respectively, the total number of triangles, triangle-edges, and triangle-vertices defining the surface. Of these, $N_V$ is the same as the number of points at which the surface intersects the grid. Thus, this number is available *even before* the triangulation begins. Every time a triangle-edge is deleted (the corresponding memory is freed), $N_E$ is incremented by 1. Similarly, every time a triangle is deleted, $N_T$ is increased by 1. Thus, we can estimate genus of the triangulated surface once it is fully constructed.

To summarise, in order to perform an online calculation of the MFs, our surface modelling scheme adheres to the following requirements:

- The vertices of all triangles are stored in an anticlockwise order. This enforces a uniform prescription on normals which always point outward from the surface being modelled. Information regarding the directionality of normals is of great importance since it is used both for calculating the volume as well as the mean curvature.

- For online calculation of the mean curvature, information concerning a given triangle must be supplemented with information concerning the triplet of triangles which are its neighbors. This allows us to unambiguously determine the contribution to the mean curvature from a local neighborhood using Eq.2.8.

- The total number of triangles, triangle-edges and triangle-vertices of a triangulated surface are counted which enable us to determine its genus using Eq.2.12.

Our surface modelling code therefore incorporates the triangulation of the entire set of 256 possible density-configurations of a cube within a *single* scheme and makes possible online computation of the area, the volume, the integrated mean curvature and the genus of a triangulated surface. The code has been tested on a variety of standard density distributions and eikonal morphologies. In the next section, we present results based on this analysis.

## 2.3   Results for Standard Eikonal Surfaces

In order to test the accuracy of our ansatz we generate triangulated surfaces whose counterpart continuum surfaces have known (analytically calculable) Minkowski functionals.

### 2.3.1   Spherical structures

In the first exercise, we consider spherically symmetric density distributions and generate surfaces of constant density for a variety of density-thresholds starting from a chosen maximum radius down to grid-size. A convenient density distribution for this exercise is

$$\rho(i,j,k) = \begin{cases} \frac{\rho_0}{R}, & (i,j,k) \neq (i_0, j_0, k_0), \\ \rho_0, & (i,j,k) = (i_0, j_0, k_0), \end{cases} \tag{2.13}$$

where R is the distance between $(i,j,k)$ and the centre.

$$R = \sqrt{(i-i_0)^2 + (j-j_0)^2 + (k-k_0)^2}. \tag{2.14}$$

Since the density field falls off as the inverse of the distance from the centre at $(i_0, j_0, k_0)$, any threshold $\rho_{TH}$ is associated with a sphere of radius $R = \frac{\rho_0}{\rho_{TH}}$. Thus larger spheres correspond to surfaces of lower constant density in this model. We assume $i_0 = j_0 = k_0 = 15$, for the centre of the sphere in our numerical calculations.

Applying the Shapefinders (see Section 1.4) to a sphere of radius $R$ one finds the simple result $\mathcal{T} = \mathcal{B} = \mathcal{L} = R$ and $\mathcal{F} = \mathcal{P} = 0$. Figure 2.15 compares the area, volume, mean curvature and Shapefinders measured for a triangulated sphere against known analytical values for these quantities. This figure clearly demonstrates that exact values, and values computed using triangulation, match exceedingly well down to the very lowest scale.

We should mention that the Genus evaluated using Eq.2.12 is identically zero for spheres of all radii and for all possible deformations of an ellipsoid (to be discussed next). This provides an excellent independent endorsement of our methodology by demonstrating that our triangulations of a sphere and ellipsoid indeed result in closed, continuous surfaces.

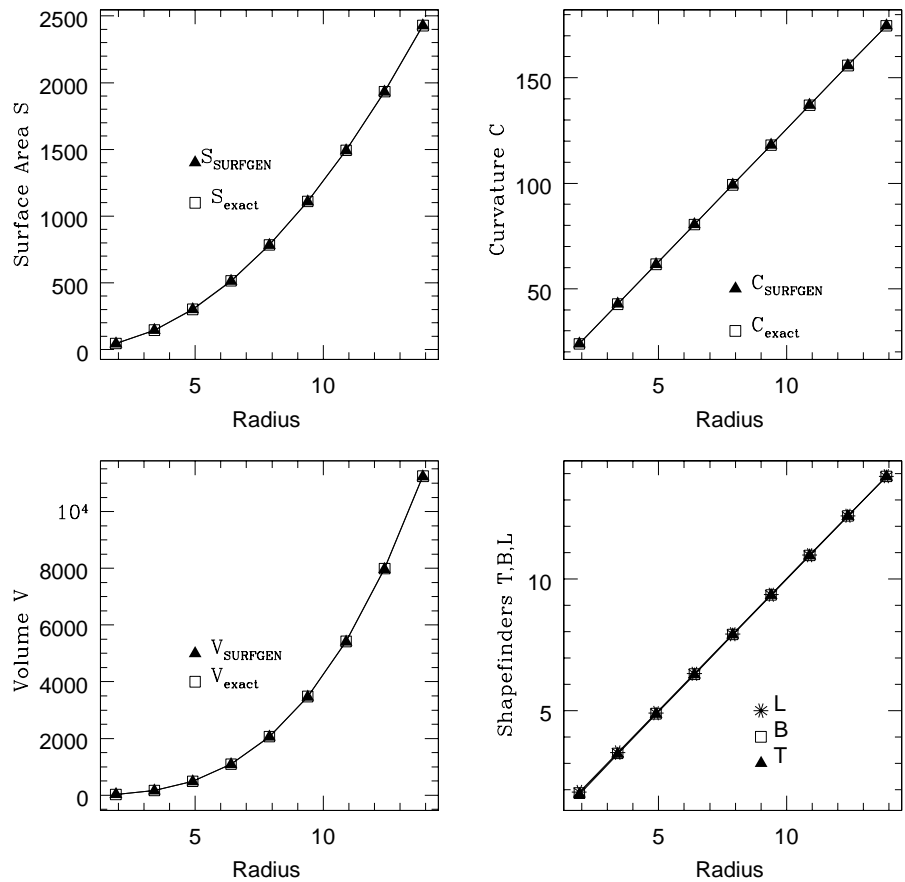

Figure 2.15: Minkowski Functionals and dimensional Shapefinders for a sphere. Note that the results for the Minkowski Functionals evaluated using our surface modelling scheme (SURFGEN) virtually coincide with the exact values.

## 2.3.2   Triaxial ellipsoid

A triaxial ellipsoid is an excellent shape with which to test a morphological statistic. This is because, depending upon the relative scales of the three axes, a triaxial ellipsoid can be oblate, prolate or spherical. We saw previously that MFs and Shapefinders give extremely accurate results for spherical surfaces. We now demonstrate that this remains true even for surfaces which are highly planar or filamentary.

The parametric form of an ellipsoid having axes $a$, $b$, $c$ and volume $V = \frac{4\pi}{3}abc$ is (Sahni et al. 1998)

$$\mathbf{r} = a(\sin\theta\cos\phi)\hat{x} + b(\sin\theta\sin\phi)\hat{y} + c(\cos\theta)\hat{z} \qquad (2.15)$$

where $0 \leq \phi \leq 2\pi$, $0 \leq \theta \leq \pi$. For the purposes of our study, we systematically deform a triaxial ellipsoid and study how its shape evolves in the process. When shrunk along a single axis a triaxial ellipsoid becomes planar. Simultaneous shrinking along a second axis makes it cigar-like or filamentary.

### 2.3.2.1   Oblate spheroids

We first study the accuracy of our triangulation scheme for planar configurations by considering oblate deformations of an ellipsoid. In this case two axes $a$ and $b$ are held fixed ($a \sim b$), while the third axis $c$ is slowly shrunk leading to an increasingly planar surface. Our results for this case are compiled in three figures. Figure 2.16 shows MFs as they evolve with the dimensionless variable $c/a$. (The exact results for the Minkowski functionals which we quote are based on the analytical expressions for MFs given in Sahni et al.(1998),we refer the reader to that paper for more details.) Figure 2.16 clearly demonstrates that the values of MFs obtained using triangulation match the exact values to a remarkable degree of accuracy. Indeed, for a wide range in $c/a$ the two distinctly different calculational algorithms give virtually indistinguishable results, thereby indicating that the triangulation ansatz is, for all practical purposes, exact for the oblate spheroids!

Figure 2.17 shows the evolution of the dimensional Shapefinders $\mathcal{T}, \mathcal{B}, \mathcal{L}$ together with the percentage errors in their estimation. We find that $\mathcal{L}$ is estimated to greatest accuracy with maximum error of $\sim 0.4\%$, while $\mathcal{T}$ and $\mathcal{B}$ can be determined to an accuracy of $\sim 0.8\%$.

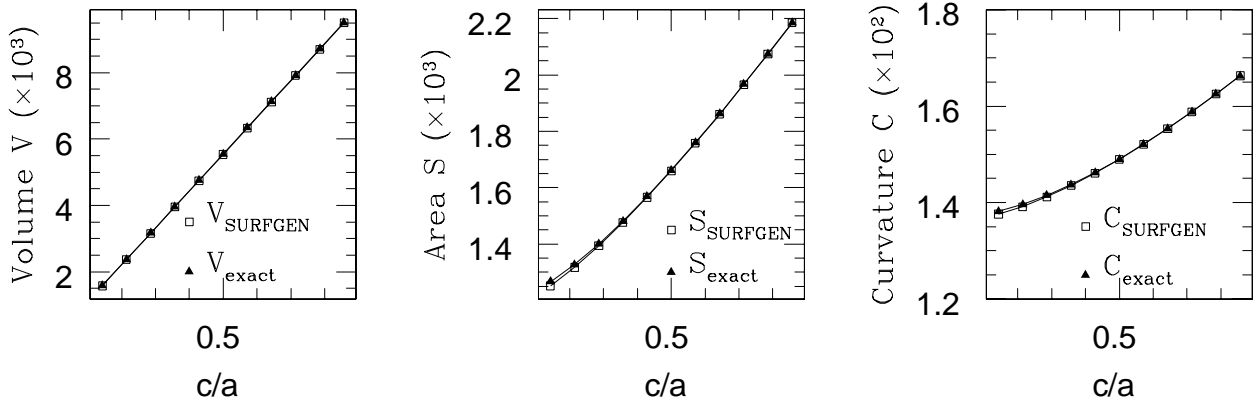

Figure 2.16: Values of Minkowski functionals determined by the surface modelling scheme SURFGEN are shown along with their exact values for oblate deformations of an ellipsoid. Note that for $c/a > 0.2$, values for Minkowski Functionals evaluated using SURFGEN virtually coincide with the exact results.

Figure 2.17 also shows the Planarity and Filamentarity for our triangulated surface alongside exact analytic values. As expected, Planarity grows from an initially low value $\sim 0.0$ to a large final value $\sim 0.4$ as the ellipsoid becomes increasingly oblate. Filamentarity, on the other hand remains small at $\sim 0.09$. Both Filamentarity and Planarity are determined to great accuracy by SURFGEN.

### 2.3.2.2  Prolate spheroids

Next we study prolate deformations of our ellipsoid. We start with $b \simeq a$, $c \ll a$, and shrink the second axis $b$ while keeping $a$ and $c$ fixed, so that finally $c \sim b \ll a$ and our initially oblate ellipsoid becomes prolate. Our results are again summarized in three figures.

Figure 2.18 reveals very good agreement between measured and true values of MFs, with the former tending to be slightly smaller than the latter.

Turning to the Shapefinders we find that, with the possible exception of extremely prolate figures, the Shapefinders are remarkably well determined. Indeed, even the extremely prolate ellipsoid with axis ratio $b/a < 0.2$ has a largest error in $\mathcal{T}$ of only $\sim 1.7\%$ while errors in $\mathcal{B}$ and $\mathcal{L}$ never exceed $\sim 1\%$ (Fig. 2.19).

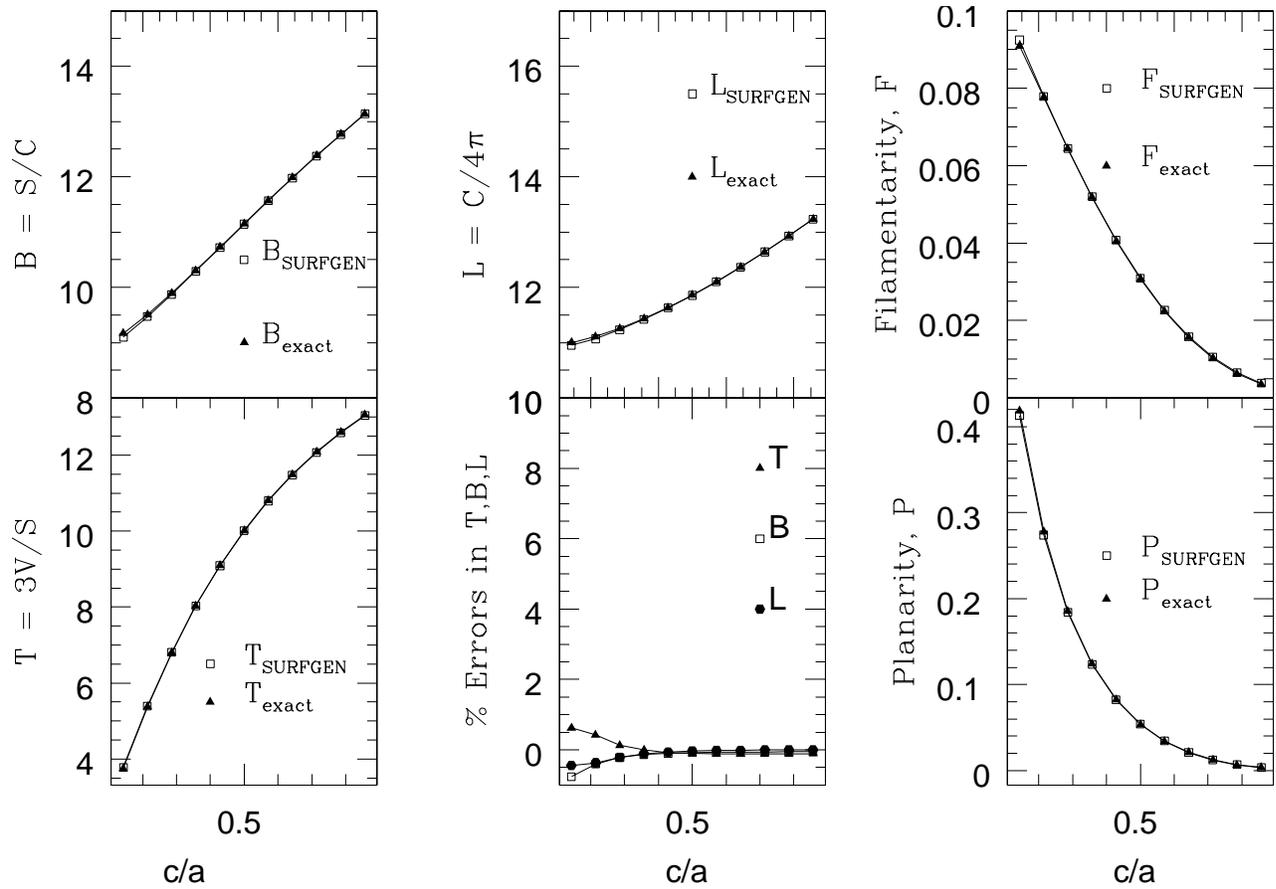

Figure 2.17:  Values of the Shapefinders derived using the surface modelling scheme SURFGEN are shown along with their exact values for oblate deformations of an ellipsoid.  The lower middle panel shows percentage errors in the estimation of the dimensional Shapefinders. We note that the errors are all in the range of $\pm 0.9\%$.

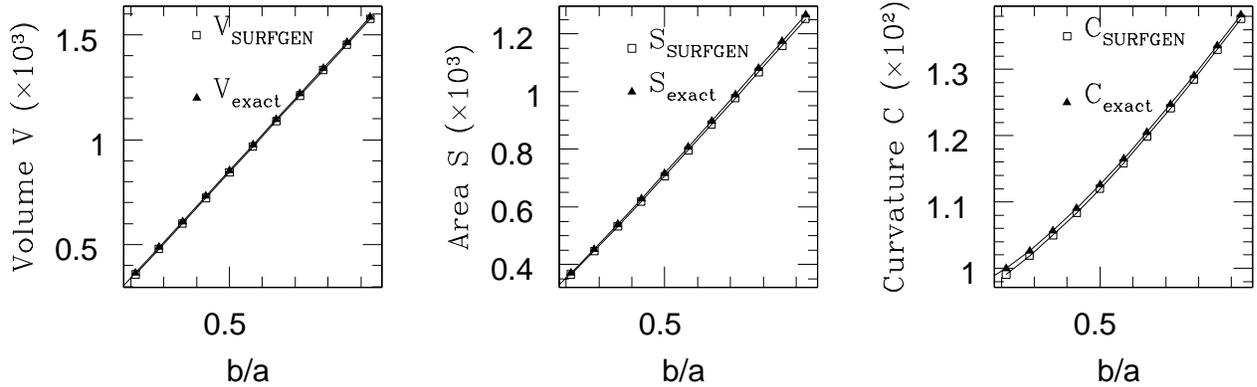

Figure 2.18: Values of Minkowski functionals determined by the surface modelling scheme SURFGEN are shown along with their exact values for prolate deformations of an ellipsoid. Note that for $b/a \geq 0.3$, the results for the Minkowski Functionals evaluated using SURFGEN virtually coincide with the exact values.

Figure 2.19 also shows the evolution of planarity ($\mathcal{P}$) and filamentarity ($\mathcal{F}$) as our ellipsoid becomes increasingly more prolate. In keeping with expectations $\mathcal{F}$ steadily increases from a small initial value $\sim 0.1$ to $\sim 0.44$. Planarity drops from its large initial value to a small final value $\mathcal{P} \sim 0.1$ and is slightly underestimated for $b/a \gtrsim 0.3$. We therefore conclude that both $P$ and $F$ are determined by SURFGEN to a sufficient accuracy for prolate ellipsoidal figures.

### 2.3.3 The Torus and its deformations

Next we extend our analysis to manifolds which are multiply connected by considering the deformations of a torus which has an elliptical cross section and which we shall refer to as an elliptical torus. A torus is an important surface on which to test SURFGEN for two reasons: it is multiply connected and, unlike an ellipsoid, it contains regions which are convex (on its outside) as well as concave (on its inside). The elliptical torus can be described by three parameters $a, b, c$. In this case the elliptical toroidal tube has diameter $2\pi b$ and $a$ & $c$ are its radii of curvature in two mutually orthogonal directions. The elliptical torus reduces to the more familiar circular torus when $a = c$. We choose to work with the elliptical torus because changing the values of $a, b, c$ can give rise to a large variety of surfaces all of which (by definition) are multiply connected but which have very different shapes. Thus our surface modelling scheme SURFGEN can be put

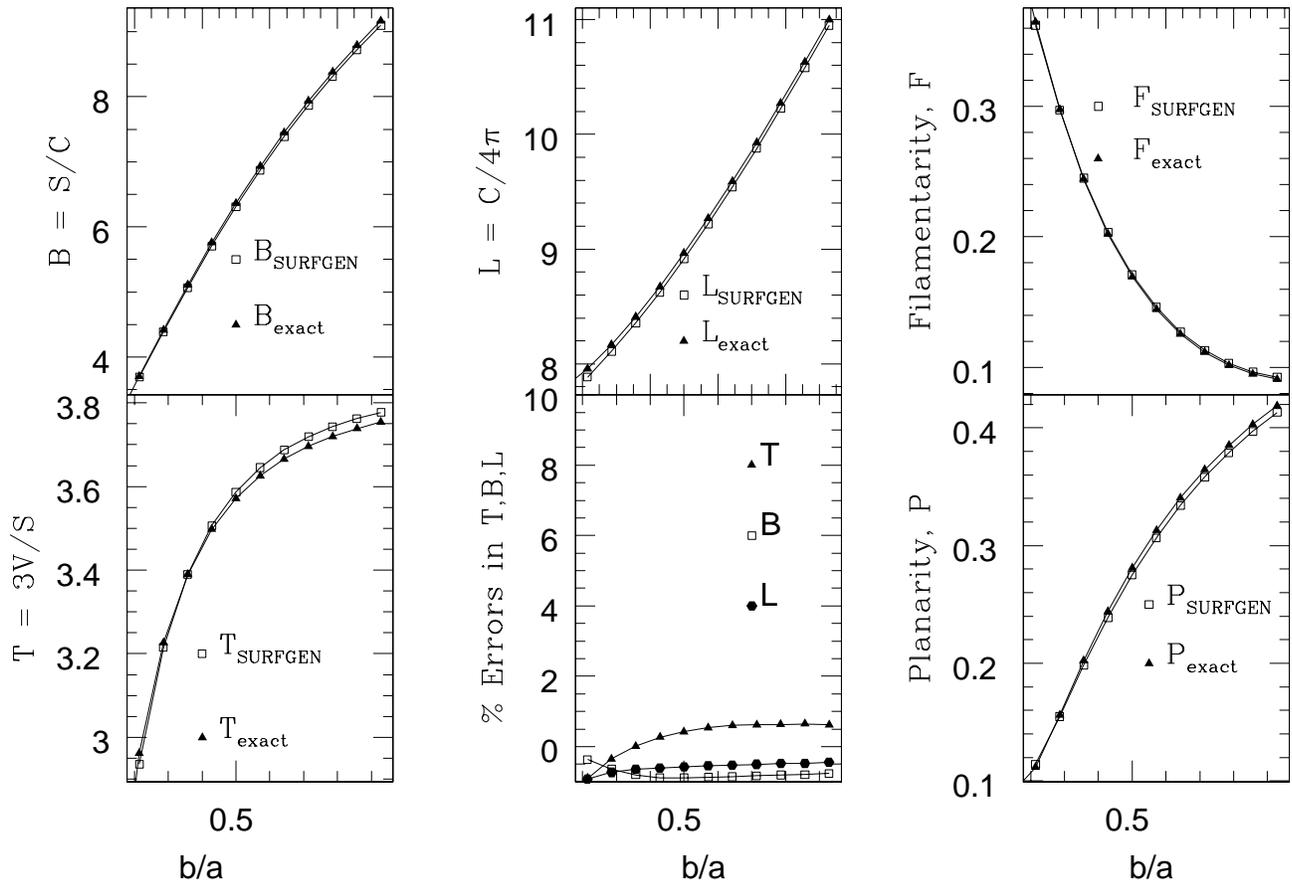

Figure 2.19: Values of the Shapefinders derived from the surface modelling scheme SURFGEN are shown along with their exact values for prolate deformations of an ellipsoid. The lower middle panel shows percentage errors in the estimation of these quantities. We note that the error in the estimation of all the quantities lies within +1% to -2%

Table 2.1: Minkowski Functionals for some extreme deformations of an elliptical torus. $\frac{\Delta C}{C}, \frac{\Delta S}{S}$ and $\frac{\Delta V}{V}$ give the percentage error in the determination of the Minkowski functionals (MFs) using SURFGEN. An accuracy of better than 1% is achieved for the three MFs: Curvature (C), Surface area (S) and Volume (V), while the genus is determined exactly.

|  | Sphere with hole | Ribbon | Pancake | Filament |
|---|---|---|---|---|
| $(b, a, c)$ | (40.0, 37.9., 37.9) | (140.0, 19.9, 1.99) | (60.0, 57.9, 3.86) | (50.02, 3.52, 3.52) |
| C | $7.93 \times 10^2$ | $2.8 \times 10^3$ | $1.2 \times 10^3$ | $9.92 \times 10^2$ |
| $\frac{\Delta C}{C}\%$ | +0.47 | +0.03 | +0.40 | +0.52 |
| S | $5.98 \times 10^4$ | $7.1 \times 10^4$ | $8.8 \times 10^4$ | $6.95 \times 10^3$ |
| $\frac{\Delta S}{S}\%$ | −0.01 | −0.42 | −0.18 | +0.18 |
| V | $1.13 \times 10^6$ | $1.1 \times 10^5$ | $2.7 \times 10^5$ | $1.21 \times 10^4$ |
| $\frac{\Delta V}{V}\%$ | −0.001 | −0.830 | +0.590 | −0.580 |
| Genus | 1 | 1 | 1 | 1 |

to a rigorous test. The parametric form for the elliptic torus is (Sahni et al.1998)

$$\mathbf{r} = (b + c \sin \phi) \cos \theta \hat{x} + (b + c \sin \phi) \sin \theta \hat{y} + a \cos \phi \hat{z}, \qquad (2.16)$$

where $a, c < b$, $0 \leq \phi < 2\pi$ and $\theta < 2\pi$. We shall compare results due to SURFGEN with the exact analytical results for four kinds of tori - a nearly spheroidal torus, a ribbon, a pancake, and a filament.

Tables 2.1, 2.2 and 2.3 refer to these four surfaces; relevant figures can be found in Sahni et al.(1998). Table 2.1 shows the estimated values of the Minkowski Functionals and the percentage error in their estimation. Table 2.2 shows all the Shapefinders for these surfaces while Table 2.3 shows the percentage errors in the estimation of the Shapefinders.

We should emphasize that the genus estimated by SURFGEN for all these deformations has the correct value of unity. This gives an important independent check on the self-consistency of the triangulation scheme since, the absence of even a single triangle out of several thousand, would create an artificial 'hole' in our surface and give the wrong value for the genus – a situation that has *never* occurred for any of the several

Table 2.2:  SURGEN determined values of the Shapefinders $\mathcal{T}$ (Thickness), $\mathcal{B}$ (Breadth), $\mathcal{L}$ (Length), $\mathcal{P}$ (Planarity) and $\mathcal{F}$ (Filamentarity) describing extreme deformations of an elliptical torus.

| Surface | $\mathcal{T}$ | $\mathcal{B}$ | $\mathcal{L}$ | $\mathcal{P}$ | $\mathcal{F}$ | Genus |
|---------|------|------|--------|-------|--------|-------|
| Sphere with hole | 56.86 | 75.44 | 31.57 | 0.140 | −0.410 | 1 |
| Ribbon | 4.67 | 25.63 | 109.96 | 0.692 | 0.622 | 1 |
| Pancake | 9.09 | 74.19 | 47.11 | 0.782 | −0.223 | 1 |
| Filament | 5.23 | 7.01 | 39.49 | 0.145 | 0.698 | 1 |

Table 2.3: Percentage errors in SURFGEN determined values of the Shapefinders describing extreme deformations of an elliptical torus. (The already small errors become even smaller for more moderate configurations.)

| Surface | $\frac{\Delta\mathcal{T}}{\mathcal{T}}\%$ | $\frac{\Delta\mathcal{B}}{\mathcal{B}}\%$ | $\frac{\Delta\mathcal{L}}{\mathcal{L}}\%$ | $\frac{\Delta\mathcal{P}}{\mathcal{P}}\%$ | $\frac{\Delta\mathcal{F}}{\mathcal{F}}\%$ |
|---------|------|-------|------|-------|------|
| Sphere with hole | 0.02 | −0.48 | 0.48 | −1.41 | 0.95 |
| Ribbon | 1.08 | −0.47 | 0.03 | −0.57 | 0.32 |
| Pancake | 0.74 | −0.59 | 0.41 | −0.26 | 2.19 |
| Filament | −0.95 | −0.43 | 0.46 | 1.40 | 0.28 |

dozen deformations of either the ellipsoid or the torus.

Our results clearly demonstrate that the surface modelling scheme SURFGEN provides values for the Minkowski functionals which are in excellent agreement with exact analytical formulae. Thus our ansatz for studying isodensity contours and their morphology works exceedingly well both for simply connected as well as multiply connected surfaces.

## 2.4 Gaussian Random Fields and the Role of Boundary Conditions

This section demonstrates the great accuracy with which SURFGEN determines Minkowski functionals for Gaussian random fields (hereafter GRFs).

Before embarking on our discussion, we briefly describe the samples that we use and summarise the analytical results against which the performance of SURFGEN is to be tested.

We work with three realizations of a Gaussian random field with a power law power spectrum (n = −1) on a $128^3$ grid. Each realization of the field is smoothed with a Gaussian kernel of length $\lambda = 2.5$ grid-units. All the fields are normalised by the standard deviation. This leaves them with zero mean and unit variance. The MFs are evaluated at a set of equispaced levels of the density field which coincide with the parameter $\nu$ on account of $\sigma$ being unity ($\rho_{\mathrm{TH}} = \nu\sigma = \nu$). $\nu$ is used to label the levels and is related to the volume filling fraction through the following equation:

$$FF_V(\nu) = \frac{1}{\sqrt{2\pi}} \int_\nu^\infty e^{-\frac{t^2}{2}} dt. \qquad (2.17)$$

Finally, the 3 realizations of the GRF will be used for ensemble averaging.

The Minkowski Functionals (MFs) of a GRF are fully specified in terms of a length-scale $\lambda_c$,

$$\lambda_c = \sqrt{\frac{2\pi\xi(0)}{|\xi''(0)|}}; \qquad \sigma^2 = \xi(0). \qquad (2.18)$$

$\lambda_c$ can be analytically derived from a knowledge of the power spectrum. It can also be estimated numerically by evaluating the variance $\xi(0)$ of the field and the variance $\xi''(0)$ of any of its first spatial derivatives (for more details see Matsubara 2003).

For a GRF in three dimensions the four MFs (per unit volume) are (Tomita 1990, Matsubara 2003)

$$V(\nu) = \frac{1}{2} - \frac{1}{2}\Phi(\frac{\nu}{\sqrt{2}}); \quad \Phi(x) = \frac{2}{\sqrt{\pi}} \int_0^{x'} \exp\left(-\frac{x'^2}{2}\right) dx', \qquad (2.19)$$

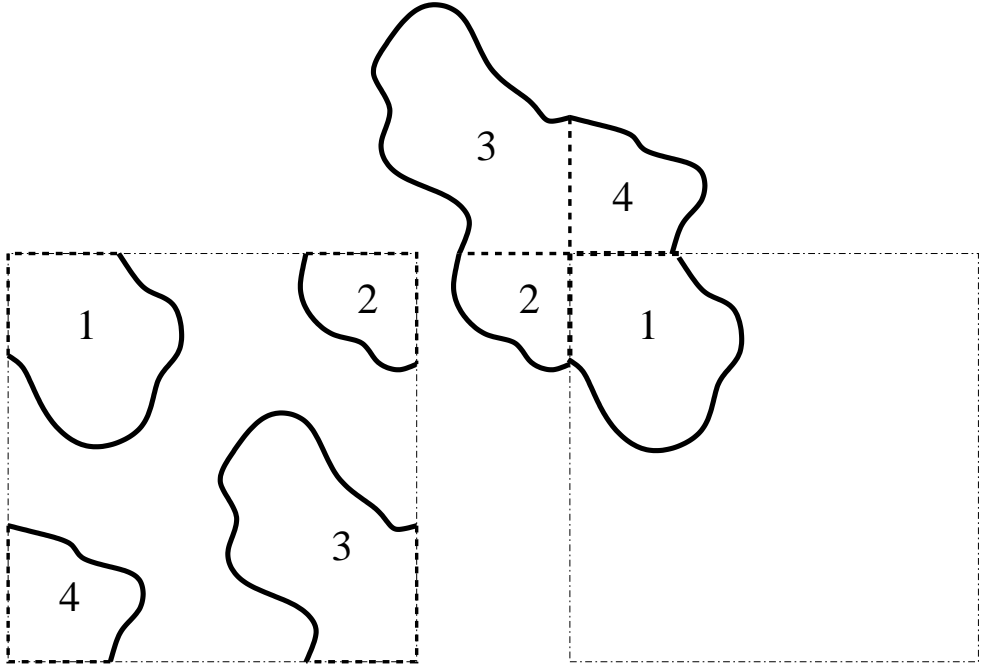

Figure 2.20: A frequent feature of a 2D Gaussian random field generated by FFT on a periodic grid is when 4 clusters touch the boundaries of the box (left panel). SURFGEN-like algorithm in 2D would ordinarily treat these as four distinct contours. However, due to the periodic nature of the GRF, these four clusters in fact constitute a single bigger cluster and the *real* boundaries of the cluster are as shown by bold lines in the right panel of the figure. The dashed solid lines which were boundaries in the left panel, now lie wholly within. In 3D the dashed lines are in fact 2-dimensional planes, which should be ignored when determining the area and mean curvature.

where $\Phi(x)$ is the error function.

$$S(\nu) = \frac{2}{\lambda_c} \sqrt{\frac{2}{\pi}} \exp\left(\frac{-\nu^2}{2}\right), \tag{2.20}$$

$$C(\nu) = \frac{\sqrt{2\pi}}{\lambda_c^2} \nu \exp\left(\frac{-\nu^2}{2}\right), \tag{2.21}$$

$$G(\nu) = \frac{1}{\lambda_c^3 \sqrt{2\pi}} (1 - \nu^2) \exp\left(\frac{-\nu^2}{2}\right). \tag{2.22}$$

Let us now describe the boundary conditions which we adopt and which incorporate the periodic nature of the GRFs generated using a FFT routine. GRFs defined on a grid are, by construction, periodic in nature. As a result, two overdense excursion sets visibly separated from each other inside the box and touching two opposite sides of the box would constitute a single cluster. To make this more clear, let us turn to Fig. 2.20 which shows four clusters (labelled 1, 2, 3 and 4 respectively) in the left panel. The boundaries of these clusters – as drawn by an analog of SURFGEN in 2-D – are shown in bold, and we note that the sides of the box (bold dashed lines) are incorporated when we define the clusters. However, due to the periodicity in the density field, these four clusters are in fact one single cluster (right panel). We note that the portion of the contours due to the boundaries of the box occurs in the interior of the actual cluster (shown with a bold dashed line).

It is easy to extrapolate this situation to 3D. Note that the interpolation method employed by SURFGEN will correctly estimate the global volume of the excursion set even if we worked with the field *as it is*. In 2D this is the same as estimating the area enclosed by the *full* cluster without having to piece the four "clusters" (labelled by 1, 2, 3 and 4) together. Our estimation of global genus will also turn out to be exact, and as it can be shown, this will hold true even if either of these four constituent clusters exhibits a nontrivial topology. On the other hand, if we work with the same set of contours for estimating global area and mean curvature, we will wrongly be adding an excess contribution due to the sides of the box. (In 2−D, this translates to adding the contribution of the dashed line segments to the perimeter of the contour in the right panel of Figure 2.20). The actual area and mean curvature contribution comes from the interface between overdense and underdense regions (such as the 2−D contours drawn in solid bold line in both the panels) as opposed to the artificial boundaries of the clusters (shown dashed).

We note that incorporating periodic boundary conditions in order to construct a total contour as shown in the right panel of Fig. 2.20 is nontrivial from the numerical point of view. When the excursion set touches all the boundaries (as, for example, at a very low threshold of density, below the percolation threshold), the corresponding algorithm could further run into an infinite loop, which makes it impractical for implementation. In the light of this, it would be more manageable if we restricted ourselves to using contours which are closed at the boundaries of the box (like the ones in the left panel of Fig. 2.20), but devise a way to exclude the *excess* contribution to surface area and mean curvature from the boundaries of the box. A uniform prescription of this type, when applied to all the appropriate clusters at all the thresholds of density will lead to correct estimation of all the global MFs. This then, is the choice of boundary conditions incorporated in our calculation of a GRF. A further cautionary note concerns the familiar W-shape of the genus-curve.  Recall that the genus-curve peaks at $\nu = 0$ signifying a sponge-like topology of the medium at the mean density threshold.  Note further that $G(\nu = \pm\sqrt{3}) = -2G(0)\exp\left(-\frac{3}{2}\right)$ are the two negative minima of the genus curve for a GRF. In fact for $|\nu| > 1$, the genus is *always* negative, and approaches zero from below when $|\nu| \gg 1$. For $\nu < -1$, this negative genus refers to isolated underdense regions (bubbles) in the overdense excursion set. Similarly, the negative genus for $\nu > 1$ should be interpreted as being caused by isolated overdense regions (meatballs) in the underdense excursion set. To summarize, in order to determine the genus for a GRF we need to estimate the genus of the *overdense* excursion set for $\nu \leq 0$ and the genus of the *underdense* excursion set for $\nu > 0$, thereby making use of the symmetry property of the GRFs.

Having elaborated on the nature of boundary conditions used in our evaluation of MF's for a GRF, we now present our results.

Figure 2.21 shows the global MFs averaged over 3 realizations of the density field (solid lines with $1\sigma$ error bars) along with the exact analytical results (shown dotted). There are in all 40 levels for every realization. Figure 2.21 clearly shows the remarkable agreement between exact theoretical results (2.19) - (2.22) and numerical estimates obtained using SURFGEN. We therefore conclude that SURFGEN determines the MF's of a GRF to great precision.

The above discussion demonstrated the important role played by boundary conditions in reproducing the analytical predictions for GRFs. Having tested the performance of SURFGEN against GRFs, we are faced with the following choice of boundary conditions when dealing with $N-$body simulations and mock/real galaxy catalogues: (i) we could either correct for the boundaries when dealing with the overdense regions which encounter the faces of the survey-volume, or, (ii) we could avoid doing this and treat all measurements with a tacit understanding that area and mean curvature contain an excess contribution which arises because of boundary effects. Of the two options, the former is computationally more demanding. In addition, its effect is appreciable only significantly after the system has percolated, i.e., when there are a large number of clusters touching the boundaries of the box. The corresponding value of the filling factor ($FF_V > FF_{perc}$) is useful only when one wishes to compare two samples through the trends in their global MFs. Since the structural elements of the cosmic web (superclusters, voids) are usually identified near the percolation threshold, and are therefore not very sensitive to boundary effects, option (ii) is a natural choice in our analysis of N-body simulations in the following chapter. Option (ii) is also more suited for dealing with real galaxy catalogues which do not satisfy periodic boundary conditions. We should also point out that since we shall compare two samples under the same conditions − the same volume and dimensions of the box, same resolution, etc.− the excess contribution to the area and mean curvature caused by box boundaries enters in identical fashion for both the samples. This is additional reason to prefer option (ii) to (i) in our study of simulations to be presented next in Chapter 3. These simulations are thus analysed without correcting for contributions from the boundaries. (It may also be noted that the contribution from the boundaries becomes less important as we deal with larger survey volumes.)

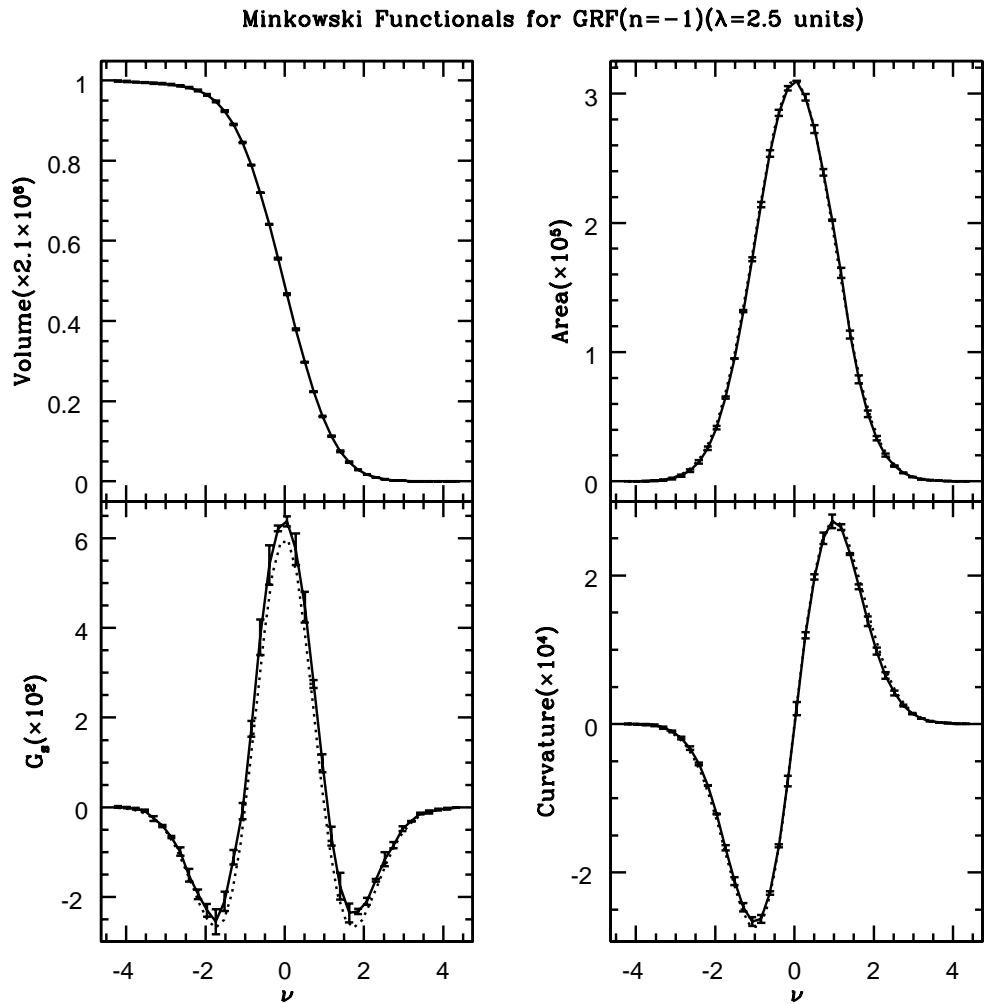

Figure 2.21: Minkowski Functionals for a GRF ($P(k)\sim k^{-1}$), smoothed with a Gaussian filter with $\lambda=2.5$ grid-units. Values of the four Minkowski functionals determined using SURFGEN are shown as solid lines together with the $1\sigma$ scatter. Exact analytical results are shown as dotted lines.

# Chapter 3

# Morphological Study of N−Body Cold Dark Matter Simulations

## 3.1   Introductory Remarks

One of the great observational discoveries of recent times is the realization that we live on a 'cosmic web' which is embedded in an accelerating Universe. Evidence for the cosmic web goes back to the late 1970's when the first redshift survey's revealed the existence of superclusters and voids on scales $\gtrsim 20 \ h^{-1}$ Mpc (Gregory & Thompson 1978; Einasto et al. 1980; Zel'dovich, Einasto & Shandarin 1982; Einasto et al., 1984). The discovery by Kirshner et al. (1980) of a large empty region in Boötes having a diameter of 50 $h^{-1}$ Mpc focussed dramatically on the importance of voids in the Universe. Perhaps the first really convincing demonstration of the ubiquity of voids and superclusters and of the existence of a supercluster-void network resembling a cosmic web was given in de Lapparent, Geller & Huchra (1986) which was soon followed by data from the Southern Sky Redshift Survey (da Costa et al. 1988, 1994). These results have been borne out by more recent studies (Colless et al., 2003, Tegmark et al. 2003), which convincingly demonstrate that the supercluster-void distribution in the Universe is 'foam-like' and has tantalizing geometrical and topological characteristics.

Since the 1970s, theory, N-body simulations, and, most importantly, galaxy redshift surveys have strongly suggested that the components of the supercluster-void network (or 'cosmic web') can be roughly divided into three classes: compact quasi-spherical or slightly elliptical structures like Abell clusters, long filaments like the famous 'bridge' connecting the Coma cluster and A1367 (Gregory & Thompson 1978) and voids. There have also been claims that pancake-like concentrations of galaxies have been observed (Fairall 1998; Martinez & Saar 2001; Gott et al., 2003). From the theoretical side, the fact that gravitational instability is going to function 'anisotropically' giving rise to pancake-like and filamentary features was first discussed in the works of Zel'dovich and collaborators, first within the context of the Zel'dovich approximation (Zel'dovich 1970, 1982; Arnol'd, Shandarin & Zel'dovich 1982, Zel'dovich, Einasto & Shandarin 1982, Shandarin & Zel'dovich 1983) and later extended to the strongly non-linear regime of gravitational instability through the adhesion model (Gurbatov et al., 1989; Weinberg & Gunn 1990; Kofman et al., 1992; Sahni, Sathyaprakash & Shandarin 1994). The existence of a cosmic web has also been confirmed by detailed N-body simulations of gravitational clustering in the distribution both of dark matter (Jenkins et al., 1998; Evrard et al. 2002; Frenk 2002; Smith et al., 2003) as well as baryons (Cen et al., 1995, Dáve et al. 1999, 2002). Interesting results have also been obtained through semi-analytical approaches of galaxy formation (Kauffmann 1999a, 1999b; Benson et al., 2001, 2003; Mathis & White 2002). An excellent overall discussion of the dynamics and geometry of the cosmic web may be found in (van de Weygaert 2002).

Fully three dimensional large scale galaxy catalogues reveal that the cosmic web consists of an interpenetrating network of superclusters and voids. It therefore becomes important to understand and quantify the geometrical and topological properties of large scale structure in a $\Lambda$CDM cosmology in a deep and integrated manner.

In this chapter, our main aim is to study the supercluster-void network in $\Lambda$CDM cosmology with emphasis on the sizes, shapes and topologies of individual superclusters and voids. We shall also study the percolation properties of the full excursion set sampling the entire density field and quantify our results in terms of Minkowski functionals (hereafter, MFs). A study such as the present one will help in setting the stage for a more detailed, subsequent comparison of theory against observations. In order to demonstrate the discriminatory power of our approach, here we shall make a preliminary comparison of three cold dark matter rival cosmological models ($\Lambda$CDM, $\tau$CDM and SCDM). We shall show that morphological properties of superclusters are indeed different in these CDM models. The corresponding results are presented in the latter part of this chapter (see Section 3.8).

It may be appropriate to mention that this is the first comprehensive analysis of large scale structure geometry and morphology in which overdense (superclusters) and underdense (voids) regions are treated on a completely equal footing. Earlier studies have emphasized either over-densities (clusters, superclusters) or under-densities (voids) as a result of which the methods used for the analysis of these two complemen-tary entities (superclusters/voids) often vary greatly in the literature. Thus overdense regions have been studied using correlation functions, minimal spanning trees, shape functions etc., whereas voids have been constructed from point processes using elaborate boundary and volume filling techniques (see for example, Sahni & Coles 1995; Martinez & Saar 2001 and references therein)*.

We shall follow an alternative route in this chapter and study the properties of the supercluster-void network using the density field as our main starting point and fundamental physical quantity. The reason for this is two fold, firstly, density fields can be easily constructed from point data sets using, for example, cloud in cell (CIC) techniques (this is true both for data from N-body simulations, which we examine in this chapter, as well as for galaxy distributions in three dimensions). Secondly, our elaborate surface modeling scheme, SURFGEN, allows us to determine the geometry, morphology and topology of excursion sets defined on a density field in a very comprehensive manner. Applying SURFGEN to the density distribution in various CDM models allows us to develop deep insights into the distribution of large scale structure in these models, which can be quantified using MFs and Shapefinders. Studying the density field at many thresholds also permits us to determine the morphological properties of the *full excursion set* describing the supercluster-void network. It leads to two natural thresholds associated with percolation in overdense and underdense excursion sets. The sizes of individual superclusters and voids change dramatically at the corresponding percolation thresholds revealing the largest scales characterizing the cosmic web.

Concretely, we study the large scale structure of the Universe by considering the geometry and topology of isodensity surfaces $\delta(\mathbf{x}) \equiv \delta\rho(\mathbf{x})/\bar{\rho} = const$. At a given threshold $\delta_{\mathrm{TH}}$, regions having higher than threshold density ($\delta > \delta_{\mathrm{TH}}$) will be called "superclusters", while regions with $\delta < \delta_{\mathrm{TH}}$ will be called "voids". Thus, we define superclusters and voids as overdense and underdense connected regions bounded by one (or several) surfaces of constant density. This definition broadly corresponds to other definitions of superclusters and voids used in the literature but differs in detail. Apart from obvious differences with superclusters and voids of galaxies in redshift space our approach specifies neither a particular density threshold nor the shapes of the structures. Despite these differences we call the objects of our study superclusters and voids mainly because they are nonlinear structures having sizes, volumes and masses roughly corresponding to superclusters and voids of galaxies.

---

*In addition, voids have frequently been claimed to have quasi-spherical or slightly elliptical shapes. One might add a cautionary note at this point since many of these claims were based on visual impressions and the statistics used in the studies of voids often tacitly assumed that voids are either spherical or close to being spherical thereby precluding any other possibility.

We study superclusters and voids at a large number of density thresholds and construct the isodensity surfaces at every threshold to best accuracy. In contrast to many studies (for example, Blumenthal et al., 1992; Goldberg & Vogeley 2003), we do not 'cook up' voids or superclusters with predefined shapes but isolate individual objects from the dark matter density field (obtained from N-body simulations) by constructing isodensity surfaces. Apart from uniformly smoothing the density field with a Gaussian filter, we do not introduce any factors that may affect the shapes or substructure in superclusters and voids of the cosmic density field. Filtering the density field obviously erases some small-scale features but it certainly does not introduce any new structures. Thus, we know beforehand that the real structure can be only richer and more elaborate compared to what we study after smoothing. Filtering high frequency modes is virtually implied in every physics study. This approach can also be viewed as an application of the standard excursion set technique to non-Gaussian three-dimensional fields.

The large–scale structure in the dark matter density field is certainly different from the one observed in galaxy redshift surveys. Galaxies are at best pointwise tracers of the parent continuous dark matter density field. It is likely that they are biased tracers (see Chapter 1 for a discussion of the observational evidences), and that they do not display real physical structures but structures which are strongly distorted by the mapping into redshift space. Thus, the observed superclusters and voids are not physical but only apparent objects, just as the retarded motion of some planets is only apparent and not real motion. At present there are only two methods of investigating real large–scale objects: by reconstructing the real space density field from peculiar velocities of galaxies and by investigating the density fields in the models that are consistent with observations such as $\Lambda$CDM.

We consider this study of the dark matter density field in real space as a necessary step in a systematic study of the morphology of large-scale structure. It is naturally complimented by studies of mock (Sheth 2004) and real galaxy catalogues. Understanding the morphology of the dark matter distribution in real space is an important component in understanding the physical processes determining the formation of galaxies and their motions as well as for building theoretical models of superclusters and voids.

Due to the fairly large smoothing scale adopted here, the over-density in superclusters ranges from $\delta \sim 1$ to $\delta \sim 10$, which makes them more extended ($\gtrsim$ few Mpc) and considerably less dense than galaxy clusters in (for instance) the Abell catalogue.

The rest of this chapter is organized as follows. We begin by giving in Section 3.2, a historical account on the approaches followed by various groups in studying the superclusters and voids. In Section 3.3 we briefly describe the Virgo simulations which are used in this study. Section 3.4 describes our method of generating a density field by smoothing the particle distributions supplied by simulations. Here we also describe the calculations which we have carried out on the density fields. Section 3.5 provides the relation of the filling factors used in this study with the probability density function. Section 3.6 discusses the morphological properties of individual superclusters and voids. Section 3.7 describes substructure in superclusters and voids. In Section 3.8 we compare three rival cosmological models based on their global MFs and morphological properties of superclusters. We conclude this chapter in Section 3.9.

## 3.2 Voids and Superclusters

Both superclusters and their complements – voids – are spectacular features in 3D galaxy catalogues as well as in N-body simulations of gravitational clustering. The first attempts to understand and study voids, considered them to be isolated, quasi-spherical expanding underdense regions embedded in an overdense background (Bertschinger 1985; Hoffman & Shamam 1982). It was shown by Bertschinger (1985) for instance, that matter within a spherical underdensity (void) expands faster than matter which lies outside, leading to shell crossing and a gradual piling up of matter at the void boundary. In contrast to collapsing overdensities, which grow more anisotropic with time, isolated expanding underdensities were shown to become more spherical (Icke 1984; Bertschinger 1985) and this result carried over even when more complex and realistic void configurations were considered (van de Weygaert & van Kampen 1993).

Studies which addressed the role of voids in more general circumstances included Regös & Geller (1991); Dubinski et al., (1993); van de Weygaert & van Kampen (1993); Sahni et al., (1994) and Arbabi-Bidgoli & Müller (2002). In particular, the adhesion model based studies of Sahni et al.(1994) showed that the void distribution in a realistic cosmology evolves with time, and that at any given time, there is a given distribution of void sizes. Sahni et al.(1994) also concluded that the number of voids evolved with epoch and although some voids grew (in comoving volume) others collapsed and disappeared as overdense regions started encroaching upon them. That voids of galaxies need not be empty and must contain a significant amount of mass in undetectable (at that time) form was firstly stressed by Zel'dovich & Shandarin (1982) and Zel'dovich,

Einasto & Shandarin (1982) and was a generic prediction of the adhesion model (Gurbatov, Saichev & Shandarin 1985, 1989; Kofman et al., 1990, 1992). Subsequently it has been confirmed both by high resolution N-body simulations (Göttlober et al., 2003) as well as by observations (Peebles 2001). (The latter gives an excellent discussion of the significance of voids in the context of cosmology.) The expansion of some voids and the shrinkage of others has been confirmed by the study of Sheth and van de Weygaert (2004) who suggested a model describing this phenomenon in many respects similar to the excursion set model used in calculation of the mass function (Bond et al., 1991).

The question of how to define voids within a galaxy distribution has been addressed in a number of papers (Kauffmann & Fairall 1991; El-Ad & Piran 1996, 1997; Hoyle & Vogeley 2002; Plionis & Basilakos 2002). One should mention that there is no unambiguous answer to this question and there have been attempts to define voids either by fitting cubes inside empty regions (Kauffmann & Fairall 1991) or by assuming that walls surround voids, and then searching for spheres which are empty of any wall galaxies and defining voids to be super-impositions of spheres (El-Ad & Piran 1996, 1997; Hoyle & Vogeley 2002). (An alternative, and quite useful way to probe voids is using the void probability function (VPF) which is defined as the probability that a randomly placed sphere of radius $R$ is devoid of galaxies/points (White 1979). This method has been used for analyzing voids in N-body simulations (Little & Weinberg 1994) and in mock galaxy catalogues (Benson et al. 2003). It may be worth mentioning that the void-finder algorithms derived from the methods of Kauffmann & Fairall (1991); El-Ad & Piran (1996, 1997); Hoyle & Vogeley (2002); Plionis & Basilakos (2002) are 'void-specific' in that they implicitly treat voids as entities which are fundamentally different from superclusters and which therefore require tools of analysis specific to them and different from those used to analyse overdense regions – clusters and superclusters. By contrast our method makes no such assumptions. On the contrary, since the density field is viewed as a fundamental characteristic of the matter distribution, voids and superclusters are defined symmetrically, as underdense or overdense regions in the density field, and the same tools (Minkowski functionals) are used to study both superclusters as well as voids.

One of the goals of this work is to quantitatively and objectively test some of the stereotypes routinely used in cosmology. Here are a few examples: both N-body simulations and redshift surveys are characterized by filamentary and sheet-like or pancake-like structures; voids are quasi-spherical bubbles in the density and/or galaxy distributions; and expand faster than the Universe as a whole; voids occupy most of the volume in the Universe. The first problem arises when one tries to address these issues in the absence of conventional definitions. Frequently the context of the origin and usage of a particular assertion is either obscure or forgotten. Proving that some of the above cliches cannot be true does not require elaborate N-body simulations or sophisticated analysis. For instance, voids cannot *both* occupy most of the Universe and all expand significantly faster than the Hubble expansion. Expansion faster than the Universe means that voids expand in the *comoving space* described by the comoving coordinates $x_i = r_i/a(t)$, where $a(t)$ is the scale factor (Peebles 1980). But this is impossible since the comoving volume is conserved! Therefore, after voids have occupied most of the space the expansion of some voids must be at the expense of the others because there is not much of free space (not already occupied by any void) anymore. This is one of trivial (but important) conclusions arising from the adhesion model. At the same time it is not necessary for the interior of the squeezed voids to shrink. In most cases the interior of the void continue to expand faster than the Hubble flow but the void eventually vanishes because its boundaries move *inwards* (Sahni et al., 1994; Sheth & van de Weygaert 2004).

There are arguments for voids to become more spherical with time (Icke 1984; Bertschinger 1985). However, the effect is quite weak quantitatively. The numerical simulations by Bertschinger (1985) have shown that an oblate void with initial axial ratio $\epsilon = c/a = 0.5$ reached $\epsilon = 0.7$ after the Universe expanded by a factor of 162. Expecting that a void lives from a redshift $z = 150$ until present time alone is quite unrealistic and obviously contradicts the visual impression from N-body simulations of realistic cosmological models (Jenkins et al., 1998). Thus, the model of a single expanding void (Bertschinger 1985) may be reasonable for a crude estimate of substructure growth rate and for some particular voids but may be completely misleading for estimating the sizes and shapes in a realistic ensemble of voids. We try to address this issue in our study.

Pancakes or sheet–like structures remain an unsolved controversy since their theoretical prediction by Zel'dovich in 1970. On the one hand they were clearly detected in the simulations with generic smooth initial conditions (Shandarin et al. 1995). There have also been made numerous assertions that pancakes had been detected in the cosmological N-body simulations and redshift surveys. However, most of that were based on visual impressions. On the other hand, many believe that pancakes are not clearly seen in any realistic cosmological N-body simulations and have not been detected objectively. It is worth noting that the simulation of the structure in the hot dark matter model did not show the existence of pancakes (Klypin & Shandarin 1983) and this was interpreted as being the result of the comparative weakness of singularities corresponding to pancakes when compared to the considerably stronger singularities which correspond to filaments (Arnol'd, Shandarin and Zel'dovich 1982).

The problem of pancakes and filaments is even more complex in the hierarchical clustering scenarios like $\Lambda$CDM. According to these models practically all dark matter is in halos that have roughly spherical or somewhat ellipsoidal shapes. The anisotropic distributions so conspicuous in modern N-body simulations actually represent the large-scale organization in the halo distribution in space. It can be approximately reproduced by the truncated Zel'dovich approximation (Coles, Melott & Shandarin 1993; Melott, Pelman & Shandarin 1994; Melott, Shandarin & Weinberg 1994) and therefore the explanation suggested by Klypin and Shandarin is relevant in this more complex situation. Filaments and pancakes obviously correspond to the tendencies in this organization to concentrate halos toward some lines and surfaces respectively. In the cases when some of these tendencies are not strong enough the limited resolution (especially mass resolution) of current N-body simulations may result in complete erasing of the most fragile structures. Thus, the coarse graining unavoidable in N-body simulations may erase pancakes while filaments survive. In our study, we do not detect pancakes after having smoothed the density fields at 2 to $5h^{-1}$Mpc.

## 3.3   Virgo Simulations

In the late 90s of the last century, a group of cosmologists from across Europe formed Virgo Consortium with an aim to simulate the formation and evolution of large scale structure. Their simulations included evolving $256^3$ dark matter particles in a comoving volume of 239.5 $[h^{-1}\text{Mpc}]^3$. The spatial resolution achieved in these simulations was $\sim$10s' of kiloparsecs. To begin with, we shall briefly summarise the cosmological models which were simulated.

Simulations of the cosmological models require us to fix the shape and amplitude of the power spectrum of the initial density fluctuations and choose the value of the matter-density parameter $\Omega_0$ in addition to the total density parameter $\Omega$.

The shape of a CDM-type power spectrum is characterised by the so called shape-parameter $\Gamma$. During mid 90s, having access to the angular positions of more than a million galaxies through APM catalogue, Maddox, Efstathiou & Sutherland(1996) computed the angular correlation function and the corresponding 3$-$dimensional power spectrum of these galaxies. The shape of the power spectrum was constrained using these measurements. This led to $\Gamma \simeq 0.2$, under a crucial assumption that galaxies *trace* mass. Conventionally, $\Gamma = \Omega_0.h$. Reconciling the above value of $\Gamma$ with a *flat* model with $\Omega_0 = 1$ is difficult, for it leads to prohibitively small values $H_0$, which contradicts the estimate of $H_0$ from CMBR measurements. However, it is possible to have $\Omega_0 = 1$ *flat* model with lower value of $\Gamma$, provided the massive $\tau-$neutrinos decay lately into $e-$ and $\mu-$neutrinos to produce an extra superthermal background (White, Gelmini & Silk 1995). This has the effect of delaying the era of matter domination, leading to a decrease in the effective value of $\Gamma$.

For reasons explained in detail by Jenkins et al.(1998), the amplitude of the power spectrum was fixed so as to reproduce the observed abundance of rich galaxy clusters in our local neighbourhood. This was achieved by fixing the value of $\sigma_8$ which refers to the r.m.s. mass fluctuations in spheres of radius $8h^{-1}$Mpc. The $\sigma_8$ normalization is 0.51 for $\tau$CDM and SCDM, 0.9 for $\Lambda$CDM and 0.85 for OCDM. For $(\Lambda, \tau)$CDM models, this normalization matches well with the corresponding COBE normalization, which is an attractive feature of these models. Other parameters of these models are discussed below.

A comparison of the baryon fraction in rich clusters with the universal baryon fraction required by big bang nucleosynthesis (White et al. 1993, White & Fabian 1995, Evrard 1997) favours the value of $\Omega_0 = 0.3$. The measured abundance of hot X-ray emitting clusters at z$\simeq$0.3 by Henry (1997) also suggested a similar value of $\Omega_0$. Two of the models ($\Lambda$CDM and OCDM) simulated by Virgo group are motivated by the above estimates of $\Omega_0$. These measurements were, however, considered controversial. Hence, Virgo Consortium also simulated models with $\Omega_0 = 1$ (SCDM and $\tau$CDM).

Out of four models simulated, three adopted power spectrum shape parameter $\Gamma = 0.21$. One of these ($\Lambda$CDM) has $\Omega_0 = 0.3$ and achieves *flat* geometry required by standard inflationary hypothesis, by fixing $\lambda \equiv \frac{\Lambda}{3H^2}$=0.7. The second model (OCDM) is an open model with $\Omega_0 = 0.3$. Both these models have $h$=0.7. The third model ($\tau$CDM) with $\Gamma = 0.21$ has $\Omega = 1$ and $h$=0.5; it corresponds to the decaying neutrino model mentioned above. Finally, the fourth model is the standard CDM (SCDM) model with $\Omega = 1$, $\Gamma = 0.5$ and $h$=0.5.

We shall briefly mention the various studies which have been carried out using Virgo simulations. This list tentatively serves to illustrate the various ways in which simulations have been utilised *so far* to learn about LSS. Further, this summary also brings out attractive features of some of the *less* favoured cosmological models such as $\tau$CDM. While $\Lambda$CDM concordance model is the most preferred cosmological model consistent with *almost* all the observations, there are several fundamental questions regarding the nature of dark matter and the physics of type-Ia supernovae, yet unanswered. Hence, the future may be expected to hold some surprises in store, and the methods developed here may provide scope of comparing the predictions of alternative cosmogonies with the observed LSS. We should take the findings on several cosmogonies presented below in *this* spirit.

1. Jenkins et al.(1998) measured the two−point correlation function $\xi_{\mathrm{dm}}(r)$ for $r \in [0.1, 20]h^{-1}$Mpc, and compared it with the galaxy-galaxy correlation function $\xi_{gg}(r)$ from the APM catalogue. Their preliminary study indicated that *all* the models require a scale-dependent bias in order to reproduce the observed, power-law behaviour of $\xi_{gg}(r)$. In particular, the $\tau$CDM model has an advantage that the required bias is *almost* scale-independent as against $\Lambda$CDM, which requires a sizeable anti-bias on small scales ($< 1 - 2h^{-1}$Mpc).

2. The same authors measured the pair-wise velocity dispersion. For low $\Omega_0$ models, they found it to be higher than that observed for galaxies from LCRS at a pair separation of $\sim 1h^{-1}$Mpc. This implies stronger velocity-bias for these models when compared with flat models like $\tau$CDM.

3. Springel et al.(1998) extensively studied the topology of LSS in the Virgo simulations, and used a principal components analysis based approach to compare the topology of all the four models with that of the IRAS 1.2-Jy redshift survey. These authors ruled out SCDM at the 99% confidence level. The rest of the three models were found to have their topology consistent with 1.2-Jy redshift survey.

   This work is very close in spirit to the work reported here. However, we evaluate the complete set of MFs, which in addition to topology, also describe the geometry and morphology of LSS. Indeed, with a fixed smoothing window at the scale of $2h^{-1}$Mpc, we succeed in identifying morphological differences between $\Lambda$CDM and $\tau$CDM.

4. Scoccimarro, Couchman & Frieman (1999) used Virgo simulations to test an analytic ansatz for the bispectrum in the strongly nonlinear regime. Their ansatz, fully developed in redshift space, can lift the degeneracy between $\Omega_0$ and the bias factor and can help one constrain the bias from redshift surveys.

It is interesting to note here that most studies have focussed on the lower order clustering statistics like 2−point correlation function, pair-wise velocity dispersion and the bispectrum (except [3] above). Valuable as these studies are, an integrated understanding of the formation and evolution of the large scale structure is still awaited. In particular, relating the developing non-trivial geometric patterns in large-scale structure to their analytic predictions would be a considerable step towards understanding the gravitational instability in N−body systems. Owing to their extensivity property, MFs can be evaluated for the entire overdense and underdense excursion sets, and through them one can probe the geometry and topology of dynamically evolving superclusters and voids. In addition, since MFs can be expressed through the full hierarchy of correlation functions (Schmalzing 1999), the knowledge of MFs helps us complement studies which focus on the lower order correlation functions. In this regard, the work by Matsubara (2003) holds much promise, for he predicts the behaviour of Minkowski functionals for smoothed cosmic density fields which are in the weakly nonlinear regime. With access to large N−body simulations and deep, wide-angle coverage redshift surveys like 2dFGRS and SDSS, it should be possible to confront theoretical predictions for MFs with the observed Universe, and issues related to bias may also be effectively addressed on all scales.

In this chapter we shall dwell upon the *first* steps we have taken towards the overall program which has been outlined in the previous paragraph. The attention has been focussed on three out of four models simulated by the Virgo group, namely ΛCDM, SCDM and τCDM (we do not consider OCDM because the CMBR maps conclusively rule out the open geometry of the Universe). Given a distribution of matter (or galaxies), which morphological and physical quantities most effectively quantify the cosmic web? This is the subject of the subsequent sections. For illustrative purposes only ΛCDM model has been analysed in detail. In Section 3.8, all three models will be compared against one another.

## 3.4 Preparing density field in the ΛCDM Cosmogony

Superclusters of galaxies appear to be in the weakly nonlinear regime. They are dynamically evolving at the present epoch and are unrelaxed. The large scale structure in our immediate neighbourhood is richly populated by several superclusters. Prominent among them are

- The Great Wall located in the northern hemisphere, centered around the Coma cluster. This supercluster indeed has a wall-like morphology, and is quite large. According to Fairall (1998), its typical dimensions are $(\mathcal{L}, \mathcal{B}, \mathcal{T}) = (120, 50,$ $10)h^{-1}$Mpc, where $\mathcal{L}$ stands for the length, $\mathcal{B}$ for the breadth or depth and $\mathcal{T}$ for the thickness of the supercluster. (Note that these dimensions are *not* measured using Shapefinder statistics).

- The Sculptor Wall is located in the southern hemisphere, and is often termed as the *Southern Great Wall*. Its lateral dimensions are indeed huge. The dimensions are $(\mathcal{L}, \mathcal{B}, \mathcal{T}) = (200, 30, 10)h^{-1}$Mpc (Fairall 1998).

We also have the Virgo supercluster, the Centaurus Wall, the Fornax Wall, the Hydra Wall, the Perseus-Pisces region, the Cetus Wall etc. The visual impression of these superclusters, as reported by Fairall is that of pancakes or ribbons. It is as yet a matter of debate as to how planar these objects are, since their morphology requires to be quantified mathematically [†]. Doroshkevich et al.(2003) quantified the morphology of LSS in SDSS (EDR) using the technique of minimal spanning trees (MST). According to these workers, the high density regions which are necessarily confined in smaller volumes of space appear sheet-like, and these are then randomly connected to one another through filaments. However, these sheet-like structures are at best a few tens of Mpc across, i.e., of the size of the Virgo supercluster. Although a set of smaller wall-like structures visually evident in our surroundings conform with this notion, the Great Wall, the Sculptor Wall and the newly discovered *SDSS Great Well* (Gott et al. 2003) are considerably larger by these standards. If confirmed, the existence of these pancake-like superclusters could pose a considerable challenge to models of structure formation, for in N−body simulations, the matter appears to be conglomerating along filaments rather than sheets (Sathyaprakash et al. 1996, Bond et al. 1996), as will be demonstrated in this chapter.

Voids also appear to be dynamic entities (Sahni et al. 1994, Sheth & van de Weygaert 2004). Being interwoven with superclusters, they occupy $\geq 70\%$ of the volume of the Universe. Voids complement superclusters and are equally important in understanding the cosmic web. Our own extragalctic neighbourhood exhibits prominent voids, e.g., the *local void*, the Sculptor void, the Eridanus void, the Microscopium void, the Pegasus void, etc. None of these voids are larger than about 50 Mpc across.

In order to determine the morphological parameters of superclusters and voids in $\Lambda$CDM, we first generate the density field from the distribution of dark matter particles. SURFGEN would operate on such three-dimensional pixelized density-maps. The process of density-reconstruction, described in detail by Sheth et al.(2003), is summarised here: The data consist of $256^3$ particles in a box of size 239.5 $h^{-1}$Mpc. We fit a $128^3$ grid to the box. Thus, the size of each cell is 1.875 $h^{-1}$Mpc. Here we follow the smoothing technique used by Springel et al.(1998) which they adopted for their preliminary topological analysis of the Virgo simulations. In the first, we apply a Cloud in Cell (CIC) technique to construct a density field on the grid. Next we smooth this field with a Gaussian kernel which offers us an extra smoothing length-scale. We study the field smoothed with $L_s = 5$ $h^{-1}$Mpc but also present the global MFs for the field smoothed with 10 $h^{-1}$Mpc for comparison. The scale of $L_s = 5$ $h^{-1}$Mpc is a fiducial smoothing scale in many studies of both density fields in N-body simulations and galaxy fields from redshift surveys; see for example, Grogin & Geller (2000). The Gaussian kernel for smoothing that we adopt here is

$$W(r) = \frac{1}{\pi^{\frac{3}{2}} L_s^3} \exp\left(-\frac{r^2}{L_s^2}\right). \tag{3.1}$$

---

[†]Schmalzing & Diaferio (2000) find the Great Wall to be an appreciably planar supercluster.

Since the kernel is isotropic, it is likely to diminish the true extent of anisotropy in filaments and pancakes. Figure 3.1 illustrates the effect of smoothing the density field by a Gaussian, isotropic kernel. The smoothed density field has been projected onto XY-plane for a Z-thickness of 8 $h^{-1}$Mpc. The black regions correspond to high density, which are overlayed by the contours. We also show the projected distribution of particles, to give an idea of the underlying mass distribution. We note that the small-scale anisotropy (on scales $\leq 5$ $h^{-1}$Mpc) in the matter-distribution is washed out in the final density field, and the contours are rounder because of the isotropic smoothing. The voids with a typical diameter of a few 10's of Mpc, however can still be fully captured at a suitably low density contrast.

The effect of isotropic smoothing could be minimized by considering anisotropic kernels and/or smoothing techniques based on the wavelet transform. An even more ambitious approach is to reconstruct density fields using Delaunay tesselations using a technique discussed by van de Weygaert (2002). Density fields reconstructed in this manner appear to preserve anisotropic features and may therefore have some advantage over conventional 'cloud-in-cell' technique followed by an isotropic smoothing (Schaap and van de Weygaert 2000). Thus, as far as one of the goals is to utilize the geometry of the patterns to discriminate between the models, such smoothing schemes should prove more powerful.

We scan the density fields at 99 values of the density threshold, all equispaced in the filling factor $FF_C$ defined as

$$FF_C(\delta_{TH}) = \frac{1}{V_{tot}} \int \Theta(\delta - \delta_{TH}) d^3x, \tag{3.2}$$

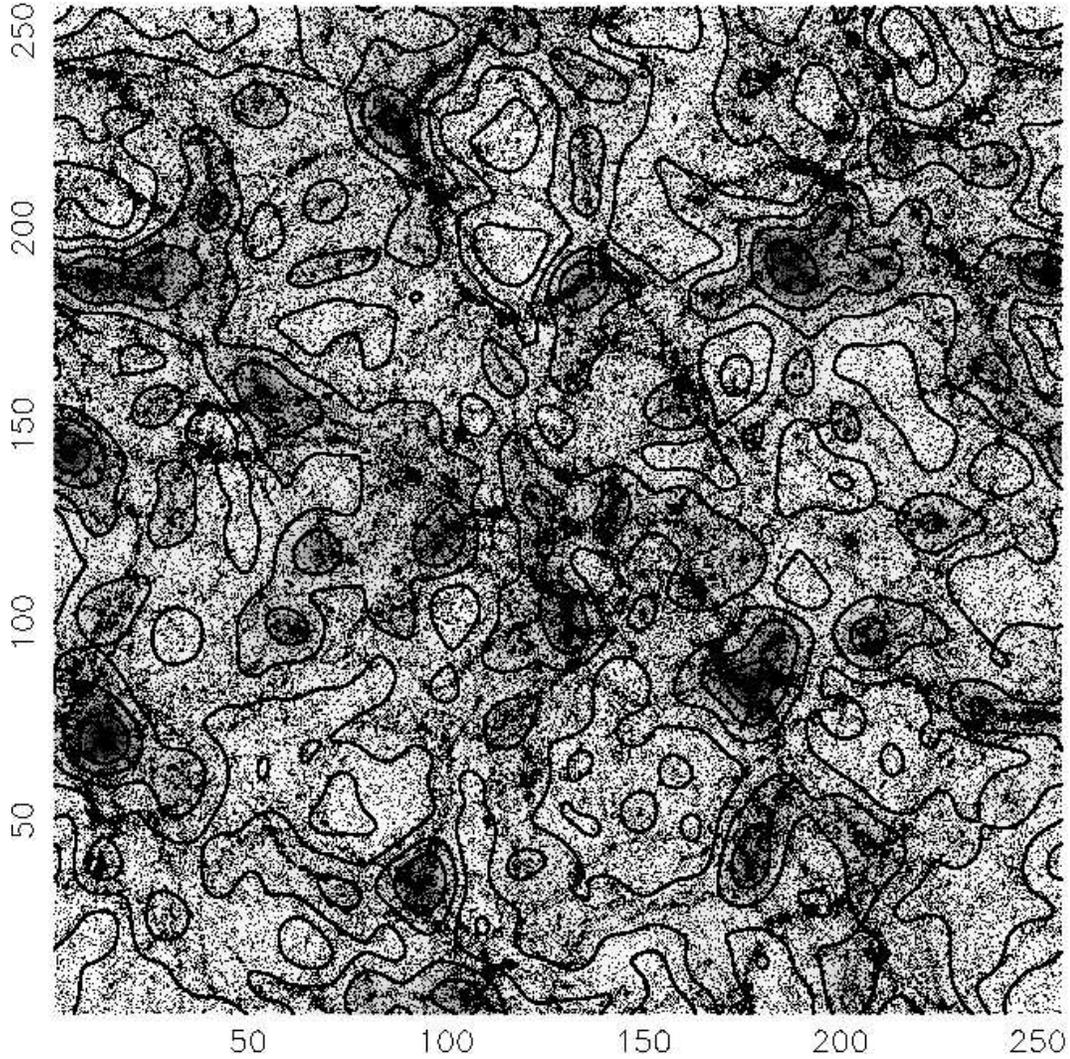

Figure 3.1: The figure illustrates the effect of an isotropic Gaussian kernel. The CIC-generated density field has been smoothed at 5 $h^{-1}$Mpc, and has been projected onto XY-plane. Overlayed on the density field are the contours which are clearly rounder on scales $\leq 5h^{-1}$Mpc than the underlying mass distribution represented by particles. We note however, that the morphology of larger voids, having a typical diameter between 30 to $50h^{-1}$Mpc can be studied with confidence. The anisotropy in structures − both superclusters and voids − is but underestimated in our investigation due to isotropic smoothing.

where $\Theta(x)$ is the Heaviside Theta function and $\delta = (\rho - \bar{\rho})/\bar{\rho}$ is the density contrast. The supercluster filling factor $FF_{\rm C}$ measures the volume-fraction in regions which satisfy the 'supercluster' criterion $\delta_{\rm sc} \geq \delta_{\rm TH}$. In the following, we use $FF_{\rm C}$ along with the void filling factor $FF_{\rm V} \equiv 1 - FF_{\rm C}$ as a parameter to label the density contours. The relation between $FF_{\rm C}$ and density contrast threshold can be seen in Figure 3.2.

At each level of the density field (labeled by $FF_{\rm C}$ or $FF_{\rm V}$), we construct a catalogue of superclusters (overdense regions) and voids (underdense regions) based on a grid realization of the Friends of Friends (FOF) algorithm.

Next we (i) run the SURFGEN code on each of these superclusters/voids to model surfaces for each of them and (ii) determine the Minkowski Functionals (MFs) for all superclusters/voids at the given threshold (these are referred to in the literature as partial MFs). Global MFs are partial MFs summed over all superclusters. Thus, at each level of the density, we first compute the partial MFs and then the global MFs. In addition, we evaluate the 5 Shapefinders for the superclusters and voids.

## 3.5 Percolation and global topology

### 3.5.1 Filling factors and one-point function

A characteristic of the density field which is both simple and useful is the one-point probability density function, $p(\delta)$, where $\delta \equiv (\rho - \bar{\rho})/\bar{\rho}$. In this study we use the integral

$$FF_C(\delta) = P(\delta) = \int\limits_{\delta}^{\infty} p(\delta')d\delta' \tag{3.3}$$

known as the cumulative probability function (cpf) as a quantity parameterizing the excursion sets. It measures the fraction of volume in the excursion set, $\delta > \delta_{\rm TH}$. In order to compare supercluster and void parameters we also use the under-density filling factor

$$FF_V(\delta) = 1 - FF_C(\delta) = \int\limits_{-1}^{\delta} p(\delta')d\delta'. \tag{3.4}$$

The filling factor $FF_C = FF_C(\delta)$ is shown in Figure 3.2 (thick solid line) for $\Lambda$CDM smoothed on the length scale of $L_s = 5\ h^{-1}$Mpc. In addition, the thick dashed line shows the fraction of mass in the excursion set for the same smoothing scale. From Figure 3.2 we find that the density field is 'nonlinear' ($\delta > 1$) in a relatively small fraction ($\lesssim 15\%$) of the total volume. However, the difference between these two curves is the first clear demonstration of nonlinearity of the density field. Convergence to the Gaussian distribution with the growth of the smoothing scale is demonstrated by the two thin lines corresponding to the smoothing scale of $L_s$=10 $h^{-1}$Mpc.

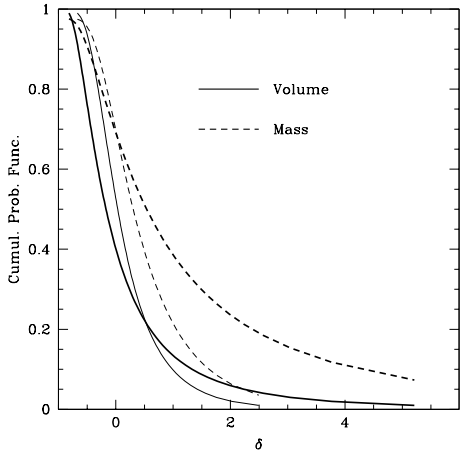

Figure 3.2: Cumulative probability functions of the density contrast in the Virgo simulations of the $\Lambda$CDM model smoothed with $L_s$=5 $h^{-1}$Mpc and 10 $h^{-1}$Mpc are shown by the thick and thin solid lines respectively. The fractions of mass in the excursion sets are shown by the dashed lines. The cumulative probability function equals the filling factor $FF_C$.

### 3.5.2   Percolation

Understanding percolation is essential for understanding the morphology of the supercluster-void network. Percolation is important because the properties of superclusters and voids radically change at the percolation transitions (see Figure 3.3). The left panel shows: (i) the fractional volumes in the largest supercluster and void (dashed lines) and (ii) the total volume in all superclusters and voids after the largest object has been removed from the sample (solid lines). The right panel shows mass fractions in all four components. At relatively high thresholds $\delta_C \gtrsim 1.8$ corresponding to small filling factors $FF_C \lesssim 0.07$ the largest supercluster has insignificant volume and mass compared to the total volume or mass contained in the overdense excursion set, $\delta > \delta_{\text{TH}}$. During the percolation transition at $FF_C \approx 0.07$ corresponding to $\delta_C \approx 1.8$, both volume and mass in the largest supercluster rapidly grow, overtaking the volume and mass in the entire excursion set, and completely dominating the entire sample from this point onwards. The largest void behaves in a qualitatively similar manner if plotted versus $FF_V$. At $FF_V \lesssim 0.22$, $\delta_V \lesssim -0.5$ its volume is small compared to the volume of the underdense excursion set, $\delta < \delta_{\text{TH}}$, but at the percolation transition $FF_V \approx 0.22$, $\delta_V \approx -0.5$, it takes over and from then on remains the dominant structure in the underdense excursion set. Since $FF_C \equiv 1 - FF_V$ the void percolation transition takes place at $FF_C \approx 0.78$ as shown in Figure 3.3.

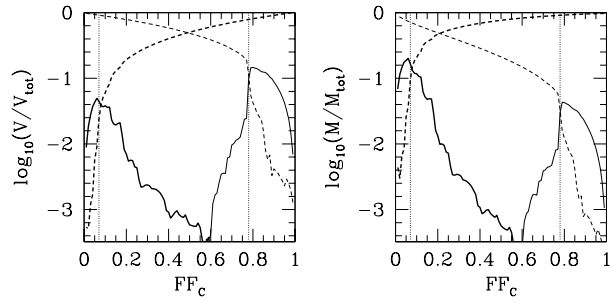

Figure 3.3: The figure refers to ΛCDM model. *Left panel*: the fractions of the total volume occupied by the largest supercluster (thick dashed line), all superclusters but the largest one (thick solid line), largest void (thin dashed line), and all voids but the largest one (thin solid line) are shown for the density field in the ΛCDM model smoothed with $L_s$=5 $h^{-1}$Mpc as a function of the filling factor, $FF_C$. *Right panel*: the y-axis shows the fraction of mass in the components shown in the left panel. Vertical dotted lines show the percolation threshold for superclusters ($FF_C \approx 0.07$) and voids ($FF_C \approx 0.78$; note: $FF_V = 1 - FF_C \approx 0.22$).

Two obvious conclusions can be drawn from the above discussion. First, at percolation the object having the largest volume becomes very different from all remaining objects, therefore it must be studied separately. Second, individual objects – both superclusters and voids – must be studied on a one-by-one basis *before percolation occurs* in the corresponding phase. Both superclusters and voids reach their largest sizes, volumes and masses just before percolation sets in.

Figure 3.3 also shows that at $FF_C > 0.5$ superclusters dominate by volume while at $FF_C < 0.5$ voids dominate. In the range $0.2 \lesssim FF_C \lesssim 0.7$ corresponding to $1.6 \gtrsim \delta \gtrsim -0.43$, no more than 10% of volume is occupied by non-percolating objects, while the remaining more than 90% of entire volume is occupied by just two largest objects: percolating supercluster and the percolating void. In the range between the two percolation thresholds $FF_C \approx 0.07$ and $FF_C \approx 0.78$, both largest objects percolate and therefore the density field has a sponge like structure. The interval of sponge like structure in a Gaussian field is between $FF_C \approx 0.16$ and $FF_C \approx 0.84$. Therefore nonlinear gravitational evolution has shifted it toward smaller $FF_C$ and increased its range by a little bit from 0.68 (=0.84−0.16) to 0.71 (=0.78−0.07).

It is worth stressing that the shift and length of the sponge like interval are determined by the percolation thresholds which in general are two independent parameters.

As a result of these shifts the interval of the so called "meat-ball" topology (when no supercluster percolates) has reduced compared to the Gaussian case from $FF_C \approx 0.16$ to $FF_C \approx 0.07$ and the interval of the "bubble" topology (when no void percolates) is increased from $FF_V \approx 0.16$ to $FF_V \approx 0.22$. All the numbers obviously depend on the adopted smoothing scale but the sign and type of change must be universal for the $\Lambda$CDM model.

For both superclusters and voids $FF = FF_{max} + FF_{ind}$, where $FF_{ind}$ stands for the fractional volume occupied by all objects excluding the largest one, and $FF_{max}$ is the fractional volume in the largest object. Further since, $FF_{max}/FF_{ind} \gg 1$ in the most part of the range between two percolation thresholds for both superclusters and voids, it is not surprising that $FF_{max}^C = FF_{max}^V$ almost exactly at $FF_C = FF_V = 0.5$. A similar result was found in the models with power law initial spectra (Yess & Shandarin 1996).

Percolation is characterized by many features, the most conspicuous being the rapid merger of disjoint parts of the excursion set into one connected structure spanning the entire volume. Merging of superclusters occurs when the density threshold is reduced whereas merging of voids correspond to the growth of the threshold. Spanning of the largest supercluster or void throughout the whole volume results in connection of the opposite faces of the cubic volume by this structure which explains the term *percolation*. Although in principle the percolation transition can be determined by checking if the opposite faces of the cube are connected, in practice it is more robust to identify percolation using other properties of the excursion set (Klypin 1987; Klypin & Shandarin 1993). As the estimators of the percolation threshold we use the following four ratios

$$m^{(i)} = \frac{MF_{max}^{(i)}}{MF_{ES}^{(i)}}, \quad i = 1, 2, 3, 4, \tag{3.5}$$

where $MF^{(i)}$ is one of four quantities: volume, area, integrated mean curvature, or mass ($MF^{(1)} = V$, $MF^{(2)} = A$, $MF^{(3)} = C$, $MF^{(4)} = M$). The subscript "max" is self-explanatory, and "$ES$" stands for the entire excursion set. At percolation these ratios grow extremely rapidly from very small values to unity. The maximum rate of growth $\delta m^{(i)}/\delta(FF)$ can be used as a reliable estimator of the percolation threshold as shown in Figure 3.4. All four parameters detect the percolation transitions at $FF_C \approx 0.07$ for superclusters and at $FF_V \approx 0.22$ for voids. These transitions correspond to the density contrast $\delta_C \approx 1.8$ for superclusters and $\delta_V \approx -0.5$ for voids.

Figure 3.5 compares the largest supercluster with the largest void. Three MFs and the fraction of the mass in the largest objects are shown as a function of the corresponding filling factor. The difference of two curves in every panel is a significant indication of non-Gaussianity of the density field. All Gaussian curves must coincide in every panel and show maximum growth at $FF_C^{SC} = FF_V^V \approx 0.16$. The nonlinear gravitational evolution in the $\Lambda$CDM model shifts the percolation transition in the overdense excursion set toward smaller filling factors ($FF_C : 0.16 \rightarrow 0.07$) and in the underdense excursion set toward larger filling factors ($FF_V : 0.16 \rightarrow 0.22$). Again the particular numbers depend on the choice of the smoothing scale but the sign of the effects is independent of the smoothing scale. At smaller smoothing scales as well as for any adaptive smoothing having a better resolution in high density regions the supercluster percolation threshold must decrease ($FF_C < 0.07$) and void percolation threshold must increase ($FF_V > 0.22$). On the other hand, increasing the smoothing scale would result in a continuous reduction of differences between superclusters and voids and their ultimate convergence to Gaussian curves in every panel (not shown).

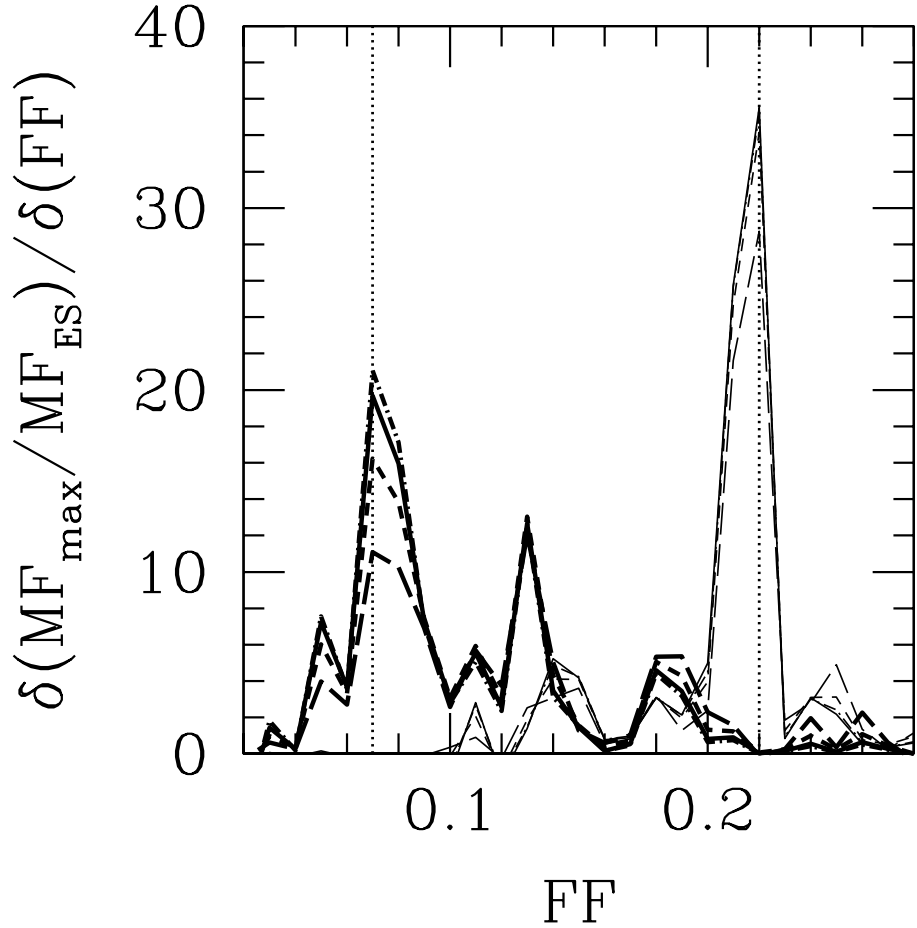

Figure 3.4: Estimates of the percolation thresholds for the $\Lambda$CDM model. The rate of growth $\delta m^{(i)}/\delta(FF)$ for the four estimators listed in Eq. 3.9 is shown as a function of $FF$. Thick lines show results for superclusters. All four curves consistently peak at $FF = FF_C = 0.07$. Thin lines show similar quantities for voids with a distinct peak at $FF = FF_V = 0.22$. Solid, short dashed, long dashed, and dot-dashed lines show the volume, area, curvature, and mass estimators respectively. Vertical dotted lines mark the percolation thresholds.

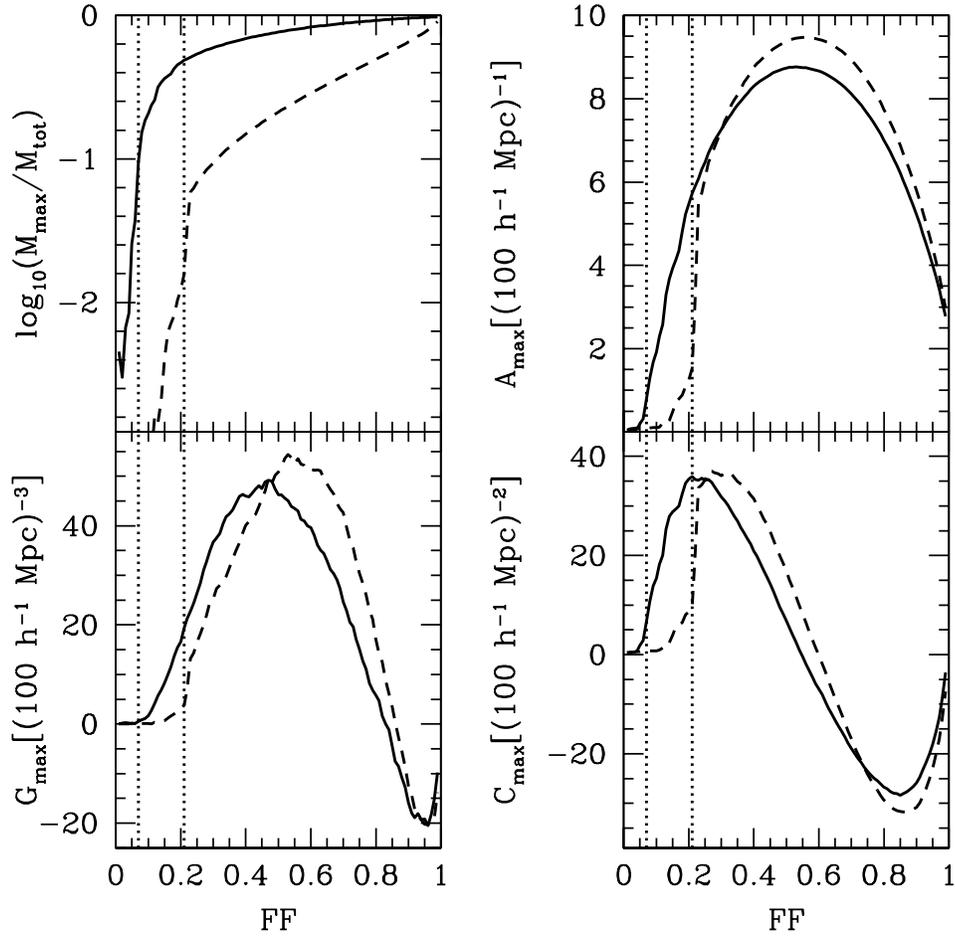

Figure 3.5: The mass fraction and Minkowski functionals of the largest (by volume) supercluster and void in the $\Lambda$CDM model smoothed with the $L_s$=5 $h^{-1}$Mpc window as a function of corresponding filling factor ($FF = FF_C$ for the largest supercluster; $FF = FF_V$ for the largest void). Solid and dashed lines show the parameters of the largest supercluster and void respectively. Vertical dotted lines mark the percolation transitions.

It is worth noting that the percolation transition in voids is more conspicuous than that in superclusters. The transition is particularly clearly marked by $A_{\max}$ and $C_{\max}$ curves. All curves look differently after percolation as well. However, in order to precisely evaluate the significance level of these differences one needs to analyze more than one realization and/or have larger simulation volume.

### 3.5.3   Global topology

It is interesting to compare the percolation and genus statistics. Both were suggested as tests for assessing the connectedness of the large-scale structure. In a series of papers Zel'dovich and Shandarin (Zel'dovich 1982; Shandarin 1983; Shandarin & Zel'dovich 1983) raised the question of topology of large-scale structure and suggested percolation statistic as a discriminator between models. The percolation test was first applied to a redshift catalogue (compiled by J. Huchra) by Zel'dovich et al.(1982) and then Einasto et al.(1984) who found that the connectivity between galaxies in this catalogue was significantly stronger than for a Poisson distribution. In contrast, a non-dynamical computer model having approximately correct correlation functions up to the fourth order (Soneira & Peebles 1978) showed significantly weaker connectivity than observed. Thus, percolation was able to detect connectedness in the galaxy distribution. It was also demonstrated that three lowest order correlation functions (two-, three- and four-point functions) are not sufficient to detect the connectedness in the galaxy distribution.

A few years later Gott, Melott and Dickinson (1986) (see Melott 1990 for a review)

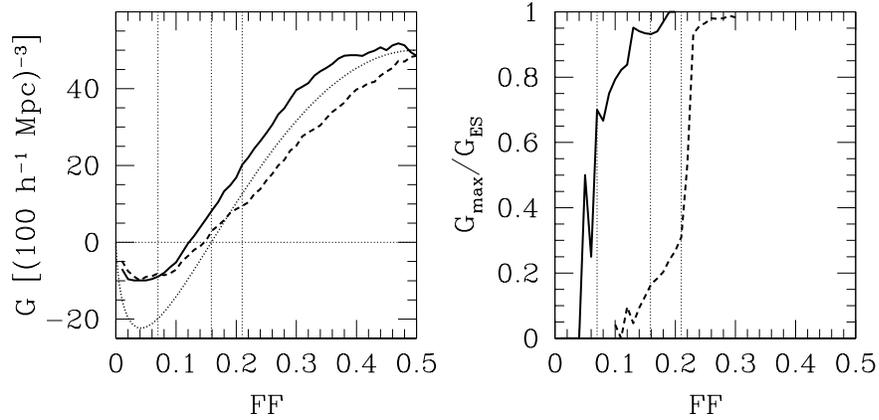

Figure 3.6: The figure refers to the $\Lambda$CDM model. *Left panel*: the global genus is shown as a function of the filling factor for the density field smoothed with $L_s = 5\ h^{-1}$ Mpc. The half of the curve corresponding to high density thresholds is plotted as a function of $FF_C$ (solid line) while the other half corresponding to low density thresholds is plotted as a function of $FF_V$ (dashed line). For comparison, the dotted line shows the Gaussian genus curve having the same amplitude. The vertical dotted lines mark two percolation thresholds in the $\Lambda$CDM ($FF = FF_C \approx 0.07$ and $FF = FF_V \approx 0.22$) and Gaussian field ($FF = FF_C = FF_V \approx 0.16$). *Right panel*: the percolation transitions in the same density field as indicated; the genus of the largest supercluster (solid line) and largest void (dashed line). The vertical dotted lines mark the percolation thresholds similar to left panel.

suggested the genus statistic as a discriminator between various models of large-scale structure. Although both percolation and genus statistics characterize density fields and are sensitive to the connectedness of the large scale structure, each carries significantly different information. It is important to remember that the genus refers to a surface which is the *interface* between overdense and underdense regions (defined at a given density threshold). An interpretation of genus as a characteristic of a three dimensional object (supercluster or void) bounded by this surface is not unique and certainly non-trivial if the region has a complicated shape. For instance, both a full sphere and a doughnut with a bubble in its body have genus of 0.

Figure 3.6 shows the genus curve in a slightly unusual form (see also Sahni et al., 1997). The half of the curve corresponding to high density thresholds is shown as a function of the over-density filling factor $FF_C$ (solid line) while the other half corresponding to low thresholds is plotted as a function of the under-density filling factor $FF_V$ (dashed line). This allows to better illustrate the deformations of the curve due to nonlinear effects. The Gaussian genus curve is symmetric for positive and negative thresholds. Thus both parts of it overlap in Figure 3.6 (dotted line). The vertical dashed lines mark three thresholds: the supercluster percolation threshold at $FF = FF_C \approx 0.07$, the void percolation threshold at $FF = FF_V \approx 0.22$ and both the percolation thresholds in a Gaussian field at $FF = FF_C = FF_V \approx 0.16$.

A marked decrease in the amplitude of the genus curve compared to the Gaussian curve at small $FF$ is noticeable for both overdense and underdense excursion sets. (Small $FF \equiv$ high density for superclusters and low density for voids.) The global genus curve has no significant features at either percolation threshold $FF_C \approx 0.07$ or $FF_V \approx 0.22$. The right panel of Figure 3.6 shows the ratios of the genus of the largest object $G_{max}$ to the global genus of the excursion set $G_{ES}$ for both superclusters (solid line) and voids (dashed line). Both percolation curves shown in the right panel would overlap in the case of a Gaussian field (not shown) and demonstrate the percolation transition at $FF \approx 0.16$. The other indicators of the percolation transitions (Eq. 3.9) are in excellent agreement with the right panel of Figure 3.6 which can be seen by comparing Figure 3.6 with Figs. 3.4 and 3.5. The splitting of the percolation curves shown in Figure 3.6 (right panel) as well as in Figure 3.5 clearly demonstrates non-Gaussianity of the density field. After percolation takes place the genus of the largest object considerably increases (Figure 3.7) which manifests its complex shape. The length and therefore filamentarity of the largest object radically changes after percolation takes place (see Figure 3.7 and 3.8) and cannot be easily interpreted. On the other hand, the breadth and thickness, (and consequently also the planarity), grow much more gradually with the growth of the corresponding filling factor and in this sense are similar to non-percolating objects. The *mean density* of the percolating supercluster at the percolation transition is $\bar{\rho}_{\mathrm{SC}} \approx 4\bar{\rho}$ so that $\delta_{\mathrm{SC}} \approx 3$. The corresponding value for the percolating void is $\bar{\rho}_{\mathrm{V}} \approx 0.3$ and $\delta_{\mathrm{V}} \approx -0.7$. Both $\delta_{\mathrm{SC}}$ and $\delta_{\mathrm{V}}$ are significantly different from the percolation threshold for superclusters ($\delta_{\mathrm{C}} \simeq 1.8$) and voids ($\delta_{\mathrm{V}} \simeq -0.5$) respectively (see Figure 3.7).

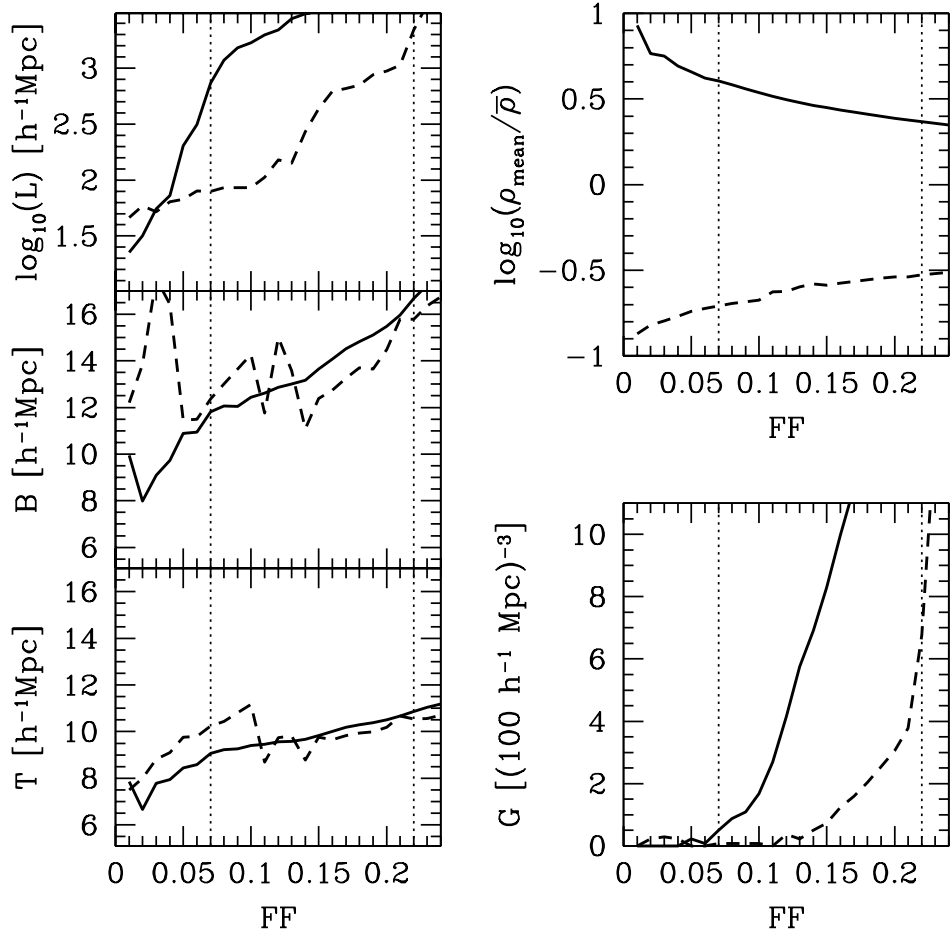

Figure 3.7: The length, breadth, thickness, mean density, and genus of the largest (by volume) supercluster (solid lines) and void (dashed line) in the $\Lambda$CDM model smoothed with $L_s = 5\ h^{-1}$Mpc are shown as a function of of $FF_C$ and $FF_V$ respectively. Vertical dotted lines show the supercluster ($FF = FF_C \approx 0.07 \delta_C \approx 1.8$) and void ($FF = FF_V \approx 0.22$, $\delta_V \approx -0.5$) percolation thresholds.

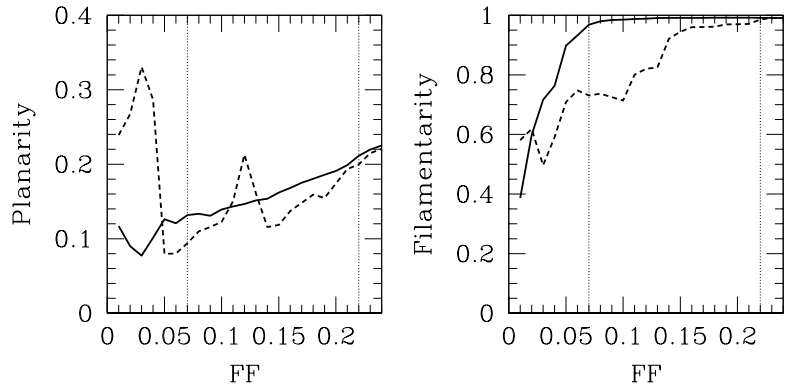

Figure 3.8: Planarity and filamentarity of the largest structures of $\Lambda$CDM model are shown as a function of the filling factor. The largest supercluster and void are shown by solid and dashed lines respectively. Vertical dotted lines show the percolation thresholds for superclusters ($FF = FF_C \approx 0.07$, $\delta_C \approx 1.8$) and voids ($FF = FF_V \approx 0.22$, $\delta_V \approx -0.5$) respectively.

## 3.6  Individual Superclusters and Voids

In this section we carry out comparative statistical analysis of physical and geometrical parameters of superclusters and voids. We stress that the largest supercluster and the largest void both are excluded from this analysis. Before percolation transition they both are extreme outliers and after percolation they have nothing in common with the other objects. The total number of individual objects is shown in Figure 3.9. There are numerous small objects among both superclusters and voids which dominate by numbers at every threshold (right panel of Figure 3.9). Including them into the statistical analysis along with large objects would seriously affect all the parameters. One way to deal with this problem would be the computation of weighted parameters, i.e., by mass for superclusters and by volume for voids. This actually corresponds to computing one or a few moments of the probability distribution function which also may be misleading because the distribution functions are strongly non-Gaussian. We analyze only the most massive superclusters contributing 90% of all mass in non-percolating overdense objects and most voluminous voids contributing 90% of total volume in non-percolating underdense objects. The smallest overdense and underdense objects are excluded from the analysis. Figure 3.9 shows the number density in $(100\ h^{-1}\ \mathrm{Mpc})^{-3}$ units for both large (left panel) and small (right panel) objects. Despite the fact that the smallest objects contribute only 10% of mass or volume they are much more numerous than the larger objects.

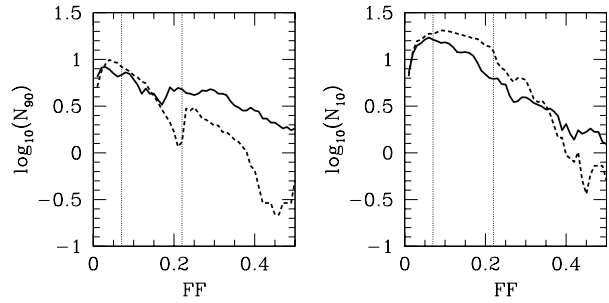

Figure 3.9: *Left panel*: the number density (in $[100\ h^{-1}\mathrm{Mpc}]^{-3}$ units) of the most massive superclusters contributing 90% of all *mass* contained in non-percolating superclusters as a function of the filling factor $FF_C$ is shown by solid line. The number density of the largest voids contributing 90% of all *volume* contained in non-percolating voids as a function of the filling factor $FF_V$ is shown by dashed line. *Right panel*: the number density of the least massive superclusters contributing 10% of all *mass* contained in non-percolating superclusters as a function of the filling factor $FF_C$ is shown by solid line. The number density of the smallest voids contributing 10% of all *volume* contained in non-percolating voids as a function of the filling factor $FF_V$ is shown by dashed line. Two vertical dotted lines mark percolation thresholds for superclusters $(FF_C \approx 0.07)$ and voids$(FF_V \approx 0.22)$

### 3.6.1   Masses, volumes and mean densities

We begin with the analysis of masses, volumes and mean densities

$$\rho_{mean} \equiv \frac{M}{V} \tag{3.6}$$

which are probably the most important factors among those that determine the visual impression. They are also the least ambiguous.

Figure 3.10 shows the masses, volumes and mean densities of superclusters (left) and voids (right) at various thresholds parameterized by the corresponding filling factor. The thick solid lines show the median of the distribution, the dashed lines show the 75% interval (25% largest and 25% smallest are beyond this interval), and thin solid lines show the 95% interval. The intervals are highly asymmetric with respect to the median which indicates the non-Gaussianity of all distribution functions. Every distribution function has a long tail at large values. Both superclusters and voids have largest masses and volumes around the corresponding percolation threshold although the maximum is more distinct in the case of voids. As one might anticipate, voids are less massive and more voluminous than superclusters. However, the difference is not huge at the chosen smoothing scale. The median mass reaches a maximum about $10^{15}$ $M_\odot$ for superclusters and $10^{14.8}$ $M_\odot$ for voids. The median volume in the corresponding maximum is about $3 \times 10^3$ $(h^{-1}Mpc)^3$ for superclusters and about $3 \times 10^4$ $(h^{-1}Mpc)^3$ for voids. The mean densities of the superclusters are well above the mean density before percolation and gradually decrease after percolation takes place however remaining above the mean density in the Universe. As expected the mean density of voids increases with the growth of $FF_V$ corresponding to the growth of the density threshold. It is also somewhat surprising that the volumes of superclusters do not change much with the threshold, the masses change more but not greater than by an order of magnitude.

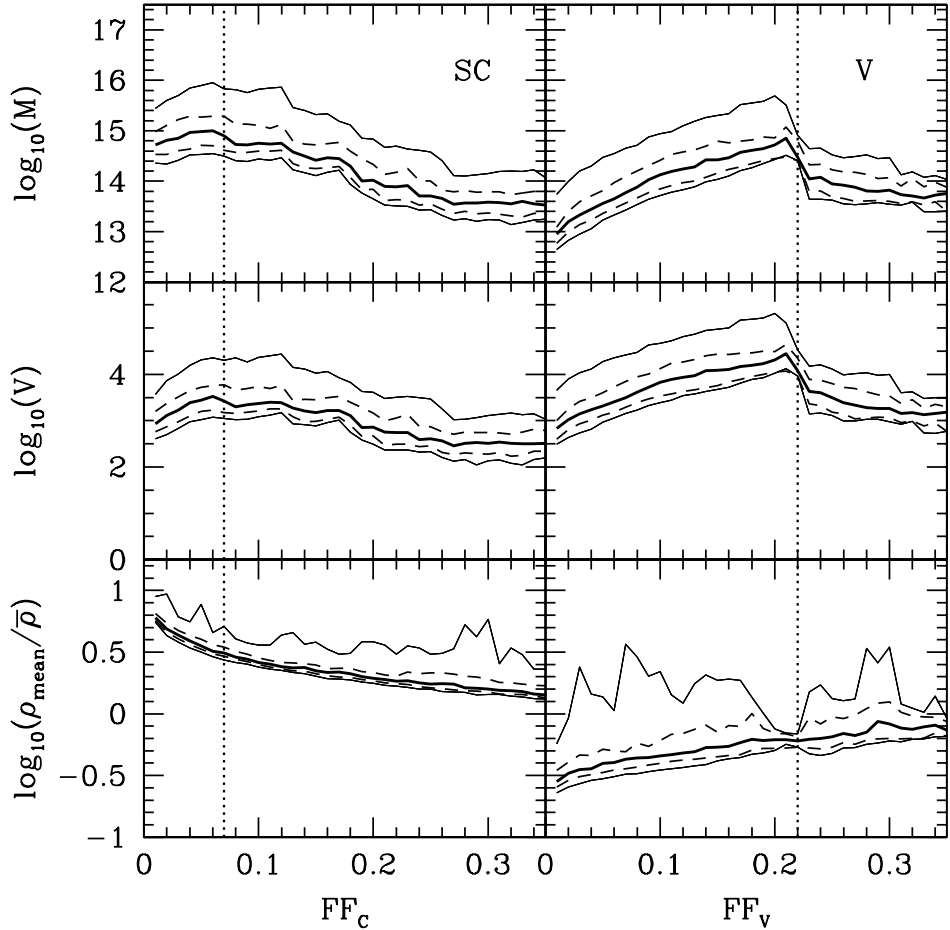

Figure 3.10: Mass (in solar masses), volume (in $(h^{-1} \text{ Mpc})^3$), and mean density (i.e., $\rho_{mean} = (M/V)/\bar{\rho}$) of all structures with the exception of the largest are shown as a function of the volume filling factor $FF_C$ for the $\Lambda$CDM model. Thick solid line is the median of the distribution, dashed lines mark 25-75% interval, and thin solid lines show the third largest and third smallest value in the distribution. Vertical dotted lines show the percolation thresholds in overdense excursion sets (superclusters) ($FF = FF_C \approx 0.07$, $\delta_C \approx 1.8$) and in underdense excursion sets (voids) ($FF = FF_V \approx 0.22$, $\delta_V \approx -0.5$). The supercluster and void parameters are shown in the left and right panels respectively.

### 3.6.2   Sizes and Shapes

Three characteristic sizes and shapes of superclusters and voids can be estimated from MFs of every object (Eqs. 1.8−1.12). Figure 3.11 shows the median and 50% and 95% intervals for the length, breadth and thickness of the superclusters and voids. The sizes of superclusters are shown as a function of $FF_C$, while the sizes of voids as a function of $FF_V$. It is surprising that the median thickness of superclusters depends on the threshold so weakly; it is within 4−6 $h^{-1}$Mpc interval for a range of thresholds between $0 \lesssim \delta \lesssim 6$. This may indicate that the actual thickness of superclusters is significantly smaller and the measured values reflect the width of the smoothing window. The breadth of superclusters is not much larger than the thickness and it is likely that this quantity is also affected by the width of the filtering window. Voids are a little fatter than superclusters and their median thickness reaches about 9 $h^{-1}$Mpc at the percolation threshold. Interestingly voids are also wider and longer than superclusters (please note the logarithmic scale used for length). Recalling that the size parameters are normalized to the radius of the sphere rather than to diameter we conclude that the longest 25% of superclusters are longer than about 50 $h^{-1}$Mpc and 25% of voids are longer than about 60 $h^{-1}$Mpc. It is interesting to reiterate here that the voids in our neighbourhood are *at best* 50 Mpc across.

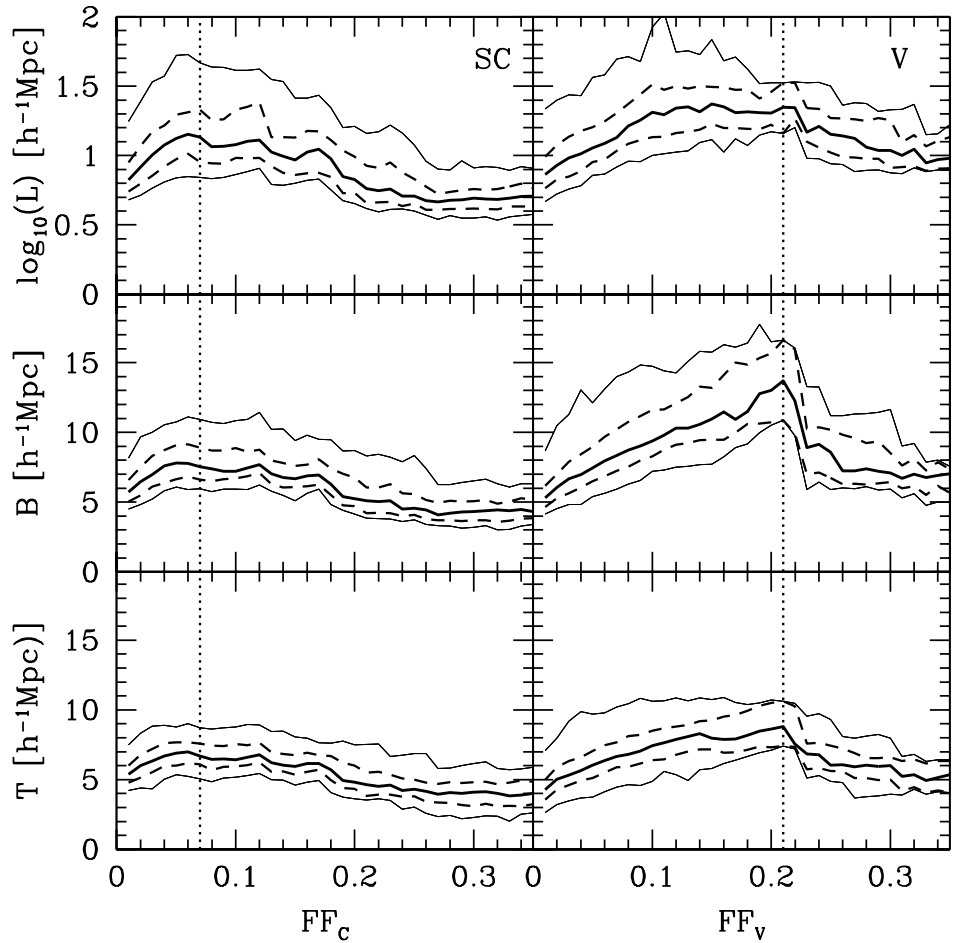

Figure 3.11: The length $\mathcal{L}$, breadth $\mathcal{B}$, and thickness $\mathcal{T}$ of all structures with the exception of the largest is shown as a function of the corresponding filling factor for the $\Lambda$CDM model. The notations are as in Figure 3.10.

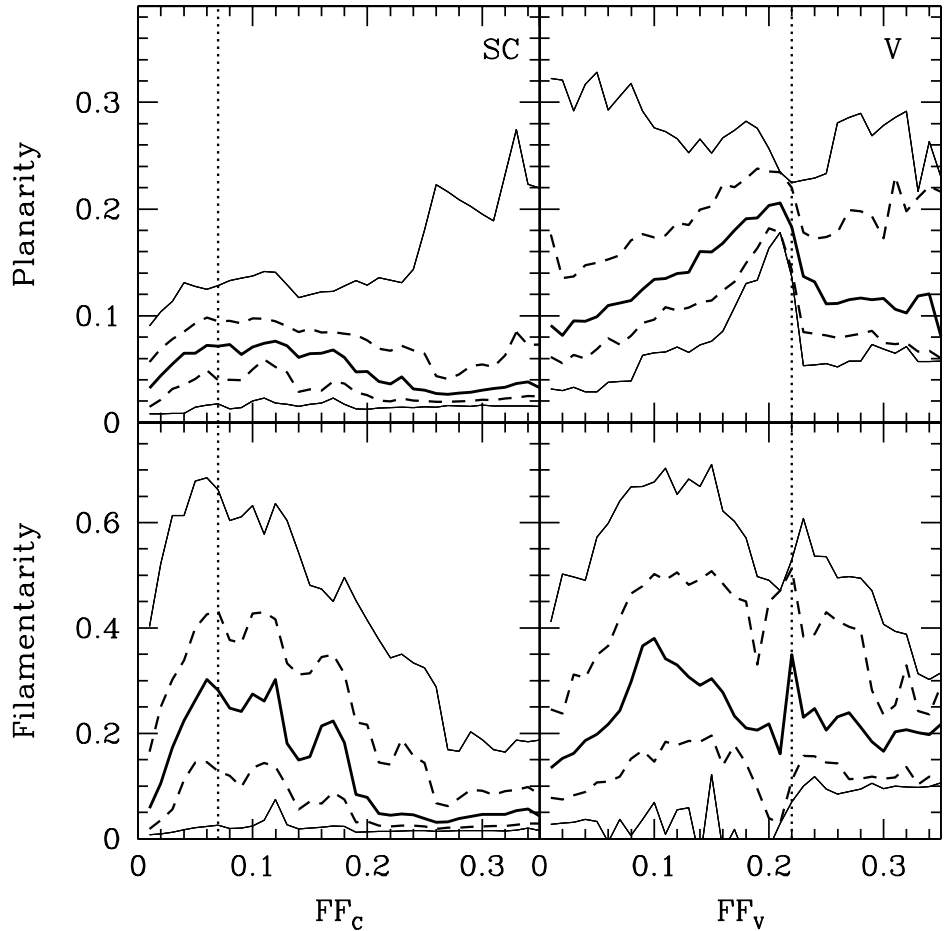

Figure 3.12: The planarity, $\mathcal{P}$ and filamentarity, $\mathcal{F}$ of all but the largest object is shown as a function of the corresponding filling factor. The notations are as in Figure 3.10.

The shape parameters of superclusters and voids are shown in Figure 3.12. The median planarities of superclusters are small which means that in the $\Lambda$CDM Universe the dark matter density field smoothed with $5h^{-1}$Mpc typically has *no* pancake-like superclusters with the diameters greater than about $10h^{-1}$Mpc. The outliers can reach planarity $\mathcal{P} \approx 0.3$ corresponding to the ratio $\mathcal{B}/\mathcal{T} \approx 1.4$ which is not large either. In addition, it happens only at quite large filling factors where the density threshold is quite low and only very few superclusters are left (see Figure 3.9).

Voids show an opposite trend to superclusters: voids are significantly more planar than superclusters and the largest planarities in voids occur at small filling factors. Filamentarities are significantly higher for both overdense and underdense objects:

median values peak at about 0.35 ($\mathcal{L}/\mathcal{B} \approx 2$). The outliers could be considerably more elongated $\mathcal{F} > 0.7$ (*i.e.* $\mathcal{L}/\mathcal{B} > 6$). As we shall see below, the more massive a supercluster or more voluminous a void, the greater is its tendency to be filamentary.

### 3.6.3 Correlations between morphological and physical parameters

Both superclusters and voids reach their largest sizes near their respective percolation thresholds: $FF_C \approx 0.07$ for superclusters and $FF_V \approx 0.22$ for voids. Figure 3.13 shows the scatter plots of three characteristic sizes ($\mathcal{L}$, $\mathcal{B}$, and $\mathcal{T}$) versus mass for superclusters and versus volume for voids. The combined plots are made for $FF_C = 0.06$, 0.07 and 0.08 for superclusters and for $FF_V = 0.21$, 0.22 and 0.23 for voids. The solid lines show the radius of a sphere having the same volume as a given object ($R = (3V/4\pi)^{1/3}$). All three sizes show a significant correlation with the mass: the greater the mass the greater the thickness, breadth and length. The thickness and breadth approximately double their values and length grows by over an order of magnitude when the mass increases from about $10^{14.5}$ $M_\odot$ to $10^{16.5}$ $M_\odot$. Both the thickness and breadth are considerably smaller than the radius $R$ of a sphere having similar volume for large superclusters ($M \gtrsim 10^{15}$ $M_\odot$) as well as for large voids ($V \gtrsim 10^4$ ($h^{-1}$ $Mpc$)$^3$). On the other hand the length is considerably greater than $R$. This is another clear manifestation of the anisotropy of large-scale objects (see Figures 3.14, 3.15, 3.16 and 3.17 for illustrations).

The corresponding plots of planarities and filamentarities show a similar correlation: the larger the mass of a supercluster or the larger the volume of a void, the greater are its planarity and filamentarity (Figure 3.18). One can clearly see that the largest objects (superclusters with $M > 10^{15}$ $M_\odot$ and voids with $V > 10^4$ ($h^{-1}Mpc$)$^3$) are the most anisotropic large-scale objects. Note that voids display a higher level of planarity when compared to superclusters. Indeed, one of the most noticeable results of this analysis is the evidence that voids defined near the onset of percolation of the underdense excursion set are *significantly* non-spherical. Finally, the larger the mass of a supercluster the greater its mean density and the more voluminous a void the lower its mean density (Figure 3.19).

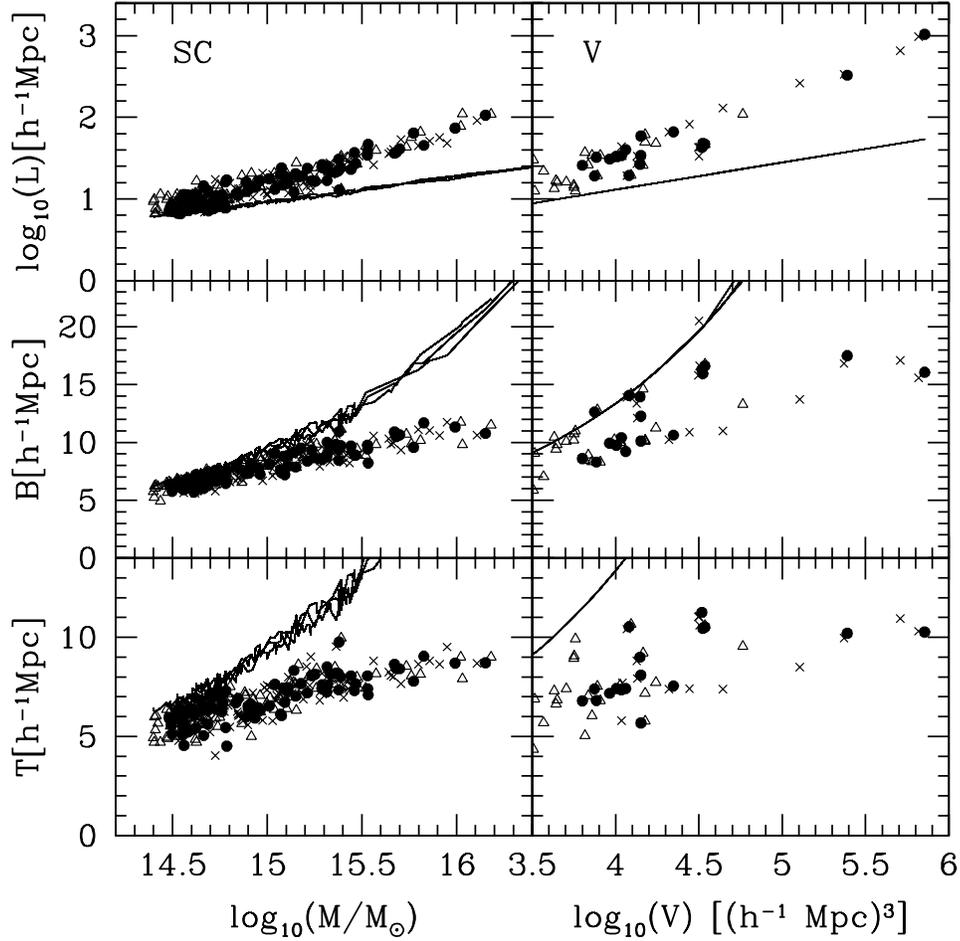

Figure 3.13: The length, breadth, and thickness versus mass for superclusters and versus volume for voids at percolation. Solid circles show the relation at percolation thresholds: $FF_C = 0.07$ for superclusters and $FF_V = 0.22$ for voids. Crosses show the parameters before percolation ($FF_C = 0.06$ for superclusters and $FF_V = 0.21$ for voids) and empty triangles after percolation ($FF_C = 0.08$ for superclusters and $FF_V = 0.23$ for voids). Solid lines show the radius of the sphere having the same volume as the corresponding object. Note the logarithmic scale used for the length. Three lines correspond to three different thresholds.

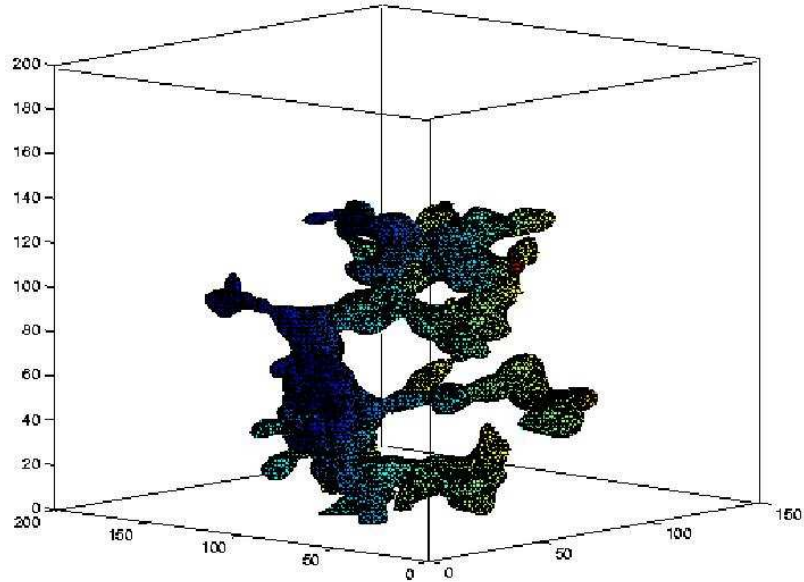

Figure 3.14: Shown here is the largest supercluster at a density threshold corresponding to $FF_C \sim 6\%$, $\delta \sim 2.5$. This is a threshold of density just before the onset of percolation. The supercluster shows genus value of 2, and its size is characterised by $(\mathcal{T}, \mathcal{B}, \mathcal{L})$=(8.61, 10.95, 358.82)$h^{-1}$Mpc. Through the pair of Shapefinders $(\mathcal{P}, \mathcal{F})$=(0.12,0.94), we infer that the supercluster is an extremely filamentary object, which is in agreement with its visual appearance. The mass enclosed by the supercluster is M = $1.26 \times 10^{16} h^{-1} M_\odot$, which is an order of magnitude larger than the mass of a typical rich Abell cluster.

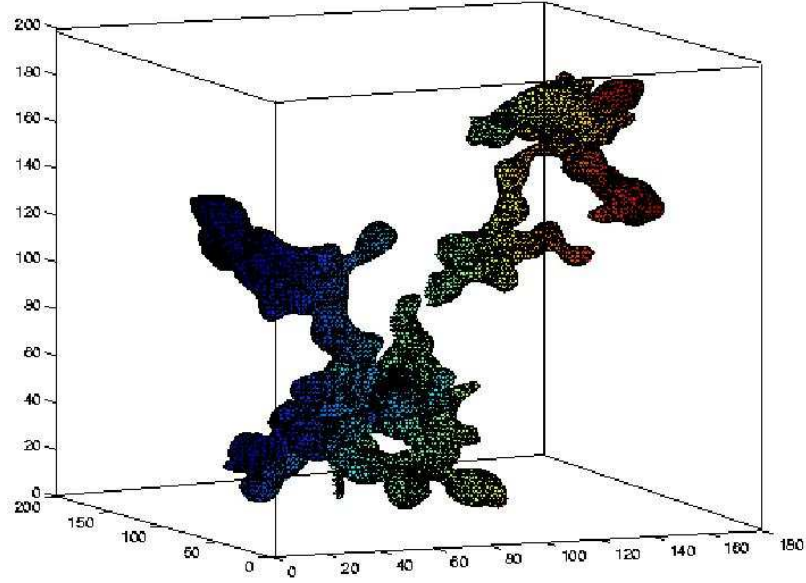

Figure 3.15: At the same threshold of density as Figure 3.14, we show the next two largest superclusters in the list. Both of them show genus $G \geq 1$ and are topologically nontrivial. Both are inferred to be quite filamentary, with F~0.87. Moreover both of them individually enclose mass similar to that shown in Figure 3.14.

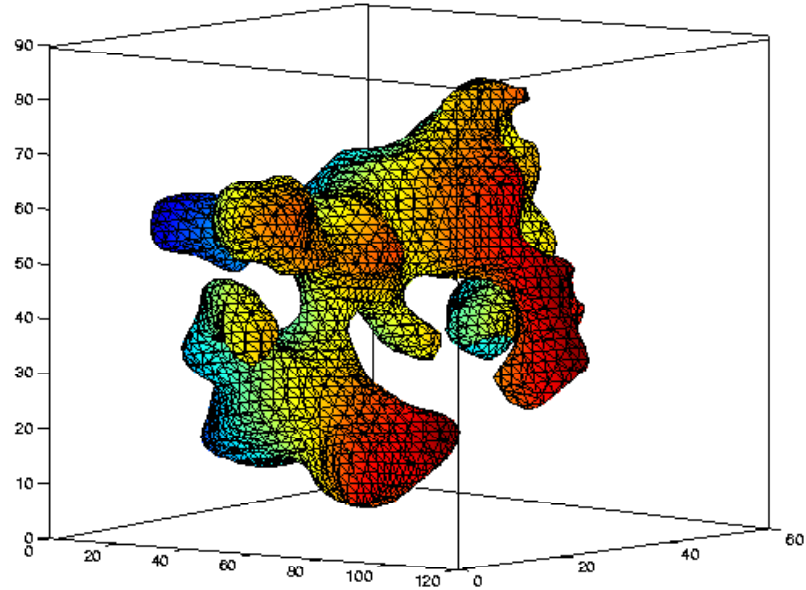

Figure 3.16: Here we show one of the largest 5 voids at the threshold of density corresponding to $FF_V$=0.1, or $\delta_V \simeq -0.7$. This is a generic, deep void which shows genus value of 1, and occupies volume of $5.7 \times 10^4 [h^{-1} \mathrm{Mpc}]^3$. Its dimensions are given by $(\mathcal{T}, \mathcal{B}, \mathcal{L}) = (11.15, 14.28, 85.44)$ $h^{-1}$Mpc. The morphology of the void is mildly planar and dominantly filamentary, as inferred by $(\mathcal{P}, \mathcal{F})$=(0.12, 0.71).

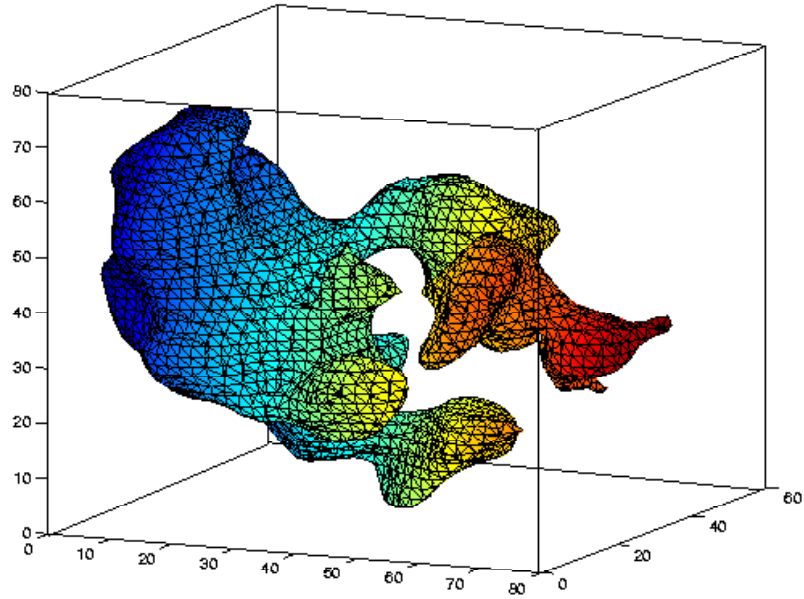

Figure 3.17: The figure shows another generic, deep void with $\delta_V \simeq -0.7$, which is characterised by genus value of 1. It occupies a volume of $4.9 \times 10^4 [h^{-1} \mathrm{Mpc}]^3$ and its dimensions are given by $(\mathcal{T}, \mathcal{B}, \mathcal{L}) = (11.66, 15.03, 66.90)$ $h^{-1}$Mpc. The void is mildly planar but has relatively larger filamentary, as inferred by $(\mathcal{P}, \mathcal{F}) = (0.12, 0.63)$.

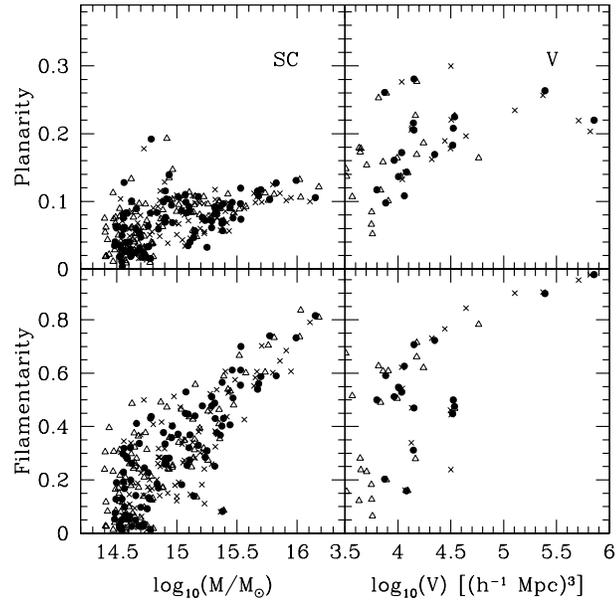

Figure 3.18: The planarity and filamentarity vs mass (for superclusters) and vs volume (for voids) at percolation. Notations are as in Fig 3.13. Note that at percolation $\delta_C \simeq 1.8$ for superclusters and $\delta_V \simeq -0.5$ for voids.

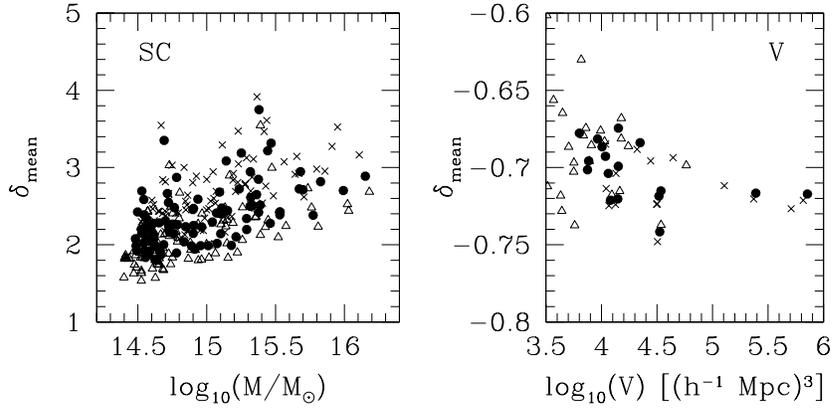

Figure 3.19: *Right panel*: correlation of the mean density contrast ($\delta_{mean} = M/(\bar\rho V) - 1$) of a supercluster with its mass at percolation. *Left panel*: correlation of the mean density contrast of a void with its volume at percolation. Notations are as in Fig 3.13.

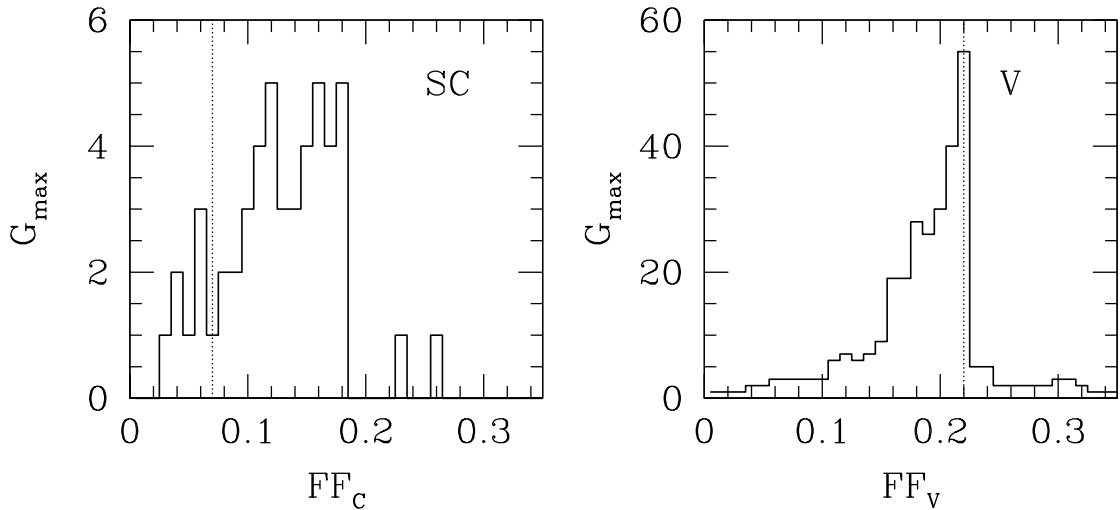

Figure 3.20: The maximum of genus as a function of the filling factor. *Left panel*: the maximum genus in isolated superclusters. Right panel: the maximum genus in isolated voids. Note the difference of the scales on the vertical axis.

## 3.7 Substructure in Superclusters and Voids

Visual inspection shows that superclusters and voids have noticeable substructure. Isolated voids inside superclusters and isolated clusters and superclusters inside voids are obvious examples. Theory (Kofman et al. 1992; Sahni et al., 1994) and targeted N-body simulations (Beacom et al., 1991; Gottlöber et al., 2003) show that voids are also filled with smaller filaments of high density. In the $\Lambda$CDM Universe these filaments are strong enough to survive smoothing on a scale $L_s = 5 \ h^{-1}Mpc$. Figure 3.20 shows the maximum of the genus of superclusters and voids.

Figure 3.20 shows that the genus of an isolated supercluster can be as large as five. It means that there are *at least* five tunnels through the supercluster. The number of tunnels could be even greater if the supercluster harbors a few small isolated voids within itself (isolated voids within the supercluster can decrease its genus and compensate for the presence of tunnels which increase the genus value – see the discussion in Section 1.4). The vast majority of superclusters are topologically isomorphic to a sphere i.e., $G = 0$, but the most massive ones have genus greater than unity. A supercluster with genus of unity is homeomorphic to a doughnut and one with genus of two to a pretzel. The substructure of voids is considerably more complex. The largest genus of voids detected in this simulation is 55, therefore at least 55 filaments span through the void !

Figure 3.21 shows the correlation between genus and mass (for superclusters) or volume (for voids). Largest superclusters $M \gtrsim 10^{15} \ M_\odot$ have a nontrivial topology. Generally the more massive the supercluster the greater its chance of having a complex topology. Voids display even stronger correlation between genus and volume.

## 3.8 Minkowski Functionals as Discriminatory Statistics

In the last section, we discussed in detail the morphology of the dark matter cosmic web in the $\Lambda$CDM cosmological model. In this section we shall utilise the morphology of the *superclusters* of cosmic web to compare three cosmological models, namely $\Lambda$CDM, $\tau$CDM and SCDM. We have already discussed the parameters used to simulate these models. We employ identical smoothing techniques as used in the earlier investigation, save for one difference: in order to detect the difference between the cosmological models, we prefer to minimise on the amount of smoothing. Following Springel et al.(1998) we have chosen to work with a modest scale of smoothing of 2 $h^{-1}$Mpc.

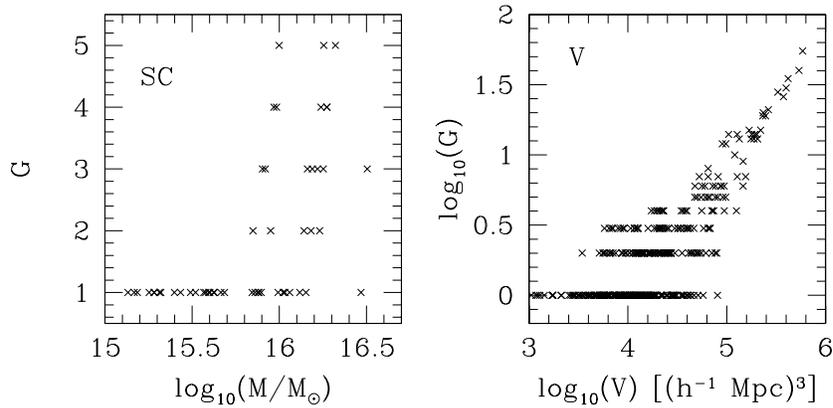

Figure 3.21: The genus-mass relation is shown for isolated superclusters in the left hand side panel. The genus-volume relation is shown for voids on the right. Every isolated supercluster and void having genus greater than zero at all density thresholds is shown. The mass is given in the solar units and the volume in $(h^{-1}\,\mathrm{Mpc})^3$. Note the difference of the scales on the vertical axis.

Another subtle difference is in the form of the Gaussian kernel used for smoothing. The kernel used here is

$$W(r) = \frac{1}{(2\pi)^{3/2}L_s^3}\exp\left(-\frac{r^2}{2L_s^2}\right).$$  (3.7)

After smoothing, we have 3 density fields defined on a $128^3$ grid pertaining to the three cosmological models considered here.

We first study the global MFs for all the three models. Next we investigate the morphology of the percolating supercluster network. This is followed by a statistical study of the morphology of smaller structures. We summarize results reported in this chapter in the next section.

### 3.8.1  Global Minkowski Functionals

We scan the density fields at 100 values of the density threshold $\rho_{TH}$, all equispaced in the filling factor

$$FF_C = \frac{1}{V}\int \Theta(\rho - \rho_{\mathrm{TH}})d^3x,$$  (3.8)

where $\Theta(x)$ is the Heaviside Theta function and $V$ is the simulation box volume. $FF_C$ measures the volume-fraction in regions which satisfy the 'cluster' criterion $\rho_{\mathrm{cluster}} \geq \rho_{\mathrm{TH}}$ at a given density threshold $\rho_{\mathrm{TH}}$. In the following, we use $FF_C$ as a parameter to label the density contours.

At each level of the density field (labelled by $FF_C$), we construct a grid-catalogue of clusters (overdense regions) based on a Friends of Friends (FOF) algorithm.[‡] Next we run the SURFGEN code on each of these clusters to model surfaces for them and determine the MFs for all clusters at the given threshold (these are referred to in the literature as partial MFs). Global MFs are obtained by summing over the partial MFs. Thus, at each level of the density, we first compute the partial MFs and then the global MFs. The plots of global MFs as functions of $FF_C$ are shown in Figures 3.22. This figure shows some interesting features. The volume-curve is the same for all models and grows linearly with the volume fraction; this is simply a restatement of the definition of 'filling factor'. Notice that the amplitude of the remaining three MFs (Area, Curvature and Genus) is substantially greater in SCDM than in $\Lambda$CDM; with $\tau$CDM falling between the two. This could mean that large scale structure is much more 'spongy' in SCDM with percolating structures in this model showing many more 'holes' (or tunnels) and resulting in a large value for the genus. A relative shift towards left, of the position of the peak of the genus curve in $\Lambda$CDM from $FF_C$ =0.5 is indicative of the bubble shift, implying more amount of clumpyness in this model compared to other two models. It should be noted that for a Gaussian random field, the peak occurs at $FF_C = 0.5$. Gravitational clustering modulates the genus curve by lowering its peak and producing a shift to the left (the 'bubble shift') or to right (the 'meatball shift') depending upon the other model parameters. The relatively large value of the surface area and the curvature may indicate that moderately overdense superclusters have many more 'twists and turns' in SCDM than in either $\tau$CDM or $\Lambda$CDM, at identical values of the filling factor. In all models the genus curve has a large negative value at large values of the volume fraction. The reason for this is that the percolating supercluster, at extremely low density thresholds (high $FF_C$), occupies most of the volume. Voids exist as small isolated bubbles in this vast supercluster and lead to a large negative genus whose value is of the same order as the total number of voids. In the opposite case, at very high density thresholds associated with small filling fractions, we probe the morphology of isolated and simply connected clusters. Since the genus for these clusters vanishes, the global topology of large scale structure at high thresholds of the density generically approaches zero for all the models. The vertical lines in the lower right panel show the percolation threshold in SCDM (dashed), $\tau$CDM (dot-dashed), and $\Lambda$CDM (solid) which occurs at intermediate density thresholds corresponding to density contrast $\delta \sim 1$. It is clear from this figure that the percolating supercluster has a relatively simple topology at the onset of percolation. The rapid increase in genus value as one moves to lower density thresholds (larger $FF_C$) reflects the progressive increase in the 'sponge-like' topology of the percolating supercluster which is more marked in SCDM than in $\Lambda$CDM. Similarly to the percolation statistic (Dominik & Shandarin 1992) the difference between models is brought out much better if, instead of plotting the MFs against the volume fraction (equivalently 'volume filling factor'), we choose the 'mass fraction' defined as $FF_M = M_{\text{total}}^{-1} \int \rho \; (\geq \rho_{\text{TH}}) \; dV$. (The 'mass fraction' can be thought of as the 'volume or mass filling factor' [§] in (initial) Lagrangian space, while the 'volume fraction' is the filling factor in (final) Eulerian space.) By employing this parameter for studying the global MFs, we essentially probe the morphology of the iso-density contours (which may refer to different thresholds of density or density contrast, but) which enclose the same fraction of the total mass. Since clustering tends to pack up the mass into progressively smaller regions of space, such a study connects to aspects of gravitational clustering in a direct way. Thus, we see that at all the values of $FF_M$ the volume that encloses the same fraction of the total mass is least for $\Lambda$CDM and most for $\tau$CDM. A further advantage of employing this parameter is that the behaviour of the rest of the global MFs is no longer restricted to follow the same pattern (like it is in case of studying global MFs with the volume filling factor as the labelling parameter). Thus, not only the peak, but also the peak-position and the shape of the MF-curves are sensitive to the models being investigated. Thus, the mass fraction parametrisation could be more useful in discriminating the models from one another as well as comparing the models with the observations. Before we begin our analysis of individual objects (connected overdense/underdense regions) we would like to reiterate that the the 'clusters' and 'superclusters' referred to further in the following discussion are elements constituting the 'cosmic web' and defined on the basis of isodensity contours in N-body simulations. They should not be mistaken for the clusters and superclusters of galaxies seen in the sky.

---

[‡]The clusters/superclusters discussed in the present paper are defined as connected overdense regions lying above a prescribed density threshold. Due to the large smoothing scale adopted, the overdensity in clusters ranges from $\delta \sim 1$ to $\delta \sim 10$, which makes them more extended ($\gtrsim$ few Mpc) and less dense than the galaxy clusters in (for instance) the Abell catalogue.

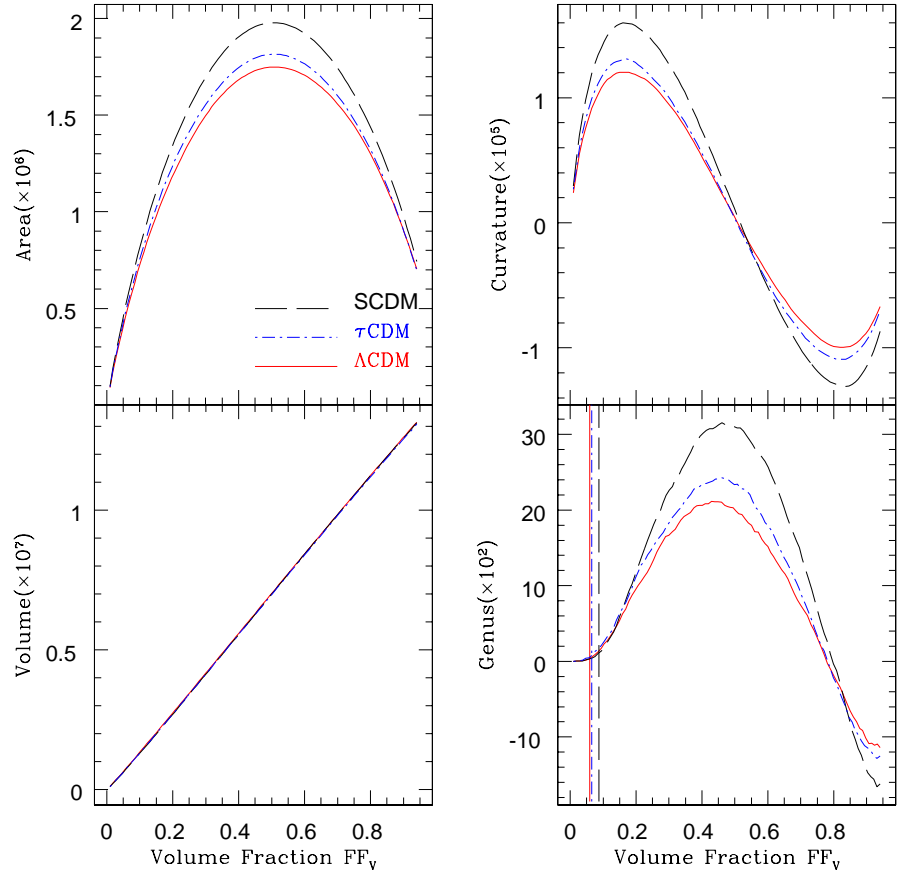

Figure 3.22: Global Minkowski functionals for $\Lambda$CDM, $\tau$CDM and SCDM are shown as functions of the volume fraction ($\equiv$ volume filling factor, $FF_C$). The three cosmological models have appreciably different morphology and hence can be distinguished from one another on the basis of morphological measures. For further discussion, please refer to the text.

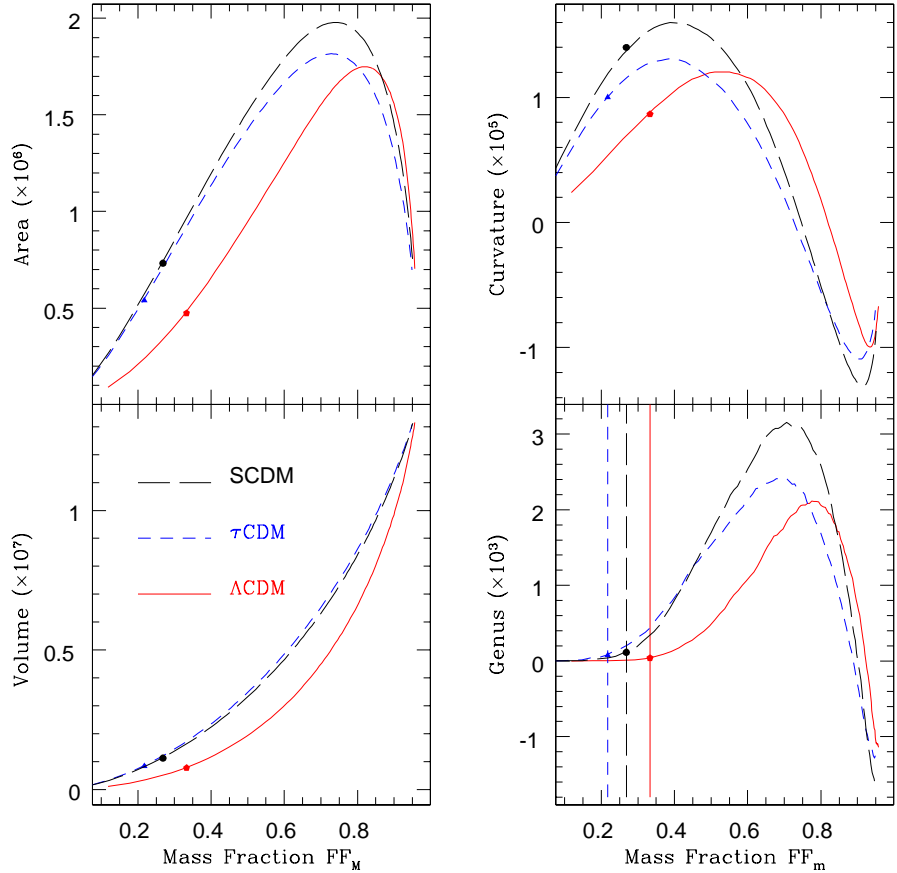

Figure 3.23: Global Minkowski functionals for $\Lambda$CDM, $\tau$CDM and SCDM are shown as functions of the mass fraction defined as $FF_{\rm M} = M_{\rm total}^{-1} \int \rho \, (\geq \rho_{\rm TH}) \, dV$. We see that near-degeneracy between different models seen in Figure 3.22 is broken when one plots the MFs in terms of the mass fraction $FF_{\rm M}$ instead of the volume fraction $FF_{\rm C}$. The markers on the three curves show the mass fraction at the onset of percolation in each of the three cosmological models. (For the sake of greater clarity, vertical lines through the markers are shown in the lower right panel.)

## 3.8.2　Cluster abundance and percolation

We have made a detailed study of the percolation properties of the $\Lambda$CDM model earlier. However, for the sake of completeness and because much of the nomenclature to be used in the next course of discussion is to be introduced here, we shall work out the percolation properties of the three models and compare them on this basis.

---

[§]Since the density in the Lagrangian space is uniform there is no difference between the volume and mass fractions.

In this context we study how the total number of clusters and the fractional volume in the largest cluster vary as we scan through a set of density levels corresponding to equispaced fractions of volume and total mass. Figure 3.24 shows the volume fraction and mass fraction as a function of the density contrast. It is interesting to note that SCDM and $\tau$CDM show the same pattern of behaviour which differs from $\Lambda$CDM. It is further to note that at high density thresholds ($\delta \gtrsim 1$) the filling factor (at a constant density threshold) is greater in $\Lambda$CDM than in SCDM or $\tau$CDM, while exactly the reverse is true for underdense regions. This figure, which relates $FF_{M,C}$ to $\delta$, serves as a reference point to all of our subsequent percolation studies. Figure 3.24 also shows that, relative to other models, *more mass occupies less space in $\Lambda$CDM*. Thus almost 67% of the total mass in the $\Lambda$CDM Universe resides in just 25% of the volume. (In case of $\tau$CDM and SCDM $\sim$ 62% mass occupies $\sim$ 34% volume.)

A statistical pair which help quantify the geometry of large scale structure and its morphology are the 'Number of Clusters Statistic' (NCS) and the 'Largest Cluster Statistic' (LCS); both are shown in Figure 3.25. The Number of Clusters Statistic shows cluster abundance as a function of the density contrast (lower left), mass fraction (lower middle) and volume fraction (lower right). The Largest Cluster Statistic measures the fractional cluster volume occupied by the largest cluster:

$$LCS = \frac{V_{LC}}{\sum_i V_i},\qquad(3.9)$$

where $V_i$ is the volume of the $i-$th cluster in the sample and $V_{LC}$ is the volume occupied by the largest cluster. Summation is over all clusters evaluated at a given threshold of the density and includes the largest cluster. In Figure 3.25 LCS is shown as a function of the density contrast (upper left), mass fraction (upper middle) and volume fraction (upper right).

The square, triangle and circle in NCS denote critical values of the parameters $\delta = \delta_{cluster-max}$, $FF_C = FF_{cluster-max}$ and $FF_M$ at which the cluster abundance peaks in a given model. Similar etching on the LCS curves denote values of $\delta = \delta_{perc}$, $FF_C = FF_{perc}$ and $FF_M$ at which the largest cluster first spans across the box in at least one direction. This is commonly referred to as the 'percolation threshold'. It is important to note that the number of clusters in SCDM is considerably greater than the number of clusters in $\Lambda$CDM at most density thresholds. It is also interesting that percolation takes place at higher values of the density contrast (and correspondingly lower values of the volume filling factor) in the case of $\Lambda$CDM. ($\delta_{perc} \simeq 2.3$ for $\Lambda$CDM, $\delta_{\mathrm{perc}} \simeq 1$ for SCDM and $\tau$CDM.) Furthermore, while in the case of $\Lambda$CDM $\delta_{\mathrm{cluster-max}} \geq \delta_{\mathrm{perc}}$, in the case of SCDM and $\tau$CDM, $\delta_{\mathrm{cluster-max}} \lesssim \delta_{\mathrm{perc}}$. Thus in the latter two models, as the density threshold is lowered, the cluster abundance initially peaks, then, as the threshold is lowered further, neighboring clusters merge to form the percolating supercluster. In $\Lambda$CDM on the other hand, the threshold at which the cluster abundance peaks also signals the formation of the percolating supercluster. Having said this we would like to add a word of caution: the analysis of one realization alone does not allow us to asses reliably the statistical fluctuations in the threshold of percolation. Therefore the final conclusion can be drawn only after a study of many realizations of each model.

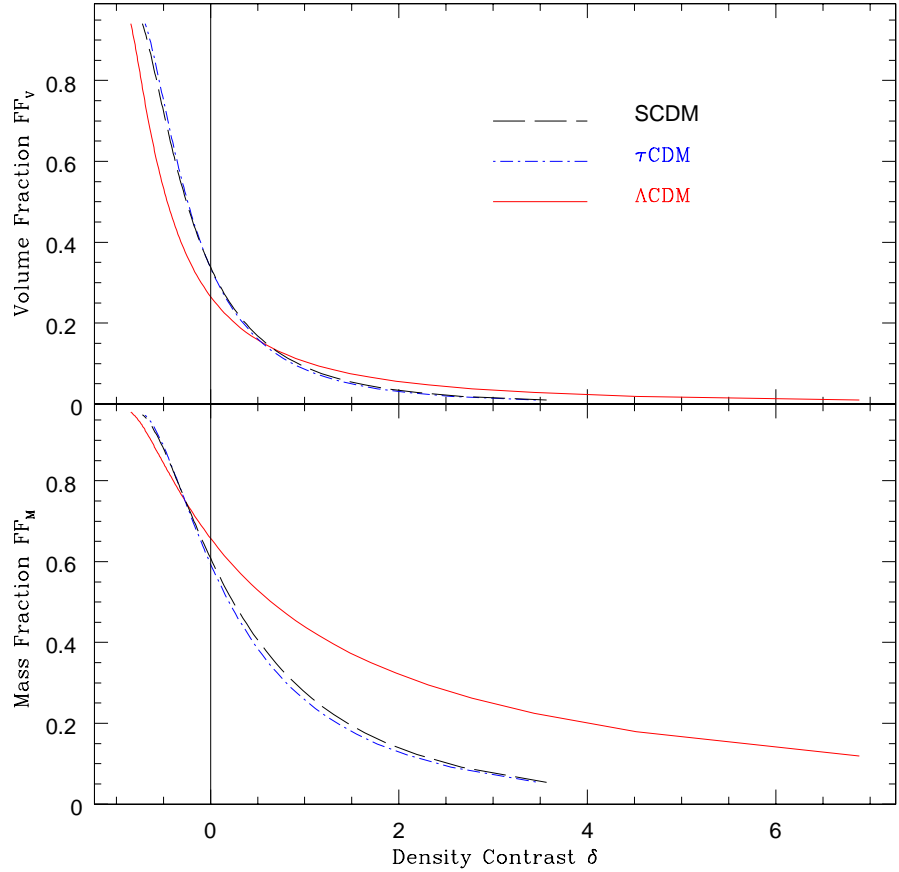

Figure 3.24: Here we study the evolution of the volume fraction and the mass fraction w.r.t. the density contrast for all the three models. $\Lambda$CDM shows maximum density contrast. SCDM and $\tau$CDM form another class of models which have similar density contrasts but smaller than $\Lambda$CDM. The vertical line across both the panels refers to the mean density, which can be used as a marker to study the mass fraction which is contained in the overdense volume (the volume occupied by regions above the mean density). We note that the overdense volume in $\Lambda$CDM is $\sim$25% which is about 10% smaller than that occupied by the other two models. At the same time, the mass that this volume encloses is $\sim$67% and is about 5% larger than that enclosed by the overdense volume in the other two models.

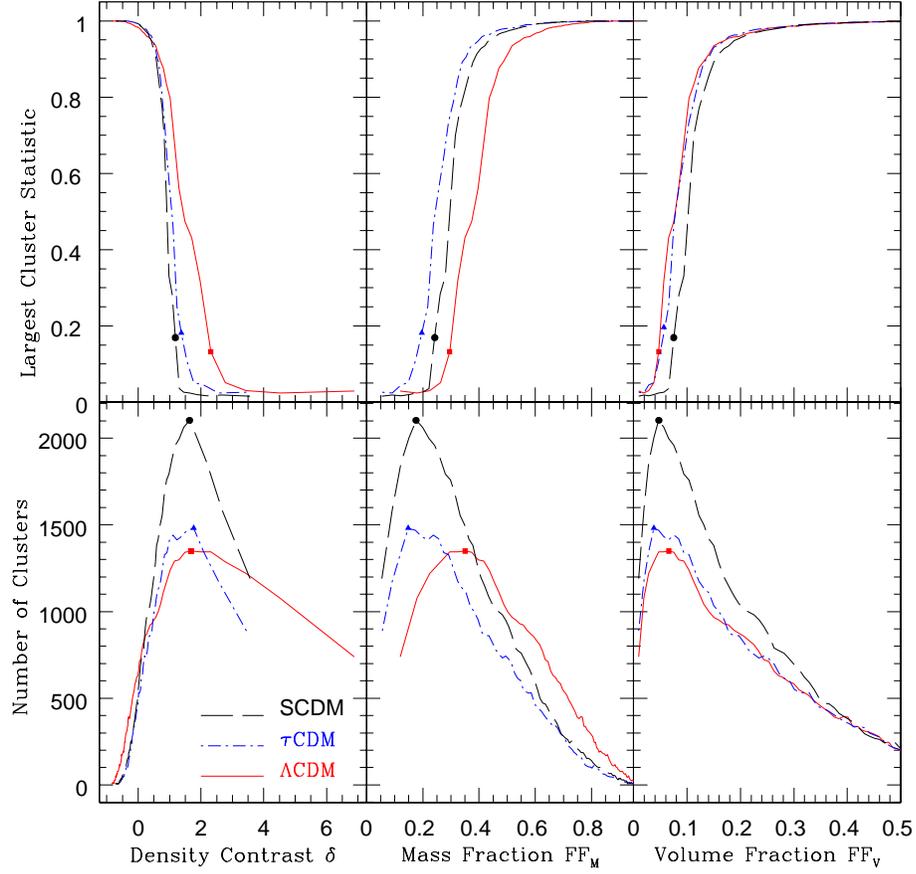

Figure 3.25: Number of Clusters statistic (NCS) and the Largest Cluster Statistic (LCS) are studied as functions of the density contrast $\delta$ (left panels), mass fraction $FF_M$ (middle panels) and volume fraction, (equivalently volume filling factor), $FF_C$ (right panels). The markers on the three curves, in the case of NCS, show values of $\delta_{\text{cluster}-\text{max}}$ (left), $FF_{\text{M,max}}$ (middle) and $FF_{\text{C,max}}$ (right) at which the cluster abundance peaks. In the case of LCS, the markers indicate percolation values of $\delta_{\text{perc}}$ (left), $FF_{\text{M,perc}}$ (middle) and $FF_{\text{C,perc}}$ (right) at which the largest cluster percolates by running through the entire simulation box.

Clearly both the LCS threshold $\delta_{\mathrm{perc}}$ and the NCS threshold $\delta_{\mathrm{cluster-max}}$ contain important information and most of our subsequent description of supercluster morphology will be carried out at one of these two thresholds. Table 3.1 shows the Shapefinders for the 5 largest (most voluminous) superclusters compiled at the percolation threshold for the three cosmological models $\Lambda$CDM, $\tau$CDM and SCDM. Shown also are the mass and genus for these superclusters.

Our $(239.5 h^{-1}\mathrm{Mpc})^3$ $\Lambda$CDM Universe contains 1334 clusters & superclusters at the percolation threshold. Of these the smallest is quasi-spherical with a radius of a few Megaparsec while the largest is extremely filamentary and percolates through the entire simulation box (see figure 3.26). From Figures 3.26 & 3.25 we find that the percolating $\Lambda$CDM supercluster is a slim but massive object.[¶] It contains 4.5% of the total mass in the Universe yet occupies only 0.6% of the total volume. In SCDM the percolating supercluster contains 4.4% of the total mass and occupies 1.2% of the total volume, i.e., a similar mass occupies two times greater volume than in the $\Lambda$CDM. The $\sim 10^3$ $\Lambda$CDM clusters with $\delta \geq \delta_{\mathrm{perc}}$ occupy 4.4% of the total volume and contain 33% of the total mass in the Universe. Gravitational clustering thus ensures that most of the mass in the $\Lambda$CDM Universe is distributed in coherent filamentary regions which occupy very little volume but contain much of the mass. (Of the 1334 $\Lambda$CDM clusters and superclusters identified at the percolation threshold, the ten most voluminous are extremely filamentary and contain close to 40% of the total cluster mass. The remaining 60% of the mass is distributed amongst other 1324 objects !)

The percolating supercluster is more filamentary and less planar in $\Lambda$CDM than in SCDM $^{\parallel}$. The percolating supercluster of the $\tau$CDM is least filamentary and most planar with $(\mathcal{P}, \mathcal{F})_\tau$=(0.15,0.70). It is interesting that superclusters evaluated at $\delta_{\mathrm{perc}}$ are more 'porous' (*i.e.* have a larger genus) in SCDM than their counterparts in $\Lambda$CDM. For instance the percolating supercluster in $\Lambda$CDM has only 6 'holes' which tunnel all the way through it, while 20 tunnels pass through the percolating supercluster in SCDM; see Table 3.1. (This number increases as the density threshold is lowered as demonstrated in Figure 3.27.) Another interesting feature of gravitational clustering appears to be that although the volume occupied by clusters at $\delta_{\mathrm{cluster-max}}$ does not vary much between models, the fractional mass contained in $\Lambda$CDM clusters is *almost twice* that in either $\tau$CDM or SCDM; see Figure 3.25.

---

[¶]Surface visualization is a difficult task especially if we wish to follow tunnels through superclusters to check whether our visual impression of the genus agrees with that calculated using the Euler formula (2.12). MATLAB has been used for surface plotting and the 'reality' of tunnels is ascertained by rotating surfaces and by viewing the supercluster at various angles.

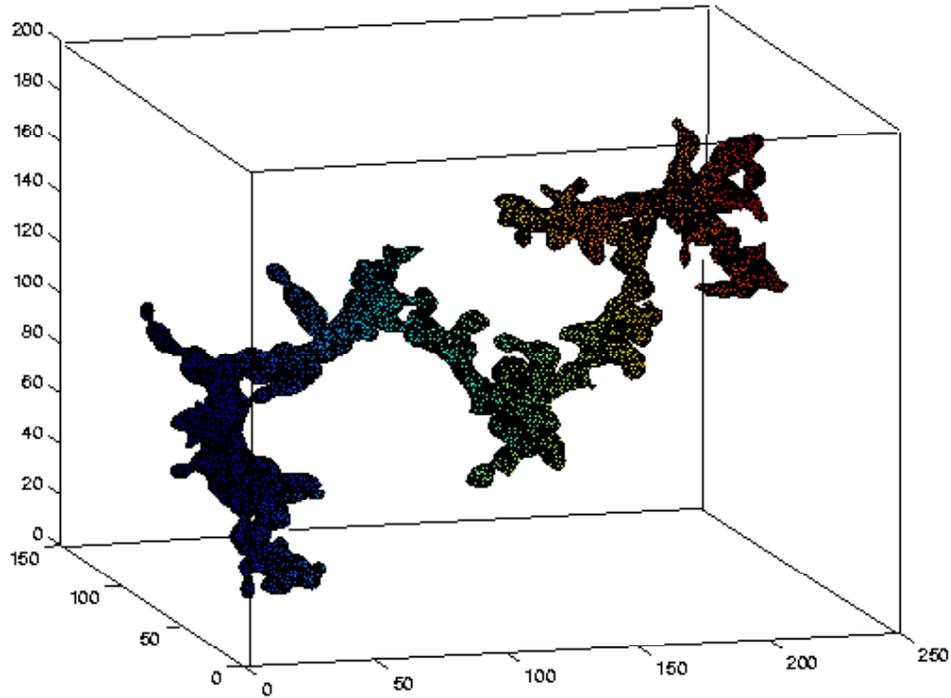

Figure 3.26: The largest (percolating) supercluster in $\Lambda$CDM. This cluster is selected at the density threshold which marks the onset of percolation ($\delta_{\text{perc}} = 2.3$). As demonstrated in the figure, the cluster at this threshold percolates through the entire length of the simulation box. It is important to note that the percolating supercluster occupies only a small fraction of the total volume and its volume fraction (filling factor) is only 0.6%. Our percolating supercluster is a multiply connected and highly filamentary object. Its visual appearance is accurately reflected in the value of the Shapefinder diagnostic assigned to this supercluster: $(\mathcal{T}, \mathcal{B}, \mathcal{L}) = (5.63, 7.30, 70.03)$ $h^{-1}$Mpc and $(\mathcal{P}, \mathcal{F}, G) = (0.13, 0.81, 6)$.

### 3.8.3   Cluster Morphology

Cosmic density fields contain a wealth of information. As demonstrated in Figure 3.25, at density thresholds significantly lower than $\delta_{\mathrm{perc}}$ most clusters merge to form a single percolating supercluster. On the other hand, the slightly larger NCS density contrast, $\delta_{\mathrm{cluster-max}}$, provides an excellent threshold at which to study the morphology of individual objects since it is precisely at $\delta_{\mathrm{cluster-max}}$ that the cluster abundance peaks. We study the morphology and topology of large scale structure in a two-fold manner: (i) properties of all superclusters are analysed at one of the two thresholds: $\delta_{\mathrm{perc}}, \delta_{\mathrm{cluster-max}}$; (ii) the morphology of the largest supercluster is extensively probed as a function of the density contrast.

Figure 3.27 shows the evolution of the morphology and topology of the largest cluster as we scan through different threshold levels of the $\Lambda$CDM density field. The density contrast, volume fraction and genus for a few of these levels have been labelled and are given in the right corner of the figure. We note that at high density thresholds the largest cluster occupies a very small fraction of the total volume and its shape is characterized by significant filamentarity ($\sim$0.67) and negligible planarity ($\sim 0$). We note a sharp increase in the filamentarity of the largest cluster as we approach the percolation threshold (shown in the figure as a solid triangle). At smaller values of the density contrast, $\delta \lesssim 3$, supercluster filamentarity rapidly drops while its planarity considerably increases. The drop in filamentarity of the supercluster is accompanied by a growth in its complexity with the result that, at moderately low thresholds ($\delta \sim 0.5$), the percolating supercluster can contain several hundred tunnels. At very low thresholds ($\delta \lesssim 0.5$) the percolating supercluster is an isotropic object possessing negligible

---

$^{\parallel}(\mathcal{P}, \mathcal{F})_\Lambda$=(0.13,0.81);  $(\mathcal{P}, \mathcal{F})_S$=(0.14,0.78)

| Model | Mass (M/M$_\odot$) | Genus | $\mathcal{T}$ ($h^{-1}$Mpc) | $\mathcal{B}$ ($h^{-1}$Mpc) | $\mathcal{L}$ ($h^{-1}$Mpc) | $\mathcal{P}$ | $\mathcal{F}$ |
|---|---|---|---|---|---|---|---|
| $\Lambda$CDM | $5.17 \times 10^{16}$ | **6** | **5.63** | **7.30** | **70.03** | **0.13** | **0.81** |
| $\delta = 2.31$ | $1.87 \times 10^{16}$ | 1 | 5.63 | 7.34 | 92.74 | 0.13 | 0.85 |
| | $1.30 \times 10^{16}$ | 1 | 5.47 | 6.93 | 70.00 | 0.12 | 0.82 |
| | $1.41 \times 10^{16}$ | 0 | 6.27 | 7.90 | 102.7 | 0.12 | 0.86 |
| | $7.68 \times 10^{15}$ | 0 | 5.71 | 7.22 | 69.95 | 0.12 | 0.81 |
| $\tau$CDM | **$1.45 \times 10^{17}$** | **19** | **5.42** | **7.36** | **40.94** | **0.15** | **0.70** |
| $\delta = 1.37$ | $4.04 \times 10^{16}$ | 2 | 5.07 | 6.57 | 94.93 | 0.13 | 0.87 |
| | $3.16 \times 10^{16}$ | 1 | 5.20 | 6.94 | 98.53 | 0.14 | 0.87 |
| | $2.45 \times 10^{16}$ | 1 | 5.16 | 6.71 | 78.47 | 0.13 | 0.84 |
| | $1.99 \times 10^{16}$ | 1 | 5.66 | 7.26 | 51.44 | 0.12 | 0.75 |
| SCDM | **$1.66 \times 10^{17}$** | **20** | **5.14** | **6.88** | **54.57** | **0.14** | **0.78** |
| $\delta = 1.19$ | $4.15 \times 10^{16}$ | 2 | 4.90 | 6.38 | 107.94 | 0.13 | 0.89 |
| | $3.43 \times 10^{16}$ | 3 | 5.19 | 6.96 | 58.27 | 0.15 | 0.79 |
| | $2.02 \times 10^{16}$ | 3 | 5.47 | 7.21 | 29.47 | 0.14 | 0.61 |
| | $1.85 \times 10^{16}$ | 2 | 5.19 | 7.08 | 42.05 | 0.15 | 0.71 |

Table 3.1: For the sake of brevity, only five out of 10 most voluminous superclusters (determined at the percolation threshold) are shown (for the full list, see Sheth et al., 2003). Listed are mass, genus and Shapefinders $\mathcal{T}, \mathcal{B}, \mathcal{L}, \mathcal{P}, \mathcal{F}$. The first row in each cosmological model describes the percolating supercluster and appears in boldface. It is interesting that in all three cosmological models the top ten superclusters contain roughly 40% of the total mass in overdense regions with $\delta \geq \delta_{\mathrm{perc}}$. (The precise numbers are 40% for $\Lambda$CDM, 37.8% for SCDM and 45.6% in $\tau$CDM.) It should also be noted that the mass of a typical supercluster in $\Lambda$CDM is somewhat smaller than that in the other two models mainly on account of the fact that the adopted particle mass in $\Lambda$CDM simulations is smaller than in simulations of SCDM and $\tau$CDM. (The particle mass in $\Lambda$CDM is $6.86 \times 10^{10} h^{-1} M_\odot$ and that in SCDM and $\tau$CDM is $2.27 \times 10^{11} h^{-1} M_\odot$; see Jenkins et al., 1998). It should be noted that the interpretation of $\mathcal{L}$ as the 'linear length' of a supercluster can be misleading for the case of superclusters having a large genus. In this case $\mathcal{L} \times (G + 1)$ provides a more realistic estimate of supercluster length since it allows for its numerous twists and turns. The morphology of the objects is conveyed through their planarity $\mathcal{P}$ and filamentarity $\mathcal{F}$. We note that the most voluminous and massive structures are highly filamentary in all the models. The largest supercluster is most filamentary in case of $\Lambda$CDM and least so in case of $\tau$CDM.

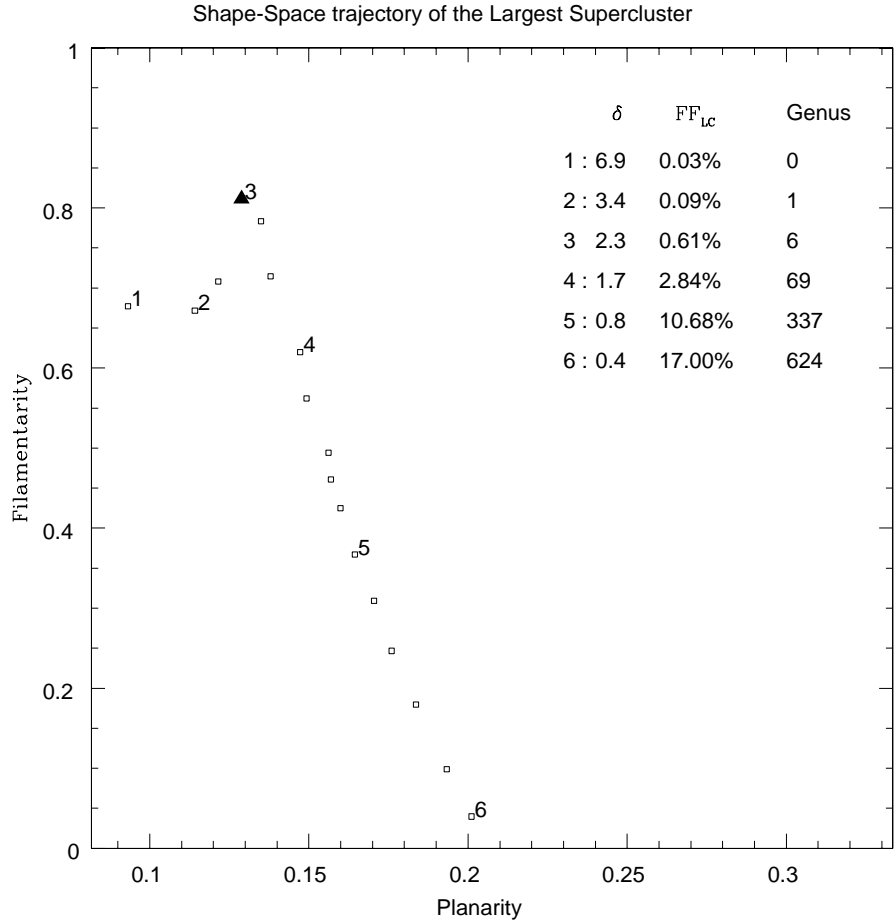

Figure 3.27: The morphological evolution of the largest supercluster in $\Lambda$CDM is shown as a series of open squares in shape-space $\{\mathcal{F}, \mathcal{P}\}$. Each square corresponds to a different value of the density threshold which is progressively lowered from a large initial value ($\delta \simeq 6.9$; left most square) until the mean density level ($\delta = 0$; lower most square). The legend lists the density contrast, the associated volume fraction and the genus of the largest supercluster at six monotonically decreasing values of the density contrast ($1 \rightarrow 6$). At the highest ($\delta \simeq 6.9$), the largest cluster appears to have a large filamentarity and a small planarity. The solid triangle (labelled 3) refers to the percolation threshold at which the largest cluster first spans across the simulation box. After percolation, the filamentarity rapidly decreases from a maximum value of 0.81 to 0.0 as the threshold is lowered from $\delta \simeq 2$ to $\delta \simeq 0$. The decline in filamentarity of the largest cluster is accompanied by growth in its complexity as revealed by its genus value, e.g., the fractional volume occupied by the largest cluster at the mean density is $\sim 26\%$ and its genus exceeds a thousand.

| Models | M | $\mathcal{T}$ | $\mathcal{B}$ | $\mathcal{L}$ | $\mathcal{P}$ | $\mathcal{F}$ |
|---|---|---|---|---|---|---|
| $\Lambda$CDM - $\tau$CDM | 0.093 | 0.082 | 0.070 | 0.037 | 0.15 | 0.101 |
| | $\mathbf{9.4 \times 10^{-4}}$ | $5.0 \times 10^{-3}$ | 0.026 | 0.56 | $\mathbf{6.5 \times 10^{-9}}$ | $\mathbf{2.2 \times 10^{-4}}$ |
| $\Lambda$CDM - SCDM | 0.077 | 0.079 | 0.067 | 0.049 | 0.12 | 0.074 |
| | $\mathbf{4.6 \times 10^{-3}}$ | $\mathbf{2.9 \times 10^{-3}}$ | 0.018 | 0.16 | $\mathbf{3.2 \times 10^{-7}}$ | $6.3 \times 10^{-3}$ |
| $\tau$CDM - SCDM | 0.034 | 0.021 | 0.023 | 0.042 | 0.052 | 0.046 |
| | 0.54 | 0.97 | 0.92 | $\mathbf{0.29}$ | $\mathbf{0.096}$ | $\mathbf{0.19}$ |

Table 3.2: The Kolmogorov-Smirnov statistic $d$ (first row for each pair of models) and the probability (second row) that the two data sets are drawn from the same distribution. The models $\Lambda$CDM, $\tau$CDM and SCDM are compared on the basis of mass (M) and the Shapefinders: thickness ($\mathcal{T}$), breadth ($\mathcal{B}$), length ($\mathcal{L}$), planarity ($\mathcal{P}$), and filamentarity ($\mathcal{F}$). Boldface highlights the three *smallest probabilities* for each pair of models.

values of both planarity ($\mathcal{P}$) and filamentarity ($\mathcal{F}$).

Since $\delta_{\mathrm{perc}}$ and $\delta_{\mathrm{cluster-max}}$ contain information pertaining to morphology and connectivity, we compile partial MFs for individual clusters at both these thresholds for the three cosmological models $\Lambda$CDM, SCDM and $\tau$CDM. We find it convenient to work with the cumulative probability function (CPF) which we define as the normalized count of clusters with the value of a quantity Q to be greater than q at a given value 'q'. The quantity Q could be one of the MFs or one of the Shapefinders. We shall study the dependence of CPF with Q on a log-log scale.

Our results are presented in Figures 3.28, and 3.29 for clusters compiled at the NCS threshold $\delta_{\mathrm{cluster-max}}$. For large values of the MFs $(V, A, C)$ the CPF declines rapidly. Although the curves may look similar, some of them are statistically very different from their peers. The various columns of Table 3.2 show the results of the Kolmogorov-Smirnov test applied to the distribution of mass and the Shapefinders. In particular the CPF of the masses and planarity clearly distinguish the $\Lambda$CDM model from SCDM & $\tau$CDM, in agreement with Figure 3.28.

Figure 3.29 shows the Cumulative Probability Function for planarity and filamentarity in our cluster sample. The last two columns of Table 3.2 show the results of the KS test for these statistics. For all three pairs they are among three best discriminators of the models. The value of KS statistic $d$ suggests that the signal comes from relatively small values of $\mathcal{P}$ ($< 0.1$) and $\mathcal{F}$ ($< 0.25$). The more conspicuous differences in the tail of the distribution is not statistically significant due to poor statistics. Clusters in $\Lambda$CDM are the least anisotropic, a fact which corresponds to their relatively early formation and therefore longer evolution. From Figure 3.29 we also find that clusters at the NCS threshold are significantly more filamentary than they are planar. This appears to be a generic prediction of gravitational clustering as demonstrated by Arnol'd, Shandarin & Zel'dovich (1982); Klypin & Shandarin (1983); Sahni, Sathyaprakash and Shandarin (1996) and Bond et al.(1996).

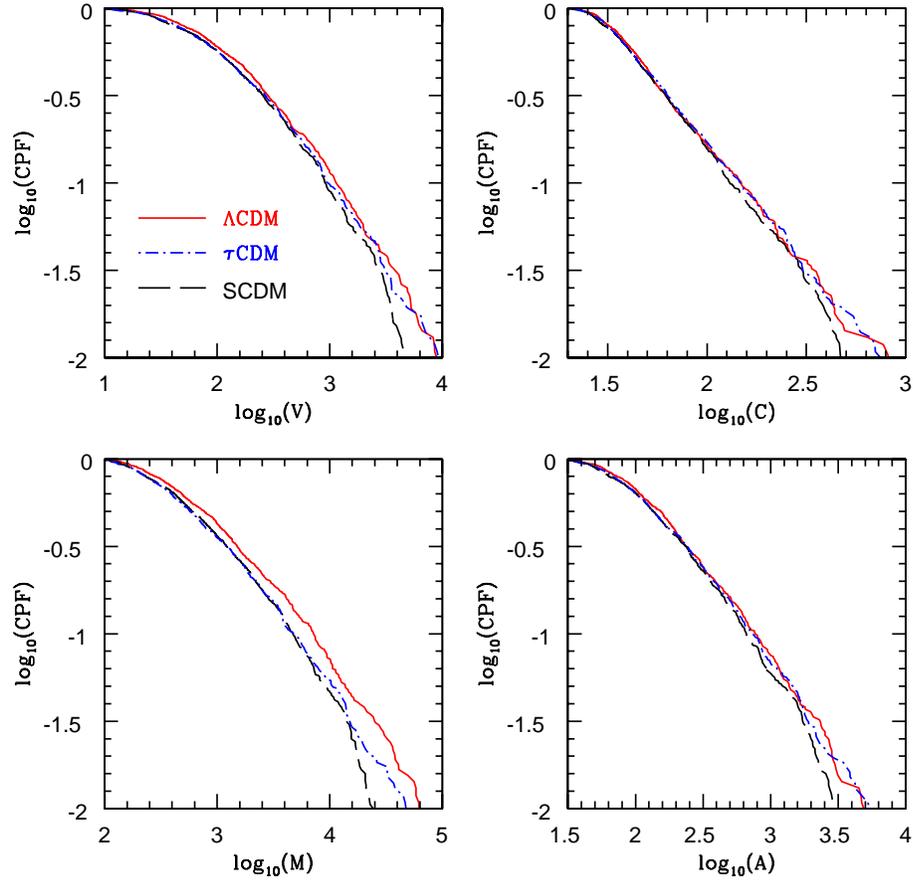

Figure 3.28: Cumulative Probability Function (CPF) is shown as a function of total mass in superclusters (lower left) and as a function of the three Minkowski functionals volume (V), surface area (A) and mean extrinsic curvature (C). Superclusters are defined to be connected overdense regions lying above the NCS threshold $\delta_{\mathrm{cluster-max}}$. (The cluster abundance peaks at $\delta = \delta_{\mathrm{cluster-max}}$ (see figure 3.25) which makes this a convenient threshold at which to probe morphology.) The colour type and line type hold the same meaning across all the panels.

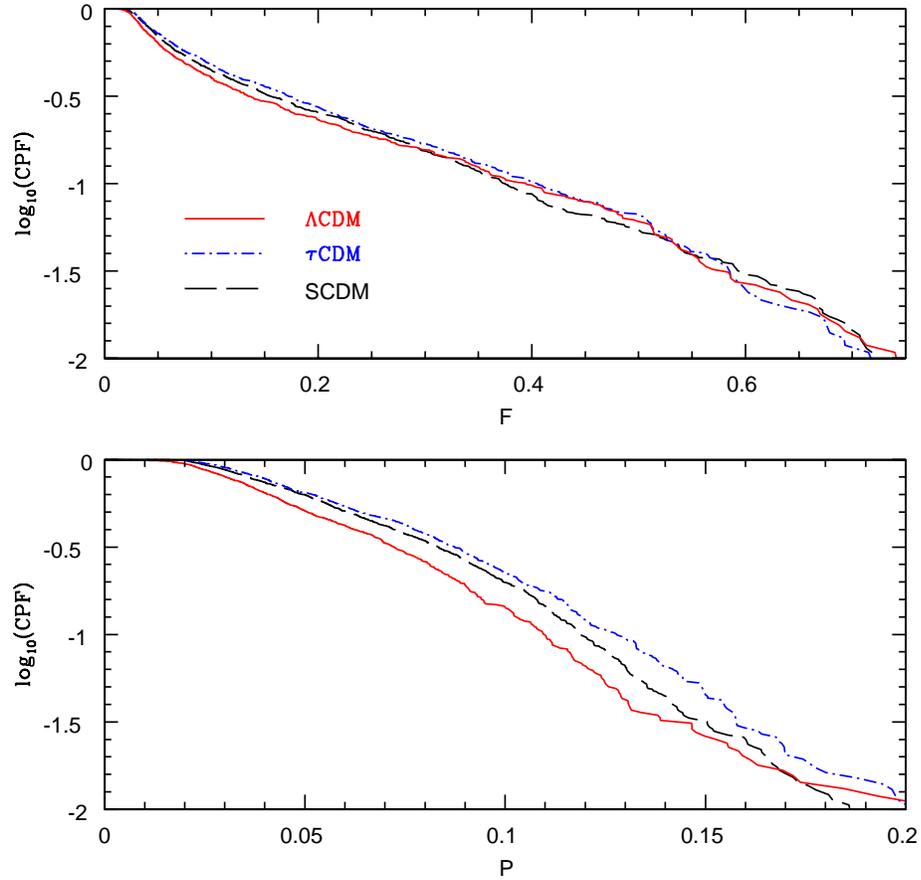

Figure 3.29: Cumulative Probability Function (CPF) for filamentarity ($\mathcal{F}$) and the planarity ($\mathcal{P}$) of superclusters selected at the density threshold $\delta_{\mathrm{cluster-max}}$. The colour type and line type hold the same meaning for both the panels.

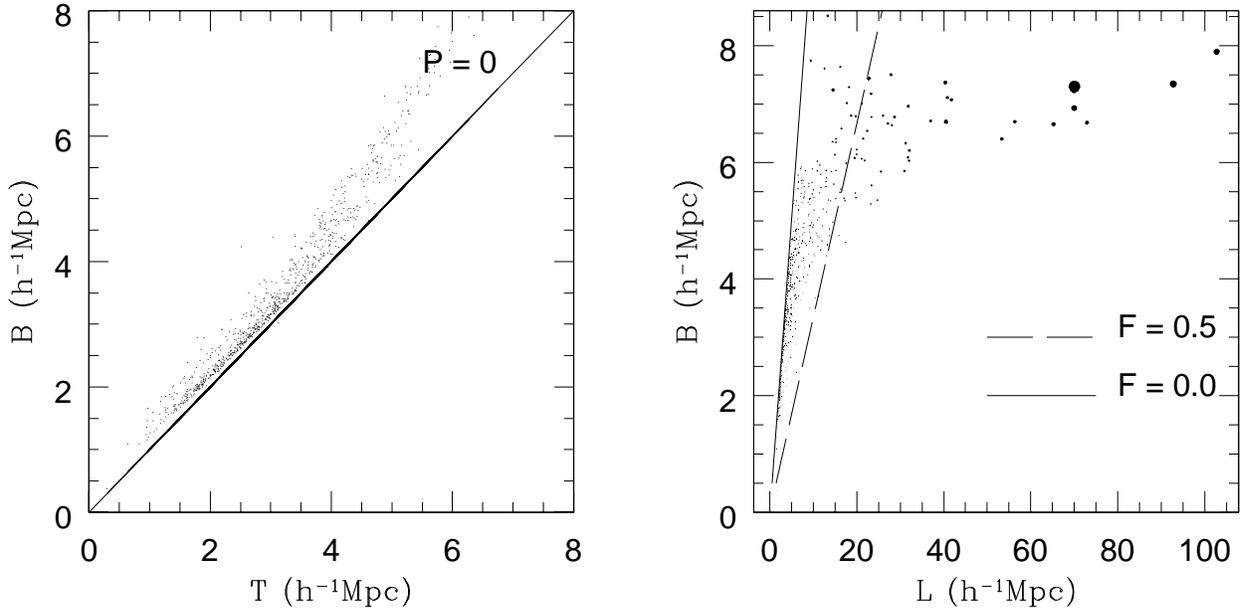

Figure 3.30: Scatter plot for the pair of Shapefinders $\mathcal{T}$ , $\mathcal{B}$ (left panel) and $\mathcal{B}$ , $\mathcal{L}$ (right panel) defining the morphology of clusters/superclusters in the $\Lambda$CDM model. The strong correlation between $\mathcal{T}$ and $\mathcal{B}$ in the left panel near the line $\mathcal{P} = 0$ indicates that two of the three dimensions defining a cluster are equal and of the same order as the correlation length ($\simeq$ few Mpc.). Judging from the left panel we find that clusters/superclusters in $\Lambda$CDM are either quasi-spherical or filamentary (since both satisfy $\mathcal{T}{\simeq}\mathcal{B}{\Rightarrow} \mathcal{P} \simeq 0$). The degeneracy between spheres and filaments is lifted by the right panel which is a mass-weighted scatter plot for the Shapefinders $\mathcal{B},\mathcal{L}$. Each dot in this panel refers to a cluster and its area is proportional to the fraction of mass in that cluster. The concentration of points near the line $\mathcal{F} = 0$ ($\mathcal{B}{=}\mathcal{L}$) reflects the fact that a large number of smaller clusters are quasi-spherical. The more massive structures, on the other hand, tend to be filamentary and the largest and most massive supercluster has $\mathcal{F} = 0.81$. All objects are determined at the percolation threshold.

Figure 3.30 is a scatter plot of Shapefinders $\mathcal{T}$, $\mathcal{B}$, $\mathcal{L}$ for clusters in $\Lambda$CDM defined at the percolation threshold. The strong correlation between $\mathcal{T}$ and $\mathcal{B}$ in the left panel indicates that two (of three) dimensions defining any given cluster assume similar values and are of the same order as the correlation length. (Note that $\mathcal{T} \simeq \mathcal{B} \simeq 5h^{-1}$Mpc for the largest superclusters in Table 3.1.) The clustering of objects near $\mathcal{T} \simeq \mathcal{B}(\mathcal{P} \simeq 0)$ in this panel suggests that our clusters/superclusters are either quasi-spherical or filamentary. The scatter plot between $\mathcal{B}$ & $\mathcal{L}$ in the right panel of Figure 3.30 breaks the degeneracy between spheres and filaments. The mass-dependence of morphology is highlighted in this panel in which larger dots denote more massive objects. This figure clearly reveals that more massive clusters/superclusters are, as a rule, also more filamentary, while smaller, less massive objects, are more nearly spherical. The concentration of points near the 'edge' of the scatter plot in the right panel, arises due to an abundance of low mass compact quasi-spherical objects with $\mathcal{L} \simeq \mathcal{B}(\mathcal{F} \simeq 0)$. The large number of massive superclusters with $\mathcal{F} > 0.5$ serves to highlight the important fact that the larger and more massive elements of the supercluster chain consist of highly elongated filaments as much as $\sim 100h^{-1}$Mpc in length, with mean diameter $\sim 5h^{-1}$Mpc (see Table 3.8.2). It is important to reiterate that almost 40% of the total overdense mass resides in the ten largest objects (the largest 5 of these are listed in Table 3.1), while the remaining 60% is distributed among 1324 clusters.

We also find short filaments to be generally thinner than longer ones in agreement with predictions made by the adhesion model with regard to the formation of hierarchical filamentary structure during gravitational clustering (Kofman et al. 1992).

To further probe the morphology of clusters and superclusters we define the notion of shape-space in Figure 3.31. Shape-space is two dimensional with the planarity (of a cluster) plotted along the x-axis and its filamentarity along the y-axis. (One can also incorporate a third dimension showing the genus.) The first panel in Figure 3.31 is a scatter plot of $\mathcal{P}$ and $\mathcal{F}$ for clusters in $\Lambda$CDM. The sizes of dots in the middle panel are proportional to cluster mass. We note that the most massive structures are also very filamentary. In the right panel we try to relate the shape of the structures with their topology by scaling the size of the dots with the genus value of clusters having a given morphology $(\mathcal{P}, \mathcal{F})$. As shown here, clusters which are multiply connected (larger dots are indicative of more complicated topologies) are also more filamentary. Together, the three panels show us that more massive superclusters are frequently very filamentary and often also topologically quite complex. We also see that a large number of less massive superclusters are simply connected and prolate. These structures are a few Mpc across along their two shorter dimensions and $\sim 20$ Mpc along the third, and therefore have appreciable filamentarity ($\mathcal{F} \sim 0.3$).

## 3.9    Conclusions

In this chapter, we have applied our method of quantifying the LSS using MFs through SURFGEN to a set of cosmological simulations of large scale structure performed by the Virgo consortium. In particular, we pay special attention to $\Lambda$CDM and quantify the supercluster-void network in this model. In addition, we study the geometry and topology of large scale structure in three cosmological scenarios $-\Lambda$CDM, $\tau$CDM and SCDM. All three cosmologies are analysed at the present epoch (z=0) using global MFs, partial MFs and Shapefinders. Our main conclusions of this study are summarised below:

- Individual superclusters totally occupy no more than about 5% of the total volume and comprise no more than 20% of mass if the largest (*i.e.* percolating) supercluster is excluded (Figure 3.3).

- The maximum of the total volume and mass comprised by all superclusters except the largest one is reached approximately at the percolation threshold: $\delta \approx 1.8$ corresponding to $FF_C \approx 0.07$.

- Individual voids totally occupy no more than 14% of volume and contain no more than 4% of mass if the largest void is excluded (Figure 3.3).

- The maximum of the total volume and mass comprised by all voids except the largest one is reached at about the void percolation threshold: $\delta \approx -0.5$ corresponding to $FF_V \approx 0.22$.

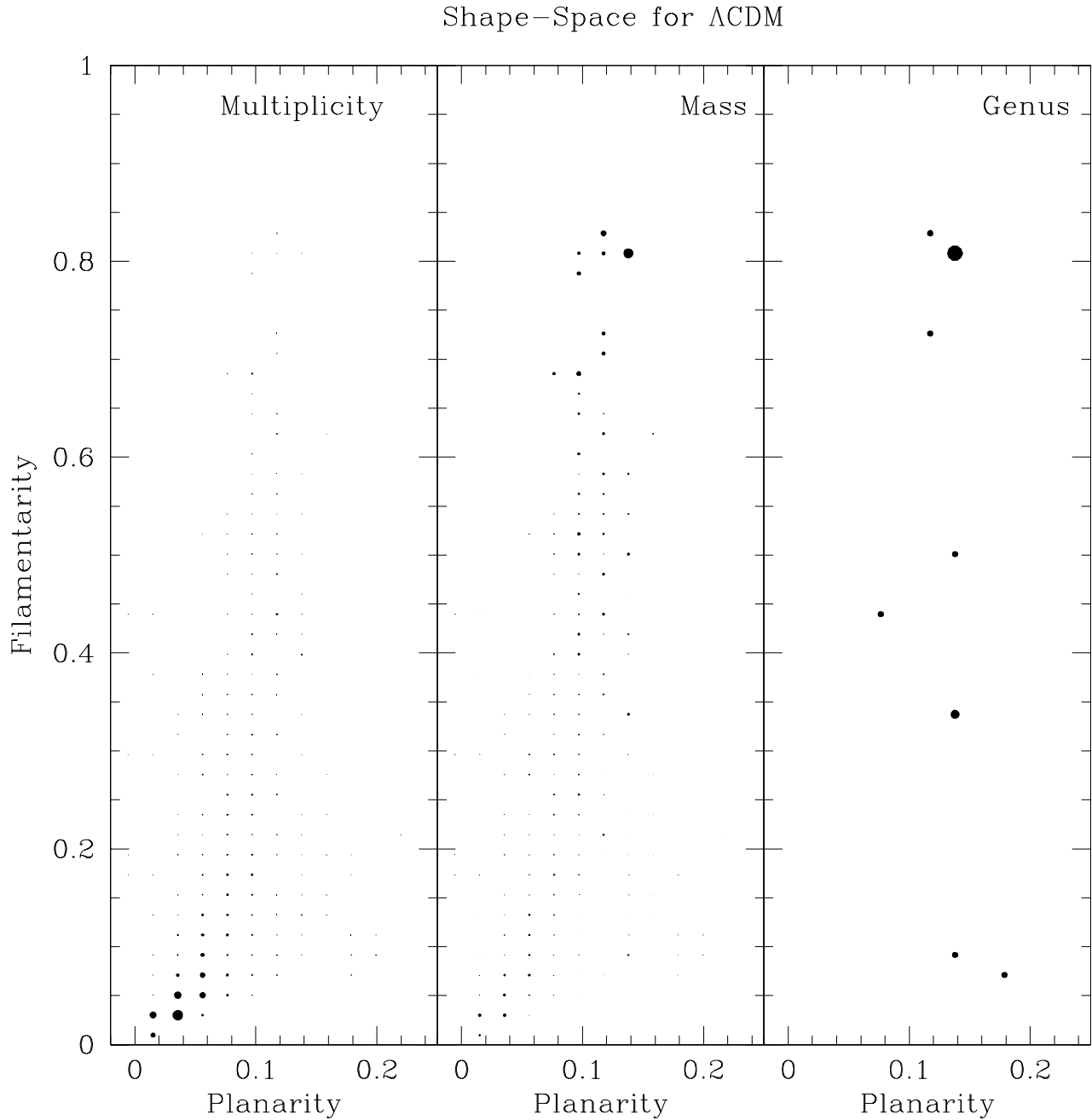

Figure 3.31: Shape-Space for ΛCDM. *Left panel* (multiplicity): The dots in this panel have area proportional to the number of clusters with a given (binned) value of Filamentarity and Planarity. Center panel (mass): The dots have area proportional to the total mass contained in clusters having a given (binned) value of Filamentarity and Planarity. Note that more massive superclusters are also more filamentary. *Right panel* (Genus): The dots have area proportional to the total genus value of superclusters with a given (binned) value of Filamentarity and Planarity. This figure demonstrates the correlation between the mass of a supercluster, its shape and its genus. More massive superclusters are, as a rule, very filamentary and also topologically multiply connected. As in the previous figure, all objects are determined at the percolation threshold.

- Between these two percolation thresholds all superclusters and voids except the largest ones take up no more than about 10% of volume and mass. Both largest supercluster and void span throughout the whole space and have a very large genus. Therefore they have no well defined sizes, volumes, masses or easily defined shapes.

- Although superclusters are more massive and voids are more voluminous the difference in maximum volumes and masses reached at the corresponding percolation thresholds is not much greater than an order of magnitude (see Figure 3.10).

- The volumes, masses and geometrical sizes of superclusters increase as the density threshold decreases and reach maximum values at about the percolation threshold ($\delta \approx 1.8$). At lower thresholds all parameters decrease as the threshold continues to decrease.

- The sizes of voids are significantly larger than those of superclusters even in the density field smoothed with $L_s = 5\ h^{-1}\ Mpc$.

- The length of a quarter of the most massive superclusters exceeds 50 $h^{-1}$ Mpc. The most voluminous voids are even longer: 25% of them are longer than 60 $h^{-1}$ Mpc. The longest non-percolating supercluster is as long as 100 $h^{-1}$ Mpc and the longest non-percolating void is as long as 200 $h^{-1}$ Mpc. Both are comparable to the size of the box (239.5 $h^{-1}\ Mpc$) and therefore may be affected by the boundaries.

- The genus value of individual superclusters can be $\sim 5$ while the genus of individual voids can reach $\sim 55$. This implies significant amount of substructure in superclusters and voids.

- Voids have considerably more developed substructure than superclusters. This is in a qualitative agreement with studies of voids of galaxies (Grogin & Geller 2000; Peebles 2001).

- One of our main results is that voids, as defined through the density field can be distinctly non-spherical. Whether this result carries over to voids in galaxy surveys will depend upon the nature of the baryon-dark matter biasing and also on whether the density field is sampled in real or in redshift space. Since gravitational lensing probes the density field directly, our results are likely to be of some relevance both for ongoing as well as future weak lensing surveys of large scale structure.

- The planarities of both superclusters and voids are quite low $\mathcal{P} \lesssim 0.3$. This implies that the pancake-like structures in the dark matter density in real space are not typical in the $\Lambda$CDM model. We stress that this conclusion may be affected by the size of the smoothing window.

- The percolation thresholds as well as some other parameters depend on the smoothing scale and for smaller smoothing scales or adaptive filtering windows the supercluster percolation threshold must decrease ($FF_C^{perc.} < 0.07$) and the void percolation threshold increase ($FF_V^{perc.} > 0.22$).

- Using the Minkowski functionals we show that, like other diagnostics of clustering, supercluster morphology too is sensitive to the underlying cosmological parameter set characterising our Universe. Although the three cosmological models considered by us, $\Lambda$CDM, $\tau$CDM and SCDM display features which are qualitatively similar, the geometrical and morphological properties of these models are sufficiently distinct to permit differentiation using MFs. We demonstrate that studying Minkowski Functionals using mass-parameterization significantly enhances their discriminatory power.

- A study of percolation and cluster abundance reveals several interesting aspects of the gravitational clustering process. For all models, on decreasing the density threshold to progressively smaller values one reaches the critical percolation threshold at which the largest supercluster runs through (percolates) the simulation box. Percolation is reached at moderate values of the density contrast ranging from $\delta_{\mathrm{perc}} \simeq 2.3$ for $\Lambda$CDM to $\delta_{\mathrm{perc}} \simeq 1.2$ for SCDM. The abundance of clusters reaches a maximum value at (or very near) the percolation threshold and the percolating supercluster occupies a rather small amount of space in all three cosmological models. Thus the fraction of total simulation-box volume contained in the percolating supercluster is least in $\Lambda$CDM (0.6%) and greatest in SCDM (1.2%). When taken together, all overdense objects at the percolation threshold occupy 4.4% of the total volume in the $\Lambda$CDM model. For comparison, the volume fraction in overdense regions at the percolation threshold is $\sim 16\%$ in an idealised, continuous Gaussian random field (Shandarin & Zel'dovich 1989). This fraction can increase upto $\sim 30\%$ for Gaussian fields generated on a grid (Yess

& Shandarin 1996; Sahni et al., 1997).  The fact that clusters and supercluster occupy a very small fraction of the total volume appears to be a hallmark of the gravitational clustering process which succeeds in placing a large amount of mass ($\sim 30\%$ of the total, in the case of $\Lambda$CDM) in a small region of space ($\sim 4\%$). The low filling fraction of the percolating supercluster in $\Lambda$CDM (0.006) strongly suggests that this object is either planar or filamentary (Sahni et al., 1997) and a definitive answer to this issue is provided by the Shapefinder statistic.

- Shapefinders were introduced to quantify the visual impression one has of the supercluster-void network of being a cosmic web of filaments interspersed with large voids (Sahni, Sathyaprakash and Shandarin 1998).  By applying the Shapefinder statistics via SURFGEN to realistic N-body simulations we have demonstrated that: (i) Most of the mass in the Universe is *indeed* contained in large superclusters which are also *extremely* filamentary; the vast abundance of smaller clusters and superclusters tends to be prolate or quasi-spherical.  (ii) Of the three cosmological models the percolating supercluster in $\Lambda$CDM is the most filamentary ($\mathcal{F} \simeq 0.81$) and the supercluster in $\tau$CDM the least ($\mathcal{F} \simeq 0.7$).  (iii) The percolating supercluster in $\Lambda$CDM is topologically a much simpler object than its counterpart in $\tau$CDM, with the former having only $\simeq 6$ tunnels compared to $\simeq 19$ in the latter.  Other differences between the models are quantified in Figures $3.22 - 3.31$.  We also show that among various morphological parameters, the Planarity and Filamentarity of clusters and superclusters is one of the most powerful statistics to discriminate between the models (see Table 3.2).

# Chapter 4

# Morphological Study of Mock Galaxy-Catalogues

## 4.1 Introduction

During the past decade, CMB experiments such as WMAP, type-Ia supernova searches, measurements of the Sunyaev-Zel'dovich effect and measurements of galaxy clustering by SDSS and 2dFGRS have all helped in making cosmology into a precision science. The combined set of measurements appear to indicate that we live in a low matter-density, cosmological constant dominated Universe with $\Omega_\Lambda \simeq 0.7$. A variety of observations such as, the power spectrum study of the MBR, the clustering properties of the galaxy-distribution and the rotational velocity measurements in normal-size galaxies as well as in clusters of galaxies (Kneib et al. 2003), lend support to the standard picture which whereby indicate that the matter density in the Universe is in the form of nonrelativistic and non-baryonic cold dark matter. Thus, the cosmological constant dominated, cold dark matter model ($\Lambda$CDM) seems to provide an excellent description of the Universe that we inhabit. However despite its many impressive successes, several anomalies have surfaced which cannot be easily reconciled with the hypothesis of cold dark matter. For example, the observed number density of dwarfs and low surface brightness galaxies is an order-of-magnitude lower than that predicted by numerical experiments. This and other anomalies relating to the paradigm of cold dark matter are summarised by Tasitsiomi (2002); Sahni(2004). On the other hand, redshift surveys like 2dFGRS, SDSS and 2MASS provide us a window to the distribution of galaxies in our neighbourhood. Whether or not the $\Lambda$CDM model reproduces the clustering properties of the observed galaxy-distributions remains to be confirmed.

Clearly, the issues related to structure formation and background cosmology are likely to have a vigorous interplay with the physics of galaxy formation, and this may bring deeper insights into our ideas about the cosmological model(s) of the Universe. It suffices, hence, to devise algorithms and statistics which enable us to compare our theoretical predictions with different kinds of observations of our Universe. In this chapter, the basic thrust is to develop a methodology to confront theoretical predictions for large-scale structure with data from redshift surveys. For the sake of completeness, we begin with a brief summary of the advances, both on theoretical and observational fronts.

The 2 degree field galaxy redshift survey (2dFGRS) containing about 250,000 galaxies and covering about 1500 square degree of the sky is complete (Colless et al., 2003). At the same time, the Sloan Digital Sky Survey (SDSS) team has recently made its second data release (SDR) reporting the redshifts of more than 360,000 galaxies (Abazajian et al., 2004). The two surveys map out large $3-$dimensional volume $\simeq 0.1 - 1[h^{-1}\mathrm{Gpc}]^3$, which is large enough for a reliable measurement of clustering properties of galaxies. A detailed account of these redshift surveys and their scientific importance has already been presented in Chapter 1.

On the theoretical front, different teams have developed novel ways in which to approach the problem of baryonic structure formation in dark matter dominated cosmogonies. See e.g., Benson et al.(2002) for semi-analytic methods and Weinberg et al.(2002, 2004) for the SPH simulations to study galaxy formation. *Mock* catalogues of "galaxies" prepared using some of these methods are to be eventually tested against galaxy redshift surveys such as 2dFGRS and SDSS. For example, see Yang et al.(2004). The two-point correlation function (hereafter, TPCF), the three-point correlation function, the pair-wise velocity dispersion, etc. are the most widely explored statistics which can be used for this purpose. These statistics reflect the dynamical history and clustering properties of the matter-distribution.

It is important to note that, although TPCF and the three-point function do provide very useful insights into gravitational clustering, they do not provide a complete description of this very complex process. It is well known that gravitational dynamics introduces phase correlations in the initially *random phase*, primordial Gaussian density field (see for example, Chiang & Coles 2000). Because of this, the entire hierarchy of correlation functions gradually becomes nonzero. A full quantification of the cosmic density field as seen in, say 2dFGRS, would ideally require knowledge of all the $n-$point functions. It is important to note that mock catalogues constructed from N−body simulations are usually *required* to reproduce the observed TPCF (Cole et al., 1998, Yang et al., 2003). Thus, mock catalogues of various cosmological model(s) agree with each other and with the large scale structure (hereafter, LSS) of the Universe *by construction*, and therefore cannot be discriminated from one another or from the observed LSS on the sole basis of TPCF. We might reiterate that higher order $n$−point functions (beyond, say, the 3-point function) are cumbersome to determine for a system consisting of a large number of particles (such as in an N−body simulation). It would be useful if a complementary set of statistics will be used to supplement the insight obtained using TPCF and other lower order cummulants. It is also desirable that these statistics simultaneously offer us an intuitively clear interpretation as well as permit numerically robust estimation. In this chapter, we use the Minkowski Functionals (hereafter, MFs) and the Shapefinders – to quantify and characterise the large scale distribution of galaxies. The motivation of using MFs for this purpose has been clarified in earlier chapters. We shall employ SURFGEN to analyse the LSS in the SDSS mock catalogues constructed from two models of structure formation – $\Lambda$CDM and $\tau$CDM. The mock catalogues, prepared by Cole et al.(1998) (hereafter, CHWF), invoke various biasing schemes to prescribe the "formation-sites" of the galaxies for a given cosmological density field. The catalogues fulfill several kinematic as well as dynamical constraints. For example, the survey geometry and the $b_J$ band luminosity functions for both 2dFGRS and SDSS have been incorporated, ensuring an accurate reproduction of the expected radial selection function in both the cases. Further, the mock catalogues due to *all* the models and due to *all* the various biasing schemes were constrained to reproduce the TPCF (Baugh 1996) and the power spectrum (Baugh & Efstathiou 1993) derived from the APM catalogue. The library provides 10 realizations for the $\tau$CDM model, which could be useful in assessing the effect of cosmic variance on the statistics used to quantify the LSS. These catalogues are therefore ideal to test the performance of MFs and their ability in distinguishing between rival scenarios of structure formation.

In the earlier chapter, we studied the geometry, topology and morphology of the dark matter distribution in SCDM, $\tau$CDM and $\Lambda$CDM models simulations of the Virgo group. In this chapter, we discuss the effects of biasing on the morphology of excursion sets in two of these 3 models. By studying constrained "galaxy distributions" in conic volumes with mock catalogues, we hope to confront at least, some of the complications which one might anticipate in dealing with real data. The present exercise provides a theoretical framework with which one might compare theory to observations once full SDSS data sets become available. Our present study is similar in intent to that carried out by Colley et al.(2000). These workers studied the topology of the mock SDSS catalogues. Our thrust will be to test MFs (which include topology) against mock datasets constructed from N−body simulations.

This chapter is structured as follows. Section 4.2 is devoted to a brief summary of the simulations which were used by CHWF to generate their mock catalogues. Here we also describe a subset of the mock catalogues which *we* use in this work. In Section 4.3, we elaborate upon the method used to extract volume limited samples from these mock catalogues for further analysis. Section 4.4 outlines our method of analysis and presents our main results. We conclude in Section 4.5.

## 4.2   The SDSS Mock Catalogues

### 4.2.1   Cosmological Models and their Simulations

Cole, Hatton, Weinberg and Frenk (1998) (CHWF) prepared a comprehensive set of mock catalogues within a variety of theoretically motivated models of structure formation. Each of the models was completely specified in terms of a set of parameters defining the background cosmology and a set of parameters fixing the power spectrum of the initial density fluctuations. The cosmological models studied in this chapter along with their respective simulation parameters are tabulated below:

| Model | $\Omega_0$ | $\Omega_\Lambda$ | $n$ | $\Gamma$ | $\sigma_8 = 0.55 \times \Omega_0^{-0.6}$ |
|-------|-----------|------------------|-----|----------|------------------------------------------|
| $\Lambda$CDM | 0.3 | 0.7 | 1 | 0.25 | 1.13 |
| $\tau$CDM | 1.0 | 0.0 | 1 | 0.25 | 0.55 |

Here, $n$ stands for the power spectrum index, which determines the shape of the primordial power spectrum $P_i(k)$. $\Gamma$ is the shape-parameter which governs the shape of the CDM power spectrum $P(k) = T^2(k).P_i(k)$, where $T(k)$ is the transfer function. According to the conventional definition, $\Gamma = \Omega_0.h \exp(-\Omega_b - \Omega_b/\Omega_0)$. With a choice of $h$ sufficient to explain the age of the Universe, one obtains $\Gamma \simeq 0.5$ for $\Omega_0 = 1$ models. However, the lower value of $\Gamma = 0.25$ is justified in case of $\tau$CDM which involves decay of a massive neutrino. The $\tau$CDM model is similar to standard CDM in having $\Omega_0$=1, but because $\Gamma_\Lambda = \Gamma_\tau$, it has the *same* power spectrum as the $\Lambda$CDM model studied here.

The above simulations were evolved up to the present epoch on a physical box of size 346.5 $h^{-1}$Mpc. We refer the reader to CHWF for further details.

## 4.2.2 Selecting the formation sites for galaxies

The simulations described above provide the distribution of dark matter. In order to prescribe the preferred sites for the formation of galaxies within a given mass-distribution, CHWF employed a set of models which invoke a local, density-dependent bias.

The values of the parameters in the parametric, bias-prescribing functions were obtained by constraining the distributions of "galaxies" within each of their cosmological simulations to reproduce two-point correlation function on the scales between 1 to 10 $h^{-1}$Mpc (Baugh 1996). The $\Lambda$CDM and $\tau$CDM models which *we* investigate in this paper were populated with galaxies by following a selection probability function given by

$$
\begin{aligned}
P(\nu) &\propto \exp(\alpha\nu + \beta\nu^{3/2}) \quad \text{if} \quad \nu \geq 0 \\
&\propto \exp(\alpha\nu) \quad \text{if} \quad \nu < 0,
\end{aligned}
\tag{4.1}
$$

where the dimensionless variable $\nu$ is defined to be $\nu(\mathbf{r}) = \delta_S(\mathbf{r})/\sigma_S$. $\delta_S(\mathbf{r})$ is the density contrast of the *initial* density field smoothed using a Gaussian window on a scale of 3 $h^{-1}$Mpc. $\sigma_S$ is the r.m.s. mass-fluctuation of this field. For the $\Lambda$CDM model, $\alpha_\Lambda$=2.55 and $\beta_\Lambda$=−17.75. The large negative value of $\beta_\Lambda$ indicates that this model requires large anti-bias in the high density peaks of the initial density field in order to reproduce the observed two-point correlation function. For the $\tau$CDM model, $\alpha_\tau$=1.10 and $\beta_\tau$=−0.56. The above probability function was normalised so as to produce a total of $128^3$ galaxies within the cubic box of size $346.5h^{-1}$Mpc, which corresponds to the observed average number density of galaxies with luminosity $\geq$ L$_*$/80. These galaxies are assigned luminosity in accordance with the $b_J$ band Schechter luminosity function. The mock catalogues are prepared in terms of *flux limited* samples with an apparent magnitude limit of 18.9. The SDSS samples mimic the original SDSS survey geometry. Further, these catalogues are constructed to reach a depth corresponding to $z = 0.5$ by periodically replicating the cube of the N−body simulation around the chosen location for the observer.

# 4.3    Data Reduction: Extracting mock samples

## 4.3.1    Preparing Volume Limited Samples

Our program, in this chapter, is to determine whether the geometric and morphological properties of the excursion sets can be used to compare and distinguish between the rival models of structure formation. For this purpose, we shall work with 10 realizations of the mock SDSS catalogue due to $\tau$CDM model and compare our results with those from a single realization of the $\Lambda$CDM model.

Since the superclusters of galaxies are extended objects, one should ensure that no user-specific bias is introduced while studying them. However, precisely such a bias is encountered when one deals with *flux-limited* samples. Such samples are characterised by a selection function which falls off at large radial distances. Consequently, galaxies of a given brightness belonging to the rear end of a typical supercluster are systematically suppressed in number compared to similar galaxies closer by. Thus the sizes and shapes of the superclusters are expected to be grossly distorted in a flux limited sample.

To ensure that our analysis is bias-free, we shall first construct volume limited subsamples from the flux limited mock surveys and then apply our methods to derive morphological and geometrical properties of these volume limited subsamples.

For this purpose, we discard all such galaxies at a given distance r, which fall below the detection limit at the farthest end of the sample, i.e., only those galaxies are retained from the flux-limited sample, which will be visible throughout the volume of the sample. This is equivalent to rendering the selection function spatially invariant.

The selection function can be written as

$$S(d_L) \propto \pi d_L^2 \Delta(d_L) \times \int_{L_{min}}^{\infty} \phi(L) dL, \tag{4.2}$$

where $d_L(z)$ is the luminosity distance corresponding to redshift $z$ for $\Lambda$CDM and $\tau$CDM models, respectively. The quantity $L_{min}(z)$ corresponds to the apparent magnitude limit of the survey and is a monotonically increasing function of $z$. It is obtained by setting the apparent magnitude limit ($b_J$=18.9) in

$$M_{b_J} - 5 \log_{10} h = b_J - (e + k) - 5 \log_{10}(d_L(z)/h^{-1} Mpc) - 25, \tag{4.3}$$

to solve for the absolute magnitude $M_{b_J}^{max}$ of the faintest galaxy detectable at redshift $z$. The factor $e$ accounts for the luminosity evolution of galaxies with redshift and $k$

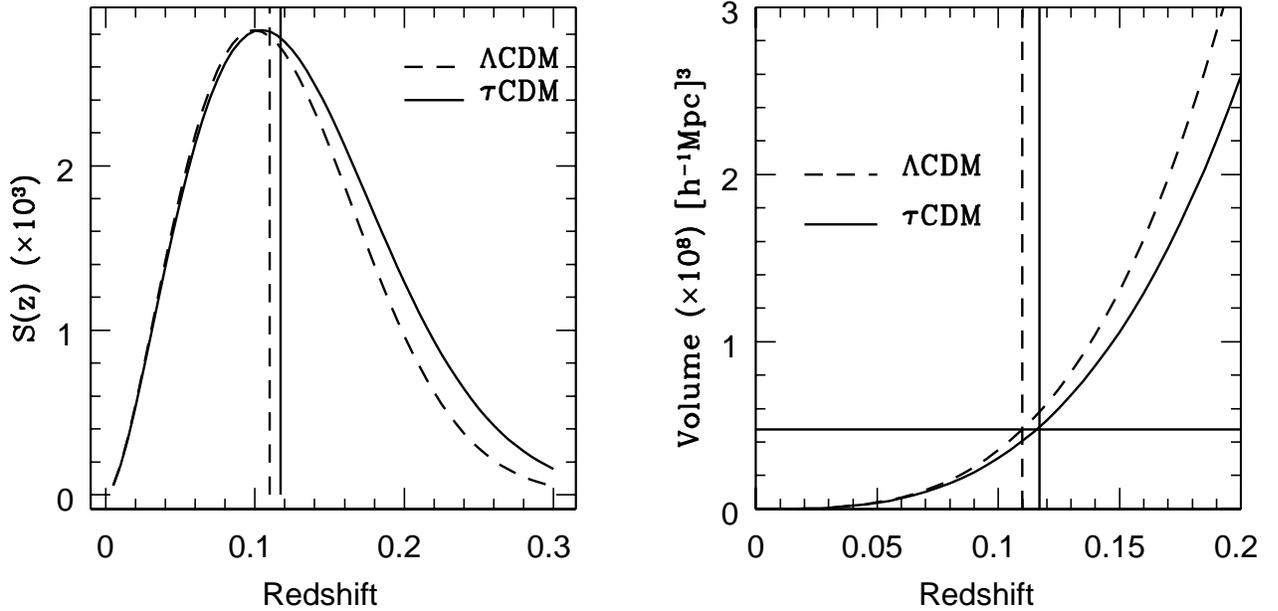

Figure 4.1: *Left panel*: The selection function is shown for both the models. S($z$) peaks around $z$=0.11 in both the cases and falls off on either side. The solid vertical line refers to the limiting redshift ($z$=0.117) for the $\tau$CDM volume limited samples. The dashed vertical line signifies the limiting redshift ($z$=0.11) for the $\Lambda$CDM model. All the samples include $\sim$2.19$\times$10$^5$ galaxies, which corresponds to a mean inter-galactic separation of $\sim$6 $h^{-1}$Mpc. *Right panel*: The survey-volume is plotted as a function of the limiting redshift $z$. We notice that the limiting redshifts (solid and dashed vertical lines) refer to similar values for the survey-volume.

stands for the k-correction*. $M_{b_J}^{max}$ is related to $L_{min}(z)$ by the following relation.

$$M_{b_J}^{max} = M_{b_J}^{\odot} - 2.5 \log_{10}\left(\frac{L_{min}(z)}{L_{b_J}^{\odot}}\right). \tag{4.4}$$

Figure 4.1 shows the selection function S($z$) along with the survey-volume in $(h^{-1}\text{Mpc})^3$ for both $\Lambda$CDM and $\tau$CDM models. The selection function S(z) stands for the number of galaxies included in the survey within a spherical shell of radius $d_L(z)$ and an infinitesimal thickness $\Delta d_L$. In both models S($z$) peaks at a redshift $\sim$0.11 and falls off on either side. From the point of view of analysis, one would prefer as large a volume as possible, and a *constant* number density of galaxies to pick out from all parts of this sample-volume. In this regard, the peak of S($z$) signifies a compromise between two competing factors, the number density and the survey volume.

---

*In mock catalogues studied here, ($e + k$) has been set to zero. It turns out that in the simplest models for the luminosity evolution, $e$ has an effect which is opposite to that of the $k-$term and cancels the k-correction.

From this analysis we expect to achieve a fair sample of LSS if we work with a volume-limited sample of depth z∼0.11. Further, care should be exercised so as to work with the same physical volume and similar number of galaxies across *both* the models and *all* the 11 samples. Considering this, we choose the limiting redshifts to be $Z^\tau_{MAX}$=0.117 and $Z^\Lambda_{MAX}$=0.11, respectively. As shown in the right panel of Figure 1, the two redshifts indeed refer to the same physical volume. All our samples have a similar number of galaxies, $N_g \simeq 2.19 \times 10^5$.

We finally have 10 $\tau$CDM samples and a $\Lambda$CDM sample, all having practically identical volume and a similar number of galaxies. Because the mean inter-galactic separation is similar in all 11 samples, we can employ the same smoothing scale in all cases.

Throughout we work with redshift-space coordinates of the galaxies, and do not attempt to correct for the redshift space distortions. The typical box enclosing all the samples is characterised by ($X_{min}$, $Y_{min}$, $Z_{min}$)≃(−322,−290.5,0.0) and ($X_{max}$, $Y_{max}$, $Z_{max}$)≃(322, 290.5, 357), where all the coordinates are measured in $h^{-1}$Mpc. Figure 4.2 shows the central slices of the $\Lambda$CDM and one of the realizations of $\tau$CDM mock catalogues, with a similar number of galaxies. These slices refer to the same set of random numbers defining the phases of the Gaussian initial density field employed in the simulations.

## 4.3.2   Generating Density Fields

Our analysis will be carried out on a continuous density field sampled on a grid with uniform resolution. As noted earlier, the size of the box enclosing a typical sample is ($\Delta$X, $\Delta$Y, $\Delta$Z)≃(644, 581, 357) $h^{-1}$Mpc. We fit a grid of size 184×166×102 onto this box. The cell-size $\ell_g$=3.5 $h^{-1}$Mpc. About 63 per cent of the cells are found to be *inside* the survey boundaries. We first carry out a Cloud-in-Cell (CIC) smoothing by attaching suitable weights to each of the eight vertices of the cube enclosing the galaxy. This procedure conserves mass. About 25 per cent of the cells inside the survey volume acquire nonzero density after CIC smoothing. The mean inter-galactic separation is $\bar{\lambda} \sim \bar{n}^{-1/3} \sim$6 $h^{-1}$Mpc $\sim 1.71 \times \ell_g$. Thus we do expect the CIC-smoothing to leave behind an appreciable fraction of cells with zero density. Ideally, a density field should be continuous, and should be defined over the entire sample volume. To achieve this, we need to smooth the CIC-smoothed density field further. We employ a Gaussian

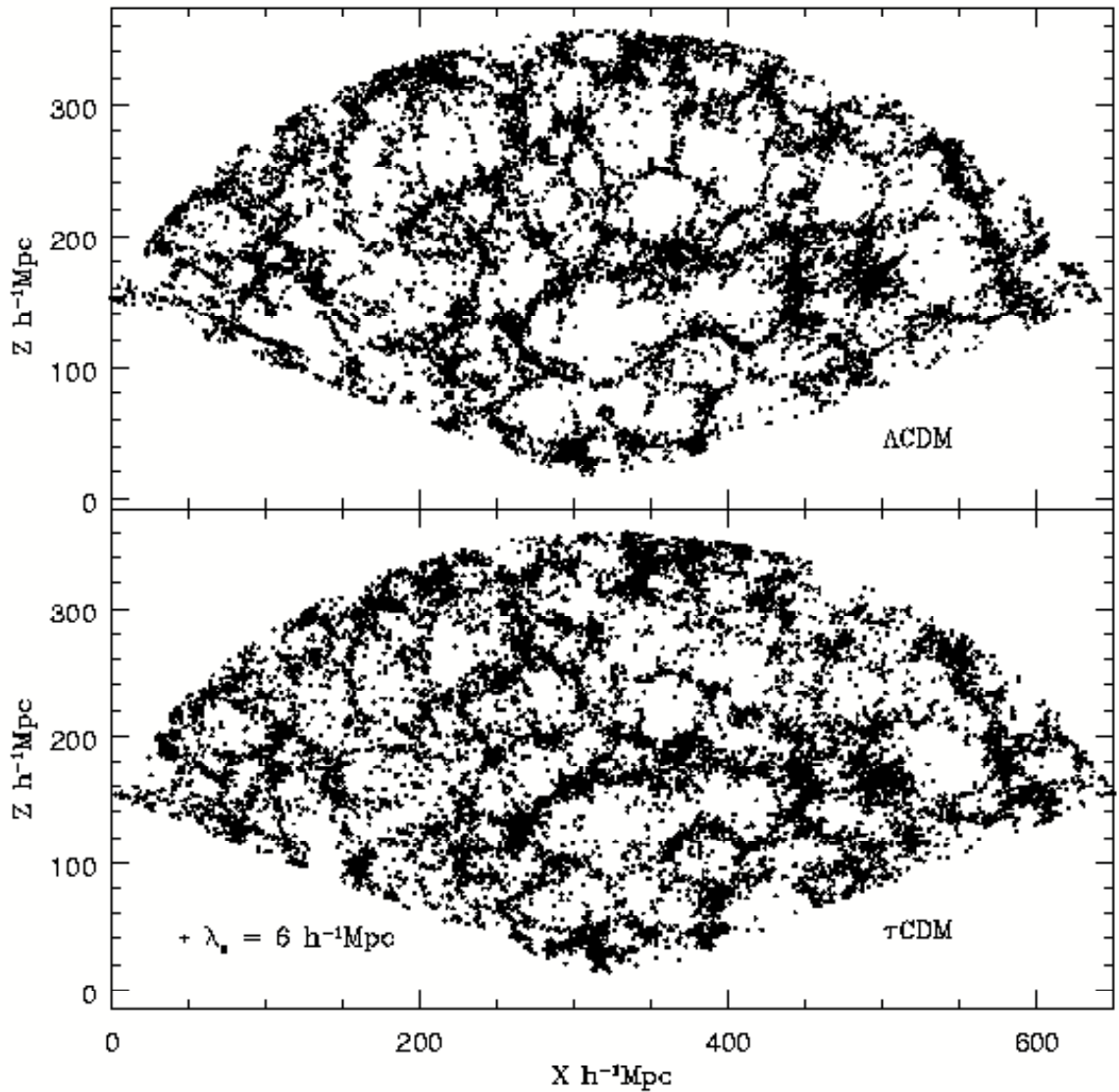

Figure 4.2: Shown here are the central slices of the ΛCDM mock SDSS catalogue and of one of the realizations of its τCDM counterpart. Both the catalogues refer to the same set of random numbers. The coordinates refer to the redshift space positions. The *fingers-of-God* effect is evident from these slices. The two slices qualitatively look quite similar. The voids in ΛCDM are however noted to be much cleaner, and the *fingers-of-God* effect stronger. In our treatment, the galaxy distributions are subsequently smoothed with a Gaussian window of size 6 $h^{-1}$Mpc. In the left corner of the bottom panel we show the window-size to give a feel for the effective smoothing done around every galaxy.

window

$$W_s(\mathbf{r}) = \frac{1}{\pi^{3/2}\lambda_s^3} \exp\left(-\frac{|\mathbf{r}|^2}{\lambda_s^2}\right) \tag{4.5}$$

for this purpose. A Gaussian window, being isotropic, effectively smooths the matter distribution within $\lambda_s \, h^{-1}\mathrm{Mpc}$ of the cell. Since the purpose of our present investigation is to quantify the morphology of coherent structures, the smoothing scale $\lambda_s$ should not be very large. It should be *just enough* to connect the neighbouring structures. While Gott et al.(1989) prescribe using $\lambda_s \geq 2.5\ell_g$ for such applications [†], we prefer to work with $\lambda_s$=6, which is comparable to the correlation length scale in the system, and reasonably smaller than $2.5\ell_g$. As we shall demonstrate below, the behaviour of MFs is remarkably regular even at this smoothing scale. The lower panel of Figure 4.2 shows the adopted smoothing scale in the overall distribution to develop a feel for the effective smoothing done with the samples. It is clear from here that the large voids should be practically devoid of matter *even after* smoothing and should be accessible for morphological analysis at the lowest, sub-zero $\delta$-values, whereas the nearby clusters and superclusters should connect up at a suitable density threshold (corresponding to $\delta \sim 1-2$, as this $\delta$ signifies a transition from linear to non-linear regime). The morphology of these superstructures should be accessible at or near the onset of percolation. In this work, we chiefly focus on the shapes, sizes and geometry of such superclusters and quantify these notions in more detail in Section 4.4.

Prior to smoothing by a Gaussian (or say, a top-hat) window, which requires utilising FFT routines, a suitable mask for the cells outside the survey boundaries has to be assumed. We also note that smoothing leads to a *spill-over* of mass from the boundaries of the survey to the outer peripheries. To conserve mass, we employ the smoothing prescription of Melott & Dominik (1993): we prepare two samples; the first containing the CIC-smoothed density field, with cells outside the survey volume set to zero density, and the second *reference grid* wherein the cells inside the survey volume are all set to unity, while the ones outside the survey boundaries are set to zero. Both density fields are smoothed with the Gaussian window (Eq.4.5) at $\lambda_s = 6h^{-1}\mathrm{Mpc}$. The *reference grid* helps us correct for the effect of having a smoothing window at the boundaries of the survey. We divide the density values of the first grid, cell-by-cell, by the entries for the density values in the reference grid. We find that this corrects for the *spill-over* of mass at the boundaries, and we indeed find mass inside the survey

---

[†]An isotropic smoothing with sufficiently large $\lambda_s$ (say, $\geq 10h^{-1}\mathrm{Mpc}$) will reduce our discriminatory power by diluting the true structures to a great extent. On the other hand, if one intends to compare MFs with analytical predictions, one should employ large scale smoothing in order to reach the quasi-linear regime, in which such predictions are available (Matsubara 2003).

volume to conserve to a very good accuracy.

We shall evaluate MFs for mock catalogues by triangulating isodensity surfaces using SURFGEN. On a dual processor DEC ALPHA machine, SURFGEN calculates MFs for *all* the clusters found at about 100 levels of density on a typical $128^3$ grid within a few hours' time. This amounts to probing the MFs for about 100,000 clusters of varying sizes. This highlights the possibility of using SURFGEN in more ambitious applications involving larger number of realizations studied at higher resolution on larger computers. Our method of analysis and results are described in the next section.

## 4.4 Geometry, Topology and Morphology of Mock Catalogues

The mock SDSS catalogues being studied here, by construction, have the same two-point correlation function in the range 1 to 10 $h^{-1}$Mpc. Since MFs are known to depend on the entire hierarchy of correlation functions (e.g., Schmalzing 1999), they may be expected to differentiate between models as similar as $\tau$CDM and $\Lambda$CDM. This will indeed be the case, as we shall show below.

One of the fascinating aspects of the LSS is related to its visual impression. The LSS as revealed in various dark matter simulations and redshift survey slices reveals a *cosmic web* of interconnected filaments running across the sample, separated by large, almost empty regions, called *voids* (see Figure 4.2 for an illustration). Our study will focus both on global MFs which characterise the full excursion sets of the cosmic density field and partial MFs which, together with Shapefinders, help us study the shapes and sizes of individual superclusters belonging to the cosmic web.

We begin by first studying the global MFs. Next we focus on how the largest cluster evolves with the volume fraction. The last two subsections deal with the morphology of individual superclusters in the two models.

### 4.4.1 Global Minkowski Functionals

We sample the 11 density fields (1 $\Lambda$CDM + 10 $\tau$CDM) at 50 levels of density. The levels uniformly correspond to the *same* set of volume fractions $FF$ for *all* the samples, which makes comparison between models easier.

We find clusters using the grid-version of the friends-of-friends (FOF) algorithm based Clusterfinder code at every level of density. Next we run SURFGEN on all the clusters and estimate their MFs by modelling their surfaces. These are the *partial MFs* and are useful for studying the morphology of individual clusters[‡]. We use the *additivity* property of the MFs and evaluate *global* MFs by summing over the contribution of *partial* MFs due to all the clusters at a given level of density. For $\tau$CDM, we use the measurements from available 10 realizations to produce average global MFs. Figure 4.3 shows the dependence of the two most effective global MFs − mean curvature and genus − on the volume fraction $FF$. The solid curves represent the mean global MFs due to $\tau$CDM model and the error-bars stand for the $1\sigma$ standard deviation. The dashed curves stand for the $\Lambda$CDM model. First, we note that the global MFs of the $\tau$CDM model are extremely well constrained, i.e., the scatter in their measurement is appreciably small. This shows the remarkable accuracy with which SURFGEN will enable us to measure the MFs of the LSS due to large datasets like SDSS and 2dFGRS. We further note that the two models *can be distinguished* from each other with sufficient confidence using the global mean curvature and genus measurements.

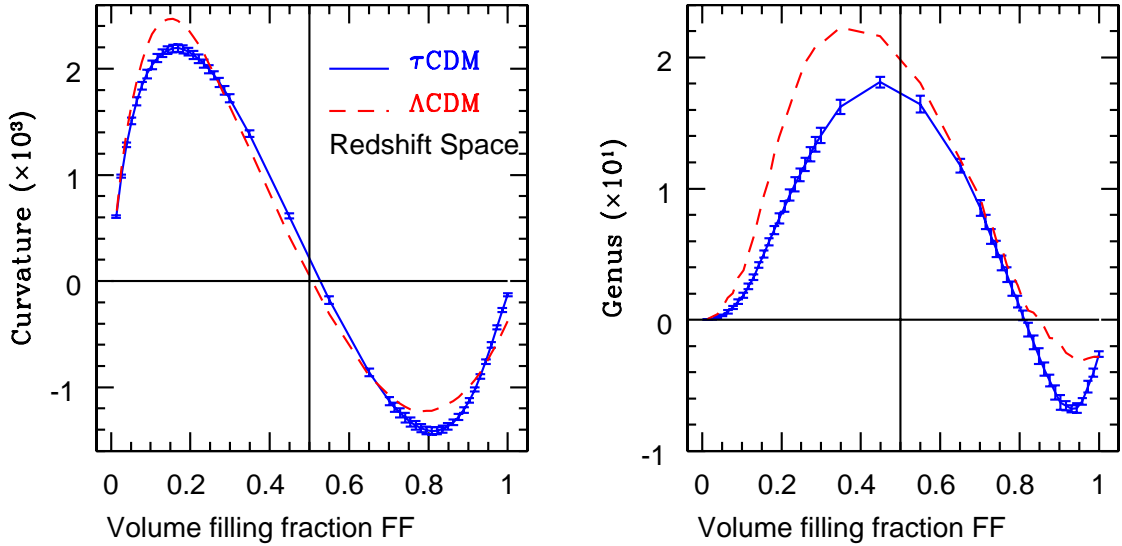

Figure 4.3: Two global MFs − curvature and genus − are evaluated at 50 density levels at a common set of volume filling fractions and are studied here with respect to $FF$. The values are normalised to the volume of $[100\ h^{-1}\mathrm{Mpc}]^3$. The global MFs for $\tau$CDM model are averaged over 10 realizations (solid lines) and the errorbars represent $1\sigma$ deviation. Assuming the same level of accuracy for $\Lambda$CDM, we may conclude that these MFs with volume parametrisation can indeed *clearly* distinguish $\tau$CDM from $\Lambda$CDM. For further discussion, please see the text.

---

[‡]It should be noted however that the "clusters" referred to in the present discussion *do not* coincide with galaxy-clusters. Our clusters are defined as connected, overdense regions and at appropriate level(s) of density may coincide with the superclusters of galaxies.

Distinctly different trends of MFs in two models are due to the contribution of all the $n$−point functions, and do indeed reflect the manner in which the galaxies cluster in $\Lambda$CDM as against $\tau$CDM. This can be called a success of MFs as robust statistics ideal for quantifying the cosmic web.

From the dynamical point of view, these results carry some surprises. As noted by Melott(1990) and Springel et al.(1998), the amplitude of the genus-curve drops as the N−body system develops phase correlations. Given two density fields, the system with larger genus-amplitude shows many more tunnels/voids which are, therefore, smaller in size. As time progresses, the voids are expected to expand and merge, leading to a drop in the genus-amplitude, while the phase correlations continue to grow. With this simple model, one could correlate the amount of phase-correlations with the relative *smallness* of the amplitude of genus, and therefore, of the area and the mean curvature. Going by this reasoning, we conclude from Figure 4.3 that the $\Lambda$CDM galaxy-distribution is relatively more porous and would show *less* coherence on large scales, compared to that due to $\tau$CDM. However, as we illustrated in Chapter 3, the dark matter distribution of $\tau$CDM model due to Virgo group shows considerably larger amplitudes for the MFs compared to the $\Lambda$CDM model , whereas, we find the reverse trend in the MFs of *the galaxy distributions* due to the same two models. Evidently biasing appears to be a source of this effect.

To establish this effect more firmly, we analysed the $\tau$CDM and $\Lambda$CDM dark matter Virgo simulations adopting the same resolution ($\ell_g$=3.5$h^{-1}$Mpc) and the smoothing scale ($\lambda_s$=6$h^{-1}$Mpc) as utilised in the present analysis. Here we randomly chose 25% of the dark matter particles, and studied 5 such realizations for both the models. We confirmed that the MFs are fairly stable even while working with this fraction of particles, and do indeed compare well with MFs computed using the full sample of particles. So as not to introduce any bias due to redshift space distortions, we further computed global MFs of the galaxy catalogues in *real space*. The effect of biasing is most dramatically seen in the global genus-curve of the dark matter and galaxies (see Figure 4.4). We notice that the amplitudes $\mathcal{G}_{DM}^{\Lambda} < \mathcal{G}_{DM}^{\tau}$, whereas $\mathcal{G}_{G}^{\Lambda} > \mathcal{G}_{G}^{\tau}$. Here subscripts $DM$ and $G$ stand for "dark matter" and "galaxies", respectively. We conclude from here that the matter distribution in $\Lambda$CDM is more porous than that in $\tau$CDM, but this trend reverses as we investigate biased distributions of galaxies due to the same two models: the galaxy distribution due to $\Lambda$CDM appears to be *less* porous than that due to $\tau$CDM. This could perhaps mean that a simple scheme of scale-dependent biasing leads to different degree of phase-correlations in the galaxy distribution as compared to the underlying mass distribution. We could interpret this as an instance of *phase-mismatch* between the galaxy-distribution and the underlying dark matter distribution and could attribute it to the biasing prescription invoked. This is supported by the fact that CHWF require large anti-bias in high density regions of the ΛCDM model. To conclude, the study of the global MFs reveals that, local, density-dependent bias could lead to an apparent *phase-mismatch* or different phase-correlations among the dark matter and the galaxies. In passing, we also note that ΛCDM preserves its unique *bubble shift* which is pronounced compared to $\tau$CDM (right panel of Figure 4.3).

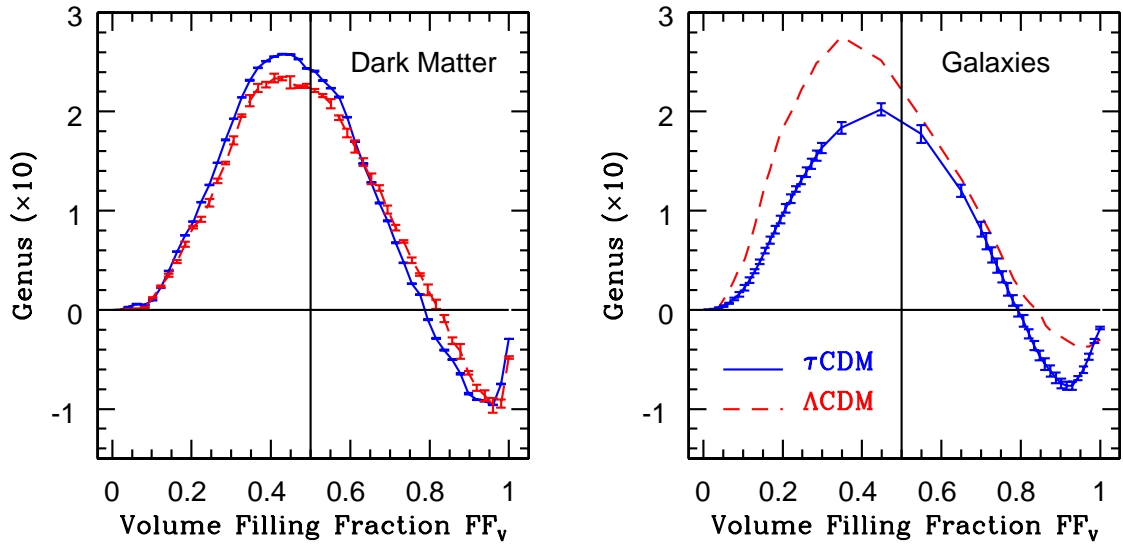

Figure 4.4: *Left panel:* Global genus curves of $\Lambda$CDM and $\tau$CDM *matter distributions* are shown. The values refer to a volume of $[100\ h^{-1}\mathrm{Mpc}]^3$. The $1\sigma$ error-bars are due to 5 realizations of both the models, each with 25% of the total number of particles. Calculations are carried out on a grid of resolution $3.5\ h^{-1}\mathrm{Mpc}$ after smoothing the density fields with $6\ h^{-1}\mathrm{Mpc}$ Gaussian kernel. *Right panel*: The galaxy catalogues due to the two models are analysed in *real* space under identical conditions. This figure illustrates the phenomenon of *phase-reversal*. For more details, please refer to the text.

### 4.4.2   Minkowski Functionals for the Largest Cluster

As we scan the density fields by lowering the density threshold, the largest cluster (hereafter, LC) of the system grows in size due to merger of smaller clusters. The manner of evolution of LC is connected with the percolation properties of the system. We have noted that LC spans most of the volume by the time the volume fraction reaches $FF \sim 0.3$. In fact, most of the features in the global MFs beyond $FF = 0.3$ pertain to the growth of LC. Hence, in this section, we study the evolution of the MFs of LC for $FF \leq 0.3$. We find that, of the four MFs, the integrated mean curvature (hereafter, IMC) and the genus are the best discriminants. In Figure 4.5, we show the IMC and genus for the largest cluster as functions of $FF$. We notice in the right panel of this figure that the genus of LC remains vanishingly small until the onset of percolation in both the models. However, once percolation sets in, $\mathcal{G}_{LC}^{\Lambda}$ grows much more rapidly than $\mathcal{G}_{LC}^{\tau}$. The larger genus is indicative of greater number of tunnels (more porosity) which the ΛCDM structure exhibits relative to its $\tau$CDM counterpart. While the genus in the case of $\tau$CDM seems extremely well-constrained (small scatter), the fluctuations in IMC, at least at smaller $FF-$values ($<0.15$), are reasonably large. This could be inferred as fluctuation in the sizes of the tunnels forming due to the merger of the smaller clusters, while their total number conserves for a given $FF$. The IMC and genus of the largest cluster in the range $FF \in [0.05, 0.3]$ could be some of the best discriminants between the two models.

### 4.4.3   Morphology of the Superclusters

Recently superclusters of galaxies have drawn a good deal of attention. For example, Baugh et al.(2004) report the existence of two massive superclusters at $z \simeq 0.11$ in 2dFGRS, one in the southern and the other in the northern galactic hemisphere (also see Erdogdu et al.2003). Further, Miller et al.(2004) have found a 200Mpc long supercluster in the recently completed 2dFQSO catalogue. On similar lines, Brand et al.(2003) have detected ~100Mpc structure. The recently released SDSS Second Data Release consists of *SDSS Great Wall* which is twice the size of the *CfA Great Wall* (Gott et al.2003).

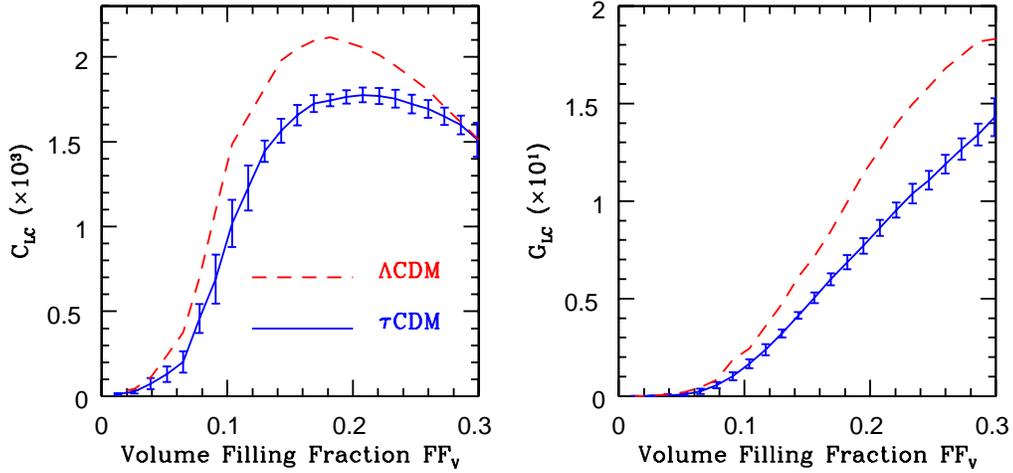

Figure 4.5: The largest cluster (LC) spans most of the sample-volume for $FF > 0.3$. Here we study the dependence of the mean curvature and genus of LC of $\tau$CDM and $\Lambda$CDM w.r.t. $FF$. Note the well-constrained behaviour of both the measures for $\tau$CDM, and the distinctly different trend observed for them in $\Lambda$CDM. Please refer to the text for the discussion.

Since superclusters are still safely within the linear to a weakly nonlinear regime, a statistical analysis of their shapes, sizes etc. may provide a conclusive test for the underlying cosmological parameter-set. In light of the above observational findings, hence, we are motivated to carry out a critical study of the morphology, i.e., shapes and sizes, of the ensemble of superclusters in our mock catalogues. By carrying out a statistical study of many realizations of $\tau$CDM, we are able to establish it *for the first time* that the supercluster shapes and their sizes are very well constrained measures, sensitive to the cosmological parameters of the model(s), and are suitable as excellent diagnostics of large scale structure.

Since we are dealing with density fields, our superclusters are defined as connected, overdense objects. By studying the percolation properties of the density fields, we have found that percolation takes place at a threshold of density corresponding to $\delta = 1.7$ in both $\Lambda$CDM and $\tau$CDM. Furthermore, $\delta_{\max} \simeq 3.5$ in these models at the scale of smoothing which we have employed in the present exercise. Thus even at the highest threshold of density, we are still within the weakly nonlinear regime, and may find sufficiently large connected objects which might act as dense progenitors of the percolating superclusters. With this anticipation, we find it advisable to study the morphology of the ensemble of superclusters as a function of the density contrast over the entire range $\delta \in [0, \delta_{max}]$. In so doing, we are also able to assess how the morphology of the LSS changes with the density-level.

The morphology of the superclusters is quantified in terms of a set of Shapefinders (Sahni et al.1998, Sheth et al.2003). For objects with fairly simple topology, the dimensional Shapefinders give us a feel for the typical thickness ($\mathcal{T}$), breadth ($\mathcal{B}$) and length ($\mathcal{L}$) of the object being studied. Since we have demonstrated that most of the superclusters are simply connected objects near or before the onset of percolation, this interpretation is valid for structures defined in this regime. Further, a combination of ($\mathcal{T}, \mathcal{B}, \mathcal{L}$) can be used to define Planarity ($\mathcal{P}$) and Filamentarity ($\mathcal{F}$) of the object.

$$\mathcal{P} = \frac{\mathcal{B} - \mathcal{T}}{\mathcal{B} + \mathcal{T}}; \quad \mathcal{F} = \frac{\mathcal{L} - \mathcal{B}}{\mathcal{L} + \mathcal{B}}. \tag{4.6}$$

Individual superclusters in a given realization of, say $\tau$CDM, may follow a distribution of shapes and sizes, not necessarily coinciding with a similar list due to another realization. Hence, to make a statistical study of a given model, we average over the shapes of individual superclusters defined at a given threshold of density, and test whether the average morphology as a function of the density-level is well constrained. For this purpose, we use the volume averaged Shapefinders defined as follows.

$$\begin{aligned} \bar{\mathcal{P}}_V(\delta_{th}) &= \frac{\sum_{i=1}^{N_C} \mathcal{V}_i \times \mathcal{P}_i}{\sum_{i=1}^{N_C} \mathcal{V}_i} \\ \bar{\mathcal{F}}_V(\delta_{th}) &= \frac{\sum_{i=1}^{N_C} \mathcal{V}_i \times \mathcal{F}_i}{\sum_{i=1}^{N_C} \mathcal{V}_i}, \end{aligned} \tag{4.7}$$

where the quantity $\mathcal{V}_i$ refers to the volume of the i$^{th}$ supercluster defined at the level $\delta_{th}$, and the summation is over all the objects. It is to be noted here that the average planarity and filamentarity are mostly contributed by the structures with large volume and/or mass. There are a finite number of such objects before the onset of percolation. Hence, in this regime, we probe the *average shape* of these objects. However, once percolation sets in, the average morphology mainly reflects the morphology of the largest object in the system.

We evaluate $\bar{\mathcal{P}}_V$ and $\bar{\mathcal{F}}_V$ for all the 11 samples at the density levels corresponding to $\delta \in [0, \delta_{max}]$. The measurements due to 10 $\tau$CDM realizations are used for averaging and for evaluating the errors. Figure 4.6 shows $\bar{\mathcal{P}}_V$ and $\bar{\mathcal{F}}_V$ studied as functions of $FF$. The solid lines refer to $\tau$CDM and the dashed lines refer to $\Lambda$CDM. The vertical bars stand for the $1\sigma$ errors. We observe that $\sqrt{}$    $\bar{\mathcal{P}}_V$ and $\bar{\mathcal{F}}_V$ are well constrained statistics and hence, can be used to reliably assess the morphology of the LSS.

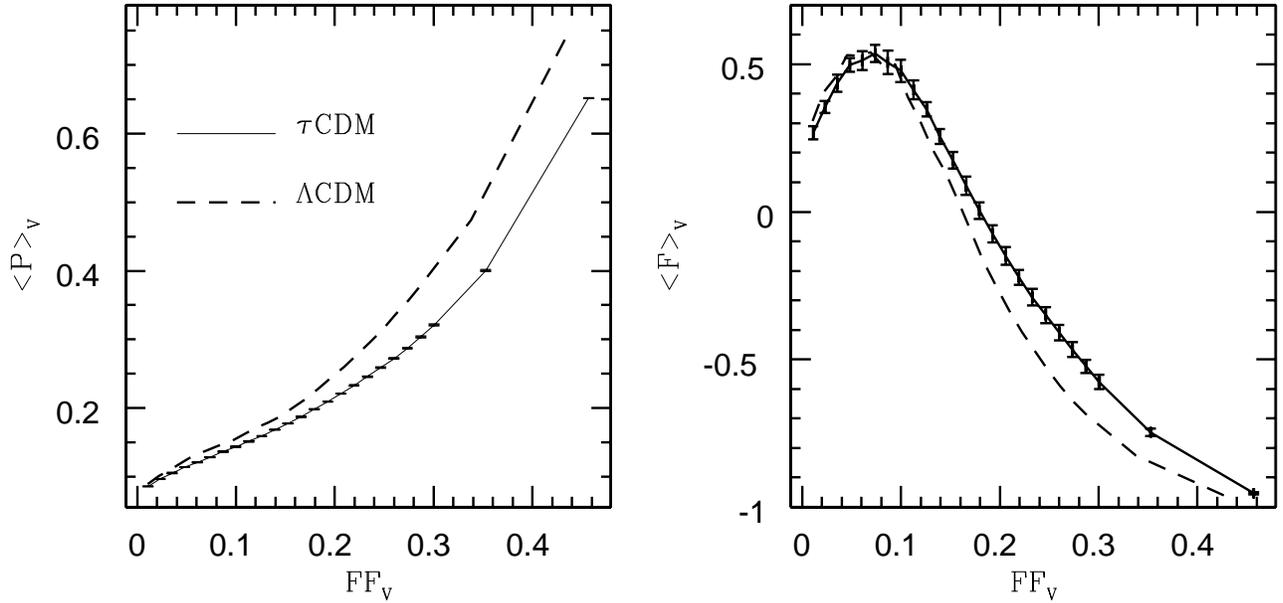

Figure 4.6: The dimensionless Shapefinders $(\mathcal{P}, \mathcal{F})$ are evaluated for all clusters at a given density level $\delta_{\text{th}}$. These Shapefinders are volume-averaged and are studied with respect to the volume filling fraction $FF$. The average shape of the cosmic web assumes increasingly planar morphology as $\delta_{\text{th}}$ is lowered. The corresponding negative filamentarity is due to the presence of large voids which lead to the development of appreciable negative curvature in the system. The average filamentarity in the system is highest near the onset of percolation, thus making it ideal for the study of structures which are evident in the visual impression of the cosmic web. The average morphology revealed in terms of $\bar{\mathcal{P}}_V$ and $\bar{\mathcal{F}}_V$ is extremely well constrained for a given model and could be a useful diagnostic for discriminating between models of structure formation.

$\sqrt{}$    The average planarity of the LSS steadily increases as we lower the threshold of density and increase in the filling factor.

$\sqrt{}$    The average filamentarity initially increases, reaches a maximum (when percolation sets in), but drops to sub-zero values after $FF{\sim}0.18$.

A steady drop in the curvature is responsible for this behaviour. We note that the system also tends to become increasingly spongy. Taken together, these observations lead to the following picture: as the density level is brought down, the overdense structures connect up across the sample-volume. This leads to the formation of large scale filaments and also to the birth of large voids in the system, signifying the onset of percolation. As revealed in the right panel, the filamentarity of the system is highest at the onset of percolation. With a further drop in $\delta$, the filaments surrounding the "voids" grow thicker and locally these assume a slab-like structure. Such an excursion set has large planarity, but also has large negative curvature due to being surrounded by large, almost spherical voids. This negative curvature is responsible for the large negative filamentarity.

Interestingly, $\bar{\mathcal{P}}_V^\Lambda \geq \bar{\mathcal{P}}_V^\tau$ and $\bar{\mathcal{F}}_V^\tau \geq \bar{\mathcal{F}}_V^\Lambda$. The $\tau$CDM cosmic web is *more filamentary* and *less planar* than that in $\Lambda$CDM. This demonstrates that the morphology of LSS, which is visually so apparent and evocative, when quantified using the $(\mathcal{P}, \mathcal{F})$ pair, is a robust statistic to compare LSS due to rival models.

We finally note that superclusters defined according to the conventional definition ($\delta_{SC} \in [1,2]$) do correspond to percolating structures in the system, and indeed exhibit the highest filamentarity (the right panel of Figure 4.6). This proves that the pair of Shapefinders $(\mathcal{P}, \mathcal{F})$ conform to our visual impression while quantifying the LSS. In the next subsection, we shall try to quantitatively asses the sizes of individual superclusters.

### 4.4.4   Cumulative Probability Functions

In this subsection we quantify the sizes of the superclusters and study the variation of these measures as functions of the density contrast. Our main focus is to make a statistical study of these measures and to develop a methodology to handle various statistics due to many realizations of a given model. We also compare the sizes of superclusters in $\tau$CDM with those in $\Lambda$CDM.

For reasons explained earlier, we find it advisable to carry out the exercise over a set of density-levels. We study the sizes of superclusters starting from the highest density level ($\delta \simeq 3.5$) down to the percolation threshold.

Since the sizes of individual structures may vary from one realization to another, it is not meaningful to prepare a list of, say, the N largest superclusters in a given realization and check its statistical significance over other realizations. There may not be any unique correspondence of the rank of the supercluster in the list and its size. This is mainly because at the density-levels above or near the percolation threshold, there may be many structures with comparable sizes and masses. So the sorted array of structures in this range of density-levels will show a large amount of Poisson fluctuations in the size of, say, the i[th] largest cluster. Hence, a more meaningful approach is to study the cumulative probability functions of the Shapefinders so as to know how many superclusters in a given ensemble of them, have their sizes greater than a given triplet $(\mathcal{T}, \mathcal{B}, \mathcal{L})$.

All the 10 samples of $\tau$CDM being studied here have comparable number of clusters at any given threshold of density, and these are only marginally smaller in number than in $\Lambda$CDM. Further, we noted earlier that the global geometry and topology of the $\tau$CDM density fields is extremely well constrained. Hence, it is justified to construct cumulative probability functions (hereafter, CPFs) of various quantities using an ensemble which contains the clusters due to *all* the 10 realizations of $\tau$CDM [§]. The CPF of a given quantity Q is a fraction of the total number of structures having value of Q greater than a chosen value $q$. In our case, the total number of clusters is determined after summing over all realizations of $\tau$CDM and is close to a few thousand. It turns out that it is beneficial to work with such large number of clusters, so that we have sufficient number of structures which are otherwise rare in a given sample. By finding these in sufficient number in a bigger ensemble, we can put an upper limit on the sizes of superclusters at any given threshold of density, without hampering our conclusions by Poisson noise.

To begin with, we study the CPFs of the dimensional Shapefinders, the so called, Thickness $(\mathcal{T})$, Breadth $(\mathcal{B})$ and Length $(\mathcal{L})$ of the superclusters with $FF$ as a parameter. Figure 4.7 shows the results. We notice that the CPFs follow a specific pattern of evolution. While a typical range of thickness and breadth of the structures is fixed ($\mathcal{T} \in (0,13)$ $h^{-1}$Mpc and $\mathcal{B} \in (0,17)$ $h^{-1}$Mpc), the fraction of clusters with a given thickness and breadth increases with increasing $FF$. Thus, the structures become thicker and broader as we lower the threshold, as anticipated. The CPF of length, on the other hand, extends to high $\mathcal{L}-$values upon increasing $FF$. This implies that the longer structures become statistically more significant with the decrease of $\delta$, while their thickness and breadth increases only marginally. Thus, as $\delta \rightarrow \delta_{\mathrm{perc}}$, the filamentarity of the individual superclusters should increase. At the highest thresholds, however, the superclusters might look only marginally filamentary ($\mathcal{F} \sim 0.25$, $\mathcal{P} \sim 0.1$) (see, Figure 4.6), and in realistic situation, may give us an impression of being more like *ribbons*. [¶]. We infer that the top 10 superclusters in $\tau$CDM should have their length $\geq 40$ $h^{-1}$Mpc at $\delta \sim \delta_{\mathrm{max}}$, and these should grow to become as long as $\mathcal{L} \geq 90 h^{-1}$Mpc at the onset of percolation. The longest superclusters at percolation, which are necessarily rare (say, a few in 1000), might be as long as $150 h^{-1}$Mpc.

---

[§]However, prior to this, we have checked that the CPFs due to individual realizations of $\tau$CDM show trends similar to CPFs due to the cumulative ensemble.

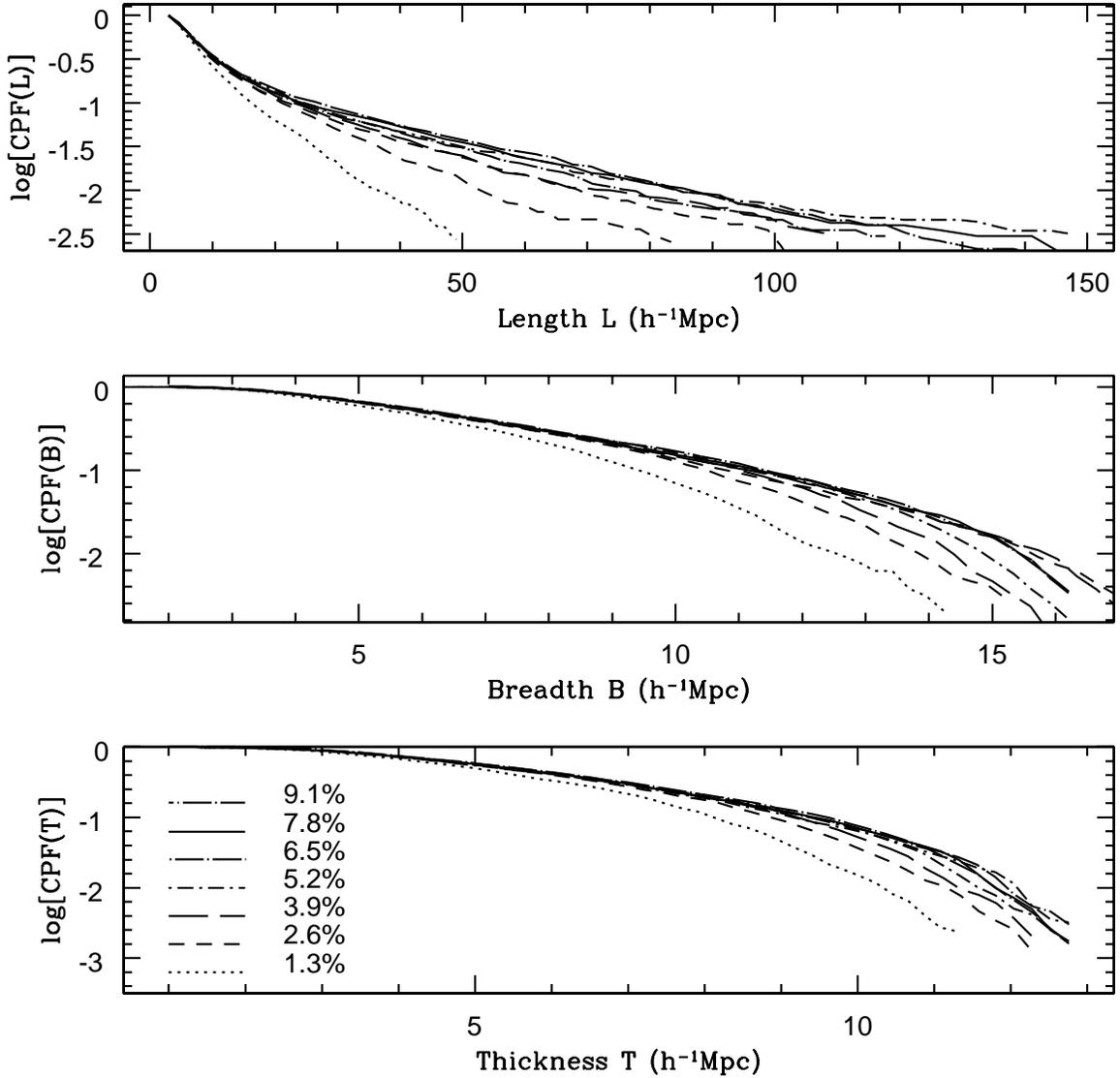

Figure 4.7: The CPFs of dimensional Shapefinders are studied as functions of the density level for $\tau$CDM. The bottom panel shows the associated volume filling fractions. The percolation curves are shown as bold, solid lines. We notice that the clusters grow thicker, broader and longer as we increase the volume fraction (at any given triplet $(\mathcal{T}, \mathcal{B}, \mathcal{L})$, there are a greater number thicker than $\mathcal{T}$, broader than $\mathcal{B}$ and longer than $\mathcal{L}$ as we decrease $\delta$).

We are limited by the availability of only one realization of $\Lambda$CDM. Hence similar conclusions cannot be reliably drawn for this model, but a study of CPFs of $\Lambda$CDM superclusters does reveal similar trends.

In order to test whether we can utilise the length of superclusters to discriminate between $(\Lambda, \tau)$CDM, we studied the CPF of length for both the models. Figure 4.8 shows the results, where the CPF($\mathcal{L}$) have been evaluated at $FF=7.8$ per cent (at the onset of percolation) and at $FF=11.7$ per cent (after the percolation). We notice that at the onset of percolation, the length of the longest $\tau$CDM superclusters could be as large as 90 $h^{-1}$Mpc, whereas their $\Lambda$CDM counterpart structures, which exhibit same degree of statistical significance, are relatively shorter with $\mathcal{L}_{\mathrm{max}} = 55$ $h^{-1}$Mpc. We conclude that the $\tau$CDM superclusters tend to be statistically longer than their $\Lambda$CDM counterpart structures. The large-scale coherence in the superclusters may be attributed to the phase-correlations in the density field; the higher the degree of phase-correlations, the larger the anticipated length-scale of coherence[‖]. Based on this, and our results obtained in Section 3.4.1, we would anticipate the $\tau$CDM superclusters to be *longer than* those in $\Lambda$CDM. As we can see, the result which we report here agrees well with this anticipation. This is a *considerable* success for the ansatz of the Shapefinder quantifying *length*, for as we noted above, it helps us capture the relative effect of phase-correlations among rival models of structure formation. After percolation, the longest structure dominates the survey volume and enters only as a Poisson fluctuation in the

---

[¶]A more detailed study of the morphology in this range of $\delta$ needs to be carried out to understand whether the super-structures with significant planarity are statistically significant or not. This question has an observational bearing because some of the super-structures in our local neighbourhood visually appear to be planar. For instance, see Fairall 1998 and Martinez & Saar 2002.

[‖]The connection between large scale coherence and the degree of phase correlations will be probed in detail in a future work.

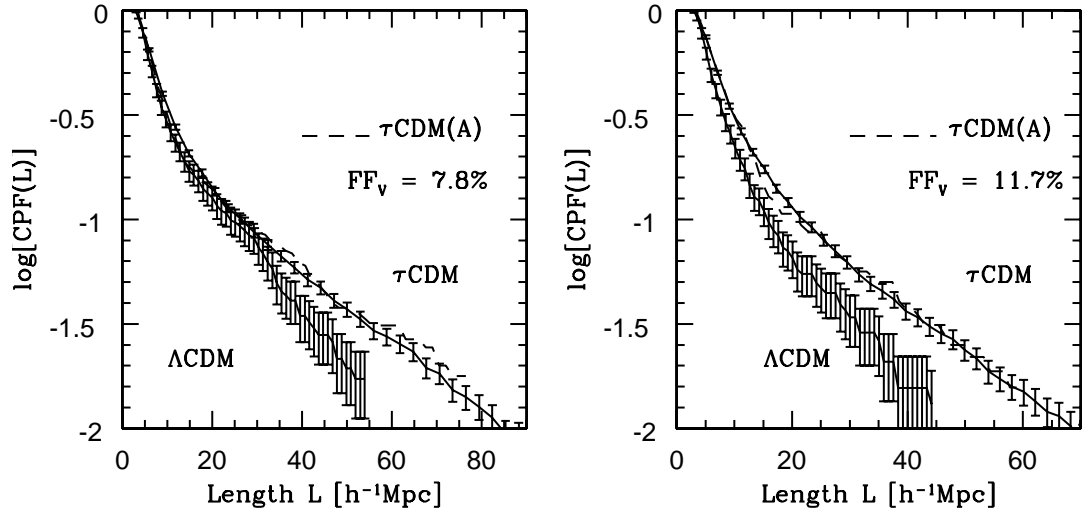

Figure 4.8: Shown here are the cumulative probability functions of length $\mathcal{L}$ for super-clusters of $\tau$CDM and $\Lambda$CDM at two thresholds of density corresponding to $FF$=7.8 per cent (just before the onset of percolation) and $FF$=11.4 per cent (after the on-set of percolation). We find the CPFs of the two models to be distinctly different at longer length-scales. The dashed line refers to the CPFs due to the first realization of $\tau$CDM. The set of random numbers used to generate random phases in the initial Gaussian density field are the same for this realization and the $\Lambda$CDM model studied here. We note that CPF($\mathcal{L}$) from this realization closely follows the CPF constructed from the total ensemble. The largest superclusters of $\tau$CDM are systematically longer than those in $\Lambda$CDM.

CPF of length. The remaining structures are evidently shorter in length compared to their longer progenitors at percolation, so that $CPF(\mathcal{L})$ are confined to lower values of $\mathcal{L}$. This effect is evident in the right panel of Figure 4.8. In both the panels, shown as dashed line is the $CPF(\mathcal{L})$ due to the first realization of $\tau$CDM which shares its set of initial random numbers with the $\Lambda$CDM simulation. We note the sharp tendency of $\tau$CDM structures to be longer than those of $\Lambda$CDM, an effect which successfully enables us to distinguish the two models.

## 4.5   Discussion and Conclusions

This chapter provides a theoretical framework and a methodology to analyse the galaxy-distribution revealed by large 3-dimensional redshift surveys. The exercise will help tackle the redshift surveys like 2dFGRS and SDSS, and consists in studying the geometry, topology and morphology of the LSS as revealed by the volume limited samples derived from 10 realizations of mock $\tau$CDM−based SDSS catalogue and a single $\Lambda$CDM catalogue. These volume limited samples are prepared so as to refer to the same physical volume, and exhibit similar number density of galaxies. The density fields used in our calculations are prepared after smoothing all the samples at 6 $h^{-1}$Mpc.

We have employed SURFGEN described in the 2$^{nd}$ chapter for calculating MFs. We find the MFs as well as the derived morphological statistics, the Shapefinders, to be well-constrained statistics, useful to indirectly probe the contribution of the higher order correlation functions in describing the clustering of the galaxy-distribution. For this purpose, we compare the measured MFs due to 10 realizations of $\tau$CDM with those due to a *single* realization of $\Lambda$CDM. All the 11 realizations, by construction, reproduce the observed two-point correlation function measured using the APM catalogue, and show similar clustering amplitude. Some of the important results of our analysis are summarised below:

- We show that the global MFs (defined as the sum of partial MFs over the set of all the clusters at a given threshold of density) of $\tau$CDM show systematically lower amplitude compared to those due to $\Lambda$CDM; this effect allows us to distinguish between the models.  be termed as a success of MFs as efficient quantifying statistics of LSS.

- The Virgo simulations of dark matter showed *higher* amplitude of MFs for $\tau$CDM compared to $\Lambda$CDM (see Sheth et al., 2003). The cosmological parameters used by the Virgo group match with those employed by Cole et al.(1998) except for the shape parameter $\Gamma$, which is 0.21 in the former case and 0.25 in the latter. We assume that this minute difference in $\Gamma$ will not alter the above trend so long as both the models start with the same power spectrum. The *galaxy distributions* due to both the models were analysed in *this* chapter, and we found a reversal in the above trend: The MF-amplitudes for mock catalogues constructed from $\tau$CDM were found to be *smaller* than that from the $\Lambda$CDM model. We confirmed this effect by analysing the dark matter simulations and galaxy-catalogues in real space, employing the same resolution and the smoothing scale (See Figure 4.4 and the associated discussion).

We speculate that the relative smallness of the amplitudes of MFs reflect the higher degree of phase correlations in the matter-distribution. If this is indeed the case, then our results show that scale-dependent biasing could lead to a phase-mismatch between the galaxy-distribution and the underlying matter-distribution. Thus, there may be *larger* degree of phase-correlations in the *matter-distribution* of $\Lambda$CDM compared to $\tau$CDM, but the biased *galaxy-distribution* of $\Lambda$CDM would appear to exhibit *smaller* phase-correlations compared to $\tau$CDM. We note however, that these results are affected by the constraint that the bias is fixed so as to reproduce the observed $2-$point correlation function (see Section 2)**.

An analysis of 2dFGRS performed by Lahav et al.(2002) concluded that the average bias on scales $0.02 \leq k \leq 0.15$ h Mpc$^{-1}$ is indeed $\sim$1 ††. In light of the above fact, it will be interesting to see how the geometry, topology and morphology of the LSS change in the presence of scale-independent, constant bias. From our analysis, however, it is clear that bias in various structure formation scenarios can play important role, and comparing the *dark matter distributions* with the SDSS and/or 2dFGRS galaxy distributions *is not* an advisable way of testing the model(s) of structure formation against the observations. A realistic treatment of bias through the proper treatment of the physics of galaxy formation while preparing a mock catalogue, is desirable in any such exercise.

- Since we had access to 10 realizations of $\tau$CDM, we made a statistical study of the shapes and sizes of the superclusters occurring in this model. We found the thickness and breadth of the superclusters to increase only marginally over a range of density threshold $\delta$ ($1\leq \mathcal{T}, \mathcal{B} \leq 17h^{-1}$Mpc; $\mathcal{T} \leq \mathcal{B}$). However, the length of the largest superclusters increases monotonically from $\sim$40 $h^{-1}$Mpc to $\sim$90 $h^{-1}$Mpc as $\delta \rightarrow \delta_{\mathrm{perc}}$. Near percolation, the $\tau$CDM superclusters with length $\geq$90 $h^{-1}$Mpc have a probability of $\sim$ 1 per cent to be found (say, the top 10 in an ensemble of 1000 structures). However, the longest (and necessarily the most massive) superclusters could be as long as 150 $h^{-1}$Mpc and are rare structures, with less than 1 per cent probability in a volume as big as that covered in this analysis. Our results show that the dominant morphology of the superclusters in mock $\tau$CDM and $\Lambda$CDM catalogues is prolate, or ribbon-like at the highest threshold of density, and it evolves to become more filamentary as the threshold is lowered toward the percolation threshold. The cosmic web in $\tau$CDM mock's appears to be less planar and more filamentary than the $\Lambda$CDM cosmic web.

---

** A detailed study of this effect lies outside the scope of the present thesis, but will be taken up elsewhere.

†† However, it should be noted that the cosmological model was *assumed* to be $\Lambda$CDM.

- A study of CPF of length of the superclusters reveals that the longest superclusters of $\tau$CDM are about 1.4 times as long as their $\Lambda$CDM counterparts at the same level of significance. In particular, whereas the longest superclusters of $\tau$CDM could be as long as 90 $h^{-1}$Mpc, the $\Lambda$CDM superclusters have $\mathcal{L} =$55 $h^{-1}$Mpc. We note that this result is in excellent agreement with the fact that the $\tau$CDM galaxy catalogue exhibits higher degree of phase-correlations compared to its $\Lambda$CDM counterpart. That the effect of relative strength of phase-correlations is captured well by the Shapefinder *ansatz* for *length*, is noted to be a remarkable success of this ansatz.

To summarise, we have established that the Minkowski Functionals estimated using SURFGEN are well constrained statistics and can be reliably used to quantify the geometry and topology of the LSS in various scenarios of structure formation. We have used these to distinguish two rival models of structure formation, namely $\Lambda$CDM and $\tau$CDM, and have explored a set of related measures and morphological statistics like Shapefinders to make such a comparison more robust. We have further measured and reported the sizes and shapes of superclusters in $\tau$CDM, and hope to extend our methodology to other models once a larger number of realizations become available.

Our present exercise does have its limitations, which are mentioned below.

- The effect of redshift space distortions has not been incorporated in the present exercise. We have throughout worked in redshift space and have not quantified the *fingers-of-God* effect.

- The mock catalogues studied in this paper are prepared by implementing *ad-hoc*

biasing schemes. Although the methodology of analysis is clearly laid out in the present work, a more realistic comparison of the mock data with the observed LSS of the Universe should be carried out using the mock catalogues prepared by incorporating the physics of galaxy-formation in as much detail as possible. Semi-analytic methods, SPH simulations of galaxy formation and the Conditional Luminosity Function based approach Yang et al.(2003) should be able to deliver such mock catalogues in the near future.

We hope to address some of the issues mentioned above and extend this work to a comparison between mock catalogues and LSS revealed by 2dFGRS and SDSS in the future work.

# Chapter 5

# The size of the Longest Superclusters in the Universe

## 5.1 Introduction

One of the most striking visual features in the distribution of galaxies in the Las Campanas redshift Survey (LCRS, Shectman et al. 1996) is that they appear to be distributed along filaments. These filaments are interconnected and form a network, with voids largely devoid of galaxies comprising the region in between the filaments. This network of interconnected filaments encircling voids extends across the entire survey and gives impression of a "Cosmic Web". Similar networks of filaments and voids are also visible in other galaxy surveys, e.g., CfA (Geller & Huchra 1989), 2dFGRS (Colless et al. 2001, Colless et al. 2003) and SDSS (EDR) (Stoughton et al. 2002, Abazajian et al. 2003, 2004).

The analysis of filamentary patterns in the galaxy distribution has a long history dating back to papers by Zel'dovich, Einasto and Shandarin (1982), Shandarin and Zel'dovich (1983) and Einasto et al. (1984). In the last paper the authors analyze the distribution of galaxies in the Local Supercluster. They use the Friend-of-Friend algorithm with varying neighborhood radius to identify connected systems of galaxies referred to as "clusters". As they increase the neighborhood radius, they find that the clusters which are initially spherical become multi-branched with multiple filaments of lengths up to several tens of $h^{-1}$Mpc extending out in different directions. Finally, as the radius is increased further, they find that the filaments get interconnected and join neighboring superclusters into a large network of superclusters and voids. A later study (Shandarin & Yess 2000) used percolation analysis to arrive at a similar conclusion for the distribution of the LCRS galaxies. The large-scale and super large-scale structures in the distribution of the LCRS galaxies have also been studied by Doroshkevich et al. (2001) and Doroshkevich et al. (1996) who find evidence for a network of sheet like structures which surround underdense regions (voids) and are criss-crossed by filaments. The distribution of voids in the LCRS has been studied by Müller, Arbabi-Bidgoli, Einasto & Tucker (2000) and the topology of the LCRS by Trac, Mitsouras, Hickson & Brandenberger (2002) and Colley (1997). A recent analysis (Einasto et al. 2003a) indicates a supercluster-void network in the Sloan Digital Sky Survey also.

Traditionally, correlation functions (Peebles 1993) have been used to quantify the statistical properties of the galaxy distribution. For the LCRS, the two-point correlation function is a power law,

$$\xi(r) = \left(\frac{r}{r_o}\right)^{-1.52} \tag{5.1}$$

with the correlation length $r_o = 6.28 \, h^{-1}\mathrm{Mpc}$ on scales $2.0 \, h^{-1}\mathrm{Mpc}$ to $16.4 \, h^{-1}\mathrm{Mpc}$. On scales larger than $30 - 40 \, h^{-1}\mathrm{Mpc}$, $\xi(r)$ fluctuates closely around zero indicating a statistically homogeneous galaxy distribution at and beyond these scales (Tucker et al. 1997). This raises the question whether the filamentary features which appear to span scales larger than $100 \, h^{-1}\mathrm{Mpc}$ are statistically significant features of the galaxy distribution or if they are mere artifacts arising from chance alignment of the galaxies.

A quantitative estimator of filamentary structure, Shapefinder, was defined (Bharadwaj et al. 2000) to provide a measure of the average filamentarity for a point distribution in 2D. (See Sahni, Sathyaprakash & Shandarin 1998 for general introduction to Shapefinders and Sheth et al. 2003 for the application of Shapefinders to 3D simulations of structure formation.) The Shapefinder statistic was used to demonstrate that the galaxy distribution in the LCRS exhibits a high degree of filamentarity compared to a random Poisson distribution having the same geometry and selection effects as the survey. This analysis provides objective confirmation of the visual impression that the galaxies *are* distributed along filaments. This, however, does not establish the statistical significance of the filaments. The features identified as filaments are essentially chains of galaxies, a crucial requirement being that the spacing between any two successive galaxies along a chain is significantly smaller than the mean inter-galaxy separation. A chain runs as long as it is possible to find another nearby galaxy which is not yet a member of the chain, and breaks when no such galaxy is to be found. The fact that the LCRS galaxies are highly clustered on small scales increases the probability of finding pairs of galaxies at small separations. This enhances the occurrence of long chains of galaxies, and we expect to find a higher degree of filamentarity aris-ing just from chance alignments in the LCRS compared to a Poisson distribution. To establish whether the observed filaments are statistically significant or if they are a result of chance alignments of smaller structural elements, it is necessary to compare the sample of galaxies (here, the LCRS slices) with a distribution of points which has the same small scale clustering properties as the original sample and for which we know that all large-scale filamentary features are solely due to chance alignments. This is achieved using a statistical technique called Shuffle (Bhavsar & Ling 1988) whereby the statistical significance of the filamentarity in a clustered dataset can be assessed.

Shuffle generates fake data-sets, practically identical in their clustering properties to the original data up to a length scale $L$, but in which all structures longer than $L$, both real and chance, of the original data, have been eliminated. In these Shuffled data, filaments spanning length-scales larger than $L$ are visually evident, even expected to be identified as a signal by the statistics used to quantify the filamentarity, but all filaments spanning length-scales larger than L have formed accidentally. The measure of the occurrence of filaments spanning length-scales larger than $L$ in the Shuffled data gives us a statistical estimate of the level at which chance filaments spanning length-scales larger than $L$ occur in the original data. Here we use Shuffle to estimate the degree of filamentarity expected from chance alignments in the LCRS and use this to determine the statistical significance of the observed filamentarity.

We present the method of analysis and our findings in Section 2. In Section 3 we discuss our results and present conclusions.

## 5.2    Analysis and Results

The LCRS contains the angular positions and redshifts of 26,418 galaxies distributed in 6 wedges, each $1.5^o$ thick in declination and $80^o$ in right ascension. Three wedges are centered around mean declinations $-3^o$, $-6^o$ and $-12^o$ in the Northern galactic cap and three at declinations $-39^o$, $-42^o$ and $-45^o$ in the Southern galactic cap. The survey has a magnitude limit $m = 17.75$ and extends to a distance of $600\,h^{-1}$Mpc. The most prominent visual feature in these wedges (Figure 5.1) is that the galaxies appear to be distributed along filaments, several of which span length-scales of $100\,h^{-1}$Mpc or more.

We extracted luminosity and volume limited sub-samples (Figure 5.1) from the LCRS data so that we have an uniform sampling of the regions that we analyze. In order to sample the largest regions that we could, with the above criterion in mind, we limited the wedges from 195 to $375\,h^{-1}$Mpc in the radial direction as shown in Figure

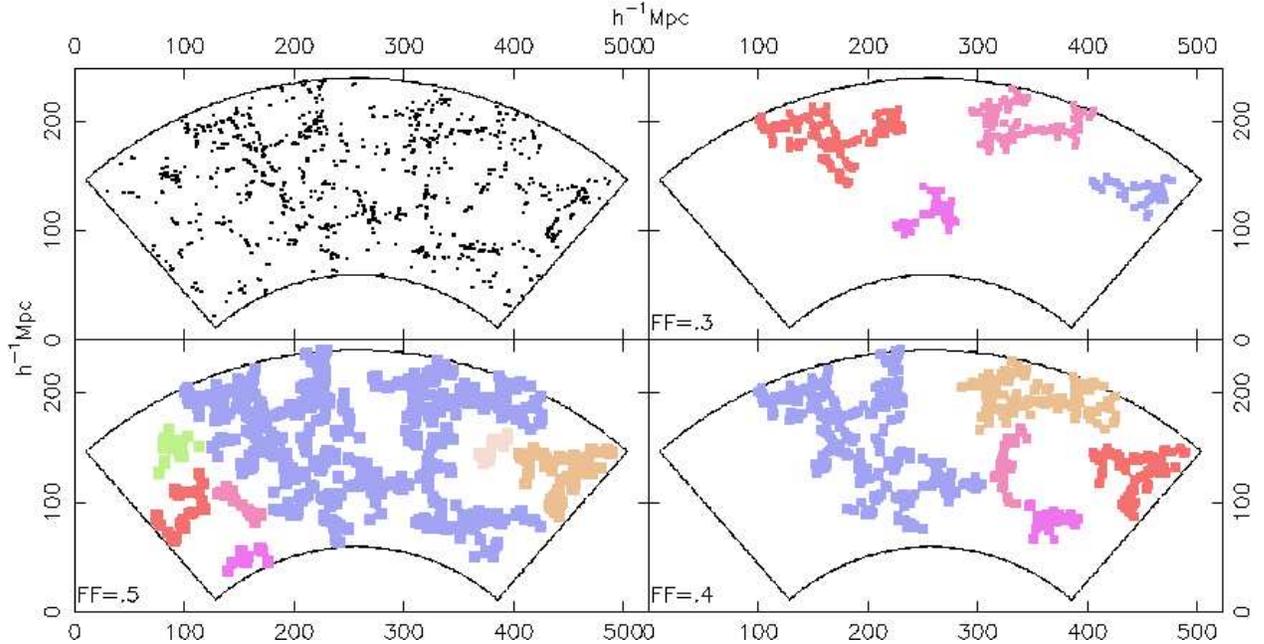

Figure 5.1: The upper left hand panel shows the distribution of galaxies in an uniformly sampled region of the $-3^o$ slice. The filamentary patterns of galaxies are evident. Going clockwise, the three other panels show some of the largest clusters identified using FOF after N=3,4 and 5 iterations of coarse-graining. Most of the clusters shown are highly filamentary ($\mathcal{F}_2 > 0.8$), with $\mathcal{F}_2 > 0.9$ for the largest cluster in each panel (the definition of $\mathcal{F}_2$ is given in the text). At each level of coarse-graining, clusters grow until we have a single, large filamentary network spanning the whole region, referred to as the "Cosmic Web".

## 5.1.

Our data consist of a total of 5073 galaxies distributed in 6 wedges. Each LCRS wedge is collapsed along its thickness (in declination) resulting in a 2 dimensional truncated conical slice which, being geometrically flat, can be unrolled onto a plane. Each slice is embedded in a $1\,h^{-1}\mathrm{Mpc} \times 1\,h^{-1}\mathrm{Mpc}$ rectangular grid. Grid cells with galaxies in them are assigned the value 1, empty cells 0. Connected regions of filled cells are identified as clusters using a "friends-of-friends"(FOF) algorithm. The geometry and topology of each cluster is described by its area $S$, perimeter $P$, and genus $G$. It is possible to utilize these measures to describe the filamentarity of the supercluster of interest. This is achievable using a set of measures termed as Shapefinders, originally defined for 2D surfaces embedded in 3D. We use a 2D version of the Shapefinder in our analysis of the superclusters in the LCRS slices.

The Shapefinder measure $\mathcal{F}$

$$\mathcal{F} = \frac{(\mathcal{P}^2 - 16\mathcal{S})}{(\mathcal{P} - 4)^2}, \qquad (5.2)$$

originally defined in Bharadwaj et al. (2000), quantifies the shape of the superclusters in the quasi-2D slices of the LCRS. By definition $0 \leq \mathcal{F} \leq 1$. $\mathcal{F}$ quantifies the degree of filamentarity of a cluster, with $\mathcal{F} = 1$ indicating a filament and $\mathcal{F} = 0$, a square (while dealing with a density field defined on a grid). The average filamentarity ($\mathcal{F}_2$), is defined as the mean filamentarity of all the clusters weighted by the square of the area of the clusters.

$$\mathcal{F}_2 = \frac{\sum_{i=1}^{N_{cl}} \mathcal{S}_i^2 \mathcal{F}_i}{\sum_{i=1}^{N_{cl}} \mathcal{S}_i^2} \qquad (5.3)$$

In the current analysis, we use the average filamentarity to quantify the degree of filamentarity in each of the LCRS slices.

The galaxy distribution in the LCRS slices is quite sparse and therefore the Filling Factor $FF$ (defined as the fraction of filled cells) is very small ($FF \sim 0.01$). The clusters identified using FOF contain at most 2 or 3 filled cells, not yet corresponding to the long filaments visually apparent in the slices. Larger structures are identified by the method of "coarse-graining". Coarse-graining is implemented by successively filling cells that are immediate neighbors of already filled cells. It may be noted that the "coarse-graining" procedure adopted by us is equivalent to smoothing successively with a top-hat kernel. The filled cells get fatter after every iteration of coarse-graining. This causes clusters to grow, first because of the growth of filled cells, and then by the merger of adjacent clusters as they overlap. The observed large scale patterns are initially enhanced as the clusters grow and then washed away as the clusters become very thick and fill up the entire region. $FF$ increases from $FF \sim 0.01$ to $FF = 1$ as the coarse graining proceeds. So as not to limit ourselves to an arbitrarily chosen value of $FF$ as the one defining filaments, we present our results showing the average filamentarity for the entire range of filling factor $FF$.

Now we describe how the Shuffle algorithm works (Figure 5.2). A grid with squares of side $L$ is superimposed on the original data slice. Square blocks of data which lie entirely within the slice are then randomly interchanged, with rotation, repeatedly, to form a new Shuffled slice. This process eliminates features in the original data on scales longer than $L$, keeping clustering at scales below $L$ nearly identical to the original data. All the structures spanning length-scales greater than $L$ that exist in the Shuffled slices are the result of chance alignments. For each value of $L$ we use different realizations of the Shuffled slices to estimate the degree of filamentarity that arises from chance alignments on scales larger than $L$. The Shuffled slices were analyzed in exactly the same way as the actual LCRS slices. At a fixed value of $L$, the average filamentarity in the *original* sample will be larger than in the *Shuffled* data only if the actual data have more filaments spanning length-scales larger than $L$, than that expected from chance alignments. We vary the value of $L$ from $10\,h^{-1}\mathrm{Mpc}$ to $100\,h^{-1}\mathrm{Mpc}$ and determine the largest value of $L$ ($L_{\mathrm{MAX}}$) such that for all $L < L_{\mathrm{MAX}}$ the values of the average filamentarity, $\mathcal{F}_2$, in the actual data are higher than the Shuffled data, indicating the presence of physical filaments of lengths greater than $L$. We use three realizations for $L = 10, 20, 30, 90$ and $100\,h^{-1}\mathrm{Mpc}$ shuffling of slices, and six realizations for the intermediate length-scales. For length-scales beyond $L_{\mathrm{MAX}}$ the average filamentarity in the Shuffled slices should continue to be the same as in the actual LCRS slice, establishing $L_{\mathrm{MAX}}$ to be the largest length-scale across which we have statistically significant filamentarity. Filaments which extend across length scales larger than $L_{\mathrm{MAX}}$ are not statistically significant and are a consequence of chance alignments. In Figure 5.3, we plot average filamentarity, $\mathcal{F}_2$, as a function of the filling factor, $FF$, for both the original sample as well as for samples generated by shuffling the patches for various values of $L$. To convey the essential aspects of the analysis, we only show the results for the slices shuffled at our lowest $L$ value, $L = 10\,h^{-1}\mathrm{Mpc}$ and at or near the length-scale of interest, $L_{\mathrm{MAX}}$. We use the $\chi^2$ test to establish $L_{\mathrm{MAX}}$ for the six individual slices. The reduced $\chi^2$ for the curves in Figure 5.3 are defined by

$$\chi^2(L) = \frac{1}{N_B - 1} \sum_{i=1}^{N_B} \left[ \frac{\mathcal{F}_2^{(i),\mathrm{LCRS}} - \mathcal{F}_2^{(i),\mathrm{Shuffled}}(L)}{\sigma_{\mathcal{F}_2}^{(i)}(L)} \right]^2, \qquad (5.4)$$

where $N_B$ is the number of data points available for comparison between the original slice and a shuffled slice for a value of $L$. The quantity $\sigma_{\mathcal{F}_2}^{(i)}(L)$ is the standard deviation in $\mathcal{F}_2$ measured at a given filling factor, $FF_i$, using all the available shuffled realizations at a given length scale $L$. We have noted that for filling factor, $FF > 0.7$, all the $\mathcal{F}_2$ curves follow the same trend, regardless of the slice (original or shuffled). This can be interpreted as the regime of $FF$ in which the coarse-graining defines structures of such large extent that they are unphysical and Shuffling does not discriminate between original or Shuffled data. Including the tiny differences in the curves in this regime ($FF > 0.7$), will give weight to an unphysical signal and determine an erroneous $\chi^2$. Hence, we only include the region of the curves for $FF < 0.7$ to determine the reduced $\chi^2$, using this as the discriminating measure between the curve for the real data and the Shuffled realizations at different $L$. The reduced $\chi^2$ quantifies how different a Shuffled slice is from the original, at various $L$. The minimum value of the reduced $\chi^2$ should correspond to the length-scale $L = L_{\mathrm{MAX}}$ at which, if slices are Shuffled, the filamentarity of the Shuffled slices and the original slice differ the least. This gives us the length scale $L_{\mathrm{MAX}}$ beyond which filaments are only chance objects and not physical.

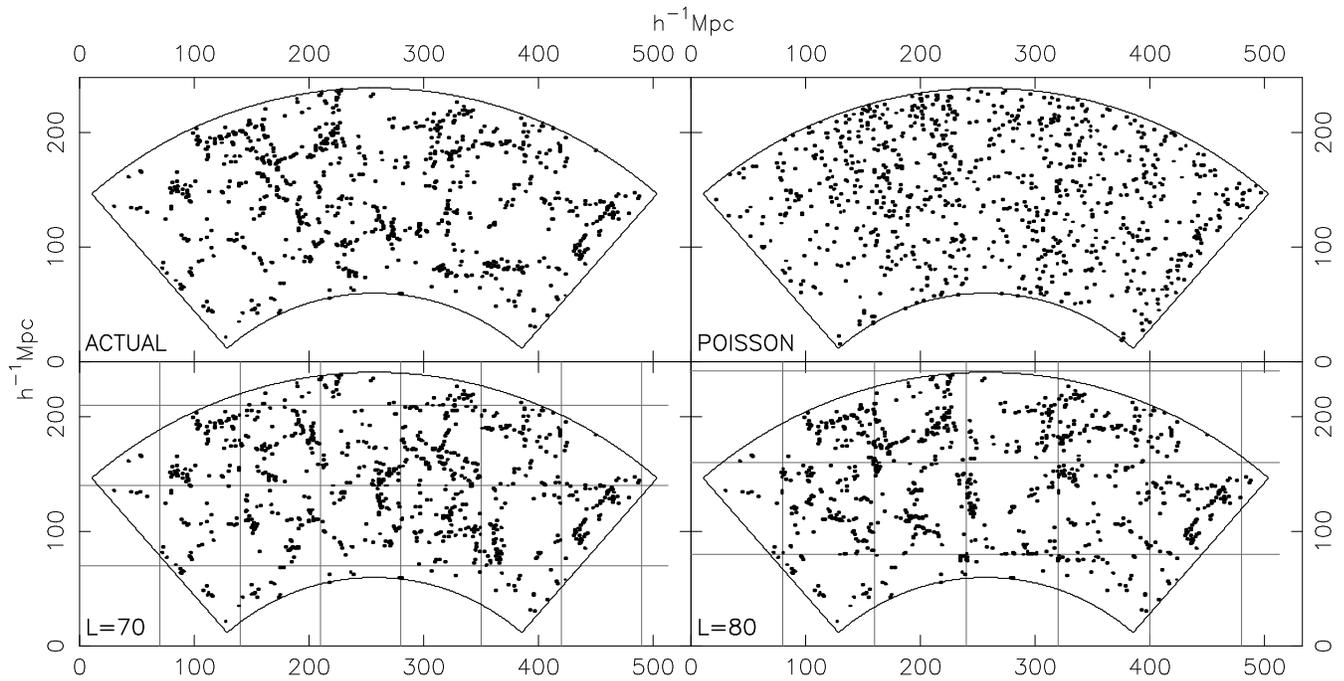

Figure 5.2: The LCRS data shown at top left, are the same as in Figure 5.1. The next panel (top right) shows a Poisson distribution of points generated over the same region. The bottom panels show Shuffled realizations of the LCRS data with $L = 70$ and $80\,h^{-1}$Mpc. The square patches show the boundaries of the Shuffled regions While it is evident that the Poisson data are much less filamentary than the LCRS galaxies, it is not possible to visually distinguish the level of filamentarity in the actual data from the Shuffled realizations. A quantitative analysis, shows that the $L = 70\,h^{-1}$Mpc Shuffled data exhibit less filamentarity than the actual data, while the $L = 80\,h^{-1}$Mpc Shuffled data are statistically identical to the original data in filamentarity (Fig. 5.3). Note that all visual features across the boundaries in the bottom two panels (Shuffled data) are chance filaments.

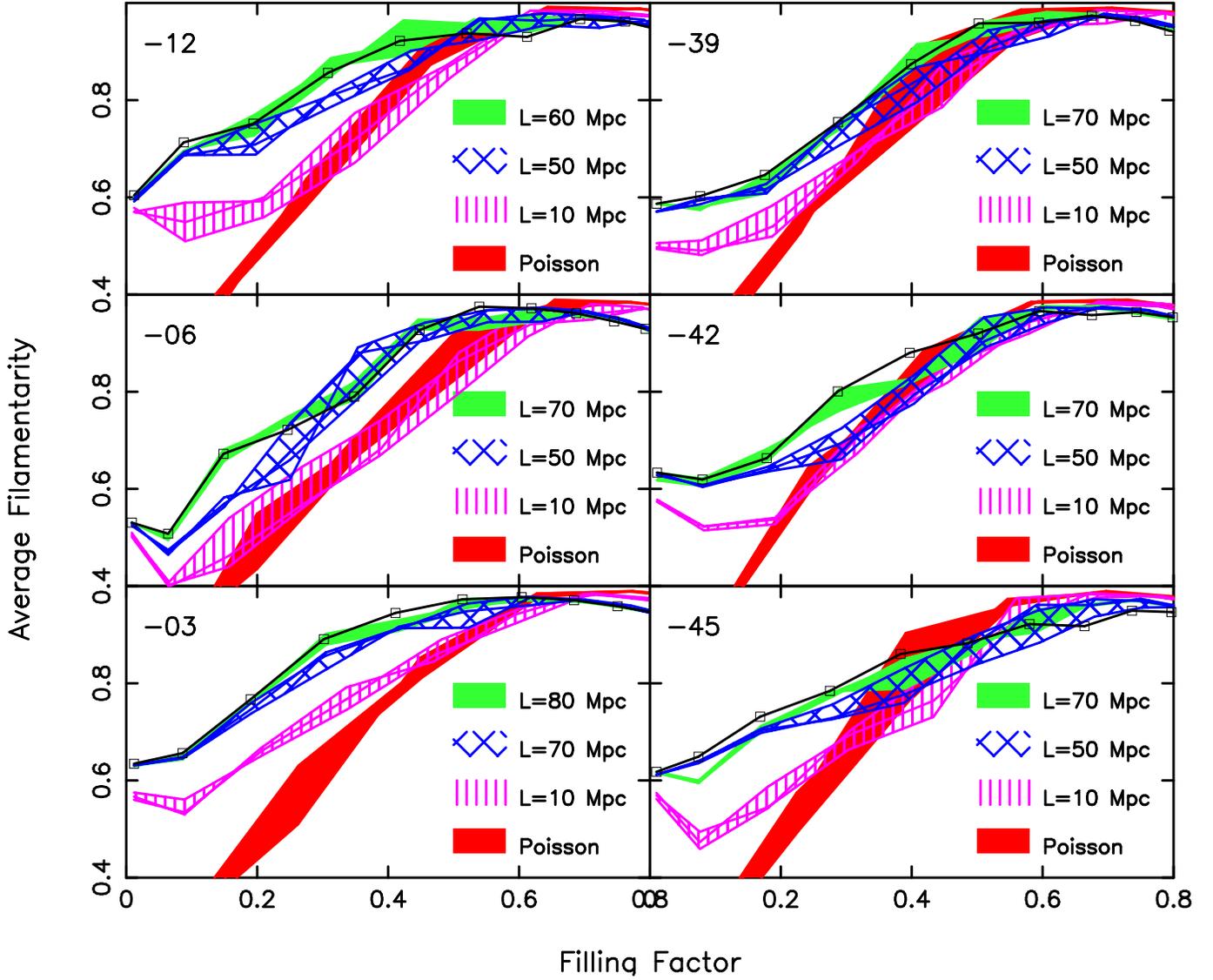

Figure 5.3: Plots of average filamentarity, $\mathcal{F}_2$, vs. Filling Factor, $FF$, for each of the LCRS slices (dark black line), three Shuffled realizations of each slice at various $L$ (hatched regions and light solid region) and three Poisson point distributions (dark solid region): We show Shuffle results for $L = 10\,h^{-1}$Mpc, the smallest patches that were Shuffled, and the two L values where a transition occurs from the Shuffled galaxy distribution being less filamentary than the original data to it being indistinguishable from the original data. For each slice, this establishes $L_{\mathrm{MAX}}$, the length beyond which filaments are just chance artifacts, somewhere between these two values of $L$.

In Figure 5.4, we show the reduced $\chi^2$ vs. $L$ plotted for the six slices. We also list the minimum values of reduced $\chi^2$ for each slice and the corresponding $L$. We see that the values of reduced $\chi^2$ are well within acceptable bounds to say that the Shuffled slices at these values of $L$ are indistinguishable from the original slice. The length-scale that corresponds to these minima is $\sim 60\,h^{-1}\mathrm{Mpc}$ for all the southern slices, whereas it is $\sim 70\,h^{-1}\mathrm{Mpc}$ for $-6^o$ slice and $80\,h^{-1}\mathrm{Mpc}$ for $-3^o$ and $-12^o$ slices. We thus establish that for the Southern slices the longest real filaments are no longer than $60\,h^{-1}\mathrm{Mpc}$ and for the Northern slices no longer than $80\,h^{-1}\mathrm{Mpc}$. Beyond $80\,h^{-1}\mathrm{Mpc}$ structure is not statistically significant.

## 5.3  Discussion and conclusions

A look at the filamentary features at different levels of coarse-graining (Figure 5.1) reveals that the size of the largest filamentary feature increases monotonically with successive iterations of coarse-graining until it spans the entire survey (Bharadwaj et al. 2000). As coarse-graining proceeds, individual filaments form and then interconnect to form the supercluster-void network, in keeping with the earlier analysis (Einasto et al. 1984; Shandarin & Yess 2000; Einasto et al. 2003a) discussed in the introduction. Although, the length of the interconnected network of filaments increases monotonically, the ratio of the length to the number of holes (Genus) stabilizes and then decreases (Bharadwaj et al. 2000). For the $-3^o$ slice, this ratio stabilizes around $140\,h^{-1}\mathrm{Mpc}$ at $FF \sim 0.4$. This ratio may be interpreted as the perimeter of the typical void in the network. This leads to a picture where there are voids of diameter $60\,h^{-1}\mathrm{Mpc}$ encircled by filaments of thickness $10\,h^{-1}\mathrm{Mpc}$ (Peebles 1993, Sheth et al. 2003, Sheth 2004) interconnected to form a large web. A void of this size, along with the filament at its perimeter, would span a length-scale $\sim 80\,h^{-1}\mathrm{Mpc}$. The results of this paper show that such voids encircled by filaments are statistically significant features. Although our analysis also finds a web of interconnected filaments which spans length-scales larger than $80\,h^{-1}\mathrm{Mpc}$ and runs across the entire survey, this is not statistically significant. The web arises from chance interconnections between the filaments encircling different voids.

Studies of the distribution of Abell superclusters (Einasto et al. 1997b) show that the mean distance between neighboring superclusters is about $50\,h^{-1}$Mpc for poor superclusters and about $100\,h^{-1}$Mpc for rich superclusters. The distribution of the SDSS superclusters (Einasto et al. 2003a) and the LCRS superclusters (Einasto et al. 2003b) shows a similar behavior. Visualizing the superclusters as being randomly distributed, we would expect filaments joining the superclusters to develop as the density field is progressively smoothed. Such filaments will arise from the chance alignments of shorter, genuine, statistically significant filaments. The filaments joining superclusters will span length-scales comparable to the mean inter-supercluster separation and the statistical properties of filaments would be stable to shuffling *i.e.*, it would not change if the superclusters were rearranged randomly. Our results may be interpreted as being indicative of the superclusters being randomly distributed on scales larger than $80\,h^{-1}$Mpc with the mean inter-supercluster separation also being of this order [*].

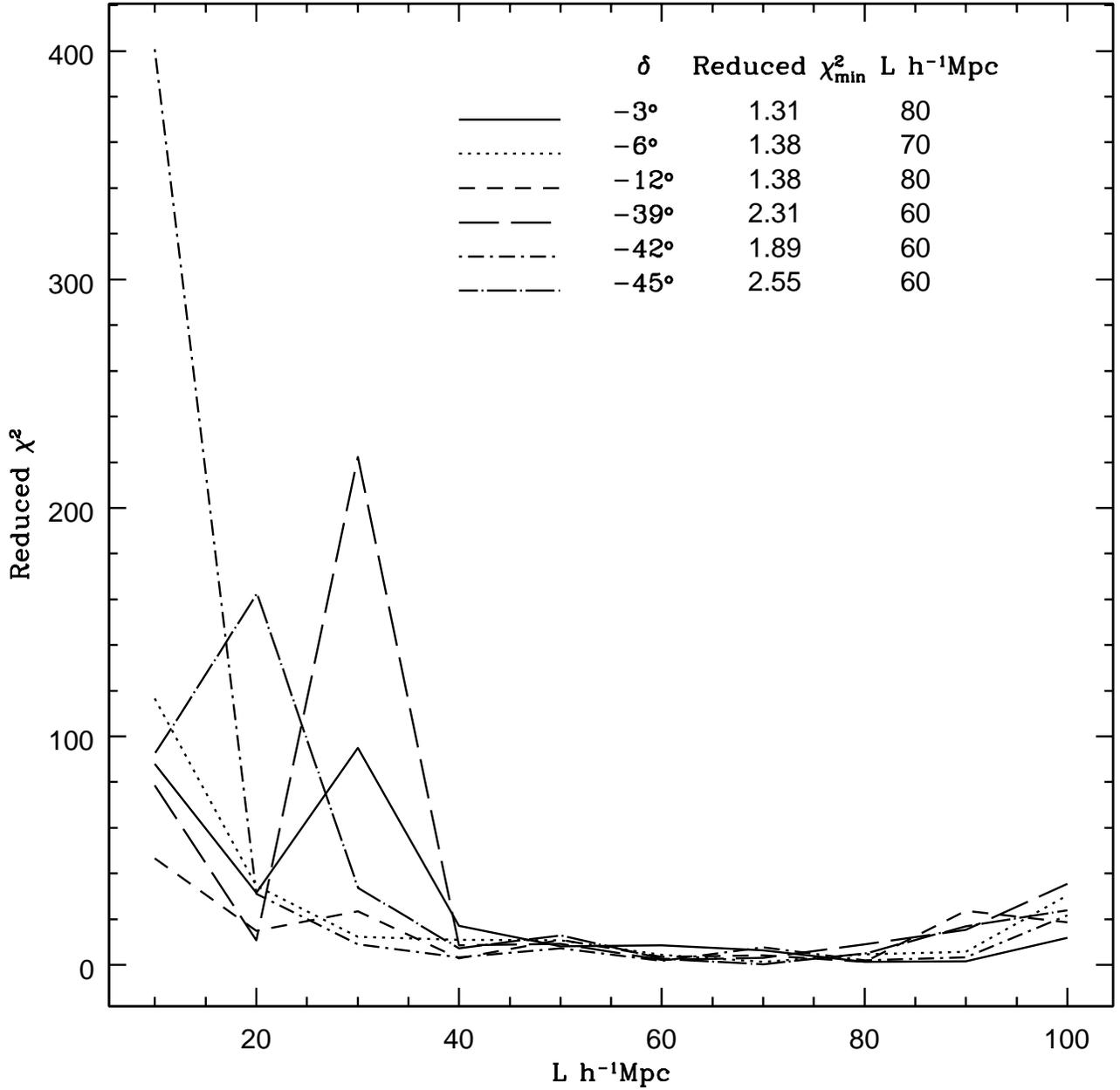

Figure 5.4: The minimum value of the reduced $\chi^2(L)$ is plotted for all the six slices. It varies from 1.4 to 2.6 and is in within the acceptable bounds. We conclude from here that the scale of longest *real* filaments is $\sim 60 \, h^{-1}$Mpc for all the three southern slices. For the northern slices this scale is $\sim 70 - 80 \, h^{-1}$Mpc. This scale can also be interpreted as the scale beyond which the LSS in the Universe is homogeneous.

The presence of statistically significant features on scales as large as 70 to $80\,h^{-1}$Mpc may seem surprising given the fact that the correlation analysis fails to detect any clustering on scales beyond 30 to $40\,h^{-1}$Mpc. This is due to the inability of the two point (and higher order) correlation functions in detecting coherence at large-scales. Pattern specific methods (like Shapefinders) are necessary to detect and quantify coherent large scale features in the galaxy distribution. It is interesting to note that the two dimensional power spectrum for the LCRS (Landy et al. 1996) exhibits strong excess power at wavelengths $\sim 100\,h^{-1}$Mpc, a feature which may possibly be related to the filamentary patterns studied here. The analysis of the three dimensional distribution of Abell clusters (Einasto et al. 1997a, 1997b) reveals a bump at $k = 0.05\,h\,\text{Mpc}^{-1}$ in the power spectrum. Also, the recent analysis of the SDSS shows a bump at $k = 0.05\,h\,\text{Mpc}^{-1}$ in the power spectrum (Tegmark et al. 2003). While it is interesting to conjecture that these features may be related to the presence of filaments, we should also note that the filaments are non-Gaussian features and cannot be characterized by the power spectrum alone.

A point which should be noted is that the filaments quite often run in a zig-zag fashion (Figure 5.1), and the length of a filament which spans a length-scale of $80\,h^{-1}$Mpc may be significantly larger than $80\,h^{-1}$Mpc. Also, the present analysis is two dimensional whereas the filaments actually extend in all three dimensions. The length of the filaments may be somewhat larger in three dimensions, and a little bit of caution may be advocated in generalizing our results.

---

[*]It is interesting to note that a study of the SDSS(EDR) superclusters conducted by Doroshkevich et al. (2003) using Minimal Spanning Trees concludes that the large-scale filaments appear to randomly connect the sheet-like structures in denser environments.

In the gravitational instability picture, small disturbances in an initially uniform matter distribution grow to produce the large-scale structures presently observed in the universe. It is possible to interpret the filaments in terms of the coherence of the deformation (or strain) tensor (Bond, Kofman & Pogosyan 1996) of the smoothened map from the initial to the present positions of the particles which constitutes the matter. Our analysis shows that the deformation tensor has correlation to length-scales up to $80\,h^{-1}$Mpc and is uncorrelated on scales larger than this. The ability to produce statistically significant filamentarity on scales up to $80\,h^{-1}$Mpc will be a crucial quantitative test of the different models for the formation of Large Scale Structure in the universe.[†]

We next address the question of the length-scale beyond which the distribution of galaxies in the LCRS may be considered to be homogeneous. The analysis of Kurokawa, Morikawa & Mouri (2001) shows this to occur at a length-scale of $\sim 30\,h^{-1}$ Mpc, whereas Best (2000) fails to find a transition to homogeneity even on the largest scale analyzed. The analysis of Amendola & Palladino (1999) shows a fractal behavior on scales less than $\sim 30h^{-1}$ Mpc but is inconclusive about the transition to homogeneity. The results presented in this paper set a lower limit to this length-scale at around $80\,h^{-1}$Mpc, in keeping with estimates based on the multi-fractal analysis of LCRS (Bharadwaj, Gupta & Seshadri 1999) who find that the LCRS exhibits homogeneity on scales 80 to $200\,h^{-1}$ Mpc. In a separate approach based on the analysis of the two point correlation applied to actual data and simulations Einasto & Gramann (1993) find that the transition to homogeneity occurs at about $130\,h^{-1}$Mpc. For the LCRS, the scale of the largest coherent structure is at least twice the length-scale at which the two-point correlation function becomes zero. Beyond this scale the filaments interconnect statistically to form a percolating network. This filament-void network of galaxies is not distinguishable, in a statistical sense, beyond scales of $80\,h^{-1}$Mpc. If the LCRS slices can be considered a fair sample of the universe then this suggests the scale of homogeneity for the universe.

---

[†]It is interesting to note that in the previous chapter devoted to analysis of mock SDSS catalogs based on $\Lambda$CDM model, we find the length-scales of the largest superclusters to be $\sim 60\,h^{-1}$Mpc.

# Chapter 6

# Discussion and Conclusions

Over the last few years, our knowledge about the large-scale structure (LSS) has increased at a great pace; thanks to the wide-angle, deep redshift surveys like 2 degree Field Galaxy Redshift Survey (2dFGRS) and Sloan Digital Sky Survey (SDSS). LSS can be perceived in several ways. For example, if we concentrate on the distribution of *galaxies*, the LSS appears web-like, due to presence of coherent, interconnected filaments and sheets. On the other hand, if we focus on voids, LSS appears frothy. The cosmological volume covered by the above redshift surveys is now large enough to sample large-scale objects like superclusters and voids in sufficient number. Perhaps, this is the best time to test our theoretical predictions about LSS, and to understand the Cosmic Web of filaments and sheets in an objective and quantifiable manner.

In this thesis, we have tried to respond to this global need and have developed techniques and methods to analyse LSS in an integrated manner. We have adopted an approach based on Minkowski Functionals to study the supercluster-void network in our Universe. To this end, a new technique for studying the geometrical and topological properties of LSS has been developed which can be uniformly applied both to the N−body experiments as well as to data from redshift surveys.

Given a density field, constructed by smoothing a point data set consisting (say) of a distribution of galaxies or dark matter particles, our method

1. constructs closed polyhedral surfaces of constant density corresponding to excursion sets of a density field, using for this purpose a surface generating triangulation scheme (SURFGEN). Discussed in Chapter 2, SURFGEN forms a major component of this thesis.

2. SURFGEN evaluates the Minkowski functionals (volume, surface area, extrinsic

curvature and genus) thus providing full morphological and topological information of three dimensional iso-density contour surfaces corresponding to a given data set. Evaluated in this manner, the Minkowski functionals can be used to study the properties of individual objects lying either above (clusters, superclusters) or below (voids) a given density threshold. They can also be used to study the morphological properties of the full supercluster-void network at (say) the percolation threshold. The ratio's of Minkowski functionals (Shapefinders) are used to probe the shape of iso-density surfaces which sample the distribution of large scale structure at different thresholds of density. (The highest density thresholds correspond to galaxies and clusters of galaxies, moderate thresholds correspond to superclusters while the lowest density thresholds characterize voids.) The performance of our ansatz has been tested against both simply and multiply connected surfaces such as the sphere, triaxial ellipsoid and triaxial torus. These three eikonal bodies can be smoothly deformed to give surfaces which are spheroidal, pancake-like (oblate) and filament-like (prolate) etc. Analytically known values of the Minkowski functionals for these surfaces allow us to test both our surface modelling scheme and our evaluation of Minkowski functionals from triangulation. Remarkably, we find that volume, area and integrated extrinsic curvature are determined to an accuracy of better than 1% on all length-scales for both simply and multiply connected surfaces. The integrated intrinsic curvature (genus) is evaluated exactly. SURFGEN also reproduces the analytic predictions for MFs of a Gaussian random field remarkably well.

3. SURFGEN employs a new contour-based method for determining topology. This method evaluates the genus for an iso-density contour surface directly, by applying the Euler formula to the triangulated surface (see Chapter 2), and presents a significant improvement over earlier grid-based methods of determining topology.

Having validated the performance of SURFGEN on eikonal surfaces and GRFs, we applied SURFGEN to study the supercluster-void network in various rival cosmological models, both in dark matter as well as in galaxies.

In Chapter 3, we studied the morphology of superclusters and voids in N−body dark matter simulation of $\Lambda$CDM cosmology. We consider this exercise as a *first* step in our overall programme of studying the LSS in redshift surveys and eventually testing theoretical models against observations. Systematic methodology would be absolutely essential in such a long term programme. This exercise has been carried out keeping this in mind, in a preparatory spirit.

Some interesting conclusions are summarised below. (Throughout this discussion, the quantities refer to superclusters and voids constructed as iso-density contours above or below a given threshold of density in a density field created by smoothing the dark matter distribution. Physical quantities like volume, mass, sizes and shapes have been evaluated using SURFGEN. The symbols for various quantities follow from Chapter 1.)

- At percolation ($\delta_{\mathrm{perc}}^{C} \sim 1.8$) Percolating superclusters together occupy only $\simeq 7\%$ of the volume of the simulation box, and yet contain about 26% of the total mass in the $\Lambda$CDM Universe.

- Voids *too* become connected with one another at a certain critical density threshold. We studied their growth in the simulation box while raising the density threshold starting from a minimum density value. We found that underdense regions (voids) totally occupy no more than $\simeq 22\%$ of the total volume at percolation ($\delta_{\mathrm{perc}}^{V} \sim -0.5$), and yet percolate the volume. At this threshold voids contain about 8% of the total mass in the $\Lambda$CDM Universe.

- It is important to understand the morphology of these percolating structures. The planarities of both superclusters and voids are quite low $\mathcal{P} \lesssim 0.3$. This implies that pancake-like structures in the dark matter density in real space are not typical in the $\Lambda$CDM model. On the other hand, an interesting trend is noted: the more massive a supercluster, or more voluminous a void, the stronger is its tendency to be filamentary. We confirmed that our conclusions drawn using Shapefinder ansatz conform well with the visual impression of these objects: objects inferred to be filamentary due to significantly high values of $\mathcal{F}$ indeed *look* filamentary.

- We studied the sizes and shapes of individual superclusters and voids at a set of density threshold values while excluding the largest structure from the analysis. The length of a quarter of the most massive $\Lambda$CDM superclusters exceeds 50 $h^{-1}$ Mpc (at percolation). The most voluminous voids at percolation are even longer: 25% of them are longer than 60 $h^{-1}$ Mpc. The longest non-percolating supercluster is as long as 100 $h^{-1}$ Mpc and the longest non-percolating void is as long as 200 $h^{-1}$ Mpc. Both are comparable to the size of the box (239.5 $h^{-1}$Mpc) and therefore may be affected by the boundaries.

- There is a significant amount of asymmetry in the topology of these superclus-

ters and voids. The genus value of individual superclusters can be $\sim 5$. The genus of individual voids can be as large as 55. This implies significant amount of substructure in superclusters and especially in voids. This is in qualitative agreement with recent theoretical and observational studies of voids of galaxies (Gottlöber et al. 2003; Peebles 2001).

- One of our main results is that voids, as defined through the density field (read dark matter distribution) can be distinctly non-spherical.

Because MFs depend on the entire hierarchy of correlation functions, it should be possible to use them as discriminatory diagnostics of LSS in its highly nonlinear regime, i.e., we may use MFs to compare LSS of rival cosmological models. We showed that not only MFs, but also the morphology of superclusters computed using Shapefinders provided useful insights into various cosmic scenarios. Dark matter distributions of three cosmological models (SCDM, $\tau$CDM and $\Lambda$CDM) were studied at $z = 0$. Some of the main results are described below.

- Using the Minkowski functionals we showed that, like other diagnostics of clustering, supercluster morphology too is sensitive to the underlying cosmological parameter set characterizing our universe. Although the three cosmological models considered by us, $\Lambda$CDM, $\tau$CDM and SCDM display features which visually appear to be quite similar, the geometrical and morphological properties of these models are sufficiently distinct to permit differentiation using MFs. We demonstrated that studying Minkowski Functionals using mass-parameterization significantly enhances their discriminatory power.

- A study of percolation and cluster abundance reveals several interesting aspects of the gravitational clustering process. For all models, on decreasing the density threshold to progressively smaller values one reaches the critical percolation threshold at which the largest supercluster runs through (percolates) the simulation box. Percolation is reached at moderate values of the density contrast ranging from $\delta_{\rm perc} \simeq 2.3$ for $\Lambda$CDM to $\delta_{\rm perc} \simeq 1.2$ for SCDM. The abundance of clusters reaches a maximum value at (or very near) the percolation threshold and the percolating supercluster occupies a rather small amount of space in all three cosmological models. Thus the fraction of total simulation-box volume contained in the percolating supercluster is least in $\Lambda$CDM (0.6%) and greatest in SCDM (1.2%). When taken together, all overdense objects at the percolation threshold occupy 4.4% of the total volume in the $\Lambda$CDM model. For comparison, the volume fraction in overdense regions at the percolation threshold is $\sim 16\%$ in an idealised, continuous Gaussian random field. This fraction can increase up to $\sim 30\%$ for Gaussian fields generated on a grid. The fact that clusters & supercluster occupy a very small fraction of the total volume appears to be a hallmark of the gravitational clustering process which succeeds in placing a large amount of mass ($\sim 30\%$ of the total, in the case of $\Lambda$CDM) in a small region of space ($\sim 4\%$). The low filling fraction of the percolating supercluster in $\Lambda$CDM (0.006) strongly suggests that this object is either planar or filamentary and a definitive answer to this issue is provided by the Shapefinder statistic.

- Shapefinders were introduced to quantify the visual impression one has of the supercluster-void network of being a cosmic web of filaments and/or sheets interspersed with large voids. By applying the Shapefinder statistics via SURFGEN to realistic N-body simulations we demonstrated that: (i) Most of the mass in the Universe is contained in large superclusters which are also extremely filamentary; the vast abundance of smaller clusters and superclusters tends to be prolate or quasi-spherical. (ii) Of the three cosmological models the percolating supercluster in $\Lambda$CDM is the most filamentary ($F \simeq 0.81$) and the supercluster in $\tau$CDM the least ($F \simeq 0.7$). (iii) The percolating supercluster in $\Lambda$CDM is topologically a much simpler object than its counterpart in $\tau$CDM, with the former having only $\simeq 6$ tunnels compared to $\simeq 19$ in the latter. We also show that among various morphological parameters, the Planarity and Filamentarity of clusters and superclusters is one of the most powerful statistics to discriminate between cosmological models.

As a next step in our programme of confronting theory with observations, it was necessary to test the performance of SURFGEN against *mock* catalogues of galaxies. We worked with mock SDSS catalogues of galaxies based on two rival cosmological models $-\Lambda$CDM and $\tau$CDM. In Chapter 4 we presented a detailed analysis of these two sets of catalogues. By construction, *all* the galaxy catalogues agree with the observed LSS *up to* the two$-$point correlation function. MFs were shown to distinguish these rather similar-looking models. Several important results regarding morphology, bias and length-scales of superclusters were obtained from this analysis.

- We showed that the global MFs of $\tau$CDM show systematically lower amplitude compared to those determined for $\Lambda$CDM, an effect which enabled us to distinguish between the models. We view this as a demonstration of the success of MFs as discriminatory diagnostics of LSS.

- The Virgo simulations of dark matter showed *higher* amplitude of MFs for $\tau$CDM compared to $\Lambda$CDM. This result was obtained in Chapter 3. We found a reversal in the above trend in case of galaxies selected as biased tracers of mass: The MF-amplitudes for $\tau$CDM were found to be *smaller* than $\Lambda$CDM. We confirmed this effect by analysing the dark matter simulations and galaxy-catalogues in real space, employing the same resolution and the smoothing scale.

  We speculate that the relative smallness of the amplitudes of MFs reflects the higher degree of phase correlations in the matter-distribution. If this is indeed the case, then our above results show that scale-dependent biasing could lead to a phase-mismatch between the galaxy-distribution and the underlying matter-distribution. Thus, there may be *larger* degree of phase-correlations in the *matter-distribution* of $\Lambda$CDM compared to $\tau$CDM, but the biased *galaxy-distribution* of $\Lambda$CDM would appear to exhibit *smaller* phase-correlations compared to $\tau$CDM. We note however, that these results are affected by the constraint that the bias is fixed so as to reproduce the observed two−point correlation function (see Section 2).

  From our analysis it is clear that bias in various structure formation scenarios can play important role, and comparing the *dark matter distributions* with the SDSS and/or 2dFGRS galaxy distributions *is not* an advisable way of testing the model(s) of structure formation against the observations. A realistic treatment of bias through the proper treatment of the physics of galaxy formation in preparing a mock catalogue, is vital in any such exercise.

- Since we had access to 10 realizations of $\tau$CDM, we made a statistical study of the shapes and sizes of the superclusters occurring in this model. We found the thickness and breadth of the superclusters to increase only marginally over a range of $\delta$ ($1 \leq \mathcal{T}, \mathcal{B} \leq 17 h^{-1}$Mpc; $\mathcal{T} \leq \mathcal{B}$). However, the length of the largest superclusters was found to be increasing monotonically from $\sim 40$ $h^{-1}$Mpc to $\sim 90$ $h^{-1}$Mpc as $\delta \rightarrow \delta_{\mathrm{perc}}$. Near percolation, the $\tau$CDM superclusters with length $\geq 90$ $h^{-1}$Mpc have $\sim 1$ per cent probability of occurrence (say, the top 10 in an ensemble of 1000 structures). The longest (and necessarily the most massive) superclusters could be as long as 150 $h^{-1}$Mpc and are extremely rare objects, with less than 1 per cent probability of being found in a volume as big as that covered by the analysis. The dominant morphology of superclusters is prolate, or ribbon-like at highest threshold of density ($\delta \sim 3.5$); it evolves and becomes more filamentary as the threshold is lowered toward the percolation threshold

$(\delta_{\mathrm{perc}}^{\Lambda,\tau} \sim 1.7)$.

- A study of the distribution of length $\mathcal{L}$ for superclusters reveals that the longest superclusters of $\tau$CDM are $\sim$1.3 times *longer* than their $\Lambda$CDM counterpart structures at the same level of significance. In particular, whereas the longest superclusters of $\tau$CDM could be as long as 90 $h^{-1}$Mpc, the $\Lambda$CDM superclusters have $\mathcal{L}_{max}$ =55 $h^{-1}$Mpc. We note that this result is in excellent agreement with the fact that $\tau$CDM galaxy catalogues exhibit higher degree of phase-correlations compared to their $\Lambda$CDM counterpart. That the effect of relative strength of phase-correlations is captured well by the Shapefinder *ansatz* for *length*, deserves to be noted.

Finally, we applied a 2−dimensional Shapefinder, "Filamentarity" to 6 slices of Las Campanas Redshift Survey in order to study the largest length-scales of coherence over which superclusters are statistically significant. For this purpose we employed a statistical technique *Shuffle*, which would allow us to preserve the correlations in galaxy distribution at or below a given length-scale and destroy these above that length-scale. By studying "Filamentarity" in samples shuffled at a variety of length-scales and comparing the morphology of these with the original slices, we found the largest, coherent and statistically significant structures to be no longer than about $60-80h^{-1}$Mpc. This was also identified as the scale of homogeneity. Structures which are seen to extend beyond this scale result purely due to chance alignment of structures from neighbouring cells of the Universe which inidividually represent the LSS in the Universe well.

Redshift surveys such as SDSS and 2dFGRS sample $\sim$[Gpc]$^3$ volume of the Universe in a controlled fashion. The results obtained in the present thesis encourage us to make a detailed study of LSS in these redshift surveys. Due to its connection with visual appearance, our morphological approach may lend useful insights into the nature of superclusters and voids. These objects can be quantified in terms of Shapefinders and MFs and may be eventually catalogued, providing crucial constraints for theoretical models of structure formation. Our method can be potentially employed to tackle issues such as the nature of bias between visible and dark matter, cosmological parameter estimation, testing the paradigm of gravitational instability etc. We hope to extend the work reported in this thesis in these directions in future.